\documentstyle[12pt,twoside]{report}
\begin{document}
\thispagestyle{empty}
\begin{flushright}
MPI-Ph/95-24\\
hep-th/9503137\\
March 1995
\end{flushright}
\vskip 3cm
\begin{center}
{\bf \Huge{$\mbox{QED}_2$ and U(1)-Problem}}
\end{center}
\vskip 1cm
\centerline{\bf
Christof Gattringer${}^*$}
\vskip 5mm
\centerline{Max-Planck-Institut f\"{u}r
 Physik, Werner-Heisenberg-Institut}
\centerline{F\"ohringer Ring 6, 80805 Munich, Germany}
\vskip 2mm
\centerline{and}
\vskip 2mm
\centerline{Institut f\"{u}r Theoretische Physik der Universit\"at Graz}
\centerline{Universit\"atsplatz 5, 8010 Graz, Austria}
\vskip 9.5cm
\begin{flushleft} \rule{2 in}{0.03cm}
\\ {\footnotesize \ ${}^*$ This work is based on the author's thesis. }
\\ {\footnotesize \ \ \ e-mail: chg@mppmu.mpg.de }
\end{flushleft}
\newpage
\thispagestyle{empty}
\hbox{}

%
%
\begin{abstract}
\noindent
$\mbox{QED}_2$ with mass and N flavors of fermions is constructed using
Euclidean path integrals. The fermion masses
are treated perturbatively and the convergence of the
mass perturbation series is proven for a finite space-time cutoff.
The expectation functional
is decomposed into clustering $\theta$-vacua and their properties
are compared to the $\theta$-vacua of QCD for zero fermion mass. The sector
that is created by the $\mbox{N}^2$ classically conserved
vector currents is identified. The currents that
correspond to a Cartan subalgebra of U(N) are bosonized
together with the chiral densities in terms of a generalized
Sine-Gordon model. The solution of the U(1)-problem of
$\mbox{QED}_2$ is discussed and a Witten-Veneziano formula
is shown to hold for the mass spectrum of the pseudoscalars.
Evaluation of the Fredenhagen-Marcu confinement order parameter clarifies
the structure of superselection sectors.
\end{abstract}
\newpage\setcounter{page}1
\tableofcontents
%
%
\chapter{Introduction}
%
%
\section{Prologue}
It is well known that the mathematical structure of four dimensional (4d),
realistic field theories is much more involved than the world
of 2d models. Therefore there is a long history of attempts
to study physical problems in low dimensional theories. Many
concepts have first been developed in 2d toy models before they were
taken over to 4d physics.

Maybe the most prominent example is U(1)-gauge theory in two
dimensions first analyzed by Schwinger \cite{schwinger} and therefore
christened {\it Schwinger model}.
A very attractive feature of the model is its rather
simple solution
as long as there is no mass term taken into account. The construction
of the massive model \cite{coleman1}, \cite{seilerfroh}
is a little bit more subtle. Also the introduction of several
flavors \cite{belvedere}, \cite{gattringer}, \cite{joos2}
makes the model less straightforward than the original version.
Nevertheless it is rather
surprising that the case of more than one flavor has not been
analyzed systematically yet.
It is the intent of this thesis to fill this gap and to push forward
the construction of
QED in two dimensions with mass and flavor ($\mbox{QED}_2$) as far as possible.

Of course this project is inspired by some `4d mysteries', as
should be any investigation of toy models. Namely the topics that will
be attacked are the construction of the $\theta$-vacuum
in QCD, the U(1)-problem and Witten-Veneziano type formulas. Those
problems are closely related to each other.

The $\theta$-vacuum \cite{callan}, \cite{jackiw}
is supposed to be the superposition of topological sectors
in order to obtain the gauge invariant, physical vacuum.
As will be discussed below, the mathematical status of this construction
is rather vague. Nevertheless the $\theta$-vacuum
is a generally accepted concept.
In particular it was used to propose a solution of the U(1)-problem
\cite{thooft1}, \cite{thooft2}.

Due to the breaking of the axial U(1)-symmetry one could a priori
expect a corresponding Goldstone boson. The lack of experimental
evidence for this particle is refered to as the U(1)-problem
\cite{weinberg}. At first
glance this problem does not seem to be there at all, since the
U(1) axial current acquires the Adler-Bardeen anomaly
when quantizing the theory
\cite{adler}, \cite{bardeenan}, \cite{bell}, \cite{schwingeran}.
Using the fact that the anomaly can be rewritten as a total
divergence \cite{bardeen}, the current can be redefined
in such a way that it is conserved.
Ignoring the missing gauge invariance of the newly defined current,
one now can indeed expect a Goldstone particle.
{}From a less reckless point of view it has to be doubted if the
U(1)-problem is really well posed, since gauge invariance is one
of the corner-stones of QCD, and the gauge variant conserved current does not
act in the physical subspace.

Finally Witten-Veneziano type formulas
\cite{seilerstam}, \cite{smit}, \cite{witten} are another link between
the topologically nontrivial structure of the QCD vacuum and the
U(1)-problem. They relate the masses of pseudoscalar
mesons (the 'would be Goldstone bosons') to the topological
susceptibility. Unfortunately the status of those formulas
is not completely clear, or they are only formulated for massless QCD.

The three quoted problems can all be addressed rather well in
$\mbox{QED}_2$. U(1)-gauge theory in two
dimensions has a nontrivial topological structure and the formal
construction of the $\theta$-vacuum can be performed. The
axial-vector current has an anomaly, and the Schwinger
model shows mass generation. Thus the situation concerning the
U(1)-problem is equivalent to QCD. Finally Witten-Veneziano
formulas should be obeyed as well.

Of course it would not make sense to repeat the argumentation
from QCD. Here the strategy will be to construct the model
independent of poorly defined concepts like $\theta$-vacua,
and to draw the lessons for QCD afterwards. On the way also a new and
careful construction of $\theta$-vacua will be given.

%
%
\section{Overview}
Before I start to explore what has been outlined in the prologue,
a short overview will be given.

To be more explicit about what should be learned for QCD,
the announced 4d topics will be discussed in Chapter 2.
I will review the construction of the
$\theta$-vacuum, the U(1)-problem and Witten-Veneziano type formulas.
In particular it will be pointed out where criticism is advisable.

The formulation that will be used to construct $\mbox{QED}_2$
is the framework of Euclidean
path integrals. Section 3.1 is dedicated to the discussion
of the Euclidean action that describes
the model under consideration. Besides the gauge field it will be necessary
to introduce another vector field $h_\mu$. This auxiliary field generates
a Thirring term (current-current interaction)
that is needed for a proper treatment of the mass term.
In order to ensure that $\mbox{QED}_2$ is appropriate for analyzing
the above mentioned problems, the symmetry properties of the
model will be discussed in 3.2 . This is followed by Section 3.3 where
topological properties of U(1) gauge fields in two dimensions are reviewed.

In the usual approach (which is adopted here), one first integrates out the
fermions. This gives rise to the fermion determinant which is discussed in
Chapter 4. It will turn out that it does not have a simple structure when
the fermions are massive. Thus the strategy will be
mass perturbation theory.
The basic formulas needed for this enterprise are derived in Section 3.4 .

Chapter 4 is dedicated to giving a precise mathematical definition
of the so far poorly defined path integral. To this end I first
elaborate on the fermion determinant in an external field
(Sections 4.1, 4.2). It will
turn out that for massless fermions the fermion determinant is Gaussian
in the external field (compare e.g. \cite{seiler}). Together with
the action for the gauge field and auxiliary field, respectively,
this will amount to common Gaussian measures which have a precise
mathematical meanig (Section 4.3). In two dimensions gauge fixing
can be used to reduce the gauge field to one scalar degree of freedom.
It is more convenient to work with those scalar fields which are
introduced in Section 4.4 where I also rewrite the fermion propagator
in terms of those variables.

A proper field theory has to obey the cluster decomposition property
(5.1) in order to guarantee the existence of a unique
vacuum state. It turns out that for
the expectation functional so far constructed clustering is violated
by a certain class of operators which I classify in 5.2.
Using the symmetry properties (Section 5.3) of those operators
the state can be decomposed into clustering $\theta$-vacua that are
introduced in 5.4 . Furthermore it is proven that the new state
defines a proper field theory. This decomposition into clustering
states is exactly what is hoped to have been obtained in QCD by
introducing the $\theta$-vacuum. Several similarities between the
two constructs will be discussed.

In two dimensions one has the elegant technique of
bosonization at hand. This means that the vacuum expectation values of
certain operators can be expressed by vacuum expectation values
in a bosonic theory. In Section 6.1 I evaluate a generalized
generating functional in the massless model which then can be mapped
onto a theory of free bosons which is described in 6.2. In particular
the vector currents that are diagonal in flavor space
({\it Cartan type currents})
have a simple transcription in the bosonic theory.
Anyhow one can define currents for all generators of $\mbox{U(N)}_{{flavor}}$,
but it is not possible to find a local bosonization for the whole set of
vector currents as I will show in 6.4 . There I also discuss the
Hilbert space of the states described by the currents. Having established the
bosonized version of the model, it is rather easy to analyze the
status of the U(1)-problem of $\mbox{QED}_2$.

By summing up the mass perturbation series one can
construct a theory that bosonizes the states described by the
Cartan currents in terms of a generalized Sine Gordon theory.
This procedure is a generalization of the Coleman isomorphism \cite{coleman1}
and will be discussed in Section 7.1 . In Section 7.2
I prove that the mass perturbation series converges if a
space-time cutoff $\Lambda$ is imposed. Due to the presence of
massless degrees of freedom in the bosonic model for more than
one flavor the known
methods to remove $\Lambda$ termwise fail.
This rather unpleasent feature of the multiflavor model
will be discussed in Section 7.3 . Nevertheless one can extract
interesting physical information (vacuum structure,
mass spectrum) from a semiclassical approximation of the model
(Section 7.4). This semiclassical spectrum will then be used to test
Witten-Veneziano formulas (7.5).

In Chapter 8 a generalized version of the Fredenhagen-Marcu parameter
\cite{fredenhagen} will be computed, and the confinement properties of the
model will be analyzed.

In a short summary the obtained results will be discussed.

Finally I announce the appendices. Appendix A collects material
that can be found in the literature, but is included to keep
the thesis self contained. In particular propagators in two dimensions,
Gaussian measures and Wick ordering will be discussed. Also a toy
example can be found there which illustrates that the
set of fields having finite action has measure zero.
Appendix B contains formulas that are of more technical nature,
and thus were not included in the main part.

%
%
%
\chapter{The three mysteries}
The aim of this chapter is to prepare the physical playground for
the toy model. I review the three 'QCD-topics' whose analogues
will be analyzed
in $\mbox{QED}_2$ and point out where criticism is advisable.
%
%
\section{$\theta$-vacuum}
Almost twenty years ago it was realized by Belavin et al. \cite{belavin},
that classical Yang-Mills
theory in Euclidean space allows for topologically nontrivial solutions
called {\it instantons}\footnote{
The discussion below can be repeated for U(1) gauge fields in two
dimensions, where the topological objects are considerably easier to
imagine. I already announce here that this 2d material will be presented in
Section 3.3, where I will be more explicit on the
topologically nontrivial configurations and
the topological index. For the reader not so much familiar with instantons,
this will be a nice illustration of the material presented here in a rather
compressed form.}.
They were obtained by analyzing gauge field configurations with finite action.
A sufficient condition for such configurations is to
approach a pure
gauge when one
sends the space-time argument to infinity
\begin{equation}
A_\mu(x) \; \;
\stackrel{x^2 \rightarrow \infty}
{-\!\!\!-\!\!\!-\!\!\!-\!\!\!\longrightarrow} \; \;
-\frac{i}{g} \Big(\partial_\mu S(x)\Big)
\Big(S(x)\Big)^{-1} \; ,
\end{equation}
where $S(x)$ are elements of SU(N).
The instanton solutions $A^{(n)}$ can be classified with respect
to their winding number $\nu[A]$ (Pontryagin index, Chern Number, see e.g.
\cite{steenrod}),
which takes on integer
values
\begin{equation}
\nu[A^{(n)}] = n \; ,
\end{equation}
where
\begin{equation}
\nu[A] := \frac{g^2}{16 \pi^2} \int d^4x \mbox{Tr}
\left( F_{\mu \nu} \tilde{F}_{\mu \nu} \right) \; ,
\end{equation}
and
\begin{equation}
\tilde{F}_{\mu \nu} =
\frac{1}{2} \varepsilon_{\mu \nu \rho \sigma} F_{\rho \sigma} \; \; ,
\; \; F_{\mu \nu} = \partial_\mu A_\nu - \partial_\nu A_\mu
- i g \left[ A_\mu\; , \; A_\nu \right] \; .
\end{equation}
There is an important identity \cite{bardeen}, \cite{chern}
(Bardeen's identity)
\begin{equation}
\frac{1}{2} \; \mbox{Tr} ( F_{\mu \nu} \tilde{F}_{\mu \nu} ) \; \; =
\; \; \partial_\mu \; K_\mu \; ,
\end{equation}
where
\begin{equation}
K_\mu \; \; = \; \; \varepsilon_{\mu \nu \rho \sigma} \mbox{Tr}
\Big( A_\nu \partial_\rho A_\sigma \; -
\; i \frac{2g}{3} A_\nu A_\rho A_\sigma \Big)
\; .
\end{equation}
Together with the Gauss theorem, (2.5) can be used to rewrite the
winding number as a surface integral
\begin{equation}
\nu[A] \; = \; \lim_{V_4 \rightarrow \infty} \; \frac{g^2}{8 \pi^2} \;
\int_{V_4} \; d^4x \; \partial_\mu K_\mu \; = \;
\lim_{V_4 \rightarrow \infty} \; \frac{g^2}{8 \pi^2} \;
\int_{\partial V_4} \; d^3 \sigma \; \hat{n}_\mu K_\mu \; .
\end{equation}
The topologically nontrivial configurations radically change the nature of
the vacuum. The standard argument \cite{callan}, \cite{jackiw} is
formulated in temporal ($A_4 = 0$) gauge.
The instanton $A^{(n)}$ in temporal gauge
has the property
\begin{equation}
A^{(n)}_i (x) \; \longrightarrow \; \left\{ \begin{array}{ll}
-\frac{i}{g}\Big(\partial_i S^{(m)}(\vec{x})\Big)
\Big(S^{(m)}(\vec{x}\Big))^{-1} &
\mbox{for} \; \; x_4 \rightarrow - \infty \\ & \\
-\frac{i}{g}\Big(\partial_i S^{(m+n)}(\vec{x})\Big)
\Big(S^{(m+n)}(\vec{x})\Big)^{-1} &
\mbox{for} \; \; x_4 \rightarrow + \infty \; ,\end{array} \right.
\end{equation}
$i=1,2,3$ and $n,m \in \mbox{Z\hspace{-1.35mm}Z}$.
\begin{equation}
S^{(1)}(\vec{x}) = \frac{{\vec{x}\;}^2 - \lambda^2}{{\vec{x}\;}^2 + \lambda^2}
- \frac{2i \lambda \vec{\sigma} \cdot \vec{x}}{{\vec{x}\;}^2
+ \lambda^2} \; ,
\end{equation}
where $\sigma_i \; , \; i = 1,2,3$ are the Pauli matrices and $\lambda$
is a real number, the {\it instanton size}. For
$S^{(l)}(x)$ with $l \neq 1$ see e.g. $\cite{rajaraman}$.
Due to the
$\varepsilon$-tensor in Equation (2.6) only $K_4$ can contribute in
the surface integral (2.7).
Choosing the surface $\partial V_4$ to be integrated over, to be the
hypercylinder of Figure 2.1 (next page),
the winding number (2.7) takes on the form
\begin{equation}
\nu[A] \; = \; \lim_{T \rightarrow \infty} \lim_{V_3 \rightarrow \infty}
\frac{g^2}{8 \pi^2} \int_{V_3} \; d^3x \; \Big(\; K_4 \; \Big|_{x_4 = T}
\; - \; K_4 \; \Big|_{x_4 = - T} \; \Big) \; =: \; \nu_+[A] \; - \; \nu_-[A]
\; .
\end{equation}
{}From Equation (2.9)
equation one easily reads off
\begin{equation}
S^{(1)}(\vec{x}) \;
\stackrel{|\vec{x}|\rightarrow \infty}{-\!\!\!-\!\!\!-\!\!\!\longrightarrow}
\;
\mbox{1\hspace{-1.1mm}I} \; ,
\end{equation}
and the same is true for all $S^{(l)}$.
Thus 3-space can be compactified to the hypersphere $S^3$. Since the
manifold of SU(2) is homeomorphic to $S^3$, the $S^{(l)}(\vec{x})$ define
mappings $S^3 \; \rightarrow \; S^3$. Such mappings are known to fall
into homotopy classes \cite{steenrod}. \newpage
\vspace*{100mm}
\noindent
{\bf Figure 2.1 : }
The surface $\partial V_4 \; $.
\vskip5mm
\noindent
By inserting the asymptotic form (2.8) for
$A^{(n)}$ into (2.6) one obtains
(similar for $\nu_+[A^{(n)}]$)
\[
\nu_-[A^{(n)}] \; = \; \]
\begin{equation}
\frac{1}{24 \pi^2} \int_{S^3} d^3x \;
\varepsilon_{ijk} \mbox{Tr} \Big[
( \partial_i S^{(m)} ) ( S^{(m)} )^{-1}
( \partial_j S^{(m)} ) ( S^{(m)} )^{-1}
( \partial_k S^{(m)} ) ( S^{(m)} )^{-1} \Big] \; = m \; .
\end{equation}
The left hand side of (2.12) is known to be the integral
over the invariant measure of the group
(see e.g. \cite{rajaraman}), and thus $\nu_-[A^{(n)}]$ and $\nu_+[A^{(n)}]$
give the homotopy classes of the gauge field configurations $A^{(n)}$ at time
equal to plus and minus infinity. Looking at (2.10) and (2.12) one now can
interpret the instanton $A^{(n)}$ obeying (2.8) in the
following way\footnote{I did not denote the explicit form for $A^{(n)}$
(it can be found in e.g. \cite{rajaraman}) since it is rather
lengthy. Only the asymptotic form
(2.8) is quoted. For the 2d case the explicit form will be given in 3.3.}.

The instanton $A^{(n)}$ with total winding number $\nu[A^{(n)}] \; = \; n$
connects a pure gauge at time equal to minus infinity that winds
$\nu_-[A^{(n)}] \; = \; m$ times around compactified space, with a pure gauge
at time equal to plus infinity that winds
$\nu_+[A^{(n)}] \; = \; m+n$ times around compactified space.

So far the instanton has only been constructed for the gauge group
SU(2). Of course
one is interested in SU(3) when dealing with QCD. In \cite{colemani}
it is discussed that mappings from $S^3$ to SU(3) can be deformed
continuously into a mapping from $S^3$ to a SU(2) subgroup. Thus the
SU(2) discussion is already sufficient for SU(3).

The configurations
\begin{equation}
-\frac{i}{g}\; \Big(\partial_i S^{(l)}(\vec{x})\Big)\;
\Big(S^{(l)}(\vec{x})\Big)^{-1} \; ,
\end{equation}
for infinite time argument are now considered as classical vacuum states
$| l \rangle$. The reasoning therefore is the following.
The classical vacua have zero potential energy separated
by a barrier \cite{jackiw}. On the other hand the instanton
$A^{(1)}$ that connects $| m \rangle$ with $| m+1 \rangle$
has Euclidean action $8\pi^2/g^2$. Thus in the WKB sense the
$| m \rangle$ to $| m+1 \rangle$ amplitude is of order $\exp(-8\pi^2/g^2)$
which is a typical tunneling amplitude \cite{callan}.

A general transition from $| m \rangle$ to $| m+n \rangle$ can
formally be expressed in terms of functional integrals
\begin{equation}
{ }_{+\infty}\langle m+n | m \rangle_{-\infty}^J \; = \;
\frac{1}{Z} \; \int {\cal D} A \; \delta \Big( \nu[A] - n \Big) \;
{\cal D} \overline{\psi} {\cal D} \psi \; e^{-S_J} \; .
\end{equation}
The left hand side denotes the transition from the vacuum $| m \rangle$
at (Euclidean) time equal to minus infinity to the vacuum $| m + n \rangle$
at time equal to plus infinity in the presence of sources $J$. Since the
instanton $A^{(n)}$ was identified to mediate such a transition
(see (2.10)), one has to integrate over gauge field configurations within
the instanton sector with winding number $n$. This is formally expressed by the
$\delta$-functional in the path integral. Finally $S_J$ denotes the
Euclidean action plus coupling terms to the sources $J$. Obviously the
expression (2.14) does not depend on the vacuum $| m \rangle$
I started with, but only on the difference $n$.

Since transitions between the vacua $| l \rangle$ are possible, none of
them can be the correct vacuum. The crucial idea in the construction of
the {\it $\theta$-vacuum} $| \theta \rangle$ is to form a superposition
that takes into account all possible transitions. In terms of functional
integrals this can be expressed as (using (2.3) in the last step)
\[
{ }_{+\infty}\langle \theta | \theta \rangle_{-\infty}^J  \; := \;
\sum_{n=-\infty}^{n=+\infty} \; e^{i \theta n} \;
{ }_{+\infty}\langle m+n | m \rangle_{-\infty}^J \;
\]
\[
= \; \frac{1}{Z} \; \int {\cal D} A \;
\sum_{n=-\infty}^{n=+\infty}\delta\Big( \nu[A] - n \Big) \;
e^{i \theta \nu[A]} \;
{\cal D} \overline{\psi} {\cal D} \psi \; e^{-S_J}
\]
\begin{equation}
= \; \frac{1}{Z} \; \int {\cal D} A \;
{\cal D} \overline{\psi} {\cal D} \psi \; \exp \left( \;
 i \theta \frac{g^2}{16 \pi^2} \int d^4x
\mbox{Tr} ( F_{\mu \nu} \tilde{F}_{\mu \nu} ) \; \; - \; \; S_J
\; \right) \; .
\end{equation}
Formally (see e.g. \cite{peccei}) the
$\theta$-vacuum can be written as a vector in a Hilbert space
\begin{equation}
| \; \theta \; \rangle \; = \; \sum_{l=-\infty}^{l=+\infty} \; e^{- i \theta l}
| \; l \; \rangle \; .
\end{equation}
Thus $\theta$ is only defined mod(2$\pi$).
The gauge transformation $S^{(1)}$ (see (2.9)) that changes $| l \rangle $
to $| l + 1 \rangle $ can formally be implemented as an unitary
operator $\hat{S}^{(1)}$
in this Hilbert space \cite{callan}, and now leaves $| \theta \rangle$
invariant up to a phase
\begin{equation}
\hat{S}^{(1)} \; | \; \theta \; \rangle \; =
\; e^{i \theta} \; | \; \theta \; \rangle \; .
\end{equation}
As can be seen from (2.15), a new term has entered the action,
namely the winding number times
a new parameter, the {\it vacuum angle} $\theta$. This
term causes a serious difficulty, since it violates CP invariance.
This problem is known as the {\it strong CP problem}
(see e.g. \cite{peccei}) for a nice review). There is a second point in
the construction of $| \theta \rangle$ which is rather problematic.
It has been pointed out (compare the second line of (2.15)), that the
measure has to be a sum over all topological sectors.
This requirement is not really mathematically well defined.
Functional measures for gauge fields (if they are constructed at all)
do not live on continuous functions and configurations with
finite action have measure zero.
The first point can be seen in Appendix A.2 on Gaussian
measures, which are measures on the space of tempered distributions.
Gaussian measures are in fact a good illustration, since 2d, U(1)
gauge theory makes use of them. The second objection is illustrated
in Appendix A.3 where I show for a toy example that configurations with
finite action
have zero measure.

Nevertheless the $\theta$-vacuum is a widely accepted concept, and
was e.g. invoked to solve the U(1)-problem. This will be discussed in the
next two sections.

I finish this section with remarking that it is generally believed that
physics does not depend on $\theta$, whenever one of the quarks is massless.
The arguments for this make use of the anomaly and will be discussed
in the next section.
%
%
\section{U(1)-problem}
The symmetry of the QCD
Lagrangian with three flavors and vanishing quark masses
is $\mbox{SU(3)}_L \times \mbox{SU(3)}_R \times
\mbox{U(1)}_V \times \mbox{U(1)}_A$.
In the following I review the
argumentation that this symmetry is
sponteously broken down to $\mbox{SU(3)}_V \times \mbox{U(1)}_V$ and
Goldstone particles have to be expected. In the discussion below, I
use the notation of \cite{bogol}.

Consider the one parameter axial transformations
\begin{equation}
q \; \; \rightarrow \; \; g^{(a)} \; q \; \; \; \; , \; \; \; \;
\overline{q} \; \; \rightarrow \; \; \overline{q} \; g^{(a)} \; ,
\end{equation}
where $q$ is the triplet $(u,d,s)^T$ and
\[
g^{(a)} = e^{i \omega^{(a)} \tau^{(a)} \gamma_5} \; \; , \; \; \; \;
a \; \; \mbox{fixed} \; \; ,
\]
\begin{equation}
\tau^{(1)} = \mbox{1\hspace{-1.1mm}I} \; \; \; , \; \;
\; \tau^{(a)} = \frac{\lambda^{(a-1)}}{2} \; \; \; , \; \; \; a = 2,3..9 \; ,
\end{equation}
where $\lambda^{(b)} \; , \; b = 1,2 .. 8$ are the Gell-Mann matrices.
As long as the quark masses vanish, this is a symmetry of the QCD
Lagrangian. The corresponding Hermitean Noether currents are
\[
{j^{(1)}}^\mu_5 \; := \; \overline{q} \; \gamma^\mu \gamma_5 \; q \; , \]
\begin{equation}
{j^{(a)}}^\mu_5 \; := \;
\overline{q} \; \gamma^\mu \gamma_5 \; \frac{\lambda^{(a-1)}}{2} \; q \; \; ,
\; \; a=2,3..9 \; .
\end{equation}
The U(1) axial current ${j^{(1)}}_5$ acquires the anomaly when
quantizing the theory
and thus will be discussed later.
The other currents are conserved
\begin{equation}
\partial_\mu \; {j^{(a)}}^\mu_5(x) \; =
\; 0 \; \; , \; \; a = 2,3 \; .. \; 9 \; .
\end{equation}
To each conserved current one can define an operator
$D^{(a)}$ acting on a polynomial $X$ of the fields via
\begin{equation}
D^{(a)} (X) \; := \; i \; \int_{x^0 = const} \;
[ \; {j^{(a)}}^0_5(x) \; , \; X \; ] \; d^3x \; .
\end{equation}
The integrals converge due to local commutativity,
and are time independent due to conservation of the currents \cite{bogol}.
They immediately can be shown to obey
$D^{(a)} (X^*) = D^{(a)} (X)^* \; , \; \;
D^{(a)} (XY) = D^{(a)} (X) \; Y + X \; D^{(a)} (Y) \; ,$
and thus are called $\star$-derivatives. In \cite{bogol} it is
discussed how the $\star$-derivatives generate internal Lie-group
symmetries of the theory\footnote{
If one assumes that the expression
$ Q := \int j^o(x) d^3x $
exists for a conserved current $j^\mu$,
then the unitary operator that performs
the underlying symmetry transformation is given by
$V_\omega = \exp(i\omega Q)$, and $D(X)$ is the second term in an
expansion of $V_\omega \; X \; V_\omega^{-1}$ in $\omega$.
}.
Now the question is if these symmetries
can be implemented unitarily or are spontaneously broken.

First I assume that the unitary implementation is possible.
This means that there exists an unitary operator $V(g^{(a)})$ depending
continuously on the element $g^{(a)}$ of the symmetry group
and the symmetry operates via
\begin{equation}
X \; \rightarrow \; V(g^{(a)}) \; X \; V(g^{(a)})^{-1} \; .
\end{equation}
As is discussed in \cite{bogol} this then implies
\begin{equation}
V(g^{(a)}) \; = \; e^{i \omega^{(a)} \; Q^{(a)}} \; ,
\end{equation}
and
\begin{equation}
Q^{(a)} \; X \; | 0 \rangle \; = \; - i \; D^{(a)} (X) \; | 0 \rangle  \; .
\end{equation}
Since the right hand side of (2.22) does not depend on time,
\begin{equation}
[ H , Q^{(a)} ] \; = \; 0 \; .
\end{equation}
Consider now an eigenstate $| E_n \rangle = X_n \; | 0 \rangle$ of the
Hamiltonian with energy $E_n$.
Due to (2.26) $Q^{(a)} \; | E_n \rangle$ is an eigenstate
with the same energy
\begin{equation}
H \; Q^{(a)} \; | E_n \rangle \; = \; E_n \; Q^{(a)} \; | E_n \rangle \; .
\end{equation}
Since the axial vector currents have odd parity, a parity
transformation ${\cal P}$ acts on $Q^{(a)} \; | E_n \rangle$ via
\begin{equation}
{\cal P} \; Q^{(a)} \; | E_n \rangle \; = \;
{\cal P} \; (-i) \; D^{(a)} (X_n) \; | 0 \rangle  \; = \;
- P_n \; Q^{(a)} \; | E_n \rangle \; ,
\end{equation}
where $P_n$ denotes the parity of the eigenstate $| E_n \rangle$.
Thus if one of the symmetries generated by the conserved currents
${j^{(a)}}_5 \; , \; a=2,3..9$ were realized unitarily, this would imply
that the hadrons come in parity doublets. Since the parity partners are
not seen in experiment, the assumption of unitary implementability
is wrong, and all those symmetries have to be broken
spontaneously.

The breaking of the symmetry for each single one paramter group generated by
one of the conserved currents ${j^{(a)}}_5 \; , \; a=2,3..9$ implies the
existence of 8 Goldstone particles. The corresponding massless
states $|\Phi^{(a)} \rangle$ are connected to the vacuum via
\begin{equation}
\langle \Phi^{(a)} \; | \; {j^{(a)}}^0_5 \; | \; 0 \rangle \; \neq \; 0 \; .
\end{equation}
The whole discussion above made use of vanishing quark masses. Since these
masses are known to be nonzero, the Goldstone particles are only
approximate Goldstone bosons. Nevertheless the pseudoscalar mesons
\begin{equation}
\pi^0 \; , \; \pi^\pm \; , \; K^0 \; , \; \overline{K}^{\;0}
\; , \; K^\pm \; , \; \eta
\end{equation}
can be properly identified to play this role.

So far the U(1)-current was excluded since it acquires the anomaly
\cite{adler}, \cite{bardeenan}, \cite{bell}
\begin{equation}
\partial_\mu \; {j^{(1)}}^\mu_5 \; = \;
2 N_f \; \frac{g^2}{16 \pi^2} \; \mbox{Tr}
\Big( F^{\mu \nu} \tilde{F}_{\mu \nu} \Big) \; ,
\end{equation}
where $N_f$ is the number of flavors.
This implies that the action of the corresponding $\star$-derivative
on some polynomial $X$ of the fields $D^{(1)} (X)$ is not time independent.

Using Bardeen's identity (\cite{bardeen}, \cite{chern}, compare (2.5), (2.6))
\[
\frac{1}{2} \; \mbox{Tr} \Big( F^{\mu \nu} \tilde{F}_{\mu \nu} \Big) \; = \;
\partial_\mu K^\mu \; ,
\]
\begin{equation}
K_\mu \; = \; \varepsilon_{\mu \nu \rho \sigma} \mbox{Tr}
\Big( A^\nu \partial^\rho A^\sigma - i \frac{2g}{3} A^\nu A^\rho A^\sigma \Big)
\end{equation}
one can define a new current
\begin{equation}
\tilde{j}^{(1) \; \mu }_{\; \; \; \; \; \; 5} \; := \; {j^{(1)}}^\mu_5 -
2 N_f \; \frac{g^2}{8 \pi^2} K^\mu \; ,
\end{equation}
which now is conserved
\begin{equation}
\partial_\mu \; \tilde{j}^{(1) \; \mu }_{\; \; \; \; \; \; 5} \; = \; 0 \; .
\end{equation}
One can also define a $\star$-derivative $\tilde{D}^{(1)}$
\begin{equation}
\tilde{D}^{(1)} (X) \; := \; i \; \int_{x^0 = const} \;
[ \tilde{j}^{(1) \; 0 }_{\; \; \; \; \; \; 5}(x) \; , \; X ] \; d^3x \; ,
\end{equation}
corresponding to the new current
$\tilde{j}^{(1)}_{\; \; \; \; \; \; 5} $.
All the arguments (2.23) - (2.29) can be repeated and a ninth pseudoscalar
particle can be expected. One would like to interpret the $\eta^\prime$ in
that sense. The common wisdom is that $\eta^\prime$ is to heavy to
be this approximate Goldstone boson. This belief is based on a
work by Weinberg \cite{weinberg} where the case of two flavors is
considered. There the $\eta$ should play the role of the U(1)
Goldstone particle. To interpret the $\eta$ in this sence, the
mass relation $m_\eta < \sqrt{3} m_\pi$ has to be obeyed.
Inserting the experimental values for the masses
($m_{\pi^0} \sim 135 \; \mbox{MeV},\;  m_{\eta} \sim 549 \; \mbox{MeV}$),
one finds that the
$\eta$ is not the wanted Goldstone particle. The same reasoning can
be done for three flavors, and the U(1)-problem can be formulated:
\vskip3mm
\centerline{
{\it Where is the ninth, light, pseudoscalar meson ?} }
\vskip3mm
\noindent
If one reanalyzes the arguments for the U(1)-current more carefully,
one finds that $K^\mu$ defined in (2.32) is not gauge invariant, and thus
${\tilde{j}^{(1)}}_{\; \; \; \; \; \; 5}$ is not a physical operator.
As discussed in $\cite{lopu}$ it is not obvious that local commutativity
which is needed to establish the convergence of
the integral (2.35)
should and can be required for non-physical operators. So it has to be doubted
that $\tilde{D}^{(1)}$ is well defined and generates a symmetry
on the physical
Hilbert space. It is unclear if the U(1)-problem is well posed.

Ignoring this criticism, one can formally define a charge $\tilde{Q}_5$
that corresponds to the current ${\tilde{j}^{(1)}}_{\; \; \; \; \; \;5}$
(compare footnote 3 on page 12)
\begin{equation}
\tilde{Q}_5 \; := \; \int_{x^0 = const} \; d^3x \;
\tilde{j}^{(1) \; 0 }_{\; \; \; \; \; \; 5}(x) \; .
\end{equation}
This can now be used to argue that physics does not depend on $\theta$
if the quarks are massless, as has been mentioned in the last
section\footnote{
There is another way to establish this result. $\theta$ can also be
introduced by modifying one of the mass terms to  $\overline{\psi}
\exp(i \theta \gamma_5) \psi$ (see \cite{baluni}). If one of
quark masses is zero, this modification vanishes and so the
$\theta$ dependence.}.
As has already been pointed out, $\tilde{Q}_5$ is not gauge invariant. In
particular under the gauge transformation $S^{(1)}$ (see Equation (2.9))
which changes $| m \rangle \rightarrow | m+1 \rangle$
\[
\triangle \tilde{Q}_5 = - \frac{2 N_f}{12 \pi^2}
\int d^3x \;
\varepsilon_{ijk} \mbox{Tr} \Big[
( \partial_i S^{(1)} ) ( S^{(1)} )^{-1}
( \partial_j S^{(1)} ) ( S^{(1)} )^{-1}
( \partial_k S^{(1)} ) ( S^{(1)} )^{-1} \Big] \]
\begin{equation}
= \; -2N_f  \; .
\end{equation}
In the last equation I used (2.33) and (2.12). Now one can perform a
chiral rotation on $| \; \theta \; \rangle$
\begin{equation}
| \; \theta \; \rangle_\delta \; := \;
e^{i \delta \tilde{Q}_5} \; | \; \theta \; \rangle_\delta \; ,
\end{equation}
and use (2.37) to obtain
\[
\hat{S}^{(1)} \; | \; \theta \; \rangle_\delta \; =
\hat{S}^{(1)} \; e^{i \delta \tilde{Q}_5} \; ( \hat{S}^{(1)} )^{-1}
\hat{S}^{(1)} | \; \theta \; \rangle \; = \]
\begin{equation}
e^{i \delta \tilde{Q}_5 \; - \; i \delta 2 N_f } \;
e^{i \theta} | \; \theta \; \rangle \; = \;
e^{i(\theta \; - \;\delta 2 N_f)} | \; \theta \; \rangle_\delta \; .
\end{equation}
Since $\tilde{Q}_5$ stems from a current which is conserved if the
quarks are massless, it is formally time independent, and hence
commutes with the Hamiltonian. Thus the chirally rotated state
$| \; \theta \; \rangle_\delta$ can also serve as 'the vacuum'. From
Equation (2.39) it follows that
\begin{equation}
| \; \theta \rangle_\delta \; = \; | \; \theta - \delta 2 N_f \; \rangle \; ,
\end{equation}
and thus the theories are equivalent for all values of $\theta$.
The argument fails for massive quarks, since then
${\tilde{j}^{(1)}}_{\; \; \; \; \; \; 5}$
is not conserved.

Arguments at the same level of rigor were used by 't Hooft to
solve the U(1)-problem
\cite{thooft1}, \cite{thooft2}.
The idea is that the nontrivial structure of the QCD vacuum leads
to a vanishing residue of the U(1)-Goldstone pole in propagators
of physical (i.e. gauge invariant) operators.
The crucial formula (see \cite{peccei})
for the cancellation of the residue is
the following structure for the vacuum expecation value of an operator
$X$ in the $\theta$-vacuum
\begin{equation}
\langle \theta | X | \theta \rangle \; = \; C_X \;
\exp \left( i \frac{\chi_X}{2N_f} \theta \right) \; ,
\end{equation}
with the remarkable property
\begin{equation}
C_X \; = \; 0 \; \; \; \mbox{unless} \; \; \; \chi_X \; = \; 2 N_f \; n \;
\; \; , \; n \; \in \; \mbox{Z\hspace{-1.35mm}Z} \; ,
\end{equation}
where $\chi_X$ is the chiral U(1)-charge of $X$. The last equation
can be seen to hold by the following formal arguments \cite{callan}.
{}From (2.37)
there follows for the operators $\hat{S}^{(m)} = (\hat{S}^{(1)})^m$ and
$\tilde{Q}_5$ that $\hat{S}^{(m)}
\tilde{Q}_5 (\hat{S}^{(m)})^{-1} \; = \; \tilde{Q}_5 - 2 N_f m$.
If the states for different vacuum topology $|m\rangle$
are defined as $|m\rangle = \hat{S}^{(m)}|0\rangle$
with $\tilde{Q}_5|0 \rangle = 0$ then there follows
$\tilde{Q}_5 |m\rangle \; = \; 2 N_f m \; |m\rangle$. However since
$\tilde{Q}_5$ is formally conserved one concludes
${}_{+\infty}\langle m + n | m \rangle_{-\infty} \; \propto \; \delta_{n,0}$.
In general one must find
\begin{equation}
{}_{+\infty}\langle m + n | X | m \rangle_{-\infty} \; \propto \;
\delta_{n,\nu} \; ,
\end{equation}
where $X$ is an operator of chirality $2 N_f \nu \; , \; \nu \in
\mbox{Z\hspace{-1.35mm}Z}$. Inserting this into
the first line of (2.15) one ends up with (2.41).
%
%

\section{Witten-Veneziano type formulas}
In 1979 Witten proposed a formula that relates the mass of the
$\eta^\prime$ meson to the topological susceptibility of
quarkless QCD. The remarkable feature of the result is that it
does not make use of instantons.
The main ingredient of the proof is that physics does not depend on the
vacuum angle $\theta$ when massless quarks are present (compare Section 2.2).
This observation can be related to the topological susceptibility.
Consider the free energy density
\begin{equation}
F \; := \; \lim_{V \rightarrow \infty} \; \frac{ \ln(Z_V)}{V} \; ,
\end{equation}
where the partition function in finite volume $Z_V$ is formally
defined as
\begin{equation}
Z_V \; := \; \int {\cal D}A {\cal D}\overline{q}{\cal D}q \;
\exp \left( iS_V \; + \; i\frac{\theta g^2}{16\pi^2}
\int_V d^4x \mbox{Tr} (F \tilde{F}) \right) \; .
\end{equation}
The second derivative of the free energy with respect to $\theta$
gives the topological susceptibility $\chi_{top}$
\begin{equation}
- \; \frac{d^2 F}{d \theta^2} \; \Bigg|_{\theta=0} \; = \;
\chi_{top} \; ,
\end{equation}
with
\begin{equation}
\chi_{top} \; := \left( \frac{g^2}{16 \pi^2} \right)^2 \; \;
\int d^4x \;\langle T \; q(x) \; q(0) \rangle \; \; , \; \;
q(x) \; := \;
\mbox{Tr} \Big( F(x) \tilde{F} (x)\Big) \; ,
\end{equation}
where $T$ denotes time ordering.
As discussed, physics does not depend on $\theta$ if the quarks are massless.
It follows that $\chi_{top}$ has to vanish then.
More generally Witten considers the propagator $U(k)$
\begin{equation}
U(k) \; := \;
\int d^4x \; e^{ikx} \; \langle \; T \; q(x) \; q(0) \; \rangle \; .
\end{equation}
Adopting some $1/N_c$-expansion arguments
\cite{thooftN}, \cite{wittenN} the propagator is rewritten as
\begin{equation}
U(k) \; = \;
\sum_{glueballs} \frac{N_c^2 \; a_n^2}{k^2-M_n^2} \;
+ \; \sum_{mesons} \frac{N_c \; c_n^2}{k^2-m_n^2} \; .
\end{equation}
The first sum also contributes in a theory without any quarks
(massless or massive) and is denoted as $U_0(k)$ then.
To lowest order in $1/N_c$, this term does not change if the
quarks are coupled.
On the other hand
the right hand side of (2.49) has to vanish in the presence of massless quarks
at $k=0$. Ignoring the fact that both sums
have the same sign, Witten claims that they cancel each other.
The condition in lowest order of $1/N_c$ is
(which would be mathematically correct if there was
an extra minus sign)
\begin{equation}
\frac{N_c \; c^2_{\eta^\prime}}{m_{\eta^\prime}^2} \; = \; U_0(0) \; .
\end{equation}
Using the anomaly equation (2.31) to rewrite $N_c c^2_{\eta^\prime}$
in terms of the decay constant
$f_{\eta^\prime}$, Witten ends up with
\begin{equation}
m_{\eta^\prime}^2 \; = \; \frac{4 N_f}{f_{\eta^\prime}^2} \; \chi_{top}^0 \; ,
\end{equation}
where $N_f$ is the number of flavors and
$\chi_{top}^0 = \left( \frac{g^2}{16 \pi^2} \right)^2 \; U_0(0)$ is
the topological susceptibility in pure SU(3) gauge theory. It has to be
remarked that $f_{\eta^\prime}$ should be evaluated in QCD with
vanishing quark masses as can be seen from the derivation above.
{}From PCAC arguments it follows that $f_{\eta^\prime}$ varies only
very slowly in the mass variable (see e.g. \cite{treiman}) and
thus the experimental value can be inserted.

Although the derivation is problematic
(besides the sign problem, questions concerning regularization
were ignored) the formula seems to have some
truth in it. It has been reanalyzed in Euclidean space
by Seiler and Stamatescu
\cite{seilerstam}. They pointed out that $\mbox{Tr}(F\tilde{F})$ is a composite
operator and requires some subtraction procedure, leading to a
spectral representation
\begin{equation}
U(k) \; = \; P(k^2) \; - \; \int_0^\infty \;
\frac{d\rho(\mu^2)}{k^2 + \mu^2} \; .
\end{equation}
$P(k^2)$ denotes some polynomial in the momentum.
This formula has to be compared to (2.49) in the Witten derivation.
Two main differences appear. There shows up the contact term $P(k^2)$
which is necessary due to the subtractions. Furthermore there is the
negative sign in front of the spectral integral, which is
required by reflection positivity and the fact that $q$ is odd under
time reflections. Now the right hand side really can vanish for
$k \rightarrow 0$ when the quarks are massless. Assuming that
$d\rho(\mu^2)$ is dominated by the $\eta^\prime$ contribution
\begin{equation}
d\rho(\mu^2) \; = \; c_{\eta^\prime}^2 \;
\delta(\mu^2 -m^2_{\eta^\prime}) \; d \mu^2 \; ,
\end{equation}
one obtains
\begin{equation}
\frac{c_{\eta^\prime}^2}{m^2_{\eta^\prime}} \; = \; P^0(0) \; .
\end{equation}
This is now the correct expression that replaces (2.50).
In Witten's result there occurs an extra factor $N_c$ which is only an
artefact of his derivation within the $1/N_c$ framework.
It vanishes when rewriting the result in terms of physical
quantities like decay constants. The final result given by
Seiler and Stamatescu reads
\begin{equation}
m_{\eta^\prime}^2 \; = \; \frac{4 N_f}{f_{\eta^\prime}^2} \; P^0(0) \; .
\end{equation}
The topological susceptibility in the Witten result (2.51)
has been replaced by the
contact term $P^0(0)$ of the
two point function of the topological charge in the
theory with vanishing quark masses\footnote{
It will turn out that at least in the Schwinger model $P^0(0)$
can be interpreted as the quenched topological susceptibility, so
that Witten's formula is recovered.
}.
There are two more articles on Witten-Veneziano type formulas I
would like two mention. The Witten result was rederived (agreeing on
the right hand side) by Veneziano \cite{veneziano}. The approach adopted
there is the analysis of anomalous Ward identities in the $1/N_c$
expansion.

The second paper by Smit and Vink \cite{smit} is an investigation of the
problem in an Euclidean lattice formulation. In this approach the
regularization procedure is rather straightforward. The problem
is the adequate matching of the lattice quantities to their
continuum counterparts. The final result for three flavors
given by Smit and Vink reads
\begin{equation}
m_{\eta^\prime}^2 - \frac{1}{2} m_{\eta}^2 - \frac{1}{2} m_{\pi^0}^2 \; = \;
\frac{12}{f_\pi^2} \; \overline{\chi} \; ,
\end{equation}
where
\begin{equation}
\overline{\chi} \; := \; \lim_{V \rightarrow \infty} \;
\frac{\kappa_P^2 m_a m_b }{V} \;
\langle \; \mbox{Tr}( \gamma_5 G_{aa} ) \; \mbox{Tr}( \gamma_5 G_{bb} ) \;
\rangle^{pbc}_U \; .
\end{equation}
Here $G_{aa}$ is the fermion propagator for fixed flavor $a$
in an external field
$G_{aa}(x,y) = \langle \psi_a(x) \overline{\psi}_a(y)\rangle_{\psi}$.
The trace is over all indices except flavor.
$\kappa_P$ is a
renormalization factor that approaches 1 in the continuum limit.
Finally $\langle .. \rangle^{pbc}_U$ denotes the expectation value with
respect to the gauge fields in a quenched approximation.
Periodic boundary conditions are imposed for gauge invariant
quantities. $\overline{\chi}$ is formally related to the topological
susceptibility through the index theorem \cite{atiyah}
(see also \cite{seilerstam2}).
In the continuum
\begin{equation}
\mbox{Tr} \; \Big( \gamma_5 G_{aa} \Big) \; = \; \frac{Q}{m_a} \; ,
\end{equation}
where $Q$ is the topological charge (now it is also
obvious that $\overline{\chi}$
does not depend on the choice of $a$ and $b$).

%
%
%
\chapter{$\mbox{QED}_2$}
In this chapter the Lagrangian of the model and its
symmetries will be discussed.
Furthermore I elaborate on the existence of topologically
nontrivial configurations,
thus showing that $\mbox{QED}_2$ is adequate for the study of
the problems announced in Chapter 2. Finally the strategy for
the construction of the model will be outlined.

%
%
\section{Formal description of the model}
The Euclidean action of the model that will be constructed is given by
\begin{equation}
S[\overline{\psi},\psi,A,h] = S_G[A] + S_h[h] +
S_F[\overline{\psi},\psi,A,h] + S_M[\overline{\psi},\psi] \; .
\end{equation}
The gauge field action reads
\begin{equation}
S_G[A] = \int d^2x\left( \frac{1}{4} F_{\mu \nu}(x) F_{\mu \nu}(x)
+ \frac{1}{2} \lambda \Big( \partial_\mu A_\mu(x) \Big)^2 \right) \; .
\end{equation}
A gauge fixing term is included that will be considered in the limit
$\lambda \rightarrow \infty$ which ensures $\partial_\mu A_\mu(x) = 0$
({\it transverse} or {\it Landau gauge}). As usual $F_{\mu \nu}(x) =
\partial_\mu A_\nu(x) - \partial_\nu A_\mu(x)$, denotes the field strength
tensor.

In addition to the gauge field an auxiliary field $h_\mu$ with action
\begin{equation}
S_h[h] = \frac{1}{2}\int d^2x h_\mu(x)\Big( \delta_{\mu \nu} -
\lambda^\prime \partial_\mu \partial_\nu \Big) h_\nu(x) \; ,
\end{equation}
has to be
introduced which generates the announced Thirring term.
$S_h[h]$ is simply $\delta_{\mu \nu}$ plus a term that
makes $h_\mu$ transverse in the limit $\lambda^\prime \rightarrow \infty$.
I postpone the discussion of the role of $h_\mu$ until the fermion
action has been introduced.

The fermion action is a sum over N flavor degrees of freedom
\begin{equation}
S_F[\overline{\psi},\psi,A,h] = \sum_{b=1}^N \int d^2x \;
\overline{\psi}^{(b)}(x)\gamma_\mu \Big( \partial_\mu - i e A_\mu(x)
-i \sqrt{g} h_\mu(x) \Big)\psi^{(b)}(x) \; .
\end{equation}
The auxiliary field $h_\mu$ is coupled in the same way as the gauge field.
$\overline{\psi}^{(b)}$ and $\psi^{(b)}$ are independent Grassmann variables.

Since the mass term will be treated differently from the rest of the
fermion action I denote it separately
\begin{equation}
S_M[\overline{\psi}, \psi] = -\sum_{b=1}^N m^{(b)} \int_\Lambda d^2x \; t(x) \;
\overline{\psi}^{(b)}(x) \psi^{(b)}(x) \; .
\end{equation}
$m^{(b)}$ are the fermion masses for the various flavors. For technical
reasons the mass term has to be smeared with a test function $t$
with compact support $\Lambda$.
In some of the results it will be possible to set $t$
equal to one.

I would like to remark that the model with zero fermion mass
(all $m^{(b)} = 0$), is
of interest on its own. It will be refered to as the {\it massless model}.
Features of this model will be discussed during the approach to the
massive case and in particular in Chapter 8.

Now one can discuss the role of $h_\mu$. The fermion action couples
$h_\mu$ to the vector currents
\begin{equation}
j^{(b)}_\mu(x) := \overline{\psi}^{(b)}(x) \gamma_\mu \psi^{(b)}(x)
\end{equation}
via
\begin{equation}
-i\sqrt{g} \sum_{b=1}^N \int d^2x h_\mu(x) j^{(b)}_\mu(x) \; .
\end{equation}
Since $S_h[h]$ gives rise to a Gaussian measure one can integrate out the
auxiliary field and obtain a new term $S_T[\overline{\psi},\psi]$
contributing to the fermion action which replaces $S_h[h]$. It is given by
\begin{equation}
S_T[\overline{\psi},\psi] = \frac{1}{2} \; g \; \sum_{b,b^\prime=1}^N \int d^2x
j^{(b)}_\mu(x)\Bigl(\delta_{\mu \nu} -
\lambda^\prime \partial_\mu \partial_\nu \Big)^{-1} j^{(b^\prime)}(x) \; .
\end{equation}
It is easy to check that the covariance operator corresponding to
$S_h[h]$ is given by
\begin{equation}
\Big(\delta_{\mu \nu} -
\lambda^\prime \partial_\mu \partial_\nu \Big)^{-1} =
\delta_{\mu \nu} + \Big(1-\lambda^\prime \triangle \Big)^{-1}
\lambda^\prime \; \partial_\mu \partial_\nu \; .
\end{equation}
In the limit $\lambda^\prime \rightarrow \infty$ this reduces to
\begin{equation}
\lim_{\lambda^\prime \rightarrow \infty}
\Big(\delta_{\mu \nu} -
\lambda^\prime \partial_\mu \partial_\nu \Big)^{-1} \; = \;
\delta_{\mu \nu} - \frac{\partial_\mu \partial_\nu}{\triangle} \; =: \;
T_{\mu \nu} \; .
\end{equation}
$T_{\mu \nu}$ obeys the projector relation
\begin{equation}
T^2 = T \; ,
\end{equation}
and projects on the transverse direction, as was expected,
since the transverse
gauge is being used. Defining now the {\it U(1)-current}
\begin{equation}
J^{(1)}_\mu(x) := \frac{1}{\sqrt{N}} \sum_{b=1}^N j^{(b)}_\mu(x) \; ,
\end{equation}
the new term of the fermion action can be written as
\begin{equation}
S_T[\overline{\psi},\psi] = \frac{g N}{2} \int d^2x
{J^{(1)}_\mu(x)}^T {J^{(1)}_\mu(x)}^T \; ,
\end{equation}
where ${J^{(1)}_\mu}^T$ denotes the transverse projection of $J^{(1)}_\mu$
\begin{equation}
{J^{(1)}_\mu(x)}^T := T_{\mu \nu} J^{(1)}_\nu(x) \; \; .
\end{equation}
Obviously $h_\mu$ generates a Thirring term for the transverse part of the
U(N) flavor singlet current $J^{(1)}_\mu$.

The purpose of this Thirring
term is to make the short distance singularity of
\begin{equation}
\overline{\psi}^{(b)}(x) \psi^{(b)}(x) \;
\overline{\psi}^{(b)}(y) \psi^{(b)}(y) \; ,
\end{equation}
integrable. The quoted expression is a typical term showing up in
a power series expansion of the mass term (3.5). It has to be integrated over
$d^2x d^2y$ which is possible only if an ultraviolet regulator
such as the Thirring term is
included.

%
%
\section{Symmetry properties of the model}
This section is devoted to the discussion of the symmetry
properties of the model. To make my notations clear, the
symmetry generators act as follows
\[
\mbox{SU(N)}_L \; : \; \; \; \; \; \; \; \; \psi \; \; \rightarrow \; \;
\exp\left( i\sum_\alpha \; \omega^{(\alpha)} \;
\tau^{(\alpha)} P_L \right) \; \psi \; ,
\]
\[
\mbox{SU(N)}_R \; : \; \; \; \; \; \; \; \; \psi \; \; \rightarrow \; \;
\exp\left( i\sum_\alpha \; \omega^{(\alpha)} \;
\tau^{(\alpha)} P_R \right) \; \psi \; ,
\]
\[
\mbox{U(1)}_V \; : \; \; \; \; \; \; \; \; \psi \; \; \rightarrow \; \;
\exp\left( i \omega \; \mbox{1\hspace{-1.2mm}I} \right) \; \psi \; ,
\]
\begin{equation}
\mbox{U(1)}_A \; : \; \; \; \; \; \; \; \; \psi \; \; \rightarrow \; \;
\exp\left( i \omega \; \mbox{1\hspace{-1.2mm}I} \;
\gamma_5 \right) \; \psi \; .
\end{equation}
The $T^{(\alpha)}$ denote the SU(N) generators, $\omega$ and
$\omega^{(\alpha)}$ are real coefficients, and
\begin{equation}
P_L \; := \; \frac{1}{2} ( 1 - \gamma_5 ) \; \; \; , \; \; \;
P_R \; := \; \frac{1}{2} ( 1 + \gamma_5 ) \; .
\end{equation}
For vanishing fermion masses $m^{(b)}$, the Lagrangian of the model has the
symmetry
$\mbox{SU(N)}_L \times \mbox{SU(N)}_R
\times \mbox{U(1)}_V \times\mbox{U(1)}_A$
as is the case for QCD.
When quantizing the massless theory the axial U(1)-current
\begin{equation}
j_{5 \; \mu}(x) \; := \; \sum_{b=1}^N \; \overline{\psi}^{(b)} (x)
\gamma_\mu \gamma_5 \psi^{(b)}(x) \; \; ,
\end{equation}
acquires the anomaly
\begin{equation}
\partial_\mu j_{5\; \mu}(x) \; = \; 2 N \; \frac{e}{2( \pi + gN)} \;
\varepsilon_{\mu \nu} \partial_\mu A_\nu(x) \;
+  \; \mbox{contact terms} \; \; .
\end{equation}
This identity can be seen to hold if one evaluates
the functional
\begin{equation}
\Big\langle \; \exp\Big( e \partial_\mu \; \Big[ \; j_{5\; \mu} (t) \; - \;
2 N \; \frac{e}{2( \pi + gN)} \;
\varepsilon_{\mu \nu}  A_\nu(t) \; \Big] \Big) \; X \; \Big\rangle \; .
\end{equation}
Here $X$ denotes an arbitrary monomial of the field operators
and $t$ is some test function that is used to smear the
fields. The expression (3.20) can be computed with the methods
developed below, and it turns out that it does not depend on $t$ up
to the announced contact term\footnote{Contact terms do not
contribute when performing the Osterwalder-Schrader reconstruction
\cite{glimm}.}.
Since this is true for arbitrary $t$ and $X$, (3.19) holds. It has to be
remarked that the coupling constant $g$ for the Thirring term
shows up in the anomaly equation. This is due to the fact that
it is the U(1) vector current that enters the Thirring term,
leading to an extra contribution to the anomaly.
Obviously when setting $g=0$ the usual result is recovered.

As in QCD, the anomaly breaks the symmetry down to
$\mbox{SU(N)}_L \times \mbox{SU(N)}_R
\times \mbox{U(1)}_V$. Since the right hand side
of the anomaly equation (3.19) is a divergence, the formal arguments that
were applied in QCD to define a conserved current $\tilde{j}_5$ can be
repeated. Thus when considering the symmetry properties, the
toy model is adequate for studying the problematic aspects of the
formulation of the U(1)-problem.

For finite fermion masses the chiral
symmetry is already broken at the classical
level. As long as all masses are equal there is a remaining
$\mbox{U(N)}_V$
symmetry, which is further reduced to $\Big(\mbox{U(1)}_V \Big)^N$
in the case of arbitrary masses.

%
%

\section{Topologically nontrivial configurations
for U(1) gauge fields in two dimensions}
In two dimensions also the gauge group U(1) allows for topologically
nontrivial configurations. There explicit calculations are much
easier than for SU(2). Since U(1) gauge fields are
relevant for the model under consideration, I decided to include the
2d discussion as an illustration of Section 2.1.

The topological index
(Chern number) is given by
\begin{equation}
\nu [ A ] \; := \; \frac{e}{4 \pi} \;
\int d^2 x \; \varepsilon_{\mu \nu} \; F_{\mu \nu} (x) \; = \;
\frac{e}{2 \pi} \; \int d^2 x \; \partial_\mu \; K_\mu (x) \; ,
\end{equation}
where
\begin{equation}
K_\mu(x) \; := \; \varepsilon_{\mu \nu}
A_\nu(x) \; .
\end{equation}
Since $F_{\mu \nu}$ is gauge invariant, so is the index.
For gauge field configurations with the boundary
condition (this is sufficient for finite action (compare (2.1))
\begin{equation}
A_\mu(x) \; \;
\stackrel{x^2 \rightarrow \infty}
{-\!\!\!-\!\!\!-\!\!\!-\!\!\!\longrightarrow} \; \;
-\frac{i}{e} \Big(\partial_\mu S(x)\Big)
\Big(S(x)\Big)^{-1} \; ,
\end{equation}
$\nu [ A ]$ can easily be seen to be an integer. Using Stokes theorem
and choosing the integration boundary to be a circle gives
\begin{equation}
\nu [ A ] \; = \; \lim_{R \rightarrow \infty} \;
\frac{e}{2 \pi} \; \int_{0}^{2\pi} \; R d \varphi \; \hat{r}_\mu \; K_\mu \; =
\; \lim_{R \rightarrow \infty} \;
\frac{-i}{2 \pi} \; \int_{0}^{2\pi} \; R d \varphi
\; \hat{r}_\mu
\; \varepsilon_{\mu \nu} (\partial_\nu S) (S)^{-1} \; .
\end{equation}
In the last step the asymptotic form (3.23) was inserted. $\hat{r}$ denotes the
unit vector pointing in radial direction. Since
$\hat{r}_\mu \varepsilon_{\mu \nu} \; = \; \hat{\varphi}_\nu$ and
$ \nabla \; = \; \hat{r} \frac{\partial}{\partial r} +
\hat{\varphi} \frac{1}{r} \frac{\partial}{\partial \varphi} $, one ends
up with
\begin{equation}
\nu [ A ] \; = \; \lim_{R \rightarrow \infty} \;
\frac{-i}{2 \pi} \; \int_{0}^{2\pi} \; d \varphi \;
\Big(\frac{\partial}{\partial \varphi} S(R,\varphi)\Big)
\Big(S(R,\varphi)\Big)^{-1} \; .
\end{equation}
The right hand side can be seen to be an integer by inserting the
explicit parametrization $S(R,\varphi) \; = \; \exp( i f(R,\varphi))$
where $f$ is a continuous function.

The standard representative $A^{(n)}$ for a configuration with
index $\nu [ A^{(n)} ] \; = \; n$ is given by
\begin{equation}
A_\mu^{(n)} \; = \; - \; \frac{i}{e} \; \frac{x^2}{x^2 + \lambda^2} \;
\Big( \partial_\mu S^{(n)}(x) \Big) \; \Big( S^{(n)}(x) \Big)^{-1} \; ,
\end{equation}
with
\begin{equation}
S^{(n)}(x) \; = \; \Big( S^{(1)}(x) \Big)^n \; \; \; \; , \; \; \; \;
S^{(1)}(x) \; = \; \frac{x_2 \; - \; i x_1}{\sqrt{x_1^2 + x^2_2}} \; .
\end{equation}
As in QCD, the construction of
the $\theta$-vacuum will be performed in temporal
gauge ($A^{(n)}_2 = 0$).
In the discussion below I restrict myself to the configuration
$A^{(1)}$ with winding number equal to 1. Inserting (3.27) in
(3.26) then gives
\begin{equation}
A^{(1)}_1(x) \; = \; \frac{1}{e} \; \frac{- x_2}{x^2 + \lambda^2}
\; \; \;  \; , \; \; \; \;
A^{(1)}_2(x) \; = \; \frac{1}{e} \; \frac{x_1}{x^2 + \lambda^2} \; .
\end{equation}
Going to temporal gauge means to gauge transform $A^{(1)}$
\begin{equation}
A^{(1)}_\mu(x) \; \longrightarrow \; \tilde{A}^{(1)}_\mu(x)
\; = \; A^{(1)}_\mu(x) \; - \;
\frac{i}{e} \Big( \partial_\mu \tilde{S}(x) \Big)
\Big( \tilde{S}(x) \Big)^{-1} \; ,
\end{equation}
with the restriction $\tilde{A}^{(1)}_2(x) \; = \; 0$ which amounts to
the partial differential equation
\begin{equation}
\partial_2 \tilde{S}(x)  \; = \; - i \frac{x_1}{x^2 + \lambda^2} \tilde{S}(x)
\; .
\end{equation}
It is solved by
\begin{equation}
\tilde{S}(x) \; = \; \exp \left( -i \frac{x_1}{\sqrt{x_1^2 + \lambda^2}}
\left[ \mbox{arctan} \left( \frac{x_2}{\sqrt{x_1^2 + \lambda^2}} \right)
\; - \; \pi \Big(\frac{1}{2} + m\Big) \right] \; \right) \; ,
\end{equation}
where $m \; \in \; \mbox{Z\hspace{-1.35mm}Z} $
comes about by a proper choice of the integration constant.
The nonvanishing field component $\tilde{A}^{(1)}_1$ is obtained
from (3.28) and (3.29)
\[
\tilde{A}^{(1)}_1(x) \; = \;
- \; \frac{1}{e} \; \frac{x_2}{x_1^2 + x_2^2 + \lambda^2} \;
\left[ 1 + \frac{x_1^2}{(x_1^2+\lambda^2)^{\frac{3}{2}}} \right]
\]
\begin{equation}
- \; \frac{1}{e} \; \frac{\lambda^2}{(x_1^2 + \lambda^2)^{\frac{3}{2}}} \;
\left[ \mbox{arctan} \left( \frac{x_2}{\sqrt{x_1^2 + \lambda^2}} \right)
\; - \; \pi\Big(\frac{1}{2} + m\Big) \right] \; .
\end{equation}
Its asymptotic form can be written as
\begin{equation}
A^{(1)}_1 (x) \; \longrightarrow \; \left\{ \begin{array}{ll}
-\frac{i}{e}\Big(\partial_1 S^{(m)}(x_1)\Big)
\Big(S^{(m)}(x_1)\Big)^{-1} &
\mbox{for} \; \; x_2 \rightarrow - \infty \\ & \\
-\frac{i}{e}\Big(\partial_1 S^{(m+1)}(x_1)\Big)
\Big(S^{(m+1)}(x_1)\Big)^{-1} &
\mbox{for} \; \; x_2 \rightarrow + \infty \; ,\end{array} \right.
\end{equation}
with
\begin{equation}
S^{(l)}(x_1) \; = \; \Big(S^{(1)}(x_1)\Big)^l \; \; , \; \;
S^{(1)}(x_1) \; = \; \exp
\left( \; i \pi \; \frac{x_1}{\sqrt{x_1^2 + \lambda^2}} \; \right) \; .
\end{equation}
Those two equations have to be compared with (2.8) and (2.9) for
the 4d instanton.
One easily reads off
\begin{equation}
\lim_{x_1 \rightarrow + \infty} \; S^{(l)}(x_1) \; \; = \; \;
\lim_{x_1 \rightarrow - \infty} \; S^{(l)}(x_1) \; \; \; \; \;
\forall \; l \; ,
\end{equation}
and the (one dimensional) space can be compactified to $S^1$.
The winding number $\nu[A]$ can be
written as $\nu[A] \; = \; \nu_+[A] - \nu_-[A]$, by choosing
the integration boundary in (3.24) to be a rectangle which replaces
the hypercylinder of Figure 2.1. The
rest of the construction of the $\theta$-vacua can be taken over from
Section 2.1 immediately.
%
%

\section{Outline of the construction}
Vacuum expectation values of operators $P[\overline{\psi},\psi,A,h]$
are formally defined as functional integrals
\begin{equation}
\langle P[\overline{\psi},\psi,A,h] \rangle :=
\frac{1}{Z} \int {\cal D}h {\cal D}A {\cal D}\overline{\psi} {\cal D}\psi \;
P[\overline{\psi},\psi,A,h] \; e^{-S[\overline{\psi},\psi,A,h]} \; .
\end{equation}
The normalization constant $Z$ is chosen such that $\langle 1 \rangle = 1$.
Obviously this is only a formal expression which has to be given some
mathematical meaning.

As I will discuss in the next chapter it is not possible to find a
simple expression of the fermion determinant for massive $\mbox{QED}_2$.
Thus the
strategy will be to expand the mass term $\exp(-S_M[\overline{\psi},\psi])$
in a power series to
obtain
\[
\langle P[\overline{\psi},\psi,A,h] \rangle = \sum_{n=0}^\infty
\frac{(-1)^n}{n!}
\frac{1}{Z} \int {\cal D}h {\cal D}A {\cal D}\overline{\psi} {\cal D}\psi \;
P[\overline{\psi},\psi,A,h] \; \Big(S_M[\overline{\psi},\psi]\Big)^n
\]
\begin{equation}
\times \exp\Big( -S_G[A]-S_h[h]-S_F[ \overline{\psi},\psi,A,h] \Big) \; .
\end{equation}
Of course also the denominator has to be expanded (compare
Equation (3.38) below).
$P[\overline{\psi},\psi,A,h]$ and hence also
$P[\overline{\psi},\psi,A,h] (S_M[\overline{\psi},\psi])^n$
are polynomials
of the Grassmann variables with coefficients that are
functionals of $A_\mu$ and $h_\mu$. They can be generated
taking derivatives of
\[
F[\overline{\eta},\eta,a,A,h] \; := \]
\begin{equation}
\int{\cal D}\overline{\psi} {\cal D}\psi
\exp \left( \sum_{b=1}^N \Big\{
(\overline{\eta}^{(b)}, \psi^{(b)}) + (\overline{\psi}^{(b)}, \eta^{(b)})
+ ie (a^{(b)}_\mu, j^{(b)}_\mu) \Big\} \right)
e^{-S_F[\overline{\psi},\psi,A,h]} \; ,
\end{equation}
with respect to the Grassmann sources
$\overline{\eta}^{(b)},\eta^{(b)}$ and the real number
sources $a^{(b)}_\mu$.
The latter are taken transverse, i.e. they obey
$\partial_\mu a^{(b)}_\mu =0$. They are convenient if one
considers expectation values of vector currents. Since they couple in the
same way as the gauge field they can be treated by replacing
$A \rightarrow A + a^{(b)}$ in the fermion action. After the functional
derivation the sources will be set to zero.

The generating functional $F[\overline{\eta},\eta,a,A,h]$
can be expressed by means of the Berezin integral \cite{berez}
\begin{equation}
F[ \overline{\eta},\eta,a,A,h] \; = \;
c \; \prod_{b=1}^N
\mbox{det} \left[ 1-K( {\not{\!\!B}}^{(b)} ) \right]
\; \exp \left( \sum_{b=1}^N \Big(\overline{\eta}^{(b)},
G(B^{(b)}) \eta^{(b)} \Big) \right) \; .
\end{equation}
$c$ denotes a constant that will be included in $Z$.
The fermion determinant and the operator $K( {\not{\!\!B}}^{(b)} )$
will be discussed in detail in the next section.
This expression shows that polynomials of Grassmann variables
turn to new polynomials that depend on propagators $G$
and functional derivatives
$\frac{\delta}{\delta a}$ when the fermions are integrated
out
\begin{equation}
Q_n[\overline{\psi},\psi,A,h] \; := \;
P[\overline{\psi},\psi,A,h] \Big(S_M[\overline{\psi},\psi]\Big)^n \;
\longrightarrow \; \tilde{Q_n} [G,\frac{\delta}{\delta a},A,h] \; .
\end{equation}
The propagator $G(x,y;B^{(b)})$ is the inverse kernel of the fermion action
\begin{equation}
G(x,y;B^{(b)}) \; := \;
[ \not{\!\partial} - i \not{\!\!B^{(b)}} ]^{-1}(x,y) \; ,
\end{equation}
where I defined
\begin{equation}
B_\mu^{(b)} \; := \; e \Big( A_\mu + a_\mu^{(b)} \Big) + \sqrt{g} h_\mu \; .
\end{equation}
In two dimensions $G$ was found by Schwinger
\cite{schwinger}. The explicit derivation is given in
Appendix A.1 and I only quote the result here
\begin{equation}
G(x,y;B) = G^o(x-y) \; e^{i[\Phi(x) - \Phi(y)]} \; ,
\end{equation}
where
\begin{equation}
\Phi(x) = -\int d^2z D(x-z) \Big( \partial_\mu B_\mu(z) + i \gamma_5
\varepsilon_{\mu \nu} \partial_\mu B_\nu(z) \Big) \; .
\end{equation}
The free fermion propagator $G^o(x-y)$ can also be found in the appendix
\begin{equation}
G^o(x-y) = \frac{1}{2\pi} \frac{\gamma_\mu x_\mu}{x^2} \; .
\end{equation}
Thus after having integrated out the fermions the final expression
for expectation values reads
\begin{equation}
\langle P[\overline{\psi},\psi,A,h] \rangle =
\sum_{n=0}^\infty
\frac{(-1)^n}{n!}
\frac{1}{Z} \bigg\langle
\tilde{Q}_n[G,\frac{\delta}{\delta a},A,h] \;
e^{i e \big(a^{(b)}_\mu,j_\mu^{(b)} \big)}
\bigg\rangle_0
\Bigg{|}_{a=0} \; .
\end{equation}
The expectation values $\langle .. \rangle_0$ of the massless theory are
defined as
\[
\langle Q[\overline{\psi},\psi,A,h] \rangle_0 \; :=
\]
\begin{equation}
\frac{1}{Z_0} \int {\cal D}h  {\cal D}A \;
Q^\prime [G,\frac{\delta}{\delta a},A,h] e^{-S_G[A]-S_h[h]}
\prod_{b=1}^N
\mbox{det} \left[ 1-K\Big( \not{\!\!B^{(b)}} \Big) \right]
\Bigg{|}_{a=0}\; .
\end{equation}
This expression can now be given a precise mathematical meaning.
In the next section I will construct the fermion determinant
$\mbox{det} \left[ 1-K\Big( \not{\!\!B^{(b)}} \Big) \right]$ in the external
field $B^{(b)}$. It will turn out that it is Gaussian in $B^{(b)}$.
Together with $\frac{1}{Z_0} {\cal D}h {\cal D}A e^{-S_G[A]-S_h[h]}$
this will amount to Gaussian measures
for the gauge and the auxiliary field that will be derived in the third
section of the next chapter.

Note that the normalization constant $Z$ of the massive theory
also has an expansion in terms
of expectation values $\langle .. \rangle_0$
\begin{equation}
Z \; = \; \sum_{n=0}^\infty
\frac{(-1)^n}{n!}
\langle \tilde{S}_{M,n} [G] \rangle_0 \; .
\end{equation}

%
%

\chapter{Construction of the massless model}
The aim of this chapter is to give a precise mathematical meaning
to the objects defined in the last section, such that one can
start to evaluate expectation values of physical interest. In particular
I will discuss the fermion determinant and construct measures for
$A_\mu$ and $h_\mu$. Furthermore the propagator will be simplified
by rewriting it in terms of scalar fields $\varphi$ and $\theta$.
%
%
\section{The fermion determinant}
When outlining the general stragtegy in the last chapter, the fermion
determinant\footnote{A very nice introduction to fermion determinants can
be found in \cite{seilerpoi}.}
was only introduced formally when integrating out the
fermions
\begin{equation}
\int \; {\cal D } \overline{\psi} {\cal D } \psi \;
e^{-(\overline{\psi}, Q \psi)} \; = \; \mbox{det}[Q] \; .
\end{equation}
This formula is inspired by reducing the system to finitely many
degrees of freedom \cite{berez} (e.g. by putting the theory on a
finite lattice).
Nevertheless the right hand side has no mathematical meaning in the
continuum yet.

The idea is to transform the problem by formal manipulations
in such a way that one has to compute $\mbox{det}[1-K]$, where $K$ is
a trace class operator. In particular formally one obtains
\begin{equation}
\mbox{det}[\; \not{\!\partial} \; -\; i e \; \not{\!\!B} \; ]  \; = \;
\mbox{det}[\; \not{\!\partial} \; ] \;
\mbox{det}[1 \; - \; i e  \not{\!\!B} \; \not{\!\partial}^{-1} \;]
\end{equation}
and includes the infinite constant $\mbox{det}[\not{\!\partial}]$
in the normalization constant $Z$ (see (3.36)). This procedure was
performed in a mathematically rigorous setting
for $\mbox{QED}_2$ in a lattice regularization \cite{weingarten}.
\newpage
\noindent
Once approached at $\mbox{det}[1-K]$
one has at hand the well known
Fredholm determinant (see e.g. Vol. 4 of \cite{reed}), that is defined by
\begin{equation}
\mbox{det}[\; 1 \; - \; K] \;   := \;
\sum_{n=0}^\infty \; (-1)^n \; \mbox{Tr} [ \Lambda^n(K)] \; .
\end{equation}
$\Lambda^n(K)$ denotes the operator that is induced by $K$ on the n-fold
antisymmetric tensor product
\begin{equation}
\Lambda^n(K) \; \varphi_1 \wedge  \varphi_2 \wedge \; ..... \;
\wedge \varphi_n  \; := \;
K \varphi_1 \wedge  K \varphi_2 \wedge \; ..... \;
\wedge K \varphi_n  \; .
\end{equation}
(4.3) can be shown \cite{simon} to have the expansion
\begin{equation}
\mbox{det}[\; 1 \; - \; K] \; = \; \exp \; \left( \; - \sum_{n=1}^\infty
\; \frac{1}{n} \;
\mbox{Tr} [ K^n] \; \right) \; ,
\end{equation}
which converges for $\mbox{Tr}|K| \; < \; 1$.
I will have to consider operators $K$  which are
not trace class, but a suitable power is. This motivates the
consideration of trace ideals ${\cal I}_q \; , \; q \geq 1$ defined as
(see Vol. 2 of \cite{reed})
\begin{equation}
{\cal I}_q \; := \; \Big\{ \mbox{compact operators}\; \; C \; \Big| \; \;
|C|^q \; := \; (C^* C)^{\frac{q}{2}} \; \;
\mbox{is trace class} \; \; \Big\} \; ,
\end{equation}
with the norms
\begin{equation}
\| C \|_q \; := \; \Big( \mbox{Tr} |C|^q \Big)^{\frac{1}{q}} \; .
\end{equation}
{}From (4.5) there follows a natural definition of modified determinants
adapted
to the trace ideals. Let $K \; \in \; {\cal I}_q \; $. Then
\begin{equation}
\mbox{det}_q[\; 1 \; - \; K \; ] \; := \;
\exp \; \left( \; - \sum_{n=q}^\infty \; \frac{1}{n} \;
\mbox{Tr} [ K^n] \; \right) \; .
\end{equation}
To understand from a physical point of view what has been done, I go back to
$\mbox{QED}_2$ (see \cite{seilerdet}). After a similarity transformation
\begin{equation}
K(\not{\!\!B}) \; := \; \not{\!\!P} \; | P |^{-\frac{3}{2}} \;
\not{\!\!B} \; | P |^{-\frac{1}{2}} \; \; \; , \; \; \mbox{with}
\; \; P_\mu \; := \; - i \partial_\mu \; ,
\end{equation}
considered as an operator on two component, square integrable functions
on $\mbox{I\hspace{-0.62mm}R}^2$. For the external field $B$ see
equation (3.42). If
\begin{equation}
\int \; d^2x \; | B_\mu |^q  \; < \; \infty
\; \; \; \; , \; \; \forall \; \; \; q \; \geq \; \frac{1}{2} \; ,
\end{equation}
(i.e. $|B_\mu|^{\frac{1}{2}} \; \in \; \bigcap_{q \geq 1} \; L^q$ ),
then $K(\not{\!\!B}) \; \in \; {\cal I}_q \;$, for all $q > 2$
(see \cite{seilersimon}). This requirement implies the inclusion
of a cutoff for the fields entering $B$.
For the gauge field $A_\mu$  this cutoff can entirely be removed in the end
(see \cite{seiler} for a discussion), and
thus is not explicitely quoted here. For the auxiliary field $h_\mu$
this is not the case and the cutoff procedure will be made explicit
where it is necessary (see (5.23)).

{}From (4.10) there follows that the modified determinant under consideration
is
\begin{equation}
\mbox{det}_3[\; 1 \; - \; K (\not{\!\!B})\; ] \; := \;
\exp \; \left( \; - \sum_{n=3}^\infty \; \frac{1}{n}\;
\mbox{Tr} [ K(\not{\!\!B})^n] \; \right) \; .
\end{equation}
It differs from Expression (4.5) by the absence of
$-\mbox{Tr} [ K(\not{\!\!B})]$ and \\
$-\frac{1}{2} \mbox{Tr} [ K(\not{\!\!\!B})^2]$ in the exponent.
The first term is zero if it is
renormalized properly in accordance with Furry's theorem
(see below). The second one corresponds to the diagram
of Figure 4.1, which is the only divergent graph in two dimensions.
\begin{center}
\unitlength0.6cm
\begin{picture}(16,6)
\put(3,3){\oval(0.5,0.5)[t]}
\put(3.5,3){\oval(0.5,0.5)[b]}
\put(4,3){\oval(0.5,0.5)[t]}
\put(4.5,3){\oval(0.5,0.5)[b]}
\put(5,3){\oval(0.5,0.5)[t]}
\put(5.5,3){\oval(0.5,0.5)[b]}
\put(6,3){\oval(0.5,0.5)[t]}
\put(6.5,3){\oval(0.5,0.5)[b]}
\put(7.95,3){\circle{6}}
\put(9.4,3){\oval(0.5,0.5)[t]}
\put(9.9,3){\oval(0.5,0.5)[b]}
\put(10.4,3){\oval(0.5,0.5)[t]}
\put(10.9,3){\oval(0.5,0.5)[b]}
\put(11.4,3){\oval(0.5,0.5)[t]}
\put(11.9,3){\oval(0.5,0.5)[b]}
\put(12.4,3){\oval(0.5,0.5)[t]}
\put(12.9,3){\oval(0.5,0.5)[b]}
\end{picture} \\
\end{center}
\noindent
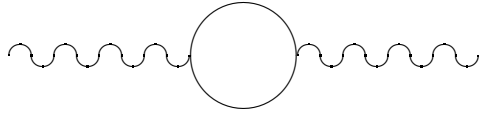
{\bf Figure 4.1 :} The diagram
corresponding to $\mbox{Tr} [ K(\not{\!\!B})^2]$.
\vskip5mm
\noindent
It is well known
how to renormalize this diagram in a gauge invariant way.
To obtain a legitimate expression for a
{\it renormalized determinant}, the renormalized trace has to be restored.
Thus the final definition of the renormalized determinant reads
\begin{equation}
\mbox{det}_{ren}[1-K(\not{\!\!B})] =  \mbox{det}_3[1-K(\not{\!\!B})] \;
\exp\Big(-\frac{1}{2} \mbox{Tr}_{ren}
\big[K^2(\not{\!\!B})\big] \Big) \; .
\end{equation}
First I discuss $\mbox{det}_3[1-K(\not{\!\!B})]$. The following
lemma holds:
\vskip3mm
\noindent
{\bf Lemma 4.1 :}\\
For $|B_\mu|^{1/2} \in \; \bigcap_{q \geq 1}  L^q\; , \;
\partial_\mu B_\mu =0$
\begin{equation}
\mbox{det}_3[1-K(\not{\!\!B})] = 1 \; .
\end{equation}
{\bf Proof:}\\
Making use of
Furry's \cite{itzykson} theorem one
finds that the determinant is even in $e,\sqrt{g}$.
In particular under a charge conjugation ${\cal C}$ ($h_\mu$ transforms in
the same way $A_\mu$ does)
\begin{equation}
K(\not{\!\!B}) \; \; \stackrel{{\cal C}}{\longrightarrow} \; \;
- K(\not{\!\!B}).
\end{equation}
Since ${\cal C}$  is a symmetry of the model only even terms have
to be taken into account, and one obtains
\begin{equation}
\mbox{det}_3[1-K(\not{\!\!B})] =
\exp\Big(-\sum_{n=2}^\infty \mbox{Tr}\big[K^{2n}(\not{\!\!B} )\big] \Big) \; .
\end{equation}
{}From the last equation and Definition (4.9)
for $K(\not{\!\!B})$ there follows immediately
\begin{equation}
\mbox{det}_3[1-K(\not{\!\!B})] = \mbox{det}_3[1-K(\gamma_5 \not{\!\!B})] \; .
\end{equation}
Define $\not{\!\!\tilde{B}} \; := \;
\gamma_5 \not{\!\!B}$ which explicitely gives
$\tilde{B}_1 = - i B_2 = i \partial_1 f$ and
$\tilde{B}_2 = i B_1 = i \partial_2 f$,  showing that $\tilde{B}_\mu$
is a pure gauge. I already used the transversality of $B_\mu$, that allows to
write $B_\mu = \varepsilon_{\mu \nu} \partial_\nu f$ for some scalar
function $f$.
{}From the definition it follows that the determinant
is invariant under the gauge transformation $\psi(x) \rightarrow
\exp(i f(x)) \psi(x),
\overline{\psi}(x) \rightarrow \overline{\psi}(x) \exp(- i f(x))$
which removes $\tilde{B_\mu}$. Now one can finish the proof
\begin{equation}
\mbox{det}_3[1-K(\not{\!\!B})] = \mbox{det}_3[1-K(\tilde{\not{\!\!B}})] =
\mbox{det}_3[1-K(0)] = 1 \; . \;\; \Box
\end{equation}
$\mbox{Tr}_{ren} (K^2(\not{\!\!B}))$
can be evaluated easily by using e.g.
dimensional regularization. To handle the infrared problem, I work
with finite mass $m$ and perform the massless limit in the end.
\[
\mbox{Tr}_{ren}(K^2(\not{\!\!B})) = \mbox{Tr}_\gamma
\int \frac{d^2 p}{(2\pi)^2}
\frac{d^2 q}{(2\pi)^2} d^2x d^2y \;
e^{-ipx} K(\not{\!\!B})(x) e^{iqx}
e^{-iqy} K(\not{\!\!B})(y) e^{ipy} \]
\begin{equation}
= \; \int d^2k \hat{B}_\mu(k) \hat{B}_\nu(-k) \hat{T}_{\mu \nu}(k) \; \; .
\end{equation}
$\hat{B}_\mu$ is the Fourier transform of $B_\mu$, and $\hat{T}_{\mu \nu}$ is
given by (now the dimensional regularization comes in;
$\omega := 1 - \epsilon$ )
\[ \hat{T}_{\mu \nu}(k) = \int \frac{d^{2\omega}p}{(2\pi)^{2\omega}}
\frac{\mbox{Tr}(\gamma_\mu \gamma_\alpha \gamma_\nu \gamma_\beta)
(p_\alpha - k_\alpha)p_\beta - m^2 \mbox{Tr}(\gamma_\mu \gamma_\nu)}
{[(p-k)^2 + m^2][p^2+m^2]}  \; \]
\[= \; \int_{x=0}^1 dx \int \frac{d^{2\omega}r}{(2\pi)^{2\omega}}
\frac{\mbox{Tr}(\gamma_\mu \gamma_\alpha \gamma_\nu \gamma_\beta)
(r_\alpha - xk_\alpha)(r_\beta + (1-x)k_\beta)- m^2 2 \delta_{\mu \nu}}
{[r^2 + m^2 + k^2x(1-x)]^2} \]
\[= \;
\mbox{Tr}(\gamma_\mu \gamma_\alpha \gamma_\nu \gamma_\beta)\int_{x=0}^1 dx
\frac{1}{(4\pi)^\omega \Gamma(2)} \Bigg( \frac{1}{2} \delta_{\alpha \beta}
\frac{\Gamma(1-\omega)}{[m^2 +k^2x(1-x)]^{1-\omega}} \]
\[ - \; x(x-1)k_\alpha
k_\beta \frac{\Gamma(2-\omega)}{[m^2 +k^2x(1-x)]^{2-\omega}} \Bigg) \]
\begin{equation}
- \; 2\delta_{\mu \nu} m^2 \int_{x=0}^1 dx
\frac{1}{(4\pi)^\omega \Gamma(2)}
\frac{\Gamma(2-\omega)}{[m^2 +k^2x(1-x)]^{2-\omega}} \; \; .
\end{equation}
In the first step a Feynman parameter $x$ was introduced and a variable
transformation $r := p - k(1-x)$ was performed
to bring the integral to standard form
where it can be solved using well known dimensional regularization
formulas (see eg. \cite{ramond}). In $2\omega = 2(1-\epsilon)$ dimensions
one has the following trace identities
\[ \mbox{Tr}(\gamma_\mu \gamma_\alpha \gamma_\nu \gamma_\beta)
k_\alpha k_\beta \; = \;
2 \Big(2 k_\mu k_\nu - \delta_{\mu \nu} k^2 \Big) \; + \;
O(\epsilon) \; , \] \\
\begin{equation}
\mbox{Tr}(\gamma_\mu \gamma_\alpha \gamma_\nu \gamma_\beta)
\delta_{\alpha \beta} \; = \;
4 \epsilon \; \delta_{\mu \nu} \; + \; O(\epsilon^2) \; .
\end{equation}
The Feynman parameter integrals over $x$ can also be expanded in $\epsilon$
and then be solved. Putting things together one ends up with
\[
\hat{T}_{\mu \nu} (k) = \frac{1}{2\pi} \delta_{\mu \nu}+
\frac{1}{2\pi}\Big( 2 \frac{k_\mu k_\nu}{k^2}\!-\!\delta_{\mu \nu} \Big)
\Bigg(-1\!+\!\Big(\sqrt{\frac{m^2}{k^2}\!+\!\frac{1}{4}}\!-\!\frac{1}{4}
\frac{1}{\sqrt{\frac{m^2}{k^2}\!+\!\frac{1}{4}}} \Big) l(k,m) \Bigg) \]
\begin{equation}
- \; \frac{1}{2\pi}\delta_{\mu \nu}\frac{m^2}{k^2}
\frac{1}{\sqrt{\frac{m^2}{k^2} + \frac{1}{4} } } l(k,m)  \; + \;
O(\epsilon) \; ,
\end{equation}
where
\begin{equation}
l(k,m) = \mbox{ln}\left( \frac{\sqrt{\frac{m^2}{k^2}
+ \frac{1}{4}} + \frac{1}{2} }
{\sqrt{\frac{m^2}{k^2} + \frac{1}{4}} - \frac{1}{2} } \right) \; .
\end{equation}
It is a remarkable feature of QED$_2$ that the terms of order $1/\epsilon$
cancel and no counterterms have to be added. $\mbox{QED}_2$ therefore
is a {\it finite theory}.

Finally I perform the limit $m \rightarrow 0$ to obtain
\begin{equation}
\hat{T}_{\mu \nu}(k) = \frac{1}{\pi}\Big( \delta_{\mu \nu} -
\frac{k_\mu k_\nu}{k^2} \Big) \Bigg( 1 +
2\frac{m^2}{k^2}\mbox{ln}\Big( \frac{m^2}{k^2}\Big)+
O\Big(\frac{m^4}{k^4}\mbox{ln}\Big( \frac{m^2}{k^2}\Big)\Big)\Bigg)  \; .
\end{equation}
In the massless limit the integral kernel $\hat{T}_{\mu \nu}(k)$ reduces to
the transversal projector already encountered in (3.10)
\begin{equation}
T_{\mu \nu} := \delta_{\mu \nu} -
\frac{k_\mu k_\nu}{k^2} \; .
\end{equation}
Using (4.12) and Lemma 4.1 one ends
up with a remarkably simple expression for the fermion determinant in
the massless case
\begin{equation}
\mbox{det}_{ren}
[1-K(\not{\!\!B})] = e^{-\frac{1}{2\pi} \| B^T \|_2^2}
\; \; , \; \;
B^T_\mu := T_{\mu \nu} B_\nu \; .
\end{equation}
It has to be remarked that the condition (4.10) on $B_\mu$ and thus
on $A_\mu$ (for $g=0$) implies zero winding. This fact will leave a trace
in the properties of the expectation functional which will be discussed in
the next chapter.
%
%
\section{Remarks on the massive case}
In the last section the following
small $m$ behaviour for the momentum space kernel
$\hat{T}_{\mu \nu}(k)$ of $\mbox{Tr}_{\mbox{ren}}[K^2(\not{\!\!B}])$
has been established
\begin{equation}
\hat{T}_{\mu \nu}(k) = \frac{1}{\pi}\Big( \delta_{\mu \nu} -
\frac{k_\mu k_\nu}{k^2} \Big) \Bigg( 1 +
2\frac{m^2}{k^2}\mbox{ln}\Big( \frac{m^2}{k^2}\Big)+
O\Big(\frac{m^4}{k^4}\mbox{ln}\Big( \frac{m^2}{k^2}\Big)\Big)\Bigg)  \; .
\end{equation}
One can try to find a similar expansion in $m$ for the traces
$\mbox{Tr}
[K^{2n}(\not{\!\!B})] , \; n > 1$ that build up
$\mbox{det}_3[1-K(\not{\!\!B})]$,
which differs from 1 in the case of finite mass. If one could show that
those higher traces decrease faster than
$\frac{m^2}{k^2}\mbox{ln}( \frac{m^2}{k^2})$, it would make sense to
perform a mass expansion of the fermion determinant directly.

As a matter of fact this optimistic scenario does not hold. By discussing
$\mbox{Tr}[K^{4}(\not{\!\!B})]$ I will make it plausible that all
$\mbox{Tr}[K^{2n}(\not{\!\!B})] \; , \; n > 1$ do not vanish faster than
$\frac{m^2}{k^2}\mbox{ln}( \frac{m^2}{k^2})$.

Using gauge invariance and the fact that
$\gamma_5 \not{\!\!B}$ is a pure gauge one obtains
\[ 0 \; =  \; \mbox{Tr}\Big[ K(\gamma_5 \not{\!\!B})^4 \Big] \]
\begin{equation}
= \; \mbox{Tr} \Big[
\frac{- \not{\!\partial}+m}{-\triangle + m^2} \not{\!\!B}
\frac{- \not{\!\partial}-m}{-\triangle + m^2} \not{\!\!B}
\frac{- \not{\!\partial}+m}{-\triangle + m^2} \not{\!\!B}
\frac{- \not{\!\partial}-m}{-\triangle + m^2} \not{\!\!B}
\Big]\; .
\end{equation}
In the last step $[\gamma_5,\not{\!\!B}] = 0 $ and $\gamma_5^2 =1$ were used.
This allows to write
\[ \mbox{Tr}\Big[ K(\not{\!\!B})^4 \Big] \; = \;
\mbox{Tr}\Big[K(\not{\!\!B})^4\Big] -
\mbox{Tr}\Big[K(\gamma_5 \not{\!\!B})^4\Big] \;
\]
\[ = \; 2m \mbox{Tr} \Big[
\frac{- \not{\!\partial}+m}{-\triangle + m^2} \not{\!\!B}
\frac{1}{-\triangle + m^2} \not{\!\!B}
\frac{- \not{\!\partial}+m}{-\triangle + m^2} \not{\!\!B}
\frac{- \not{\!\partial}-m}{-\triangle + m^2} \not{\!\!B} \Big] \] \\
\[ + \; 2m \mbox{Tr} \Big[
\frac{- \not{\!\partial}+m}{-\triangle + m^2} \not{\!\!B}
\frac{- \not{\!\partial}+m}{-\triangle + m^2} \not{\!\!B}
\frac{- \not{\!\partial}+m}{-\triangle + m^2} \not{\!\!B}
\frac{1}{-\triangle + m^2} \not{\!\!B}
\Big] \] \\
\begin{equation}
= \; 8m^2 \mbox{Tr} \Big[
\frac{1}{-\triangle + m^2} \not{\!\!B}
\frac{1}{-\triangle + m^2} \not{\!\!B}
\frac{1}{-\triangle + m^2} \not{\!\partial} \not{\!\!B}
\frac{1}{-\triangle + m^2} \not{\!\partial} \not{\!\!B}
\Big] \; ,
\end{equation}
where successively $\not{\!\partial}$ terms in the
numerator were canceled.
For the gauge field I make the following ansatz that trivially obeys
the $\partial_\mu B_\mu = 0$ gauge condition
\begin{equation}
\hat{B}_1(q) = -q_2f(q) \; \; \; \; ,
\; \; \; \; \hat{B}_2(q) = q_1f(q) \; \; ,
\end{equation}
where $f(q)$ is a $L^2(\mbox{I\hspace{-0.62mm}R})$ scalar function.
Inserting this ansatz and Fourier transforming the trace
one obtains
\[ \mbox{Tr}\Big[ K(\not{\!\!B})^4 \Big] = \frac{16m^2}{(2\pi)^8}
\int d^2q_2 \; d^2q_3 \; d^2q_4 \;
f(q_2 + q_3 + q_4) f(-q_2) f(-q_3) f(-q_4) \times
\] \\
\begin{equation}
\int d^2k \frac{-(q_2 + q_3 + q_4) \cdot q_2 \; q_3^2 \; q_4^2}
{(k^2\!+\!m^2)((k\!+\!q_2)^2\!+\!m^2)((k\!+\!q_2\!+\!q_3)^2\!+\!m^2)
((k\!+\!q_2\!+\!q_3\!+\!q_4)^2\!+\!m^2)}
\; .
\end{equation}
For non-exceptional momenta $q_2, q_3, q_4$ the $k$-integral behaves as
$\mbox{ln}(m)$ and one ends up with
\begin{equation}
\mbox{Tr}\Big[K(\not{\!\!B})^4\Big] \; \; \propto \; \; m^2 \mbox{ln}(m) \; .
\end{equation}
In the light of this result it is quite unlikely that
$\mbox{Tr}\Big[K^{2n}(\not{\!\!B})\Big] \; , \; n > 2$
vanish faster than $m^2 \mbox{ln}(m)$. Hence it is not
very promising to try a mass
expansion of the fermion determinant directly.

%
%
\section{Measures for $A$ and $h$}
As announced the determinant comes out
Gaussian in the external fields.
In this section I will construct a common Gaussian measure for
the formal expression
\begin{equation}
\frac{1}{Z_0} {\cal D}h  {\cal D}A \; e^{-S_G[A]-S_h[h]} \;
\prod_{b=1}^N \mbox{det}_{ren}
\left[ 1-K\Big( {\not\!\!B}^{(b)} \Big) \right] \; .
\end{equation}
The field combination $B_\mu$ is taken from (3.42)
\begin{equation}
B_\mu^{(b)} := e \Big( A_\mu + a_\mu^{(b)} \Big) + \sqrt{g} h_\mu \; .
\end{equation}
Inserting this into the result for the determinant (4.24) one obtains
\[
\prod_{b=1}^N \mbox{det}_{ren}
\left[ 1-K\Big( {\not\!\!B}^{(b)} \Big) \right] =
\exp\left(- \frac{1}{2\pi}\sum_{b=1}^N \| {B^{(b)}}^T \|^2_2  \right) \; =
\]
\[
\exp \left(-\frac{e^2 N}{2\pi} \Big(A,TA \Big) -
\frac{e \sqrt{g} N}{\pi} \Big( h,T A \Big) -
\frac{ g N }{2\pi}\Big( h,T h \Big) \; - \right. \]
\begin{equation} \left.
\frac{e^2}{\pi} \Big( A, T \sum_{b=1}^N a^{(b)} \Big) -
\frac{e \sqrt{g}}{\pi} \Big( h, T \sum_{b=1}^N a^{(b)} \Big) -
\frac{e^2}{2\pi} \sum_{b=1}^N \Big( a^{(b)} , T a^{(b)} \Big) \right) \; .
\end{equation}
Here the transverse projector $T$ (3.10) was used to obtain
$\| B^T \|^2_2 = (TB,TB) = (B,TB)$.

The gauge field action as well as the action for the auxiliary field
can be written
as quadratic forms (compare (3.2), (3.3))
\[
S_G[A] = \frac{1}{2}
\Big( A_\mu,\Big[ (1-\lambda)\partial_\mu \partial_\nu -
\triangle \delta_{\mu \nu}\Big]
A_\nu \Big) \, , \]
\begin{equation}
S_h[h] = \frac{1}{2}
\Big( h_\mu,\Big[ \delta_{\mu \nu} -
\lambda^\prime \partial_\mu \partial_\nu \Big] h_\nu
\Big) \, .
\end{equation}
In a first step I unify the auxiliary field action and the quadratic term in
$h$ from the determinant to one common quadratic form
\begin{equation}
\frac{1}{2}(h,C^{-1} h) := S_h[h] + \frac{g N}{2\pi} (h,Th) \; .
\end{equation}
One obtains
\begin{equation}
C^{-1}_{\mu \nu} = \delta_{\mu \nu} - \lambda^\prime
\partial_\mu \partial_\nu +
\frac{g N}{\pi}\left( \delta_{\mu \nu} -
\frac{\partial_\mu \partial_\nu}{\triangle} \right) \; .
\end{equation}
It is easy to verify (by Fourier transformation)
that $C^{-1}$ is a positive, nondegenerate
operator in
${\cal S}^2(\mbox{I\hspace{-0.62mm}R}^2)$ and hence gives rise to a
covariance operator $C$, which is given by
\begin{equation}
C_{\mu \nu} = \Big( 1 - \lambda^\prime \triangle \Big)^{-1} \delta_{\mu \nu}
- \frac{\pi}{\pi + gN} \Big( 1 - \lambda^\prime \triangle \Big)^{-1}
\left( \lambda^\prime \triangle + \frac{gN}{\pi} \right) T_{\mu \nu} \; ,
\end{equation}
and reduces in the transverse limit to
\begin{equation}
\lim_{\lambda^\prime \rightarrow \infty} C_{\mu \nu} =
\frac{\pi}{\pi + gN} T_{\mu \nu} \; .
\end{equation}
This is a proper covariance in the transverse subspace of
${\cal S}^2(\mbox{I\hspace{-0.62mm}R}^2)$.
In the following I will only need this expression.

By completing the square one can include the mixed term
$\frac{e \sqrt{g} N}{\pi} \Big( h,TA \Big)$
\[
S_h[h] + \frac{g N}{2\pi} \Big( h,Th \Big) +
\frac{e \sqrt{g} N}{\pi} \Big( h,TA \Big) =
\frac{1}{2} \Big(h,C^{-1} h \Big) +
\frac{e \sqrt{g} N}{\pi} \Big( h,TA \Big)  \]
\begin{equation}
= \; \frac{1}{2} \Big(h^\prime, C^{-1} h^\prime  \Big) -
\frac{1}{2} \frac{ge^2N^2}{\pi^2} \Big(A,TC A \Big) \; ,
\end{equation}
where
\begin{equation}
h^\prime = h + \frac{e \sqrt{g}  N}{\pi} T C A =
h + \frac{e \sqrt{g} N}{\pi + gN} T A\; .
\end{equation}
The coordinate transform $h \rightarrow h^\prime$ can be performed
without any trouble in the Gaussian integrals
and the observables and propagators can be rewritten in terms of
$h^\prime$.

There are now three pieces quadratic in $A$. One from the completion
of the square (4.40), one from the determinant (4.34)
and one from the action (4.35). They can be unified to one
common quadratic form
\begin{equation}
\frac{1}{2}(A,Q^{-1} A) := S_G[A] + \frac{e^2 N}{2\pi} \Big(A,TA \Big) -
\frac{1}{2} \frac{ge^2N^2}{\pi^2} \Big(A,TC A \Big) \; ,
\end{equation}
with
\[
Q^{-1}_{\mu \nu} = \Big( (1-\lambda)\partial_\mu \partial_\nu -
\triangle \delta_{\mu \nu} \Big) +
\frac{e^2N}{\pi}T_{\mu \nu} -
\frac{ge^2N^2}{\pi^2} \frac{\pi}{\pi + gN} T_{\mu \nu}
\]
\begin{equation}
= \; \left( -\triangle + \frac{e^2 N}{\pi + gN} \right) T_{\mu \nu}
- \lambda \partial_\mu \partial_\nu \; .
\end{equation}
Again it is easy to check that $Q^{-1}$ gives rise to a proper covariance $Q$
on ${\cal S}^2(\mbox{I\hspace{-0.62mm}R})$. $Q$ is given by
\begin{equation}
Q_{\mu \nu} = -\frac{1}{\lambda} \frac{\partial_\mu \partial_\nu}{\triangle}
\; + \;
\left( -\triangle + \frac{e^2 N}{\pi + gN} \right)^{-1} \; T_{\mu \nu}  \; ,
\end{equation}
and reduces in the transverse subspace to
\begin{equation}
Q _{\mu \nu} = \Big( -\triangle + \frac{e^2 N}{\pi + gN} \Big)^{-1}
\; T_{\mu \nu} \; .
\end{equation}
The fact that the gauge field has a massive propagator is known as the
{\it Schwinger mechanism}.

Finally I rewrite the terms in the determinant (4.34)
where the sources $a^{(b)}$
and the fields $h$ and $A$ mix in terms of $h^\prime$ and $A$
\[
\frac{e^2}{\pi} \Big( A, T \sum_{b=1}^N a^{(b)} \Big) +
\frac{e \sqrt{g}}{\pi} \Big( h, T \sum_{b=1}^N a^{(b)} \Big) \]
\begin{equation}
= \; \frac{e^2}{\pi + gN} \Big( A, T \sum_{b=1}^N a^{(b)} \Big)
+ \frac{e \sqrt{g}}{\pi} \Big( h^\prime , T \sum_{b=1}^N a^{(b)} \Big) \; .
\end{equation}
Thus one has established that the initial expression (4.32)
can be replaced by
\[
\frac{1}{Z_0} {\cal D}h  {\cal D}A
\exp\left( -\frac{1}{2} \Big(A,Q^{-1} A \Big) -
\frac{1}{2}\Big(h^\prime,C^{-1} h^\prime\Big)
\right) \]
\begin{equation}
\times
\exp\left( -\frac{e^2}{\pi + gN} \Big( A, T \sum_{b=1}^N a^{(b)} \Big)
-\frac{e \sqrt{g}}{\pi} \Big( h^\prime , T \sum_{b=1}^N a^{(b)} \Big)
-\frac{e^2}{2\pi} \sum_{b=1}^N \Big( a^{(b)} , T a^{(b)} \Big) \right) \; ,
\end{equation}
which has the clear meaning
\[
d\mu_Q[A] d\mu_C[h^\prime] \]
\begin{equation}
\times \;
\exp\left( -\frac{e^2}{\pi + gN} \Big( A, T \sum_{b=1}^N a^{(b)} \Big)
-\frac{e \sqrt{g}}{\pi} \Big( h^\prime , T \sum_{b=1}^N a^{(b)} \Big)
-\frac{e^2}{2\pi} \sum_{b=1}^N \Big( a^{(b)} , T a^{(b)} \Big) \right)  \; ,
\end{equation}
in terms of Gaussian measures\footnote{For
a short introduction to Gaussian measures see Appendix A.2}
$d\mu_Q[A]$ and $d\mu_C[h^\prime]$
with covariances $Q$ and $C$ given by (4.45) and (4.39).
The poorly defined vacuum
expectation value (3.47) of the massless model now has the precise
definition
\[
\Big\langle Q[\bar{\psi},\psi,A,h] \Big\rangle_0 \; := \;
\int d\mu_Q[A] d\mu_C[h^\prime] \; \;
Q^\prime [G,\frac{\delta}{\delta a},A,h^\prime ] \]
\begin{equation}
\exp\left( -\frac{e^2}{\pi\!+\!gN} \Big( A, T \sum_{b=1}^N a^{(b)} \Big)
-\frac{e \sqrt{g}}{\pi} \Big( h^\prime , T \sum_{b=1}^N a^{(b)} \Big)
-\frac{e^2}{2\pi} \sum_{b=1}^N \Big( a^{(b)} , T a^{(b)} \Big) \right)
\Bigg{|}_{a^{(b)}=0}\; .
\end{equation}
%
%
\section{The propagator and the fields $\varphi$ and $\theta$}
The fermion propagator in the transverse external field $B^{(b)}$
was obtained Appendix A.1 and reads (c.f. (A.15))
\begin{equation}
G(x,y;B^{(b)}) =
\frac{1}{2\pi}\frac{1}{(x\!-\!y)^2}\!\left( \begin{array}{cc}
0 & e^{-[\chi^{(b)}(x)-\chi^{(b)}(y)]} \;
\overline{(\tilde{x}\!-\!\tilde{y})} \\
e^{+[\chi^{(b)}(x)-\chi^{(b)}(y)]} \; (\tilde{x}\!-\!\tilde{y}) & 0
\end{array} \right)\; ,
\end{equation}
where
\begin{equation}
\chi^{(b)}(x) :=
\frac{\varepsilon_{\mu \nu} \partial_\mu}{\triangle} B^{(b)}_\nu(x)
\; \; \; \; ,
\; \; \; \; \tilde{x} := x_1 + ix_2 \; .
\end{equation}
The external field $B^{(b)}$ that enters the propagator was defined in (3.42).
It has to be rewritten in terms of $h^\prime$.
$\frac{\varepsilon_{\mu \nu} \partial_\mu}{\triangle}$ acts on it
to give $\chi^{(b)}$, which reads in terms of $h^\prime$
\begin{equation}
\chi^{(b)}(x) = \frac{e \pi}{\pi + gN} \varphi(x) + \sqrt{g} \theta(x)
+ e\frac{\varepsilon_{\mu \nu} \partial_\mu}{\triangle} a^{(b)}_\nu(x) \; ,
\end{equation}
where I defined the scalar fields
\begin{equation}
\varphi(x) :=
\frac{\varepsilon_{\mu \nu} \partial_\mu}{\triangle} A_\nu(x) \; \; \; ,
\; \; \; \theta(x) :=
\frac{\varepsilon_{\mu \nu} \partial_\mu}{\triangle} h^\prime_\nu(x) \; .
\end{equation}
It will turn out that it is more
convenient to work with the scalar fields $\varphi$ and $\theta$.
Obviously their measures are also Gaussian and the corresponding
covariance operators $\tilde{Q}$ and $\tilde{C}$
are immediately obtained from $Q$ and $C$
\begin{equation}
\tilde{Q} \; = \; - \frac{\varepsilon_{\rho \mu} \partial_\rho}{\triangle}
Q_{\mu \nu} \frac{\varepsilon_{\sigma \nu} \partial_\sigma}{\triangle} \; =
\; \frac{1}{-\triangle + \frac{e^2 N}{\pi + gN}} \; \frac{-1}{\triangle} \; ,
\end{equation}
and
\begin{equation}
\tilde{C} \; = \; - \frac{\varepsilon_{\rho \mu} \partial_\rho}{\triangle}
C_{\mu \nu} \frac{\varepsilon_{\sigma \nu} \partial_\sigma}{\triangle} \; =
\; \frac{\pi}{\pi + gN} \; \frac{-1}{\triangle} \; .
\end{equation}
$\tilde{C}$ behaves $\propto 1/p^2$ in momentum space which
causes an ultraviolet problem. As was discussed above (see page 28) the
construction of the determinant requires a cutoff.
Here I adopt the following procedure. The scalar field $\chi(x)$ at
the single space-time point $x$ will be replaced by the convolute
\begin{equation}
\chi(x) = \; \longrightarrow \;
\int \; d^2\xi \; \chi( \xi ) \; \delta_{n} ( \xi - x ) \; =: \;
\Big(\chi,\delta_n(x)\Big)\; ,
\end{equation}
where $\delta_n(x)$ denotes a $\delta$-sequence peaked at x
\begin{equation}
\delta_n(\xi-x) :=
\int \; \frac{d^2 p}{(2 \pi)^2} \; e^{ - \frac{|p|}{n} } \; e^{ip(\xi-x)} \; .
\end{equation}
Thus the propagator takes the form
\begin{equation}
G(x,y;\chi) =
\frac{1}{2\pi}\frac{1}{(x-y)^2} \left( \begin{array}{cc}
0 & e^{-\big(\chi,\delta_n(x)-\delta_n(y)\big)} \;
\overline{(\tilde{x}-\tilde{y})} \\
e^{+\big(\chi,\delta_n(x)-\delta_n(y)\big)} \; (\tilde{x}-\tilde{y}) & 0
\end{array} \right) \; .
\end{equation}
When one considers the limit $n \rightarrow \infty$, some
of the operators will have to be multiplied with a wave function
renormalization constant (see (5.53))
diverging as $n \rightarrow \infty$.

%
%

\chapter{Decomposition into clustering states and the vacuum angle}
It will turn out that for vanishing fermion masses
a certain class of operators violates the cluster
decomposition property when using the expectation functional constructed so
far. I am going to classify those operators and discuss their symmetry
properties. Finally the expectation functional will be decomposed into
clustering $\theta$-vacua giving rise to proper states.
%
%
\section{Clustering and the uniqueness of the vacuum}
If there is any truth in the $\theta$-vacuum philosophy,
then one should face some problems with the vacuum state.
The reason for this is that in the construction
of the fermion determinant and the propagator
it was assumed that the gauge fields
decrease appropriately (compare (4.10) and Appendix A.1),
and thus have zero winding.
In the $\theta$-language this means that up to now only
transitions of the $\langle n | n \rangle$ type are included, whereas
topologically nontrivial configurations enforce $\langle n + \nu | n \rangle$
(compare (2.14))
contributions as well. Thus one has to expect that the expectation
functional constructed so far does not lead to a unique vacuum
when the Osterwalder-Schrader reconstruction (see \cite{glimm}) is performed.
The ergodicity axiom which guarantees the uniquenes of the vacuum
state is not fulfilled. Ergodicity is equivalent to the
cluster property (compare \cite{glimm}, Section 19.7)
\begin{equation}
\lim_{t \rightarrow \infty} \; t^{-1} \int_0^t \;d\tau
\; \Big[ \langle A T(\tau) B \rangle -
\langle A \rangle \langle B \rangle \Big]
\; \stackrel{!}{=} \; 0 \; .
\end{equation}
$T(\tau)$ denotes time translation, and $A,B$  are arbitrary polynomials
in the fields smeared with test functions\footnote{In fact $A,B$ can be taken
from ${\cal E}$ which is the closure of the vectors
$\sum_j c_j \exp(\phi(f_j))\; , \; c_j \in \mbox{C\hspace{-1.7mm}l} \; , \;
f_j \in {\cal D}$, in the $L_2(d\mu[\phi])$ inner product.}.
To detect expectation values that violate clustering, and thus
connect different vacua, I will consider a rather general ansatz for
$A$ and $B$ in the next section.

It is more convenient to consider the $\tau \rightarrow \infty$ limit of
\begin{equation}
C(\tau) \; := \; \langle A T(\tau) B \rangle -
\langle A \rangle \langle B \rangle \; .
\end{equation}
Obviously a nonvanishing limit of $C(\tau)$ implies the violation of
clustering in the formulation (5.1).

%
%
\section{Identification of operators that violate clustering}
To identify the operators that violate clustering, I start with an
ansatz containing only the chiral densities
$\overline{\psi}^{(b)} P_\pm \psi^{(b)} \; , \;
( P_\pm := (1 \pm \gamma_5)/2 )$
and discuss the effect of
adding vector currents and other modifications later.
Define
\begin{equation}
C(\tau) \; := \; C_1(\tau) \; - \; C_2
\end{equation}
where
\[
C_1(\tau) \; := \;
\Big\langle \prod_{b=1}^N
\prod_{i=1}^{n_b} \overline{\psi}^{(b)}(x^{(b)}_i\!+ \hat{\tau})
P_+ \psi^{(b)}(x^{(b)}_i\!+ \hat{\tau})
\prod_{i=1}^{m_b} \overline{\psi}^{(b)}(y^{(b)}_i\!+ \hat{\tau})
P_- \psi^{(b)}(y^{(b)}_i\! + \hat{\tau}) \]
\begin{equation}
\times\;
\prod_{i=1}^{n^\prime_b}
\overline{\psi}^{(b)}({x^\prime}^{(b)}_i)
P_+ \psi^{(b)}({x^\prime}^{(b)}_i)
\prod_{i=1}^{m^\prime_b}
\overline{\psi}^{(b)}({y^\prime}^{(b)}_i)
P_- \psi^{(b)}({y^\prime}^{(b)}_i) \Big\rangle_0
\end{equation}
and
\[
C_2 \; := \;
\Big\langle \prod_{b=1}^N
\prod_{i=1}^{n_b} \overline{\psi}^{(b)}(x^{(b)}_i )
P_+ \psi^{(b)}(x^{(b)}_i )
\prod_{i=1}^{m_b} \overline{\psi}^{(b)}(y^{(b)}_i )
P_- \psi^{(b)}(y^{(b)}_i ) \Big\rangle_0 \]
\begin{equation}
\times \; \Big\langle \prod_{b=1}^N
\prod_{i=1}^{n^\prime_b}
\overline{\psi}^{(b)}({x^\prime}^{(b)}_i)
P_+ \psi^{(b)}({x^\prime}^{(b)}_i)
\prod_{i=1}^{m^\prime_b}
\overline{\psi}^{(b)}({y^\prime}^{(b)}_i)
P_- \psi^{(b)}({y^\prime}^{(b)}_i) \Big\rangle_0 \; .
\end{equation}
$\hat{\tau}$ denotes the vector of length $\tau$ in 2-direction.
Violation of the cluster property now manifests itself in a nonvanishing
limit
\begin{equation}
\lim_{\tau \rightarrow \infty} C(\tau) =: C  \neq 0 \; .
\end{equation}
It will be obtained for certain $n_b,m_b,n^\prime_b,m^\prime_b$.

\noindent
Since it is rather hopeless to evaluate the traces over general
products of $\gamma$-matrices, I make use of the special form (4.58)
of the propagator
\begin{equation}
G(x,y;\chi) =
\frac{1}{2\pi}\frac{1}{(x-y)^2} \left( \begin{array}{cc}
0 & e^{-\big(\chi,\delta_n(x)-\delta_n(y)\big)} \;
\overline{(\tilde{x}-\tilde{y})} \\
e^{+\big(\chi,\delta_n(x)-\delta_n(y)\big)} \; (\tilde{x}-\tilde{y}) & 0
\end{array} \right) \; .
\end{equation}
With the chosen representation of the $\gamma$-algebra
it has only off diagonal entries. For the evaluation of $C_1(\tau)$ and $C_2$
no sources $a^{(b)}$ are necessary and $\chi(x)$ (see (4.52)) reduces to
\begin{equation}
\chi(x) = \frac{e \; \pi}{\pi + gN} \; \varphi(x) \; + \;
\sqrt{g} \; \theta(x) \; .
\end{equation}
Due to the chosen repr\"asentation
of $\gamma_5$ the chiral densities are given by
\begin{equation}
\overline{\psi}^{(b)} P_+ \psi^{(b)}  =
\overline{\psi}^{(b)}_1 \psi^{(b)}_1 \; \; \; , \; \; \;
\overline{\psi}^{(b)} P_- \psi^{(b)}  =
\overline{\psi}^{(b)}_2 \psi^{(b)}_2 \; .
\end{equation}
The propagator (5.7) implies that $\overline{\psi}^{(b)} P_+ \psi^{(b)}$
and $\overline{\psi}^{(b)} P_- \psi^{(b)}$ have to come in pairs for all
flavors $b$ in order to allow complete contractions of the fermions
which is necessary for nonvanishing results.
Thus $C_1(\tau)$ does not vanish only for
\begin{equation}
n_b + n^\prime_b = m_b + m^\prime_b \; \; \; , \; \; b = 1,...N \; .
\end{equation}
After some reordering $C_1(\tau)$ reads
\[ s \; \Big\langle
\prod_{b=1}^N
\prod_{i=1}^{n_b} \psi^{(b)}_1(x_i^{(b)}\!+ \hat{\tau})
\prod_{i=1}^{m^\prime_b} \overline{\psi}^{(b)}_2({y^\prime}^{(b)}_i)
\prod_{i=1}^{n^\prime_b} \psi^{(b)}_1({x^\prime}_i^{(b)})
\prod_{i=1}^{m_b} \overline{\psi}^{(b)}_2(y_i^{(b)}\!+ \hat{\tau}) \]
\begin{equation} \times \;
\prod_{i=1}^{m^\prime_b} \psi^{(b)}_2({y^\prime}_i^{(b)})
\prod_{i=1}^{n_b} \overline{\psi}^{(b)}_1(x_i^{(b)}\!+ \hat{\tau})
\prod_{i=1}^{m_b} \psi^{(b)}_2(y_i^{(b)}\!+ \hat{\tau})
\prod_{i=1}^{n^\prime_b} \overline{\psi}^{(b)}_1({x^\prime}^{(b)}_i)
\Big \rangle_0 \; .
\end{equation}
$s$ denotes an overall sign depending on $n_b,m_b,n^\prime_b,m^\prime_b$
which is not relevant for the following.

One can make use of the simple exponential dependence of the propagator
on the external fields to factorize $C_1(\tau)$. Whenever
$\overline{\psi}^{(b)}_1(x) \psi^{(b)}_1(x)$
contracts with
$\overline{\psi}^{(b)}_2(y) \psi^{(b)}_2(y)$ this amounts to a factor
$\exp\big( -2 (\chi,\delta_n(x) - \delta_n(y) \big)$ as can be seen from (5.7).
Thus the factorization
\begin{equation}
C_1(\tau) \; = \; I(\tau) \; C_1(\tau)_{free} \; \; \; ,
\end{equation}
holds.
$C_1(\tau)_{free}$ simply denotes the replacement
$\langle ..\rangle_0 \rightarrow \langle .. \rangle_{free}$
where the latter means expectation value with respect to free,
massless fermions.

The factor $I(\tau)$ is the integral over gauge and auxiliary fields
\[
I(\tau) = \int d\mu_{\tilde{Q}}[\varphi] d\mu_{\tilde{C}}[\theta] \;
\]
\[
\times \; \exp \Bigg(- 2\sum_{b=1}^N \Big[
\sum_{i=1}^{m_b} (\chi,\delta_n(x^{(b)}_i\!+ \hat{\tau}) +
\sum_{i=1}^{m^\prime_b} (\chi,\delta_n({x^\prime}^{(b)}_i) \Big] \Bigg)
\]
\begin{equation}
\times \; \exp \Bigg(+ 2\sum_{b=1}^N \Big[
\sum_{i=1}^{n_b} (\chi,\delta_n(y^{(b)}_i\!+ \hat{\tau}) +
\sum_{i=1}^{n^\prime_b} (\chi,\delta_n({y^\prime}^{(b)}_i) \Big] \Bigg) \; .
\end{equation}
To keep my notation simple I introduce two sets
of space-time variables $\{w_j\}$ and $\{z_j\}$ which are
given by
\[
\{ w_j \}_{j=1}^M :=
\{ x_l^{(b)} +\hat{\tau} , {x^\prime}^{(b)}_k \; | \; \;
l = 1,...n_b ; \; k = 1,...n^\prime_b ; \; b = 1,...N \} \; ,
\]
\begin{equation}
\{ z_j \}_{j=1}^M :=
\{ y_l^{(b)} +\hat{\tau} , {y^\prime}^{(b)}_k \; | \; \;
l = 1,...m_b ; \; k = 1,...m^\prime_b ; \; b = 1,...N \} \; .
\end{equation}
Due to (5.10) both sets contain the same number $M$
\begin{equation}
M \; := \; \sum_{b=1}^N (n_b + n_b^\prime) \; = \;
\sum_{b=1}^N (m_b + m_b^\prime) \; ,
\end{equation}
of elements.
Using this notation and (5.13)
\[
I(\tau) = \int d \mu_{\tilde{Q}}[\varphi]
e^{-2 \frac{e \pi}{\pi + g N}\sum_{j=1}^M
\big( \varphi,\delta_n(w_j) - \delta_n(z_j) \big) } \times
\int d \mu_{\tilde{C}}[\theta]
e^{-2 \sqrt{g}\sum_{j=1}^M
\big( \theta,\delta_n(w_j) - \delta_n(z_j) \big) } \]
\begin{equation}
= \; \; \exp \left(
2 \sum_{i,j=1}^{M}
\Big( \delta_n(w_i) - \delta_n(z_i)\; , \; K \;
[ \delta_n(w_j) - \delta_n(z_j) ] \; \Big) \right) \; ,
\end{equation}
where $K$ denotes
\begin{equation}
K := \frac{ e^2 \pi^2}{( \pi + g N )^2} \; \tilde{Q} \; +
\; g \; \tilde{C} \; .
\end{equation}
The functional integral was solved using (A.22) from Appendix A.2.
Inserting the choice (4.57) for the
$\delta$-sequence, $I(\tau)$ becomes
\[
I(\tau) = \exp \left(
\int \frac{d^2p}{(2\pi)^2} \; e^{-2\frac{|p|}{n}} \; \hat{K}(p) \;
2 \sum_{i,j=1}^{M} \Big( e^{-ipw_i} - e^{-ipz_i} \Big)
\Big( e^{+ipw_j} - e^{+ipz_j} \Big) \right)  \]
\begin{equation}
= \; \exp \left( \sum_{i,j=1}^M V(w_i -  z_j)
- \frac{1}{2}\sum_{i \neq j}^M V(w_i - w_j)
- \frac{1}{2}\sum_{i \neq j}^M V(z_i - z_j)  \right) \; ,
\end{equation}
where the potential $V(x)$ is defined as (compare (4.54), (4.55)
for the covariances $\tilde{Q}$ and $\tilde{C}$)
\[
V(x) := 2 \int \frac{d^2p}{(2\pi)^2} \; e^{-2\frac{|p|}{n}} \; \hat{K}(p)
\Big( 2 - 2 \cos(px) \Big) \]
\[
= \; 4 \int \frac{d^2p}{(2\pi)^2} \; e^{-2\frac{|p|}{n}} \;
\frac{e^2 \pi^2}{( \pi + g N )^2} \; \frac{1}{p^2 + \frac{e^2 N}{\pi + gN}}
\; \frac{1}{p^2} \;
\Big( 1 - \cos(px) \Big) \]
\begin{equation}
+ \; 4 \int \frac{d^2p}{(2\pi)^2} \; e^{-2\frac{|p|}{n}} \;
g \; \frac{\pi}{\pi + gN}
\frac{1}{p^2} \;
\Big( 1 - \cos(px) \Big) \; .
\end{equation}
In both integrals the infrared problem is cured by the
$\Big( 1 - \cos(px) \Big)$ term. The first one even has no
ultraviolet problem, and it can be solved after the limit
$n \rightarrow \infty$ was taken. The other one has to be evaluated
for finite $n$. Using Appendix B.2 one obtains
\[
V(x) = \frac{2}{\pi}
\frac{e^2 \pi^2}{( \pi\!+\!g N )^2}\frac{\pi\!+\!gN}{e^2 N}
\left( \ln|x| + \mbox{K}_0\left(\sqrt{\frac{e^2 N}{\pi\!+\!gN}} |x|\right)
+ \ln\left(\frac{1}{2} \sqrt{\frac{e^2 N}{\pi\!+\!gN}}\right)+\gamma\right)\]
\[
 + \; \frac{2}{\pi} \; g \; \frac{\pi}{\pi + gN}\;
\Bigg( \ln|x| + \ln\Big(\frac{n}{4}\Big) + O\Big(\frac{1}{n}\Big) \Bigg)
\]
\begin{equation}
= \; \frac{1}{N}\ln|x| + \tilde{V}(x) +
\frac{2\pi g}{\pi\!+\!gN} \ln\left( \frac{n}{4} \right) +
O\left(\frac{1}{n}\right) \; ,
\end{equation}
where I defined
\begin{equation}
\tilde{V}(x) :=
\frac{2\pi}{N(\pi + gN)} \;
\left(\mbox{K}_0\left(\sqrt{\frac{e^2 N}{\pi\!+\!gN}} |x|\right)
+ \ln\left(\frac{1}{2}
\sqrt{\frac{e^2 N}{\pi\!+\!gN}}\right)+\gamma\right) \; .
\end{equation}
Thus one ends up with
\[ I(\tau) = \Big(\frac{n}{4}\Big)^{\frac{\pi g}{\pi + gN} \; 2 M} \;
e^{O(\frac{1}{n})} \]
\[ \times
\exp \left( \sum_{i,j=1}^M \tilde{V}(w_i -  z_j)
- \frac{1}{2}\sum_{i \neq j}^M \tilde{V}(w_i - w_j)
- \frac{1}{2}\sum_{i \neq j}^M \tilde{V}(z_i - z_j)  \right) \]
\begin{equation}
\times \prod_{i,j=1}^M \Big(w_i -  z_j\Big)^2 \;
\prod_{i<j}^M \Big(w_i -  w_j\Big)^{-2} \Big(z_i -  z_j\Big)^{-2} \; .
\end{equation}
As announced, the $n$ dependent factor can be absorbed in a
wave function renormalization constant $Z$ for the
chiral densities $\overline{\psi}^{(a)}P_\pm \psi^{(a)}$
\begin{equation}
Z \; := \left(\frac{4}{n}\right)^{\frac{\pi g}{\pi + gN}} \; .
\end{equation}
(5.22) allows a discussion of the large $\tau$ behaviour of $I(\tau)$.
$\tilde{V}(x)$ depends on $x$ only via the modified Bessel function
$\mbox{K}_0$. Since $\mbox{K}_0$ approaches zero exponentially,
$\exp\big(\tilde{V}(x)\big)$ goes to a constant for large $\tau$, and the
only remaining $\tau$ dependence for large $\tau$
of $I(\tau)$ must come from the
rational function of the space-time arguments. Inserting the
sets $\{w_j\}, \{z_j\}$ (see (5.14)) and extracting the leading power in
$\tau$ gives
\begin{equation}
I(\tau) \; = \;  F(\{w_j\}, \{z_j\}) \times \Big( \tau^2
\Big)^{ -\frac{1}{N} \sum_{b,b^\prime=1}^N
(n_b-m_b)(n^\prime_{b^\prime} - m^\prime_{b^\prime})} \;
\left( 1 + O\Big(\frac{1}{\tau}\Big) \right) \; ,
\end{equation}
where $F(\{w_j\}, \{z_j\})$ is some function depending
on the space time arguments within the clusters.

$C_1(\tau)_{free}$ factorizes with respect to the
flavors
\[
C_1(\tau)_{free} \; = \; s \; \prod_{b=1}^N \Big\langle
\prod_{i=1}^{n_b} \psi^{(b)}_1(x_i^{(b)}\!+ \hat{\tau})
\prod_{i=1}^{m^\prime_b} \overline{\psi}^{(b)}_2({y^\prime}^{(b)}_i)
\prod_{i=1}^{n^\prime_b} \psi^{(b)}_1({x^\prime}_i^{(b)})
\prod_{i=1}^{m_b} \overline{\psi}^{(b)}_2(y_i^{(b)}\!+ \hat{\tau}) \]
\begin{equation}
\times \;
\prod_{i=1}^{m^\prime_b} \psi^{(b)}_2({y^\prime}_i^{(b)})
\prod_{i=1}^{n_b} \overline{\psi}^{(b)}_1(x_i^{(b)}\!+ \hat{\tau})
\prod_{i=1}^{m_b} \psi^{(b)}_2(y_i^{(b)}\!+ \hat{\tau})
\prod_{i=1}^{n^\prime_b} \overline{\psi}^{(b)}_1({x^\prime}^{(b)}_i)
\Big \rangle_{free} \; .
\end{equation}
Using the explicit form of the free propagator $G^o$ (Appendix A.1)
it can be expressed
in terms of determinants. The general structure of a factor with
fixed flavor (flavor indices suppressed) is
\[
\Big \langle
\prod_{i=1}^n \psi_1(w_i)
\overline{\psi}_2(z_i)
\prod_{i=1}^n \psi_2(z_i)
\overline{\psi}_1(w_i) \Big\rangle_{free} \]
\[
= \; \sum_{\pi(n)} \mbox{sign}(\pi)
G^o_{12}(w_1 - z_{\pi(1)}) G^o_{12}(w_2 - z_{\pi(2)}) ...
G^o_{12}(w_n - z_{\pi(n)}) \]
\[ \times
\sum_{\pi(n)} \mbox{sign}(\pi)
G^o_{21}(z_1 - w_{\pi(1)}) G^o_{21}(z_2 - w_{\pi(2)}) ...
G^o_{21}(z_n - w_{\pi(n)}) \]
\[
= \; (-1)^n \; \left( \frac{1}{2\pi} \right)^{2n} \left|
\sum_{\pi(n)} \mbox{sign}(\pi)
\frac{1}{\tilde{w}_1 - \tilde{z}_{\pi(1)}} ...
\frac{1}{\tilde{w}_n - \tilde{z}_{\pi(n)}} \right|^2 \]
\begin{equation}
= \; \left( \frac{1}{2\pi} \right)^{2n} \left|
{ \; \atop { \mbox{det} \atop {\scriptstyle (i,j)} } }\Big(
\frac{1}{\tilde{w}_i - \tilde{z}_j} \Big) \right|^2 \; .
\end{equation}
Determinants of this type can be rewritten using Cauchy's identity
(see e.g. \cite{deutsch})
\begin{equation}
{ \; \atop { \mbox{det} \atop {\scriptstyle (i,j)} } }\Big(
\frac{1}{\tilde{w}_i - \tilde{z}_j} \Big) =
(-1)^{\frac{n(n-1)}{2}}
\frac{\prod_{1 \leq i < j \leq n} (\tilde{w}_i - \tilde{w}_j )
( \tilde{z}_i - \tilde{z}_j ) }
{\prod_{i,j = 1}^n ( \tilde{w}_i - \tilde{z}_j ) }\; .
\end{equation}
Hence one obtains
\[
C_1(\tau)_{free} \; = \; \]
\begin{equation}s^\prime \;
\left( \frac{1}{2\pi} \right)^{2M}
\prod_{b=1}^N
\prod_{i,j = 1}^{n_b+n_b^\prime}
\Big( w_i^{(b)} - z_j^{(b)} \Big)^{-2}
\prod_{1 \leq i < j \leq n_b + n_b^\prime}
\Big( w_i^{(b)} - w_j^{(b)} \Big)^2
\Big( z_i^{(b)} - z_j^{(b)} \Big)^2 \; ,
\end{equation}
where the sets $\{w_j^{(b)}\},\{z_j^{(b)}\}$ for fixed flavor $b$ are
given by
\[
\{ w_j^{(b)} \}_{j=1}^{n_b + n_b^\prime} :=
\{ x_l^{(b)} +\hat{\tau} , {x^\prime}^{(b)}_k \; | \; \;
l = 1,...n_b ; \; k = 1,...n^\prime_b \} \; ,
\]
\begin{equation}
\{ z_j^{(b)} \}_{j=1}^{m_b + m_b^\prime} :=
\{ y_l^{(b)} +\hat{\tau} , {y^\prime}^{(b)}_k \; | \; \;
l = 1,...m_b ; \; k = 1,...m^\prime_b \} \; .
\end{equation}
Note that $m_b + m_b^\prime\; = \; n_b + n_b^\prime$ due to (5.10).
Inserting the sets (5.29), one can extract the large $\tau$ behaviour
\begin{equation}
C_1(\tau)_{free} \; = \; s^\prime \;
\left( \frac{1}{2\pi} \right)^{2M}
\Big(\tau^2\Big)^{\sum_{b=1}^N (n_b - m_b)(n^\prime_b - m^\prime_b) }
\; \left( 1 + O\Big(\frac{1}{\tau}\Big) \right) \; .
\end{equation}
Thus (use (5.12), (5.24) and (5.30))
\begin{equation}
C_1(\tau) \; \; \propto \; \; \Big(\frac{1}{\tau^2}\Big)^E \;
\left( 1 + O\Big(\frac{1}{\tau}\Big) \right) \; ,
\end{equation}
where the exponent $E$ is given by
\begin{equation}
E := \frac{1}{N} \sum_{b,b^\prime=1}^N
(n_b-m_b)(n^\prime_{b^\prime} - m^\prime_{b^\prime})
- \sum_{b=1}^N (n_b - m_b)(n^\prime_b - m^\prime_b) \; .
\end{equation}
Using (5.10) one can rewrite $E$
\[
E \; = \; \sum_{b=1}^N (n_b\!-\!m_b)(n_b\!-\!m_b) -
\frac{1}{N} \sum_{b,b^\prime=1}^N
(n_b\!-\!m_b)(n_{b^\prime}\!-\!m_{b^\prime}) = \]
\begin{equation}
\frac{1}{N}  \sum_{b,b^\prime=1}^N
(n_b\!-\!m_b)R_{bb^\prime}(n_{b^\prime}\!-\!m_{b^\prime}) \; .
\end{equation}
The matrix $R$
\begin{equation}
R_{bb^\prime} \; =: \; \delta_{bb^\prime} \; N \; - \; 1 \; ,
\end{equation}
is discussed in Appendix B.3.
There the corresponding eigenvalue problem is solved.
One finds one eigenvalue 0, and $N-1$ eigenvalues $N$. The eigenvector
$x^0$ to the eigenvalue 0 is given by $x^0 = 1/\sqrt{N} (1,1,...1)^T$.
Hence the quadratic form $x^T R x$ is positive semidefinite, and
vanishes only if $x$ is a multiple of $x^0$. This implies that the
exponent $E$ is nonnegative and vanishes only for
\begin{equation}
n_b - m_b = m^\prime_b - n^\prime_b = n  \;\; \; \; \; \;
\forall \; b \; = \; 1,2, \; ..... \; N \; , \; \;  \;\;  n \in
\mbox{Z\hspace{-1.35mm}Z} \; .
\end{equation}
All those possibilities lead to a nonvanishing limit
$C_1(\infty) := \lim_{\tau \rightarrow \infty} C_1(\tau)$.
In some of the cases $C_1(\infty)$ will be cancelled by $C_2$
which is given by
\[ C_2 =
\Big\langle \prod_{b=1}^N
\prod_{i=1}^{n_b} \overline{\psi}^{(b)}(x^{(b)}_i )
P_+ \psi^{(b)}(x^{(b)}_i )
\prod_{i=1}^{m_b} \overline{\psi}^{(b)}(y^{(b)}_i )
P_- \psi^{(b)}(y^{(b)}_i ) \Big\rangle_0 \]
\begin{equation}
\times \;
\Big\langle \prod_{b=1}^N
\prod_{i=1}^{n^\prime_b}
\overline{\psi}^{(b)}({x^\prime}^{(b)}_i)
P_+ \psi^{(b)}({x^\prime}^{(b)}_i)
\prod_{i=1}^{m^\prime_a}
\overline{\psi}^{(b)}({y^\prime}^{(b)}_i)
P_- \psi^{(b)}({y^\prime}^{(b)}_i) \Big\rangle_0 \; ,
\end{equation}
Using (5.7), (5.9) it is possible to find a necessary condition
similar to (5.10) that has to be fulfilled in order to be able to
contract the fermions entirely. $C_2$ does not vanish only for
\begin{equation}
n_b = m_b \; \; \; \; \mbox{and} \; \; \; \;
n^\prime_b = m^\prime_b \; \; , \; \; \; \; b = 1,...N \; .
\end{equation}
In these cases $C_2$ then cancels $C_1(\infty)$ and the
operators cluster. Thus violation of clustering of $C_{\tau}$
is expressed in the condition
\begin{equation}
n_b - m_b = m^\prime_b - n^\prime_b = n  \;\; , \;\;
\forall \; b = 1,2,..N \;\; , \;\;  n \in
\mbox{Z\hspace{-1.35mm}Z} \setminus \{0\} \; .
\end{equation}
How does this picture change when one allows vector currents as well?
First I notice that vector currents do not contribute to the
$\varphi$ and $\theta$ integrals. Consider e.g. the term
$\overline{\psi}^{(1)}_1(x) \psi^{(2)}_2(x)$
showing up in a current that mixes flavors 1 and 2
\footnote{Together with $\overline{\psi}^{(1)}_2(x) \psi^{(2)}_1(x)$
this is generic due to the off diagonal choice of
the $\gamma$-algebra (see Appendix B.1).}.
The $\overline{\psi}^{(1)}_1(x)$
enters a propagator $G_{21}(\cdot,x)$, $\psi^{(2)}_2 (x)$ a propagator
$G_{21}(x,\cdot)$. Inspecting (5.7) immediately shows the cancellation
of the $\chi$ dependence in the product of the
propagators. Hence each vector current can only
contribute a $1/\tau$ from the free propagator.

Nonvanishing results remain only if the flavors that occur in the
vector currents can contract entirely. Thus one has to consider
only `closed cycles' like e.g.
\begin{equation}
\overline{\psi}^{(1)}(x) \gamma_\mu \psi^{(2)}(x) \;\;\;
\overline{\psi}^{(2)}(y) \gamma_\nu \psi^{(3)}(y) \;\;\;
\overline{\psi}^{(3)}(z) \gamma_\omega \psi^{(1)}(z) \; .
\end{equation}
In principle there are two possibilities to distribute the space-time
arguments $x,y,z$. If they are all in one cluster they do not bring in
any $\tau$-dependence. They do not modify the clustering, only the constant
$C$. If one distributes the closed cycle over both clusters
then the situation changes. Each vector current with a partner in the
other cluster contributes a factor $1/\tau$ from the free propagator.
This implies that any combination of vector currents alone clusters.
Nevertheless a combination
of vector currents together with the ansatz (5.3) could violate clustering.
But the gauge field integral contributions from the chiral
charges
$\overline{\psi} P_\pm \psi$ can at most compensate the $1/\tau$
of these charges, nothing else
($x^T R x$ is positive semidefinite, compare (5.34) and the Appendix B.3).
Hence adding vector currents that can only contract between the clusters
can at most enforce operators to cluster, never create extra
powers of $\tau$  that lead to violation of clustering. The same is true
when inserting currents containing only a single flavor
where one has to introduce
a point splitting regulator. The gauge field transporter that
connects
$\overline{\psi}^{(b)}(x-\varepsilon)$ and $\psi^{(b)}(x+\varepsilon)$
is only a modification within a cluster that does not change the
clustering behaviour.

%
%
\section{Symmetry properties of operators that violate clustering}

In this section the symmetry properties of the nonclustering
operators will be analyzed.
To make the notation more convenient I introduce
\begin{equation}
{\cal O}_\pm ( \{x\} ) :=
\prod_{b=1}^{N}\overline{\psi}^{(b)}(x^{(b)})P_\pm \psi^{(b)}(x^{(b)}) \; .
\end{equation}
It will turn out that the lack of clustering of ${\cal O}_\pm$
is related to the fact that they
are singlets\footnote{
The condition that operators that violate clustering should be singlets under
$\mbox{SU(N)}_L\times \mbox{SU(N)}_R$ was already discussed in
\cite{belvedere}.}
under the conserved symmetry group
$\mbox{U(1)}_V \times \mbox{SU(N)}_L\times \mbox{SU(N)}_R$, but
transform nontrivially under the explicitly broken $\mbox{U(1)}_A$.
To obtain operators that transform under a single irreducible representation
of the symmetry group namely the trivial representation,
I antisymmetrize ${\cal O}_\pm$ with respect to the
flavor indices and call the result ${\cal O}_\pm^a$.
\[
{\cal O}_\pm^a( \{x\} ) := (-1)^{\frac{N(N-1)}{2}}\left[
\frac{1}{N!} \sum_\pi \mbox{sign}(\pi) \prod_{b=1}^{N}
\overline{\psi}^{(\pi(b))}_{\stackrel{{\scriptstyle 1}}{{\scriptstyle 2}}}
(x^{(b)})
\right] \]
\begin{equation}
\left[ \frac{1}{N!} \sum_{\pi^\prime} \mbox{sign}(\pi^\prime)
\prod_{b^\prime=1}^{N}
\psi^{(\pi^\prime(b^\prime))}_{\stackrel{{\scriptstyle 1}}{{\scriptstyle 2}}}
(x^{(b^\prime)}) \right] \; .
\end{equation}
The global sign comes from shifting all $\overline{\psi}$ to the left.
Using
\begin{equation}
\prod_{b=1}^N \psi^{(b)}_\alpha(x^{(\pi(b))}) =
\mbox{sign}(\pi^{-1}) \prod_{b=1}^N
\psi^{(\pi^{-1}(b))}_{\alpha}(x^{(b)})
\; ,
\end{equation}
one can express ${\cal O}^a_\pm$ in terms of symmetrized
space-time arguments as well,
\begin{equation}
{\cal O}_\pm^a( \{x\} ) := \frac{1}{(N!)^2}\sum_{\pi,\pi^\prime}
\prod_{b=1}^{N}\overline{\psi}^{(b)}(x^{(\pi(b))})
P_\pm \psi^{(b)}(x^{(\pi^\prime(b))}) \; .
\end{equation}
Since the constant $C = \lim_{\tau \rightarrow \infty} C(\tau)$
is invariant under
the permutation of arguments within a cluster (compare (5.22), (5.28)), the
latter expression shows explicitely that the constant $C$
is the same for $\cal O_\pm$ and ${\cal O}^a_\pm$.
{}From (5.41) one easily reads off the invariance of ${\cal O}^a_\pm$ under
\begin{equation}
\mbox{U(1)}_V \times \mbox{SU(N)}_L\times \mbox{SU(N)}_R \; ,
\end{equation}
and nontrivial transformation properties under $\mbox{U(1)}_A$.
I end up with the following picture of the clustering behaviour:
The prototype of a correlation function that violates clustering is given by
\[
C(\tau) =
\Big \langle
\prod_{i=1}^n {\cal O}^a_+ ( \{ x_i + \hat{\tau} \} )
\prod_{i=1}^m {\cal O}^a_- ( \{ y_i + \hat{\tau} \} )
\prod_{i=1}^{n^\prime} {\cal O}^a_+ ( \{ x^\prime_i \} )
\prod_{i=1}^{m^\prime} {\cal O}^a_- ( \{ y^\prime_i \} ) \Big\rangle \]
\begin{equation}
- \; \Big\langle
\prod_{i=1}^n {\cal O}^a_+ ( \{ x_i \} )
\prod_{i=1}^m {\cal O}^a_- ( \{ y_i \} ) \Big\rangle \;\; \Big\langle
\prod_{i=1}^{n^\prime} {\cal O}^a_+ ( \{x^\prime_i \} )
\prod_{i=1}^{m^\prime} {\cal O}^a_- ( \{y^\prime_i \} ) \Big\rangle \; ,
\end{equation}
with the condition
\begin{equation}
n-m = -n^\prime + m^\prime \in \mbox{Z\hspace{-.8ex}Z}\backslash
\{0\} \; .
\end{equation}
Insertion of closed cycles of vector currents into a
cluster does not change the
clustering behaviour. If one inserts vector currents that can contract
only to a partner in the other cluster,
operators that violated clustering before now become operators
obeying the
cluster property. Of course it is possible to generalize
the operators ${\cal O}_\pm$
further by e.g. splitting the arguments and connect them with a
parallel transporter. Since this is a modification within a
cluster, the extra terms in the functional integral will not
depend on $\tau$ and only modify the constant.

For completeness I determine the constant $C = \lim_{\tau \rightarrow \infty}
C(\tau)$ with $C(\tau)$ defined in (5.45)
\begin{equation}
C =  {\cal F} (\{x\},\{y\}) \; \;
{\cal F} ( \{x^\prime\},\{y^\prime\}) \; .
\end{equation}
In the limit $\tau \rightarrow \infty$
the dependence on the space time arguments factorizes into two parts
that depend on the arguments in the two clusters. This function
$\cal F$ is unambiguous for an operator, if only the other operator has the
right quantum numbers to form a pair that violates clustering,
otherwise it vanishes. It is
given by
\[
{\cal F}(\{x\},\{y\}) = \prod_{b=1}^N
\frac{\prod_{1 \leq i < j \leq n} \left(x_i^{(b)} - x_j^{(b)}\right)^2
\prod_{1 \leq i < j \leq m} \left(y_i^{(b)} - y_j^{(b)}\right)^2}
{\prod_{i=1}^n \prod_{j=1}^m \left(x_i^{(b)} - y_j^{(b)} \right)}
\]
\[
\times \;
e^{\sum \tilde{\tilde{V}} \left(x_i^{(b)} - y_j^{(b^\prime)} \right)
-\frac{1}{2}\sum (1-\delta_{b b^\prime} \delta_{i j})
\left( \tilde{\tilde{V}} \left(x_i^{(b)} - x_j^{(b^\prime)} \right)
+ \tilde{\tilde{V}} \left(y_i^{(b)} - y_j^{(b^\prime)} \right)  \right)} \]
\begin{equation}
\times\;
\left( \frac{1}{2\pi} \right)^{N(n+m)}
\left[\frac{e^2 N}{4 ( \pi +gN)} e^{2\gamma} \right]^{
\frac{\pi}{\pi+ gN} \frac{N(n-m)^2}{2}} \; \;
\; .
\end{equation}
$\sum$ denotes summation over all possible values of the indices
$b,b^\prime,i,j$ within the clusters.
$\tilde{\tilde{V}}$ is defined as  (compare (5.20))
\begin{equation}
\tilde{\tilde{V}}(x) := \frac{1}{N} \ln ( x^2 ) + \tilde{V}(x) \; .
\end{equation}
%
%
\section{Decomposition into clustering states}
In the last section it was shown that operators that violate
clustering are singlets under
$\mbox{U(1)}_V \times \mbox{SU(N)}_L\times \mbox{SU(N)}_R$, but
transform nontrivially under $\mbox{U(1)}_A$. Thus the decomposition
of the vacuum state in terms of clustering
$\theta$-vacua\footnote{The term $\theta$-vacua will be
used here as well, but it has to be remarked that the
construction here is completely different from
the concept discussed in Section 2.1.} can make
use of the charge that is associated to the axial
transformation $\mbox{U(1)}_A$
\begin{equation}
\psi^{(b)} \; \longrightarrow \; e^{i \varepsilon \gamma_5}
\psi^{(b)} \; .
\end{equation}
An arbitrary product $\cal B$ of $\overline{\psi}_\alpha^{(b)}(x),
\psi^{(b^\prime)}_{\alpha^\prime}(x^\prime)$ transforms under
$\mbox{U(1)}_A$ as
\begin{equation}
{\cal B}({\scriptstyle\{x\}}) \; \longrightarrow
e^{ i m \varepsilon} {\cal B}({\scriptstyle\{x\}}) \; , \; m \in
\mbox{Z\hspace{-.8ex}Z} \; .
\end{equation}
(More generally one can consider observables that are sums of operators
with definite transformation properties under $\mbox{U(1)}_A$).
Define the corresponding charge $Q_5({\scriptstyle {\cal B}})$ as $m$.
Obviously ${\cal O}^a_\pm$ (see (5.41)) have the charge $\pm 2N$.
I will now decompose the expectation functional $\langle .. \rangle_0$
into states $\langle .. \rangle^\theta_0$ labeled by a parameter
$\theta \in [-\pi,\pi]$ defined as follows
\begin{equation}
\langle {\cal B}({\scriptstyle\{x\}}) \rangle^\theta_0 :=
e^{i \theta \frac{Q_5({\scriptstyle{\cal B}})}{2N}}
\lim_{\tau \rightarrow \infty}
\langle {\cal U}_\tau ({\scriptstyle {\cal B}}) \;
{\cal B}({\scriptstyle\{x\}}) \rangle_0 \; .
\end{equation}
The set of `test operators' ${\cal U}_\tau ({\scriptstyle {\cal B}})$
is defined by
\begin{equation}
{\cal U}_\tau ({\scriptstyle {\cal B}}) :=
\left\{ \begin{array}{l}
{\cal N}^{(n)} (\{y\})  \prod_{i=1}^n {\cal O}^a_\mp (\{y+\hat{\tau}\})
\; \mbox{for} \; \; Q_5({\scriptstyle{\cal B}}) = \pm 2nN , \; n \geq 1 \; , \\
\; \\
1 \; \; \mbox{otherwise} \; \; . \end{array} \right.
\end{equation}
Up to the requirement of being nondegenerate,
the arguments $\{y\}$ are arbitrary.
The normalizing factor ${\cal N}^{(n)} (\{y\})$ is defined such that
\begin{equation}
\lim_{\tau^\prime \rightarrow \infty}
\langle {\cal U}_{\tau^\prime} ({\scriptstyle{\cal B}^\dagger})
\; {\cal U}_\tau ({\scriptstyle{\cal B}}) \rangle_0 = 1 \; .
\end{equation}
It can be read off from (5.48)
\[
{\cal N}^{(n)} (\{y\}) =
\left( \frac{1}{2\pi} \right)^{-Nn} \!
\left[\frac{e^2 N}{4 ( \pi +gN)} e^{2\gamma} \right]^{
-\frac{\pi}{\pi + gN} \frac{Nn^2}{2}} \]
\begin{equation}
\times \;
\prod_{b=1}^N \prod_{1 \leq i < j \leq n}
\left(y_i^{(b)}\!-\!y_j^{(b)}\right)^{-2}
e^{\frac{1}{2}\sum_{b,b^\prime,i,j} (1-\delta_{bb^\prime} \delta_{ij})
\tilde{\tilde{V}}\left(y_i^{(b)} - y_j^{(b^\prime)} \right) } \; .
\end{equation}
The expectation functionals (states) $\langle .. \rangle^\theta_0$ have
the following properties:
\vskip5mm
\noindent
{\bf Theorem 5.1 :}\\
{\bf i) } The state $\langle .. \rangle_0$ constructed initially
is recovered by averaging over $\theta$
\begin{equation}
\langle \; .. \; \rangle = \frac{1}{2\pi}\int_{-\pi}^{\pi}
\langle \; .. \; \rangle^\theta_0 \; d\theta \;.
\end{equation}
{\bf ii) } The cluster decomposition property holds.\\
 \\
{\bf Proof:} \\
{\bf i):}
The averaging procedure leaves a nonvanishing result only for
operators $\cal B$ with vanishing charge $Q_5({\scriptstyle {\cal B}}) = 0$
\begin{equation}
\frac{1}{2\pi}\int_{-\pi}^\pi
\langle {\cal B} \rangle^\theta_0 d\theta =
\frac{1}{2\pi}\int_{-\pi}^\pi \;
e ^{i \theta \frac{Q_5({\scriptstyle{\cal B}})}{2N} } \;
d\theta \;  \lim_{\tau \rightarrow \infty}
\langle {\cal U}_\tau({\scriptstyle {\cal B}}) \;
{\cal B} \rangle_0 =
\delta_{Q_5({{\scriptscriptstyle \cal B}}),0} \; \langle {\cal B} \rangle
\; = \; \langle {\cal B} \rangle_0 \; .
\end{equation}
In the second step I used ${\cal U}_\tau({\scriptstyle {\cal B}}) = 1$
if $Q_5({\scriptstyle {\cal B}}) = 0$. To complete the proof,
one has to check that in the state $\langle .. \rangle_0$
constructed initially the vacuum expectation values of operators $\cal B$ with
$Q_5({\scriptstyle {\cal B}}) \neq 0$ vanish (cf. also \cite{patr}).
In the chosen representation of the
$\gamma$-matrices this can be seen immediately.
$Q_5({{\scriptstyle\cal B}}) \neq 0$
means that the number of $\overline{\psi}_1,\psi_1$ is not equal to the
number $\overline{\psi}_2,\psi_2$. Since $G_{\alpha \alpha} = 0$, for
each $\psi_1$ there has to be a $\overline{\psi}_2$ and for each
$\psi_2$ a $\overline{\psi}_1$ to give a nonvanishing contribution.
But this implies that the number of fields
$\overline{\psi}_1,\psi_1$ in $\cal B$
is equal to the number of fields $\overline{\psi}_2,\psi_2$.
Hence
\begin{equation}
\langle {\cal B} \rangle_0 = 0  \;\; \mbox{for} \;\;
Q_5({\scriptstyle {\cal B}}) \neq 0 \;,
\end{equation}
and the last equality in (5.57) holds.
\vskip5mm
\noindent
{\bf ii):}
Let $\cal A$ and $\cal B$ be arbitrary operators. Define
\begin{equation}
C_\theta(\tau) := \lim_{\tau \rightarrow \infty} \Big[ \;
\langle {\cal A}(\tau) {\cal B}(0) \rangle^\theta_0 \; - \;
\langle {\cal A}(0) \rangle^\theta_0 \;
\langle {\cal B}(0) \rangle^\theta_0 \; \Big] \; .
\end{equation}
The dependence on the
space-time arguments $\{ x \}$ is not displayed explicitely,
only the dependence on the shift variable $\tau$.
Depending on the axial charge $Q_5$ of the operators
$\cal A,B,AB$, one has to insert the different alternatives for
$\langle \cdot \rangle^\theta_0$ given in (5.53).
I introduce the following convenient notation:
An operator for which the first alternative in (5.53)
holds I call a `type {\bf I}' operator, the operators where the second
alternative holds is called `type {\bf II}'. It has to be
remarked that all operators of
type {\bf II } cannot have the problem of violating the clustering condition.
Even among the type {\bf I} operators there are examples that are not
able to violate clustering (like e.g.
$( \overline{\psi}^{(b)} P_+ \psi^{(b)} )^N$, for fixed $b$,
which does not contain
a $\mbox{SU(N)}_L\times \mbox{SU(N)}_R$ singlet part). The operators with
the structure $\prod^n {\cal O}_+ \prod^m {\cal O}_- \prod J_\mu^{(l)} \;, \;
n-m \neq 0$ I call `type {\bf V}' for violating.

The following cases have to be distinguished
 \\
\begin{center}
\begin{tabular}{c|l|l|l}
case $\#$ &
$Q_5({\scriptstyle {\cal A}})$ &
$Q_5({\scriptstyle {\cal B}})$ &
$Q_5({\scriptstyle {\cal AB}})$ \\ \hline
{\bf 1} & {\bf II} & {\bf II} & {\bf II} \\
{\bf 2} & {\bf II} $\;\;(\neq 0)$ & {\bf II}
$\;\;(\neq 0)$ & {\bf I} \\
{\bf 3} & {\bf II} $\;\;(\neq 0)$ & {\bf I} & {\bf II} \\
{\bf 4} & {\bf II} $\;\;( = 0)$ & {\bf I} & {\bf I} \\
{\bf 5} & {\bf I} $\;\;\;( = q)$ & {\bf I} $\;\;\;( = -q)$ &
{\bf II} $\;\;( = 0)$ \\
{\bf 6} & {\bf I} & {\bf I} & {\bf I} \\
\end{tabular}
\end{center}
In parenthesis $(\cdot)$ I denoted facts that necessarily follow.
For example in Case 2 the requirement
$Q_5({\scriptstyle {\cal A}}) \in \mbox{{\bf II}},
Q_5({\scriptstyle {\cal B}}) \in \mbox{{\bf II}},
Q_5({\scriptstyle {\cal AB}}) \in \mbox{{\bf I}}$ implies
$Q_5({\scriptstyle {\cal A}}) \neq 0$ and
$Q_5({\scriptstyle {\cal B}}) \neq 0$. If one of these two had
charge zero,
the other operator would be of type {\bf I},
since $Q_5({\scriptstyle {\cal AB}}) =
Q_5({\scriptstyle {\cal A}}) + Q_5({\scriptstyle {\cal B}})$. \\
 \\
{\bf Case 1:}
\begin{equation}
C_\theta(\tau) = \langle {\cal A} (\tau) {\cal B} \rangle_0 -
\langle {\cal A } \rangle_0 \langle {\cal B} \rangle_0
\stackrel{\tau \rightarrow \infty}{\longrightarrow} 0 \; ,
\end{equation}
since if $\cal A$ and $\cal B$ are of type {\bf II}, they do not
form a pair that violates clustering. \\
{\bf Case 2:}
\[
C_\theta(\tau) = \lim_{\tau^\prime \rightarrow \infty}
\langle {\cal U}_{\tau^\prime} ( {\scriptstyle {\cal AB}})
{\cal A}(\tau) {\cal B} \rangle_0 \; - \; \langle {\cal A} \rangle_0
\langle {\cal B} \rangle_0
\stackrel{\tau \rightarrow \infty}{\longrightarrow} \]
\begin{equation}
\langle {\cal A} \rangle_0 \lim_{\tau^\prime \rightarrow \infty}
\langle {\cal U}_{\tau^\prime} ( {\scriptstyle {\cal AB}})
{\cal B} \rangle_0 \; - \; \langle {\cal A} \rangle_0
\langle {\cal B} \rangle_0 \; = \; 0 \; ,
\end{equation}
where I used the fact that $\langle {\cal A} \rangle_0$ factorizes, since
$\cal A$ is type {\bf II} and $\langle {\cal A} \rangle_0 = 0$ since
$Q_5({\scriptstyle {\cal A}}) \neq 0$ (compare the note in the table).
The interchange of the $\tau , \tau^\prime$ limits is justified, since
all the functions involved are continuous in these variables and bounded
(the exponent $E$ in Equation (5.31) is nonnegative). \\
{\bf Case 3:}
\begin{equation}
C_\theta(\tau) = \langle {\cal A}(\tau) {\cal B} \rangle_0 \; - \;
\langle {\cal A} \rangle_0 \lim_{\tau^\prime \rightarrow \infty}
\langle {\cal U}_{\tau^\prime} ( {\scriptstyle {\cal B}})
{\cal B} \rangle_0 \stackrel{\tau \rightarrow \infty}{\longrightarrow} 0 \; ,
\end{equation}
for the same reasons as in the last case. \\
{\bf Case 4:}
\[
C_\theta(\tau) = \lim_{\tau^\prime \rightarrow \infty}
\langle {\cal U}_{\tau^\prime} ( {\scriptstyle {\cal AB}})
{\cal A}(\tau) {\cal B} \rangle_0 \; - \; \langle {\cal A} \rangle_0
\lim_{\tau^\prime \rightarrow \infty}
\langle {\cal U}_{\tau^\prime} ( {\scriptstyle {\cal B}})
{\cal B} \rangle_0 \stackrel{\tau \rightarrow \infty}{\longrightarrow}  \]
\begin{equation}
\langle {\cal A} \rangle_0  \lim_{\tau^\prime \rightarrow \infty}
\langle {\cal U}_{\tau^\prime} ( {\scriptstyle {\cal AB}})
{\cal B} \rangle_0 \; - \; \langle {\cal A} \rangle_0
\lim_{\tau^\prime \rightarrow \infty}
\langle {\cal U}_{\tau^\prime} ( {\scriptstyle {\cal B}})
{\cal B} \rangle_0 = 0 \; .
\end{equation}
Again $\langle {\cal A} \rangle_0$ factorizes, and
${\cal U}_{\tau^\prime} ( {\scriptstyle {\cal AB}}) =
{\cal U}_{\tau^\prime} ( {\scriptstyle {\cal B}})$ since
$Q_5({\scriptstyle {\cal AB}}) = Q_5({\scriptstyle {\cal B}})$. \\
{\bf Case 5:}
\[
C_\theta(\tau) = \langle {\cal A}(\tau) {\cal B} \rangle_0 \; - \;
\lim_{\tau^\prime \rightarrow \infty} \langle {\cal U}_{\tau^\prime}
( {\scriptstyle {\cal A}}) {\cal A} \rangle_0
\lim_{\tau^{\prime \prime} \rightarrow \infty} \langle
{\cal U}_{\tau^{\prime \prime}} ( {\scriptstyle {\cal B}}) {\cal B} \rangle_0
\stackrel{\tau \rightarrow \infty}{\longrightarrow} \]
\begin{equation}
\left\{ \begin{array}{l}
{\cal F_A  F_B - F_A  F_B}  = 0 \; \; \mbox{for} \; \; {\cal A,B} \; \;
\mbox{of {\bf V}-type} \; , \\
0 \; \; \mbox{otherwise} \; . \end{array} \right.
\end{equation}
In the first case I used the factorization of the argument function
$\cal F_{AB}$ introduced in (5.48). \\
{\bf Case 6:}
\[
C_\theta(\tau) = \lim_{\tau^\prime \rightarrow \infty}
\langle {\cal U}_{\tau^\prime} ( {\scriptstyle {\cal AB}})
{\cal A}(\tau) {\cal B} \rangle_0 \; - \;
\lim_{\tau^\prime \rightarrow \infty} \langle {\cal U}_{\tau^\prime}
( {\scriptstyle {\cal A}}) {\cal A} \rangle_0
\lim_{\tau^{\prime \prime} \rightarrow \infty} \langle
{\cal U}_{\tau^{\prime \prime}} ( {\scriptstyle {\cal B}}) {\cal B} \rangle_0
\stackrel{\tau \rightarrow \infty}{\longrightarrow} \]
\begin{equation}
\left\{ \begin{array}{l}
{\cal F_A} \lim_{\tau^\prime \rightarrow \infty}
{\cal F}_{{\cal U}_{\tau^\prime} {\cal B}} -
{\cal F_A  F_B}  = 0 \; \; \mbox{for} \; \; {\cal A,B} \; \;
\mbox{of {\bf V}-type} \; , \\
0 \; \; \mbox{otherwise} \; . \end{array} \right.
\end{equation}
To justify the first case, one has to show that
$\lim_{\tau^\prime \rightarrow \infty}
{\cal F}_{{\cal U}_{\tau^\prime} {\cal B}} = {\cal F_B}$. This can be
seen immediately from (5.48). Shifting $\tau^\prime$ in
${\cal F}_{{\cal U}_{\tau^\prime} {\cal B}}$ corresponds to shifting the
arguments of the test operator
${\cal U}_{\tau^\prime}( {\scriptstyle {\cal B}})$. For
example this could be
the set (refering to (5.48) for $\cal F$)
\begin{equation}
\left\{ x_i^{(b)} \mid n^\prime < i \leq n ; b = 1,2,...N \right\} \; ,
\end{equation}
where $n^\prime < n$ depends on the charge $Q_5( {\scriptstyle {\cal B}})$.
The $\tau^\prime$ terms cancel for $\tau^\prime \rightarrow \infty$,
and what remains is ${\cal F_B}$, since the normalization of
${\cal U}_{\tau^\prime}( {\scriptstyle {\cal B}})$ cancels exactly the
extra terms.
$\Box$

This concludes the analysis of the vacuum structure of the massless model.
The decomposition performed here is equivalent to what is hoped
to have been obtained by the superposition of topological sectors
to the $\theta$-vacuum. The procedure adopted here has the advantage
of being in a better mathematical status. Finally it has to be remarked
that it is in full agreement with the picture that is conventionally
deduced from the discussions of topological sectors on a compact manifold
\cite{joos}, \cite{wipf}.

One now can compare properties of
the vacuum state in $\mbox{QED}_2$ to the expected properties of
the instanton construction of QCD.
This opens a series of lessons for the topics discussed in
Chapter 2 that I will draw from the model.
\vskip5mm
\noindent
{\bf Lesson 1 :}
\vskip3mm
\noindent
{\it The structure
of the vacuum functional that has been suggested
within the instanton picture is recovered.}
\vskip3mm
\noindent
In particular the formulas (2.41), (2.42) are confirmed. By looking at the
prescription (5.53) one sees that only operators with chirality
$2 \mbox{N} \nu \; , \; \nu \in
\mbox{Z\hspace{-1.35mm}Z}$ have non vanishing vacuum expectation values,
as has been claimed by 't Hooft for QCD.
Also the phase of (2.41) comes out correctly, if $\theta$ is
defined mod(2$\pi$). It has to be stressed that the
vacuum state in $\mbox{QED}_2$ has been constructed without making
use of topologically nontrivial configurations
but nevertheless has the properties expected from
this picture.

%
%

\chapter{Bosonization and vector currents}
In this chapter it will be shown that expectation values
$\langle \; .. \; \rangle^\theta_0$ of certain operators
(vector currents of the Cartan subalgebra, chiral densities)
can be rewritten in terms of expectation values of a bosonic
theory. Furthermore I will comment on currents that are not of the
Cartan type. I will derive two theorems on n-point functions of
vector currents in the
model with vanishing fermion masses. This will clarify the structure of
the Hilbert space of the massless model.

\section{Evaluation of a generalized generating functional}
I evaluate the following generating functional
\[
E(n_b,m_b;a^{(b)}) := \]
\begin{equation}
\Big\langle
\prod_{b=1}^N
\prod_{i=1}^{n_b}
\overline{\psi}^{(b)}({x}^{(b)}_i)
P_+ \psi^{(b)}({x}^{(b)}_i)
\prod_{j=1}^{m_b}
\overline{\psi}^{(b)}({y}^{(b)}_j)
P_- \psi^{(b)}({y}^{(b)}_j)
e^{ie\sum_{b=1}^N (a_\mu^{(b)},j_\mu^{(b)})} \Big\rangle_0^\theta \; .
\end{equation}
Obviously this is a simple modification of expectation values already
considered in the last chapter. The insertion of the chiral densities
$\overline{\psi}^{(b)} P_\pm \psi^{(b)}$ is motivated by the
expansion (3.37) of the mass term of the action. Setting $n_b = m_b =0$ for
$b = 1,2, ... \mbox{N}$ reduces $E(n_b,m_b;a^{(b)})$ to the generating
functional for vector currents in the model with vanishing fermion masses.

To work out the dependence on the
sources $a^{(b)}$ I first consider the case
\begin{equation}
n_b - m_b = 0 \; \; \; \; \; \; b = 1,2, \; ... \; , N \; ,
\end{equation}
where the $\theta-$prescription is remarkably simple, i.e. it coincides
with the naive expectation value (compare (5.53)).

\noindent
As in the last chapter the expectation value $E$ factorizes
\begin{equation}
E(n_b,n_b;a^{(b)}) := I(n_b,n_b;a^{(b)})
\times E_{free} (n_b,n_b) \; ,
\end{equation}
where the result for the
free expectation value $E_{free}$ can immediately be taken
over from (5.28)
\[
E_{free} (n_b,n_b) :=
\Big\langle \prod_{b=1}^N
\prod_{i=1}^{n_b}
\overline{\psi}^{(b)}({x}^{(b)}_i)
P_+ \psi^{(b)}({x}^{(b)}_i)
\prod_{j=1}^{n_b}
\overline{\psi}^{(b)}({y}^{(b)}_j)
P_- \psi^{(b)}({y}^{(b)}_j) \Big\rangle_{free} \; =
\]
\begin{equation}
\Big( \frac{1}{2\pi} \Big)^{2 \sum_b n_b}
\prod_{b=1}^N
\prod_{i,j = 1}^{n_b}
\Big( x_i^{(b)} - y_j^{(b)} \Big)^{-2}
\prod_{1 \leq i < j \leq n_b}
\Big( x_i^{(b)} - x_j^{(b)} \Big)^2
\Big( y_i^{(b)} - y_j^{(b)} \Big)^2 \; .
\end{equation}
The second factor $I$ from the functional integration
over the fields $A$ and $h^\prime$ reads (compare (5.13))
\[
I(n_b,n_b;a^{(b)}) \; :=
\]
\begin{equation}
\left\langle
\exp \left( - 2 \sum_{b=1}^N \sum_{j=1}^{n_b}
\Big( \chi^{(b)}, \delta_n(x_j^{(b)})-\delta_n(y_j^{(b)}) \Big) \right)
\exp \left( ie\sum_{b=1}^N (a_\mu^{(b)},j_\mu^{(b)}) \right)
\right\rangle_0 \; ,
\end{equation}
where the second exponential collects the terms from the propagators,
and $\chi^{(b)}$ is given by (see (4.52))
\begin{equation}
\chi^{(b)}(x) \; = \;
\frac{e \pi}{\pi + gN} \; \varphi(x) \; + \; \sqrt{g} \; \theta(x)
+ e\; \frac{\varepsilon_{\mu \nu}
\partial_\mu}{\triangle} a^{(b)}_\nu(x) \; .
\end{equation}
As discussed in Section 3.4 the external sources $a^{(b)}_\mu$ can be
treated by including them into the fermion determinant.
One can take over the result (4.49) from Section 4.3 to obtain
\[
I(n_b,n_b;a^{(b)}) = \int d\mu_Q[A] d\mu_C[h^\prime] \; \;
\exp \left( - 2 \sum_{b=1}^N \sum_{j=1}^{n_b}
\Big( \chi^{(b)}, \delta_n(x_j^{(b)})\!-\!\delta_n(y_j^{(b)}) \Big) \right)
\]
\begin{equation}
\times \;
\exp\left( -\frac{e^2}{\pi\!+\!gN} \Big( A, T \sum_{b=1}^N a^{(b)} \Big)
-\frac{e \sqrt{g}}{\pi} \Big( h^\prime , T \sum_{b=1}^N a^{(b)} \Big)
-\frac{e^2}{2\pi} \sum_{b=1}^N \Big( a^{(b)} , T a^{(b)} \Big) \right) \; ,
\end{equation}
where the fields $\varphi$ and $A_\mu$ and $\theta$ and $h^\prime_\mu$
respectively are related by (c.f. (4.53))
\begin{equation}
\varphi(x) \; := \;
\frac{\varepsilon_{\mu \nu} \partial_\mu}{\triangle} A_\nu(x) \; \; \; ,
\; \; \; \theta(x) \; := \;
\frac{\varepsilon_{\mu \nu} \partial_\mu}{\triangle} h^\prime_\nu(x) \; .
\end{equation}
This can be used to rewrite everything in terms of the scalar
fields $\varphi$ and $\theta$
\begin{equation}
\Big( A_\mu , T a^{(b)}_\mu \Big) =
- \Big( \varphi, \varepsilon_{\mu \nu} \partial_\mu a^{(b)}_\nu \Big)
\; \; \; , \; \; \;
\Big( h^\prime_\mu , T a^{(b)}_\mu \Big) =
- \Big( \theta, \varepsilon_{\mu \nu} \partial_\mu a^{(b)}_\nu \Big) \; .
\end{equation}
Thus one ends up with
\[
I(n_b,n_b;a^{(b)}) = \int d\mu_{\tilde{Q}}[\varphi]
d\mu_{\tilde{C}}[\theta] \;
\exp \left( - 2 \sum_{b=1}^N \sum_{j=1}^{n_b}
\Big( \chi^{(b)}, \delta_n(x_j^{(b)})\!-\!\delta_n(y_j^{(b)}) \Big) \right)
\times\]
\begin{equation}
\exp\!\left(
\frac{e^2}{\pi\!+\!gN} \Big( \varphi ,
T \sum_{b=1}^N \varepsilon_{\mu \nu} \partial_\mu a^{(b)}_\nu \Big)
+ \frac{e \sqrt{g}}{\pi} \Big( \theta ,
T \sum_{b=1}^N \varepsilon_{\mu \nu} \partial_\mu a^{(b)}_\nu \Big)
-\frac{e^2}{2\pi} \sum_{b=1}^N \Big( a^{(b)} , T a^{(b)} \Big)
\right).
\end{equation}
The Gaussian integrals over $\varphi$ and $\theta$ can be solved and
one obtains
\[
I(n_b,n_b;a^{(b)}) = \exp \left(
-\frac{e^2}{2\pi} \sum_{b=1}^N \Big( a^{(b)} , T a^{(b)} \Big) \right)
\]
\[
\times \exp \left( \frac{1}{2} \sum_{b,b^\prime=1}^N
\Big( \varepsilon_{\mu \nu} \partial_\mu a^{(b)}_\nu,
\Big[ \Big(\frac{e^2}{\pi\!+\!g N}\Big)^2 \tilde{Q} +
\frac{e^2 g}{\pi^2} \tilde{C} \Big]
\varepsilon_{\rho \sigma} \partial_\rho a^{(b^\prime)}_\sigma \Big)
\right)
\]
\[
\times \exp \left( -2 e \sum_{b=1}^N \sum_{j=1}^{n_b}
\Big( \frac{\varepsilon_{\mu \nu} \partial_\mu}{\triangle} a^{(b)}_\nu,
\delta_n(x_j^{(b)}) - \delta_n(y_j^{(b)}) \Big) \right)
\]
\[
\times \exp \left( - 2 e \sum_{b,b^\prime =1}^N \sum_{j=1}^{n_b}
\Big( \varepsilon_{\mu \nu} \partial_\mu a^{(b^\prime )}_\nu,
\Big[ \frac{e^2 \pi}{(\pi + g N)^2} \tilde{Q} + \frac{g}{\pi} \tilde{C} \Big]
\delta_n(x_j^{(b)}) - \delta_n(y_j^{(b)}) \Big) \right)
\]
\begin{equation}
\times \exp\!\left( 2 \sum_{b,b^\prime=1}^N
\sum_{j,j^\prime=1}^{n_b,n_{b^\prime}}
\Big( \delta_n(x_j^{(b)})\!-\!\delta_n(y_j^{(b)}),
\Big[ (\frac{e \pi}{\pi\!+\!g N})^2 \tilde{Q}\!+\!g \tilde{C} \Big]
\Big[ \delta_n(x_{j^\prime}^{(b^\prime)}
)\!-\!
\delta_n(y_{j^\prime}^{(b^\prime)}) \Big] \Big) \right).
\end{equation}
The first step in the analysis of this expression is to simplify the
exponents quadratic in $a^{(b)}$. Using
\begin{equation}
\Big( a^{(b)}_\mu,T_{\mu \nu} a^{(b)}_\nu \Big) =
\Big( \varepsilon_{\mu \nu} \partial_\mu a^{(b)}_\nu, \frac{-1}{\triangle}
\; \varepsilon_{\rho \sigma} \partial_\rho a^{(b)}_\sigma \Big) \; ,
\end{equation}
one can rewrite the quadratic form for $a^{(b)}_\mu$ in a quadratic form for
$\varepsilon_{\mu \nu} \partial_\mu a^{(b)}_\nu$. The corresponding
term of the exponent now reads
\begin{equation}
-\frac{e^2}{2\pi} \sum_{b,b^\prime=1}^N
\Big(\varepsilon_{\mu \nu} \partial_\mu a^{(b)}_\nu, M_{bb^\prime}
\varepsilon_{\rho \sigma} \partial_\rho a^{(b^\prime)}_\sigma \Big) \; ,
\end{equation}
where the covariance $M$ is given by
\begin{equation}
M = \frac{1}{-\triangle  + \frac{e^2N}{\pi + gN}}
\Big[ \frac{\pi}{\pi + gN} \mbox{1\hspace{-.6ex}I}
+ \frac{g}{\pi + gN} R \Big]
\; + \; \frac{e^2}{\pi + gN} \frac{-1}{\triangle}
\frac{1}{-\triangle  + \frac{e^2N}{\pi + gN}} R \; .
\end{equation}
The numerical matrix
\begin{equation}
R_{bb^\prime} = \delta_{b,b^\prime} N - 1 \; \; \; , \; \; \; \; \; \;
b,b^\prime = 1, ... N \; ,
\end{equation}
is analyzed in Appendix B.3. There I quote explicitely the orthogonal
matrix $U$ that diagonalizes $R$
\begin{equation}
U R U^T \; = \; \mbox{diag} ( 0 , N , N , \; .... \;N ) \; .
\end{equation}
This allows to express the quadratic form in terms of a covariance
$K$ diagonal in flavor
\begin{equation}
K \; := \; U \; M \; U^T \; \; \Longleftrightarrow \; \;
M \; = \; U^T \; K \; U \; .
\end{equation}
Explicitely $K$ is given by
\begin{equation}
K = \left( \begin{array}{ccccc}
\frac{\pi}{\pi\!+\!gN} \frac{1}{-\triangle\!+\!\frac{e^2N}{\pi\!+\!gn}}
& & & & \\
 & \frac{1}{-\triangle} & & & \\
 & & . & & \\
 & & & . & \\
 & & & & \frac{1}{-\triangle} \end{array} \right) \; .
\end{equation}
Obviously the covariance $K$ describes one
massive and $\mbox{N}-1$ massless particles. Finally the quadratic form reads
\begin{equation}
-\frac{e^2}{2\pi}
\Big( \varepsilon_{\mu \nu} \partial_\mu U a_\nu, K \;
\varepsilon_{\rho \sigma} \partial_\rho U a_\sigma \Big) \; ,
\end{equation}
where matrix notation in flavor space was used.

The next step is to define new sources $A_\mu^{(I)}$ that are linear
combinations of the $a_\mu^{(b)}$
\begin{equation}
A_\mu^{(I)} := \sum_{b=1}^N U_{I b} a_\mu^{(b)} \; \;
\stackrel{U^T = U^{-1}}{\Longleftarrow \!\!=\!\!=\!\!=\!\!=\!\!=
\!\!\! \Longrightarrow}  \; \;
a_\mu^{(b)} = \sum_{I=1}^N U_{I b} A_\mu^{(I)} \; .
\end{equation}
Rewriting the coupling term in $E(n_b,n_b;a^{(b)})$
allows the identification of the
currents $J_\mu^{(I)}$ that couple to the new sources $A_\mu^{(I)}$
\begin{equation}
\sum_{b=1}^N \Big( a_\mu^{(b)},j_\mu^{(b)} \Big) =
\sum_{I=1}^N \sum_{b=1}^N
\Big( U_{I b} A_\mu^{(I)}, j_\mu^{(b)} \Big) :=
\sum_{I=1}^N \Big( A_\mu^{(I)}, J_\mu^{(I)} \Big) \; ,
\end{equation}
where I defined
\begin{equation}
J_\mu^{(I)} := \sum_{b=1}^N U_{I b} j_\mu^{(b)} \; \; \;
\; \; \; I = 1,2, ... N \;.
\end{equation}
Inspecting the explicit form of the matrix $U$ quoted in
Appendix B.3 one can express the new currents also as
\begin{equation}
J_\mu^{(I)} := \sum_{b,b^\prime=1}^N \;
\overline{\psi}^{(b)} \gamma_\mu H^{(I)}_{b,b^\prime} \; \psi^{(b^\prime)}
\end{equation}
where the $\mbox{N}\times \mbox{N}$ matrices $H^{(I)}$
are generators of a {\it Cartan subalgebra} of $\mbox{U(N)}_{flavor}$
\[
H^{(1)} = \frac{1}{\sqrt{N}} \; \mbox{1\hspace{-.6ex}I} \;  , \] \[
H^{(2)} = \frac{1}{\sqrt{N-1+(N-1)^2}} \; \mbox{diag}
(1,1, .... 1, -N+1 ) \; , \] \[
H^{(3)} = \frac{1}{\sqrt{N-2+(N-2)^2}} \; \mbox{diag}
(1,1, .....1, -N+2 , 0) \; , ...... \]
\begin{equation}
H^{(N)} = \frac{1}{\sqrt{2}} \; \mbox{diag}
(1,-1,0,0, ...... 0) \; .
\end{equation}
The currents (6.23) with the generators (6.24) will be referred to as
{\it Cartan type currents}. Note that $J^{(1)}$ is the U(1)-current
already definded in (3.12).
Later I will also discuss vector currents that correspond to generators that
do not belong to this Cartan subalgebra.

Putting things together the generating functional now
reads
\[
E(n_b,n_b;A^{(I)}) =
\Big( \frac{1}{2\pi} \Big)^{2 \sum_b n_b}
\; \exp \left( -\frac{e^2}{2\pi} \sum_{I=1}^N
\Big( \varepsilon_{\mu \nu} \partial_\mu A_\nu^{(I)} , K_{I I}
\varepsilon_{\rho \sigma} \partial_\rho A^{(I)}_\sigma \Big) \right)
\]
\[
\times \exp \left( 2 e \sum_{I=1}^N \sum_{b=1}^N \sum_{j=1}^{n_b} U_{I b}
\Big( \varepsilon_{\mu \nu} \partial_\mu A^{(I)}_\nu, \frac{-1}{\triangle}
\Big[ \delta_n(x_j^{(b)}) - \delta_n(y_j^{(b)}) \Big] \Big) \right)
\]
\[
\times \exp \left(
- 2 e \sum_{I=1}^N \sum_{b,b^\prime=1}^N \sum_{j=1}^{n_b}
U_{I b^\prime}\Big( \varepsilon_{\mu \nu} \partial_\mu A^{(I)}_\nu,
\Big[ \frac{e^2 \pi}{(\pi\!+\!g N)^2} \tilde{Q} + \frac{g}{\pi}
\tilde{C} \Big] \Big[
\delta_n(x_j^{(b)})\!-\!\delta_n(y_j^{(b)}) \Big]\Big) \right)
\]
\[
\times \exp \left( 2 \sum_{b,b^\prime=1}^N
\sum_{j,j^\prime=1}^{n_b,n_{b^\prime}}
\Big( \delta_n(x_j^{(b)})\!-\!\delta_n(y_j^{(b)}),
\Big[ (\frac{e \pi}{\pi\!+\!g N})^2 \tilde{Q} + g \tilde{C} \Big]
\Big[ \delta_n(x_{j^\prime}^{(b^\prime)})\!-\!
\delta_n(y_{j^\prime}^{(b^\prime)}) \Big] \Big) \right)
\]
\begin{equation}
\times \exp \left( -2 \sum_{b=1}^N \left[ \sum_{i,j=1}^{n_b}
\ln \Big| x_i^{(b)} - y_j^{(b)} \Big| - \frac{1}{2} \sum_{i \neq j}^{n_b}
\Big[ \ln \Big| x_i^{(b)} - x_j^{(b)} \Big|  +
\ln \Big| y_i^{(b)} - y_j^{(b)} \Big| \Big] \right] \right) \; .
\end{equation}
In order to understand the representation of the chiral densities in
a bosonized theory one has to study the terms that mix the sources
$A^{(b)}_\mu$ and the space-time arguments of the densities.
Using (see Appendix B.3, Formula B.17)
\begin{equation}
\sum_{b^\prime=1}^N U_{I b^\prime} \; = \;
\frac{1}{\sqrt{N}} N \; \delta_{I 1} \; = \;
U_{1 b} N \; \delta_{I 1} \; \; \; \; \; ( b \; \; \; \mbox{arbitrary} ) \; ,
\end{equation}
one can write the part of the exponent that corresponds to the mixing term \\
(linear in $A^{(b)}_\mu$) as
\[
2e \left( \varepsilon_{\mu \nu} \partial_\mu A^{(1)}_\nu,
\bigg[ \frac{-1}{\triangle}\!-\!
N \Big[ \frac{e^2 \pi}{(\pi\!+\!gN)^2}\tilde{Q} +
\frac{g}{\pi} \tilde{C} \Big]
\bigg] \sum_{b=1}^N \sum_{j=1}^{n_b} U_{1 b}
\Big[ \delta_n(x_j^{(b)})\!-\!\delta_n(y_j^{(b)}) \Big] \right)
\]
\[
+ \; 2e \sum_{I=2}^N \left( \varepsilon_{\mu \nu} \partial_\mu A^{(I)}_\nu,
\frac{-1}{\triangle} \sum_{b=1}^N \sum_{j=1}^{n_b} U_{I b}
\Big[ \delta_n(x_j^{(b)}) - \delta_n(y_j^{(b)}) \Big] \right)
\]
\begin{equation}
= \; 2e \sum_{I=1}^N \left( \varepsilon_{\mu \nu} \partial_\mu A^{(I)}_\nu,
K_{I I} \sum_{b=1}^N \sum_{j=1}^{n_b} U_{I b}
\Big[ \delta_n(x_j^{(b)}) - \delta_n(y_j^{(b)}) \Big] \right) \; .
\end{equation}
Putting things together one ends up with
\[
E(n_b,n_b;A^{(I)}) =
\Big( \frac{1}{2\pi} \Big)^{2 \sum_b n_b}
\times \exp \left( -\frac{e^2}{2\pi} \sum_{I=1}^N
\Big( \varepsilon_{\mu \nu} \partial_\mu A_\nu^{(I)} , K_{I I}
\varepsilon_{\rho \sigma} \partial_\rho A^{(I)}_\sigma \Big) \right)
\]
\[
\times \prod_{b=1}^N \prod_{j=1}^{n_b} \exp \left(
2e \sum_{I=1}^N \left( \varepsilon_{\mu \nu} \partial_\mu A^{(I)}_\nu,
K_{I I} U_{I b}
\Big[ \delta_n(x_j^{(b)}) - \delta_n(y_j^{(b)}) \Big] \right) \right) \]
\begin{equation}
\times \tilde{\rho}_n ( \{x^{(b)}_j \}, \{y^{(b)}_j \} ) \; .
\end{equation}
$\tilde{\rho}_n( \{x^{(b)}_j \}, \{y^{(b)}_j \} )$ denotes the
factor that depends on the space-time arguments. Furthermore
it still depends on $n$, the index of the $\delta$-sequence
The wave function renormalization procedure introduced in the last
chapter (compare(5.23))
has to be applied before the limit $n \rightarrow \infty$ is
taken. I will give the explicit form of $\tilde{\rho}$ after this
procedure in the end of this section.

The expression (6.28) now can be generalized to the case
\begin{equation}
n_a - m_a = l \; \; \; \;
l \in \mbox{Z\hspace{-.8ex}Z}
\; \; \; \; a = 1,2, \; ... \; , N \; .
\end{equation}
For those cases the $\theta-$prescription gives a nonvanishing
result which for $l \neq 0$ is different from the naive expectation
functional. The result can be obtained easily by following the
argumentation given in the last chapter. In fact the term quadratic in
the sources is not affected by the $\theta$-prescription, and
$\tilde{\rho}( \{x^{(b)}_j \}, \{y^{(b)}_j \} )$ can be read off from
(5.48) immediately. Only the term that mixes the sources with the
space time arguments of the chiral densities has to be generalized,
but this is straightforward. One ends up with
\[
E(n_b,m_b;A^{(b)}) \]
\[
= \; \Big\langle
\prod_{b=1}^N
\prod_{i=1}^{n_b}
\overline{\psi}^{(b)}({x}^{(b)}_i)
P_+ \psi^{(b)}({x}^{(b)}_i)
\prod_{j=1}^{m_b}
\overline{\psi}^{(b)}({y}^{(b)}_j)
P_- \psi^{(b)}({y}^{(b)}_j)
e^{ie\sum_{b=1}^N (a_\mu^{(b)},j_\mu^{(b)})} \Big\rangle_0^\theta
\]
\[
= \; \exp \left( -\frac{e^2}{2\pi} \sum_{I=1}^N
\Big( \varepsilon_{\mu \nu} \partial_\mu A_\nu^{(I)} , K_{I I}
\varepsilon_{\rho \sigma} \partial_\rho A^{(I)}_\sigma \Big) \right)
\]
\[
\times \prod_{b=1}^N
\prod_{j=1}^{n_b} \exp \left(
2e \sum_{I=1}^N \Big( \varepsilon_{\mu \nu} \partial_\mu A^{(I)}_\nu,
K_{I I} U_{I b} \; \delta(x_j^{(b)}) \Big) \right) \]
\[
\times \prod_{b=1}^N
\prod_{j=1}^{m_b} \exp \left(
- 2e \sum_{I=1}^N \Big( \varepsilon_{\mu \nu} \partial_\mu A^{(I)}_\nu,
K_{I I} U_{I b} \; \delta(y_j^{(b)}) \Big) \right)
\]
\begin{equation}
\times \rho ( \{x^{(b)}_j \}, \{y^{(b)}_j \} ) \; ,
\end{equation}
where $\rho ( \{x^{(b)}_j \}, \{y^{(b)}_j \} )$ is given by
(the limit $n \rightarrow \infty$ and the wave function renormalization
(compare (5.23)) has already been performed)
\[
\rho ( \{x^{(b)}_j \}, \{y^{(b)}_j \} ) = \]
\[
h(n_b,m_b) \;
\Big(\frac{1}{2\pi}\Big)^{\sum_b (n_b + m_b)} \; \;
e^{ i \frac{\theta}{N} \sum_b (n_b - m_b)}
\;
\times \; \left( e \sqrt{\frac{N}{\pi\!+\!gN}} \frac{e^\gamma}{2}
\right)^{\frac{\pi}{\pi + gN} \frac{1}{N}
\Big( \sum_b (n_b - m_b) \Big)^2} \]
\[
\times \exp \left(
\sum_{b, b^\prime =1}^N \sum_{j=1}^{n_b} \sum_{j^\prime=1}^{m_{b^\prime}}
\tilde{\tilde{V}}(x^{(b)}_j - y^{(b^\prime)}_{j^\prime}) \right) \]
\[
\times \exp \left( -\frac{1}{2}
\sum_{b, b^\prime =1}^N \sum_{j=1}^{n_b} \sum_{j^\prime=1}^{n_{b^\prime}}
\Big(1 - \delta_{b b^\prime} \delta_{j j^\prime} \Big)
\tilde{\tilde{V}}(x^{(b)}_j - x^{(b^\prime)}_{j^\prime}) \right) \]
\[
\times \exp \left( -\frac{1}{2}
\sum_{b, b^\prime =1}^N \sum_{j=1}^{m_b} \sum_{j^\prime=1}^{m_{b^\prime}}
\Big(1 - \delta_{b b^\prime} \delta_{j j^\prime} \Big)
\tilde{\tilde{V}}(y^{(b)}_j - y^{(b^\prime)}_{j^\prime}) \right) \]
\[
\times \exp \left(
- \sum_{b=1}^N \sum_{j=1}^{n_b} \sum_{j^\prime=1}^{m_{b}}
\ln (x^{(b)}_j - y^{(b)}_{j^\prime})^2 \right) \]
\[
\times \exp \left( \frac{1}{2}
\sum_{b=1}^N \sum_{j=1}^{n_b} \sum_{j^\prime=1}^{n_{b}}(1-\delta_{j j^\prime})
\ln (x^{(b)}_j - x^{(b)}_{j^\prime})^2 \right) \]
\begin{equation}
\times \exp \left( \frac{1}{2}
\sum_{b=1}^N \sum_{j=1}^{m_b} \sum_{j^\prime=1}^{m_{b}}(1-\delta_{j j^\prime})
\ln (y^{(b)}_j - y^{(b)}_{j^\prime})^2 \right) \; .
\end{equation}
The factor $h(n_b,m_b)$ is defined as
\begin{equation}
h(n_b,m_b) \; := \;
\sum_{l= -\infty}^{+\infty} \prod_{b=1}^N \delta_{n_b\!-\!m_b, l} \; \; \; \; .
\end{equation}
It is equal to one whenever
the $\theta-$prescription allows a nonvanishing result for
$E(n_b,m_b;A^{(b)})$, otherwise it is zero. $\tilde{\tilde{V}}$ is given by
(compare (5.49) and (5.21))
\begin{equation}
\tilde{\tilde{V}}(x) := \frac{1}{N} \ln ( x^2 ) +
\frac{2\pi}{N(\pi + gN)} \;
\Bigg(\mbox{K}_0\Big(\sqrt{\frac{e^2 N}{\pi\!+\!gN}} |x|\Big)
+ \ln\Big(\frac{1}{2} \sqrt{\frac{e^2 N}{\pi\!+\!gN}}\Big)+\gamma\Bigg) \; .
\end{equation}
Expression (6.30) can now be used to identify the correct bosonization.

%
%
\section{Bosonization prescription}

Bosonization means that the generating functional
$E(n_b,m_b;A^{(b)})$ can also be obtained by computing the
vacuum expectation value of a certain functional
$F(n_b,m_b;A^{(b)};\Phi^{(I)})$ in a bosonic theory
with some fields $\Phi^{(I)}$. Every operator that was used to
define $E(n_b,m_b;A^{(b)})$
will have a transcription
in terms of the $\Phi^{(I)}$ which then enters
$F(n_b,m_b;A^{(b)};\Phi^{(I)})$.
Inspecting (6.30) makes it plausible to try it with Gaussian fields
$\Phi^{(I)}$ with some covariances $K^{(I)}$ which are related to the
$K_{I I}$ (see (6.18)).
Thus one can express the idea of bosonization in the following
formula
\begin{equation}
E(n_b,m_b;A^{(b)}) \; = \;
\Big\langle F(n_b,m_b;A^{(b)};\Phi^{(I)})
\Big\rangle_{ \{K^{(I)}\} } \; ,
\end{equation}
where $\langle .. \rangle_{ \{K^{(I)} \} }$ denotes expectation value
for the fields $\Phi^{(I)}$ with respect to the covariances $K^{(I)}$.
Two steps have to be done. First define an appropriate covariance
$K^{(I)}$ and then establish the correct transcription of the fermionic
operators into bosonic ones.

The definition of the $K^{(I)}$ is rather simple. I define
\begin{equation}
K^{(1)} \; := \; \frac{1}{-\triangle + \frac{e^2 N}{\pi + gN}} \; = \;
\frac{\pi + gN}{\pi} K_{1 1} \; ,
\end{equation}
and
\begin{equation}
K^{(I)} \; := \; \frac{1}{-\triangle} \; = \; K_{I I} \; \; \; I = 2, ...N \; .
\end{equation}
Thus the $K^{(I)}$ are just the canonically normalized $K_{I I}$.
The term in (6.30) which is quadratic in the sources $A^{(b)}$ then implies
the following prescription for the bosonization of the Cartan
currents\footnote{The bosonization prescription (6.37) for the Cartan
currents was already obtained for the $g=0$ case in \cite{belvedere}.}
\begin{equation}
J^{(I)}_\nu(x) \; \longleftrightarrow \;
\left\{ \begin{array}{ll}
- \frac{1}{\sqrt{\pi\!+\!gN}} \;
\varepsilon_{\mu \nu} \partial_\mu \Phi^{(1)}(x) \; \; \; \; \; \; & I = 1 \\
\; & \;   \\
- \frac{1}{\sqrt{\pi}} \;
\varepsilon_{\mu \nu} \partial_\mu \Phi^{(I)}(x) \; \; \; \; \; \;
& I = 2,\; ... \; N \; .
\end{array} \right.
\end{equation}
With this choice the term linear in $A^{(b)}$ already fixes the
structure of the transcription of the chiral densities to
\[
\overline{\psi}^{(b)}(x) P_\pm \psi^{(b)}(x) \; \longleftrightarrow \]
\begin{equation}
\frac{1}{2\pi} c^{(b)}
: e^{\mp i 2 \sqrt{\pi} \sqrt{\frac{\pi}{\pi +gN}} U_{1 b} \Phi^{(1)}(x)}
:_{M^{(1)}} \; \prod_{I=2}^N
: e^{\mp i 2 \sqrt{\pi} U_{I b} \Phi^{(I)}(x)}
:_{M^{(I)}} e^{\pm i \frac{\theta}{N}}\; ,
\end{equation}
as can be seen from the exponentials in (6.30) linear in the
sources.
Here \\
$: .. :_{M^{(I)}}$ denotes normal ordering with respect to mass
$M^{(I)}$ (compare Appendix A.4). Those normal ordering masses as well
as the real numbers $c^{(b)}$ are free parameters that will be fixed
later.

Inserting the prescriptions (6.37), (6.38) into the definition of
$E(n_b,m_b;A^{(b)})$, one obtains
\[
E(n_b,m_b;A^{(b)}) \; \longleftrightarrow \;
\]
\[
\Big(\frac{1}{2\pi}\Big)^{\sum_b (n_b + m_b)} \; \;
e^{ i \frac{\theta}{N} \sum_b (n_b - m_b)} \; \;
\prod_{b=1}^N \Big( c{(b)} \Big)^{n_b + m_b}
\]
\[
\Bigg\langle \; \; \prod_{b=1}^N \prod_{j=1}^{n_b} \left[
: e^{- i 2 \sqrt{\pi} \sqrt{\frac{\pi}{\pi +gN}} U_{1 b}
\Phi^{(1)}(x^{(b)}_j)} :_{M^{(1)}} \; \prod_{I=2}^N
: e^{- i 2 \sqrt{\pi} U_{I b} \Phi^{(I)}(x^{(b)}_j)}
:_{M^{(I)}} \right]
\]
\[
\times \; \prod_{b=1}^N \prod_{j=1}^{m_b} \left[
: e^{+ i 2 \sqrt{\pi} \sqrt{\frac{\pi}{\pi +gN}} U_{1 b}
\Phi^{(1)}(y^{(b)}_j)} :_{M^{(1)}} \; \prod_{I=2}^N
: e^{+ i 2 \sqrt{\pi} U_{I b} \Phi^{(I)}(y^{(b)}_j)}
:_{M^{(I)}} \right]
\]
\begin{equation}
\times \; \exp \left( - \frac{i e}{\sqrt{\pi + gN}}
\Big( A_\mu^{(1)}, \varepsilon_{\nu \mu} \partial_\nu \Phi^{(1)} \Big)
- \frac{i e}{\sqrt{\pi}} \sum_{I=2}^N
\Big( A_\mu^{(I)}, \varepsilon_{\nu \mu} \partial_\nu \Phi^{(I)} \Big)
\right) \; \; \Bigg\rangle_{ \{ K^{(I)} \} } \; .
\end{equation}
The Gaussian integrals can be solved rather easily since they factorize
with respect to the $\Phi^{(I)}$. One then obtains for the right hand
side of the last equation
\[
\exp \left( - \frac{e^2}{2 \pi} \Big(
\varepsilon_{\mu \nu} \partial_\mu A_\nu^{(1)},
\frac{\pi}{\pi\!+\!gN} K^{(1)}
\varepsilon_{\rho \sigma} \partial_\rho A_\sigma^{(1)} \Big)
- \frac{e^2}{2 \pi} \sum_{I=2}^N \Big(
\varepsilon_{\mu \nu} \partial_\mu A_\nu^{(I)}, K^{(I)}
\varepsilon_{\rho \sigma} \partial_\rho A_\sigma^{(I)} \Big) \right)
\]
\[
\times \prod_{b=1}^N \prod_{j=1}^{n_b} \exp \left(
+ 2e U_{1b} \Big( \varepsilon_{\mu \nu} \partial_\mu A_\nu^{(1)},
\frac{\pi}{\pi+gN} K^{(1)} \delta(x^{(b)}_j) \Big) \right)
\]
\[
\times \prod_{b=1}^N \prod_{j=1}^{n_b} \exp \left(
+ 2e \sum_{I=2}^N U_{Ib} \Big( \varepsilon_{\mu \nu} \partial_\mu A_\nu^{(I)},
K^{(I)} \delta(x^{(b)}_j) \Big) \right)
\]
\[
\times \prod_{b=1}^N \prod_{j=1}^{m_b} \exp \left(
- 2e U_{1b} \Big( \varepsilon_{\mu \nu} \partial_\mu A_\nu^{(1)},
\frac{\pi}{\pi+gN} K^{(1)} \delta(y^{(b)}_j) \Big) \right)
\]
\[
\times \prod_{b=1}^N \prod_{j=1}^{m_b} \exp \left(
- 2e \sum_{I=2}^N U_{Ib}
\Big( \varepsilon_{\mu \nu} \partial_\mu A_\nu^{(I)},
K^{(I)} \delta(y^{(b)}_j) \Big) \right)
\]
\begin{equation}
\times \; \rho_B(\{x_j^{(b)}\}, \{y_j^{(b)}\}) \; .
\end{equation}
Comparing (6.30) and (6.40) shows immediately that the terms
quadratic and linear in the sources come out correctly. Thus it is
left to show
\begin{equation}
\rho_B(\{x_j^{(b)}\}, \{y_j^{(b)}\}) \; \; = \; \;
\rho(\{x_j^{(b)}\}, \{y_j^{(b)}\}) \; ,
\end{equation}
where $\rho(\{x_j^{(b)}\}, \{y_j^{(b)}\})$ is given by (6.31).
As mentioned before, the integral over the $\Phi^{(I)}$ factorizes
such that $\rho_B(\{x_j^{(b)}\}, \{y_j^{(b)}\})$ reads
\[
\rho_B(\{x_j^{(b)}\}, \{y_j^{(b)}\}) =
\left(\frac{1}{2\pi}\right)^{\sum_b (n_b + m_b)} \; \;
e^{ i \frac{\theta}{N} \sum_b (n_b - m_b)} \; \;
\prod_{b=1}^N \Big( c^{(b)} \Big)^{n_b + m_b}
\]
\[
\times \Bigg\langle
\prod_{b=1}^N \; \prod_{j=1}^{n_b}
: e^{- i 2 \sqrt{\pi} \sqrt{\frac{\pi}{\pi +gN}} U_{1 b}
\Phi^{(1)}(x^{(b)}_j)} :_{M^{(1)}} \;
\prod_{j=1}^{m_b}
: e^{+ i 2 \sqrt{\pi} \sqrt{\frac{\pi}{\pi +gN}} U_{1 b}
\Phi^{(1)}(y^{(b)}_j)} :_{M^{(1)}} \Bigg\rangle_{K^{(1)}}
\]
\begin{equation}
\times \prod_{I=2}^N \;
\Bigg\langle
\prod_{b=1}^N \; \prod_{j=1}^{n_b}
: e^{- i 2 \sqrt{\pi} U_{I b}
\Phi^{(I)}(x^{(b)}_j)} :_{M^{(I)}} \;
\prod_{j=1}^{m_b}
: e^{+ i 2 \sqrt{\pi} U_{I b}
\Phi^{(I)}(y^{(b)}_j)} :_{M^{(I)}} \Bigg\rangle_{K^{(I)}}
\end{equation}
The $\Phi^{(1)}$ expectation value is rather simple since the covariance is
massive. Using $U_{1b} = 1/\sqrt{N}$ one finds (compare Appendix A.4)
\[
\left( \frac{e^2 N}{\pi+gN} \frac{1}{(M^{(1)})^2}
\right)^{\frac{\pi}{\pi+gN} \frac{1}{2N} \sum_b (n_b + m_b)}
\]
\[
\times
\exp \left(
\sum_{b, b^\prime=1}^N \sum_{j=1}^{n_b} \sum_{j^\prime=1}^{m_{b^\prime}}
2 \frac{\pi}{\pi+gN} \frac{1}{N} \; \mbox{K}_0
\Big( e \sqrt{\frac{N}{\pi+gN}} |x_j^{(b)} - y_{j^\prime}^{(b^\prime)}|
\Big) \right)
\]
\[
\times \exp \left( -
\sum_{b, b^\prime=1}^N \sum_{j=1}^{n_b} \sum_{j^\prime=1}^{n_{b^\prime}}
\Big(1-\delta_{b b^\prime} \delta_{j j^\prime} \Big)
\frac{\pi}{\pi+gN} \frac{1}{N} \; \mbox{K}_0
\Big( e \sqrt{\frac{N}{\pi+gN}} |x_j^{(b)} - x_{j^\prime}^{(b^\prime)}|
\Big) \right)
\]
\begin{equation}
\times \exp \left( -
\sum_{b, b^\prime=1}^N \sum_{j=1}^{m_b} \sum_{j^\prime=1}^{m_{b^\prime}}
\Big(1-\delta_{b b^\prime} \delta_{j j^\prime} \Big)
\frac{\pi}{\pi+gN} \frac{1}{N} \; \mbox{K}_0
\Big( e \sqrt{\frac{N}{\pi+gN}} |y_j^{(b)} - y_{j^\prime}^{(b^\prime)}|
\Big) \right) \; .
\end{equation}
The evaluation of the expectaion values
$\langle .. \rangle_{K^{(I)}}$ with $I > 1$ is a little bit more involved,
since for massless fields the neutrality condition
(see Appendix A.4) is relevant. It will turn
out that the neutrality condition produces the factor $h(n_b,m_b)$ which was
defined in (6.32). The condition implies
\begin{equation}
\sum_{b=1}^N U_{Ib} (n_b - m_b) \; \stackrel{!}{=} \; 0 \; \; \; \; \; \;
\forall \; \; I = 2,3, ... N  \; \; ,
\end{equation}
for nonvanishing expectation values. Interpreting the lines of $U_{Ib}$
as vectors $\vec{r}^{\;(I)}$ (see Appendix B.3 for the definition),
the condition reads
\begin{equation}
(\vec{n} - \vec{m}) \cdot \vec{r}^{\;(I)} \; \stackrel{!}{=} \; 0 \; \; \; \;
\forall \; \; I = 2,3, ... N  \; \; .
\end{equation}
One finds that the only solution is
\begin{equation}
\vec{n} - \vec{m} \; \; \propto \; \; (1,1, \; .... \; 1) \; .
\end{equation}
Since $n_b$ and $m_b$ are integers this solution is equivalent to
multiplication with $h(n_b,m_b)$.

Using Appendix A.4 one obtains
\[
\prod_{I=2}^N \;
\Bigg\langle
\prod_{b=1}^N \; \prod_{j=1}^{n_b}
: e^{- i 2 \sqrt{\pi} U_{I b}
\Phi^{(I)}(x^{(b)}_j)} :_{M^{(I)}} \;
\prod_{j=1}^{m_b}
: e^{+ i 2 \sqrt{\pi} U_{I b}
\Phi^{(I)}(y^{(b)}_j)} :_{M^{(I)}} \Bigg\rangle_{K^{(I)}}
\]
\[
= \; h(n_b,m_b) \; \;
\prod_{I=2}^N \left( \frac{1}{M^{(I)}}\right)^{\sum_b (U_{Ib})^2 (n_b - m_b)}
\]
\[
\times \exp \left( - \sum_{b,b^\prime=1}^N \sum_{j=1}^{n_b}
\sum_{j^\prime=1}^{m_{b^\prime}} \sum_{I=2}^N U_{I b} U_{I b^\prime}
\Big[ \ln(x^{(b)}_j - y^{(b^\prime)}_{j^\prime})^2 + 2\gamma - \ln(4)
\Big] \right)
\]
\[
\times \exp \left( \frac{1}{2} \sum_{b,b^\prime=1}^N \sum_{j=1}^{n_b}
\sum_{j^\prime=1}^{n_{b^\prime}}
\Big(1-\delta_{b b^\prime} \delta_{j j^\prime} \Big)
\sum_{I=2}^N U_{I b} U_{I b^\prime}
\Big[ \ln(x^{(b)}_j - x^{(b^\prime)}_{j^\prime})^2 + 2\gamma - \ln(4)
\Big] \right)
\]
\[
\times \exp \left( \frac{1}{2} \sum_{b,b^\prime=1}^N \sum_{j=1}^{m_b}
\sum_{j^\prime=1}^{m_{b^\prime}}
\Big(1-\delta_{b b^\prime} \delta_{j j^\prime} \Big)
\sum_{I=2}^N U_{I b} U_{I b^\prime}
\Big[ \ln(y^{(b)}_j - y^{(b^\prime)}_{j^\prime})^2 + 2\gamma - \ln(4)
\Big] \right)
\]
\[
= \; h(n_b,m_b) \;
\prod_{I=2}^N \left( \frac{1}{M^{(I)}}\right)^{\sum_b (U_{Ib})^2 (n_b - m_b)}
\]
\[
\times \exp \left( \sum_{b,b^\prime=1}^N \sum_{j=1}^{n_b}
\sum_{j^\prime=1}^{m_{b^\prime}}
\frac{1}{N}
\Big[ \ln(x^{(b)}_j - y^{(b^\prime)}_{j^\prime})^2 + 2\gamma - \ln(4)
\Big] \right)
\]
\[
\times \exp \left( - \frac{1}{2} \sum_{b,b^\prime=1}^N \sum_{j=1}^{n_b}
\sum_{j^\prime=1}^{n_{b^\prime}}
\Big(1-\delta_{b b^\prime} \delta_{j j^\prime} \Big)
\frac{1}{N}
\Big[ \ln(x^{(b)}_j - x^{(b^\prime)}_{j^\prime})^2 + 2\gamma - \ln(4)
\Big] \right)
\]
\[
\times \exp \left( - \frac{1}{2} \sum_{b,b^\prime=1}^N \sum_{j=1}^{m_b}
\sum_{j^\prime=1}^{m_{b^\prime}}
\Big(1-\delta_{b b^\prime} \delta_{j j^\prime} \Big)
\frac{1}{N}
\Big[ \ln(y^{(b)}_j - y^{(b^\prime)}_{j^\prime})^2 + 2\gamma - \ln(4)
\Big] \right) \;
\]
\[
\times \exp \left( - \sum_{b=1}^N \sum_{j=1}^{n_b}
\sum_{j^\prime=1}^{m_{b}}
\Big[ \ln(x^{(b)}_j - y^{(b)}_{j^\prime})^2 + 2\gamma - \ln(4)
\Big] \right)
\]
\[
\times \exp \left( \frac{1}{2} \sum_{b=1}^N \sum_{j=1}^{n_b}
\sum_{j^\prime=1}^{n_{b}}
\Big(1-\delta_{j j^\prime} \Big)
\Big[ \ln(x^{(b)}_j - x^{(b)}_{j^\prime})^2 + 2\gamma - \ln(4)
\Big] \right)
\]
\begin{equation}
\times \exp \left( \frac{1}{2} \sum_{b=1}^N \sum_{j=1}^{m_b}
\sum_{j^\prime=1}^{m_{b}}
\Big(1- \delta_{j j^\prime} \Big)
\Big[ \ln(y^{(b)}_j - y^{(b)}_{j^\prime})^2 + 2\gamma - \ln(4)
\Big] \right) \; ,
\end{equation}
where I made use of (B.20) to remove the $U_{Ib}$ in the last step.
Putting things together one ends up with
\[
\rho_B (\{x_j^{(b)}\}, \{y_j^{(b)}\})
\]
\[ = \; C \;
\times \exp \left(
\sum_{b, b^\prime =1}^N \sum_{j=1}^{n_b} \sum_{j^\prime=1}^{m_{b^\prime}}
\tilde{\tilde{V}}(x^{(b)}_j - y^{(b^\prime)}_{j^\prime}) \right) \]
\[
\times \exp \left( -\frac{1}{2}
\sum_{b, b^\prime =1}^N \sum_{j=1}^{n_b} \sum_{j^\prime=1}^{n_{b^\prime}}
\Big(1 - \delta_{b b^\prime} \delta_{j j^\prime} \Big)
\tilde{\tilde{V}}(x^{(b)}_j - x^{(b^\prime)}_{j^\prime}) \right) \]
\[
\times \exp \left( -\frac{1}{2}
\sum_{b, b^\prime =1}^N \sum_{j=1}^{m_b} \sum_{j^\prime=1}^{m_{b^\prime}}
\Big(1 - \delta_{b b^\prime} \delta_{j j^\prime} \Big)
\tilde{\tilde{V}}(y^{(b)}_j - y^{(b^\prime)}_{j^\prime}) \right) \]
\[
\times \exp \left(
- \sum_{b=1}^N \sum_{j=1}^{n_b} \sum_{j^\prime=1}^{m_{b}}
\ln (x^{(b)}_j - y^{(b)}_{j^\prime})^2 \right) \]
\[
\times \exp \left( \frac{1}{2}
\sum_{b=1}^N \sum_{j=1}^{n_b}
\sum_{j^\prime=1}^{n_{b}}(1-\delta_{j j^\prime})
\ln (x^{(b)}_j - x^{(b)}_{j^\prime})^2 \right) \]
\begin{equation}
\times \exp \left( \frac{1}{2}
\sum_{b=1}^N \sum_{j=1}^{m_b}
\sum_{j^\prime=1}^{m_{b}}(1-\delta_{j j^\prime})
\ln (y^{(b)}_j - y^{(b)}_{j^\prime})^2 \right) \; .
\end{equation}
The constant $C$ is given by
\[
C = h(n_b,m_b)
\left(\frac{1}{2\pi}\right)^{\sum_b (n_b + m_b)} \;
e^{ i \frac{\theta}{N} \sum_b (n_b - m_b)}
\left( e \sqrt{ \frac{N}{\pi\!+\!gN}} \frac{e^\gamma}{2}
\right)^{\frac{\pi}{\pi + gN} \frac{1}{N}
\Big( \sum_b (n_b - m_b) \Big)^2} \]
\begin{equation}
\times \prod_{a=1}^N \Big[ c^{(a)} \Big]^{n_a + m_a}
\; \;
\prod_{b=1}^N \left[
\left( \frac{2e^{-\gamma}}{M^{(1)}}\right)^{\frac{\pi}{\pi+gN} \frac{1}{N}}
\prod_{I=2}^N
\Big( \frac{2e^{-\gamma}}{M^{(I)}}\Big)^{(U_{Ib})^2}
\right]^{n_b + m_b} \; .
\end{equation}
The evaluation of the constant $C$ is straightforward but lenghty.
One has to add to the Bessel functions and the logarithms in the exponent
terms proportional to
\begin{equation}
\frac{\pi}{\pi +gN} \Big[ \ln
\Big( \frac{e}{2} \sqrt{\frac{N}{\pi+gN}} \Big)
+ \gamma \Big]
\end{equation}
to obtain $\tilde{\tilde{V}}$ (see (6.33)).
This modification has to be compensated
by a factor that enters $C$. The factor picks up an exponent proportional
to $(\sum_b(n_b-m_b))^2$, which shows up also in the result
for $\rho(\{x_j^{(b)}\}, \{y_j^{(b)}\})$ (see (6.31)).
It is rather crucial since it cannot be produced by the factors
$c^{(b)}$ of the ansatz (6.38), which can only obtain exponents linear in
$n_b$ and $m_b$. Furthermore one has to use $U_{1b} = 1/\sqrt{N}$
and (B.20) to obtain (6.49).
Finally (6.41) can be fulfilled by setting
\begin{equation}
c^{(b)} = \Big(\frac{M^{(1)} e^\gamma}{2}
\Big)^{\frac{\pi}{\pi+gN} \frac{1}{N}} \;
\prod_{I=2}^N \Big(\frac{M^{(I)} e^\gamma}{2} \Big)^{(U_{Ib})^2} \; .
\end{equation}
Thus the bosonization is given by (6.37) and (6.38) together with (6.51).

\noindent
I finish this section with a discussion of the Cartan currents in the
massless model.
Up to a constant the vector currents $J^{(I)}$ are bosonized by
$- \frac{1}{\sqrt{\pi}} \;
\varepsilon_{\mu \nu} \partial_\mu \Phi^{(I)}$. The covariance $K^{(I)}$
(6.35) then implies that the U(1)-current $J^{(1)}$ describes
a particle with mass
\begin{equation}
e \; \sqrt{\frac{N}{\pi +gN}} \; .
\end{equation}
The rest of the Cartan currents are massless as can be seen from the
covariances $K{(I)}, \; I = 2,3, ... \mbox{N}$.

It has to be remarked that the U(1)-current remains massive also
in the case $g=0$. At first glance it might seem a little bit suspicious
that the U(1)-current which is treated differently from the others
(only $J^{(1)}$ has the Thirring term in the action), acquires
mass. As long as one is not interested in the massive theory,
the Thirring term is not needed, and $g$ can be set equal to zero
since the expectation values are continuous in $g$.
(6.52) shows that $J^{(1)}$ remains massive.

%
%
\section{More vector currents}

In this section the vector currents that are not
of the Cartan type will be discussed.
In analogy to the construction of meson states in QCD, one
can define vector currents for all the generators of U(N).
A convenient basis of the Lie algebra of U(N) is given by the
$\mbox{N}(\mbox{N}-1)/2$ generators $H^{(I)}$ of the form
\begin{equation}
\frac{1}{\sqrt{2}} \left( \begin{array}{cccccc}
 & & & & & \\
 & & & &1& \\
 & & & & & \\
 &1& & & & \\
 & & & & & \\
 & & & & &
\end{array} \right) \; ,
\end{equation}
and $\mbox{N}(\mbox{N}-1)/2$ of the form
\begin{equation}
\frac{1}{\sqrt{2}} \left( \begin{array}{cccccc}
 & & & & & \\
 & & & &-i& \\
 & & & & & \\
 &i& & & & \\
 & & & & & \\
 & & & & &
\end{array} \right) \; .
\end{equation}
The new generators are given indices $I = \mbox{N}+1,\mbox{N}+2, ...
\mbox{N}^2$.
Together with $H^{(1)}, ..., H^{(N)}$ defined in (6.24)
they generate U(N)\footnote{
It has to be remarked that the generators (6.24), (6.53) and (6.54) I use are
not normalized as usual (like e.g. the Gell-Mann matrices
for SU(3)). This is due to the
fact, that the generators (6.24) stem from the orthonormal matrix $U$
which diagonalizes the covariance (compare (6.17)).}.
The corresponding vector currents are (compare (6.23))
\begin{equation}
J_\mu^{(I)}(x) \; := \; \sum_{b,b^\prime=1}^N \;
\overline{\psi}^{(b)}(x)
\gamma_\mu H^{(I)}_{b,b^\prime} \; \psi^{(b^\prime)}(x) \; .
\end{equation}
Since only different flavors (which cannot contract) sit at one space-time
point no short distance singularity can emerge. For the Cartan currents
$J_\mu^{(1)}, ... J_\mu^{(N)}$ (which are diagonal in flavor)
this problem was circumvented by including
the sources that couple to these currents into the fermion determinant which
has its own renormalization.

First I notice that the set of the $\mbox{N}^2$ vector
currents generate orthogonal
states
\begin{equation}
\Big\langle J_\mu^{(I)}(x) J_\nu^{(I^\prime)}(y)\Big\rangle =
\delta_{I I^\prime} \;\; {\cal F}_{\mu \nu}^{(I)} (x,y)
\;\;\;\; I = 1,2,...N^2 \; ,
\end{equation}
where $\cal F$ is the two point function. For the Cartan currents
(6.56) follows dircectly from the bosonization. To prove it for the set of all
$\mbox{N}^2$ currents one has to take functional derivatives of
the generating functional (3.38)
with respect to the sources $\eta, \bar{\eta}$ and
$A_\mu^{(I)} \Leftrightarrow a^{(b)}_\mu$.
If $I \neq I^\prime$, either the different flavors do not contract
entirely, or terms with opposite sign cancel.

The two point function for the new currents can be obtained easily.
In the case $ I = I^\prime , I = \mbox{N}+1,\mbox{N}+2,...\mbox{N}^2$,
functional derivation of (3.38) leads to
(take e.g. $\mu = \nu = 1$)
\[
{\cal F}_{1 1} = -\!\int d\mu_{\tilde{Q}}[\varphi]
d\mu_{\tilde{C}}[\theta]  \; \Bigg(
G_{21}(x,y;\varphi,\theta)G_{21}(y,x;\varphi,\theta) \; + \]
\begin{equation}
G_{12}(x,y;\varphi,\theta)G_{12}(y,x;\varphi,\theta) \Bigg) \; .
\end{equation}
Using the explicit form (4.58) for $G$, one immediately sees that
the exponentials involving $\varphi$ and $\theta$ cancel. Integration over
$\varphi$ simply gives a factor 1. The same is true for arbitrary
$\mu,\nu$. Hence the two point function is the same as for currents
made from free, massless
fermions. It can be expressed in terms of derivatives of the
propagator of a free massless boson, giving rise to the same expression as
was obtained for the Cartan type corrents, i.e. for $I = 2,3,...\mbox{N}$.

Putting things together
one concludes for the two-point functions of the vector currents
\begin{equation}
\Big\langle J_\mu^{(I)}(x) J_\nu^{(I^\prime)}(y) \Big\rangle =
\delta_{I I^\prime} \;\left\{\!\begin{array}{ll}
\; \frac{1}{\pi\!+\!gN} \varepsilon_{\mu \rho} \partial x_\rho
\,\varepsilon_{\nu \sigma} \partial y_\sigma
\;\Big\langle \varphi^{(I)}(x) \varphi^{(I)}(y) \Big\rangle_{K^{(1)}} \;
& I = 1 \\
\; & \; \\
\; \frac{1}{\pi} \varepsilon_{\mu \rho} \partial x_\rho
\,\varepsilon_{\nu \sigma} \partial y_\sigma
\;\Big\langle \varphi^{(I)}(x) \varphi^{(I)}(y) \Big\rangle_{K^{(I)}} \;
& I = 2,...N,
\end{array} \right.
\end{equation}
where the scalar field $\varphi^{(1)}$ has the mass given in
equation (6.52) and
the $\varphi^{(I)}, I=2,3,...\mbox{N}^2$ are massless.

The $\mbox{N}^2$ vector
have the same mass as the corresponding Bose fields. Only the
U(1)-current is massive, while the others are massless. When the
model is considered as a toy model for QCD, the vector currents describe
the pseudoscalar mesons. Note that
$J^{(I)}_{\mu \; 5} = -i \varepsilon_{\mu \nu} J^{(I)}_\nu$ due to the choice
of the $\gamma$-matrices (compare Appendix B.1).
The particle related to
$J^{(1)}$ has to be identified with the
$\eta^\prime$ meson. Since it is the only massive current,
the model perfectly mimics the U(1)-problem and its solution.

The lesson that has to be learned here reads
\vskip5mm
\noindent
{\bf Lesson 2 :}
\vskip3mm
\noindent
{\it The axial U(1)-symmetry is not a symmetry on the physical
Hilbert space, and there is no  U(1)-problem.}
\vskip3mm
\noindent
This can be seen rather easily in the $\mbox{N}=2$ flavor case. The Lagrangian
for the scalar fields $\varphi^{(1)},\; \varphi^{(2)}$ that bosonize
the currents and the chiral densities then, is given by
\begin{equation}
\frac{1}{2} \Big( \partial_\mu \varphi^{(1)} \Big)^2 \; + \;
\frac{1}{2} \Big( \partial_\mu \varphi^{(2)} \Big)^2 \; + \;
\frac{1}{2} \Big( \varphi^{(1)} \Big)^2 \frac{e^2 N}{\pi+gN} \; .
\end{equation}
The bosonization prescription (6.38) for the left-handed densities gives
\[
\overline{\psi}^{(1)}(x) P_+ \psi^{(1)}(x) \; \longleftrightarrow \;
\frac{1}{2\pi} c^{(1)}
:e^{-ia \varphi^{(1)}(x)}:_{M^{(1)}}
:e^{-ib \varphi^{(2)}(x)}:_{M^{(2)}} \; e^{i\frac{\theta}{2}} \; , \]
\begin{equation}
\overline{\psi}^{(2)}(x) P_+ \psi^{(2)}(x) \; \longleftrightarrow \;
\frac{1}{2\pi} c^{(2)}
:e^{-ia \varphi^{(1)}(x)}:_{M^{(1)}}
:e^{+ib \varphi^{(2)}(x)}:_{M^{(2)}} \; e^{i\frac{\theta}{2}} \; ,
\end{equation}
with (see B.17) $a = 2\sqrt{\pi/2} \sqrt{\pi/(\pi+gN)}$ and
$b = 2\sqrt{\pi/2}$. The axial transformation (3.16) acts on the
densities via
\begin{equation}
\overline{\psi}^{(b)}(x) P_+ \psi^{(b)}(x) \; \longrightarrow \;
\overline{\psi}^{(b)}(x) P_+ \psi^{(b)}(x) \; e^{i 2 \omega} \; .
\end{equation}
In the bosonized theory this corresponds to
\[
a \; \varphi^{(1)}(x) \; + \; b \varphi^{(2)}(x) \; \longrightarrow \;
a \; \varphi^{(1)}(x) \; + \; b \varphi^{(2)}(x) \; - \; 2 \omega \; ,
\]
\begin{equation}
a \; \varphi^{(1)}(x) \; - \; b \varphi^{(2)}(x) \; \longrightarrow \;
a \; \varphi^{(1)}(x) \; - \; b \varphi^{(2)}(x) \; - \; 2 \omega \; .
\end{equation}
Obviously this is not a symmetry, since $\omega$ on the right hand side of
(6.62) cannot be transformed away, by shifting one of the fields by a
constant. $\varphi^{(1)}(x)$ cannot be shifted since it is
massive (see (6.59)). $\varphi^{(2)}(x)$ would have to be shifted by
$+2 \omega/b$ in order to remove $\omega$ in the first line of (6.62) and by
$-2 \omega/b$ to remove it in the second line. Thus $\mbox{U(1)}_A$
is not a symmetry, and no Goldstone particle exists in
the physical Hilbert space. The generalization
of the arguments to $\mbox{N} > 2$ flavors is straightforward.

%
%
\section{Theorems on n-point functions}
To gain more insight into the sector generated by the vector currents in
the massless model,
I compute explicit expressions for connected $n$-point
functions. Considering fully connected correlations
has the advantage of excluding manifestly contractions of fermions
at the same point. Therefore one can avoid using test functions as
would be required in principle by the distributional nature of the fields.
Let
\[
C_{n+1} := \Big\langle J_{\mu_0}^{(I_0)}(x_0) J_{\mu_1}^{(I_1)}(x_1) .....
 J_{\mu_n}^{(I_n)}(x_n) \Big\rangle_{0\; c}^\theta   \]
\[
= \; \sum_{a_i,b_i} H^{(I_0)}_{a_0 b_0} H^{(I_1)}_{a_1 b_1} .....
H^{(I_n)}_{a_n b_n} \sum_{\alpha_i,\beta_i}
(\gamma_{\mu_0})_{\alpha_0 \beta_0} (\gamma_{\mu_1})_{\alpha_1 \beta_1} .....
(\gamma_{\mu_n})_{\alpha_n \beta_n} \]
\begin{equation} \Big\langle
\overline{\psi}^{(a_0)}_{\alpha_0}(x_0) \psi^{(b_0)}_{\beta_0}(x_0)
\overline{\psi}^{(a_1)}_{\alpha_1}(x_1) \psi^{(b_1)}_{\beta_1}(x_1) .....
\overline{\psi}^{(a_n)}_{\alpha_n}(x_n) \psi^{(b_n)}_{\beta_n}(x_n)
\Big\rangle_{0\; c}^\theta \; ,
\end{equation}
where (6.23) was inserted.
Nonvanishing contributions occur only if the color indices can form a
closed chain (e.g. $b_0 = a_1,b_1 = a_2, ... , b_{n-1} = a_n,b_n = a_0$),
such that the fermions can contract entirely. The
corresponding factor is simply the trace over the flavor matrices $H^{(I)}$.
To find all possible contributions one has to sum over all permutations $\pi$,
keeping the first term fixed.
\[
C_{n+1} =  -\!\!\sum_{\pi(1,2,...n)}\!\mbox{Tr} \Big[
H^{(I_0)} H^{(I_{\pi(1)})} ..... H^{(I_{\pi(n)})} \Big]
\sum_{\alpha_i,\beta_i}
(\gamma_{\mu_0})_{\alpha_0 \beta_0} (\gamma_{\mu_1})_{\alpha_1 \beta_1} .....
(\gamma_{\mu_n})_{\alpha_n \beta_n} \]
\[
\int\!d\mu_{\tilde{Q}}[\varphi] d\mu_{\tilde{C}}[\theta]
G_{\beta_0 \alpha_{\pi(1)}} (x_0,x_{\pi(1)};\varphi, \theta)
G_{\beta_{\pi(1)} \alpha_{\pi(2)}} (x_{\pi(1)},x_{\pi(2)};\varphi, \theta) \;
..... \]
\begin{equation}
...... \;
G_{\beta_{\pi(n)} \alpha_0} (x_{\pi(n)},x_0;\varphi, \theta)
\end{equation}
Since in the chosen representation the $\gamma$-matrices and the propagator
$G$ have only off-diagonal entries (cf. Equation(4.58)),
one finds the following chain of implications for e.g. $\beta_0 = 1$
\begin{equation}
\beta_0 = 1 \Rightarrow \alpha_{\pi(1)} = 2 \Rightarrow \beta_{\pi(1)} = 1
\Rightarrow \alpha_{\pi(2)} =2 ..... \Rightarrow \beta_{\pi(n)} = 1
\Rightarrow \alpha_0 = 2 .
\end{equation}
When starting with $\beta_0 = 2$ one ends up with the reverse result of all
$\beta_i = 2$ and all $\alpha_i = 1$. Besides those two no other
nonvanishing terms contribute.
Again one finds that all dependence on $\varphi$ and $\theta$ cancels, and
only free propagators $G^0$ remain. Using $G^0(x) = \frac{1}{2\pi}
\frac{\gamma_\mu x_\mu}{x^2}$ and making use of the complex notation
(A.15) one obtains
\[
C_{n+1} \; = \; - \; \frac{1}{(2\pi)^{n+1}}
\sum_{\pi(1,2...n)} \mbox{Tr}\Big[H^{(I_0)} H^{(I_{\pi(1)})} .....
H^{(I_{\pi(n)})}\Big]
\]
\begin{equation}
\left\{ \prod_{i=0}^n (\gamma_{\mu_i})_{21}
\frac{1}{\tilde{x}_0 - \tilde{x}_{\pi(1)}}
\frac{1}{\tilde{x}_{\pi(1)} - \tilde{x}_{\pi(2)}} .....
\frac{1}{\tilde{x}_{\pi(n)} - \tilde{x}_0} \; + \; c.c. \right\} \; ,
\end{equation}
where $c.c.$ denotes complex conjugation.

Using this basic formula, I construct for $\mbox{N}\geq2$ and arbitrary $n$
fully connected
$n$-point functions that do not vanish; in other words it will be
shown that the
entire set of all $\mbox{N}^2$ currents is not Gaussian. For simplicity the
construction is performed for the case of $\mbox{N}=2$.
Since for arbitray $\mbox{N} > 2$
there exist generators with the same commutation relations, it is obvious how
to generalize the construction to general N.
For $\mbox{N}=2$ the generators needed in the construction
are simply the Pauli matrices
up to a normalization factor.
To distinguish them from arbitrary generators $H^{(I)}$
(which would be there for $\mbox{N}>2$), I denote the
special set of matrices needed in the construction
\begin{equation}
\tau^{(1)} := \frac{1}{\sqrt{2}} \left( \begin{array}{cc}
0 & 1 \\
1 & 0 \end{array} \right) \; , \;
\tau^{(2)} := \frac{1}{\sqrt{2}} \left( \begin{array}{cc}
0 & -i \\
i & 0 \end{array} \right) \; , \;
\tau^{(3)} := \frac{1}{\sqrt{2}} \left( \begin{array}{cc}
1 & 0 \\
0 & -1 \end{array} \right) \;.
\end{equation}
Define
\[
F_{n+1}(\tilde{x}_0,\tilde{x}_1,... \tilde{x}_n) :=
\]
\begin{equation}
\sum_{\pi(1,2,...n)} \mbox{Tr}\Big[\tau^{(I_0)} \tau^{(I_{\pi(1)})} .....
\tau^{(I_{\pi(n)})} \Big]
\frac{1}{\tilde{x}_0 - \tilde{x}_{\pi(1)}}
\frac{1}{\tilde{x}_{\pi(1)} - \tilde{x}_{\pi(2)}} .....
\frac{1}{\tilde{x}_{\pi(n)} - \tilde{x}_0}  \; ,
\end{equation}
where the set of $\tau$ matrices is given by
\[
\{\tau^{(I_0)},\tau^{(I_1)},.....\tau^{(I_n)} \} :=
\]
\begin{equation} \left\{ \;\;
\begin{array}{ll}
\{\tau^{(2)},\tau^{(3)},\tau^{(3)},...\tau^{(3)},\tau^{(1)} \} \;\;\;\;\;
\mbox{for}\ n+1 \; = 2m+1 \; \; \; \; \;\;
(n+1\ \mbox{odd)} \\ \; \\
\{\tau^{(1)},\tau^{(3)},\tau^{(3)},...\tau^{(3)},\tau^{(1)} \} \;\;\;\;\;
\mbox{for}\ n+1 \; = 2m+2 \; \; \; \; \;\;
(n+1\ \mbox{even)} \;\; .\end{array} \right.
\end{equation}
\vskip3mm
\noindent
{\bf Theorem 6.1 : } \vskip3mm
\noindent
For arbitrary $m\geq 1$ :
\begin{equation}
F_{2m+1} \not{\!\!\equiv} 0 \; , \; F_{2m+2} \not{\!\!\equiv} 0 \; .
\end{equation}
\vskip3mm
\noindent
{\bf Proof:}
\vskip3mm
\noindent
The statement will be proven by induction. \\
{\bf i:} $F_3 \not{\!\!\equiv} 0$ and $F_4 \not{\!\!\equiv} 0$ can
be checked easily.\\
{\bf ii:} Assume $F_{2m+1} \not{\!\!\equiv} 0$ and $F_{2m+2}
\not{\!\!\equiv} 0$ .\\
{\bf iii:} One has to show $F_{2m+3} \not{\!\!\equiv} 0$ and
$F_{2m+4} \not{\!\!\equiv} 0$. First I check $F_{2m+3}$. Again I
abbreviate $2m+3 := n+1$. The trick is to consider
$F_{n+1}(\tilde{x}_0,\tilde{x}_1, ... \tilde{x}_n)$ as a function of
$\tilde{x}_0$, and to compute the corresponding residues\footnote{
I acknowledge a useful discussion with
Peter Breitenlohner on this point.}.
\[
\mbox{Res}_{\tilde{x}_0} [ F_{n+1},\tilde{x}_0 = \tilde{x}_1 ] \]
\[
= \; \sum_{\pi(1,2,...n),\pi(1)=1}\!\!\!
\mbox{Tr}\Big[\tau^{(2)} \tau^{(3)} \tau^{(\pi(2))} ...
\tau^{(\pi(n))} \Big]
\frac{1}{\tilde{x}_1\!-\!\tilde{x}_{\pi(2)}}
\frac{1}{\tilde{x}_{\pi(2)}\!-\!\tilde{x}_{\pi(3)}} ...
\frac{1}{\tilde{x}_{\pi(n)}\!-\!\tilde{x}_1}
\]
\begin{equation}
-\!\!\!\!\!\sum_{\pi^\prime(1,2,...n),\pi^\prime(n)=1}\!\!\!
\mbox{Tr}\Big[\tau^{(2)} \tau^{(\pi^\prime(1))}
... \tau^{(\pi^\prime(n-1))} \tau^{(3)} \Big]
\frac{1}{\tilde{x}_1\!-\!\tilde{x}_{\pi^\prime(1)}}
\frac{1}{\tilde{x}_{\pi^\prime(1)}\!-\!\tilde{x}_{\pi^\prime(2)}} ...
\frac{1}{\tilde{x}_{\pi^\prime(n-1)}\!-\!\tilde{x}_1}
\end{equation}
For a given permutation $\pi$ choose the permutation $\pi^\prime$
such that $\pi^\prime(1) = \pi(2),\pi^\prime(2) = \pi(3), ...
\pi^\prime(n-1) = \pi(n), \pi^\prime(n) = \pi(1) = 1$. This gives for
the residue
\[
\sum_{\pi(1,2,...n),\pi(1)=1}
\frac{1}{\tilde{x}_1 - \tilde{x}_{\pi(2)}}
\frac{1}{\tilde{x}_{\pi(2)} - \tilde{x}_{\pi(3)}} ...
\frac{1}{\tilde{x}_{\pi(n)} - \tilde{x}_1}\]
\begin{equation}
\left\{
\mbox{Tr}\Big[\tau^{(2)} \tau^{(3)} \tau^{(\pi(2))} \tau^{(\pi(3))} ...
\tau^{(\pi(n))} \Big] -
\mbox{Tr}\Big[\tau^{(2)} \tau^{(\pi(2))} \tau^{(\pi(3))} ...
\tau^{(\pi(n))} \tau^{(3)} \Big] \right\}  \; .
\end{equation}
Using the cycliticity of the trace and $\{\tau^{(2)},\tau^{(3)}\} = 0$, one
finds that the second trace is the negative of the first one. Furthermore
$\tau^{(2)} \tau^{(3)} = \frac{i}{\sqrt{2}} \tau^{(1)}$.
Looking at the definition (6.69) of
$F_n$ for even $n$ one concludes
\begin{equation}
\mbox{Res}_{\tilde{x}_0} [ F_{n+1},\tilde{x}_0 = \tilde{x}_1 ]  =
i \sqrt{2} F_n(\tilde{x}_1,... \tilde{x}_n) =
i \sqrt{2} F_{2m+2}(\tilde{x}_1,... \tilde{x}_{2m+2}) \; .
\end{equation}
Since by assumption $F_{2m+2} \not{\!\!\equiv} 0$,  Res$[F_{2m+3}]
\neq 0$ and hence $F_{2m+3}$ does not vanish. The same trick can be applied to
prove that this implies $F_{2m+4} \not{\!\!\equiv} 0. \; \Box$
\vskip3mm
\noindent
A $n+1$-point function with the flavor content given by (6.69) is
simply the real part of a multiple of $F_{n+1}$ (compare (6.66)).
Hence there exist nonvanishing, fully connected $n$-point functions
for arbitrary $n$.

This result implies that it is not possible to bosonize
the whole set of all $\mbox{N}^2$ vector currents
$J^{(I)}_\mu\; , \; I = 1,2, ...\mbox{N}^2$ linearly
in terms of free bosons (this was the reason why Witten \cite{witten2}
introduced
his {\it nonabelian bosonization }).
The bosonization sketched above ({\it abelian bosonization}) can be
done only for a Cartan
subalgebra where all generators commute,
and the two traces in Equation (6.71) cancel. The last observation allows to
prove a second theorem. \vskip3mm
\noindent
{\bf Theorem 6.2 :} \vskip3mm
\noindent
Any fully connected $n$-point function
\begin{equation}
\Big\langle J^{(1)}_{\mu_0}(x_0) J^{(I_1)}_{\mu_1}(x_1) J^{(I_2)}_{\mu_2}(x_2)
..... J^{(I_n)}_{\mu_n}(x_n) \Big\rangle_{0 \; c}^\theta
\end{equation}
containing the $U(1)$-current
vanishes, with the exception of the two point function
$\langle J^{(1)}_\mu(x) J^{(1)}_\nu(y) \rangle_{0 \; c}^\theta$
(without loss of generality I shifted
the U(1)-current in Equation (6.74) to the first position). \vskip3mm
\noindent
{\bf Proof :}\vskip3mm
\noindent
To prove the statement I use the same trick as for Theorem 6.1.
Again I consider the residues of functions $F_{n+1}$ defined by
\[
F_{n+1}(\tilde{x}_0,\tilde{x}_1,... \tilde{x}_n) :=
\]
\begin{equation}
\sum_{\pi(1,2,...n)} \mbox{Tr}\Big[H^{(1)} H^{(I_{\pi(1)})} .....
H^{(I_{\pi(n)})} \Big]
\frac{1}{\tilde{x}_0 - \tilde{x}_{\pi(1)}}
\frac{1}{\tilde{x}_{\pi(1)} - \tilde{x}_{\pi(2)}} .....
\frac{1}{\tilde{x}_{\pi(n)} - \tilde{x}_0}  \; .
\end{equation}
For the residue at $\tilde{x}_0 = \tilde{x}_1$ one obtains an equation
equivalent to (6.71). After making the same choice for $\pi^\prime$ one
finds
\[
\mbox{Res}_{\tilde{x}_0} [ F_{n+1},\tilde{x}_0 = \tilde{x}_1 ] =
\sum_{\pi(1,2,...n),\pi(1)=1}
\frac{1}{\tilde{x}_1 - \tilde{x}_{\pi(2)}}
\frac{1}{\tilde{x}_{\pi(2)} - \tilde{x}_{\pi(3)}} ...
\frac{1}{\tilde{x}_{\pi(n)} - \tilde{x}_1} \]
\begin{equation}
\left\{
\mbox{Tr}\Big[H^{(1)} H^{(I_1)} H^{(I_{\pi(2)})}  ...
H^{(I_{\pi(n)})} \Big] -
\mbox{Tr}\Big[H^{(1)} H^{(I_{\pi(2)})}  ...
H^{(I_{\pi(n)})} H^{(I_1)} \Big] \right\}  \; .
\end{equation}
Since $H^{(1)}$ commutes with all generators $H^{(I)}$, the two traces
are the same and cancel. Using the same argument one can show that
all residues at $\tilde{x}_1,\tilde{x}_2,.....\tilde{x}_n$ vanish.
So $F_{n+1}$ is analytic and bounded in the entire $\tilde{x_0}$ plane. By
Liouville's theorem $F_{n+1}$ is a constant, and the
limit $\tilde{x}_0 \longrightarrow \infty$ shows that this constant
is zero. Since the $n+1$-point function defined in (6.74) is
proportional to $F_{n+1}$, it has to vanish. $\Box$ \vskip3mm
\noindent
Theorem 6.2 allows to prove the following proposition about the
structure of the Hilbert space.\vskip3mm
\noindent
{\bf Proposition 6.1 :} \vskip3mm
\noindent
The Hilbert space $\cal H$ generated by the vector currents from the
vacuum is the tensor product
\begin{equation}
{\cal H} = {\cal H}_{U(1)} \otimes {\cal H}_{\mbox{mass=0}} \; .
\end{equation}\vskip3mm
\noindent
{\bf Proof :}
\vskip3mm
\noindent
To prove this one uses the connection between untruncated and
fully connected $n$-point functions (see e.g. \cite{glimm}).
\begin{equation}
\langle \phi_1 ..... \phi_n \rangle =
\sum_{\pi \in {\cal P}_n} \prod_{p \in \pi}
\langle \phi_{i_1} ..... \phi_{i_{\mid p \mid}} \rangle_c \; ,
\end{equation}
where ${\cal P}_n$ is the set of all partitions of $\{1,2,...n\}$,
$\pi = \{ p_1, p_2, ... p_{\mid \pi \mid} \}$ denotes an element of
${\cal P}_n$, and $\{i_1,i_2,...i_{\mid p \mid} \}$ is an element $p$ of
$\pi$.
An arbitrary $n+k$-point function (without loss of
generality I write the U(1)-currents first;
$I_i \neq 1 , i = 1,2,...k$) factorizes
due to Theorem 6.2
\[
\Big\langle J^{(1)}_{\mu_1}(x_1) J^{(1)}_{\mu_2}(x_2) .....
J^{(1)}_{\mu_n}(x_n) J^{(I_1)}_{\nu_1}(y_1)
J^{(I_2)}_{\nu_2}(y_2) ..... J^{(I_k)}_{\nu_k}(y_k) \Big\rangle_0^\theta \]
\[
= \; \sum_{\pi \in {\cal P}_{n+k}} \prod_{p \in \pi}
\Big\langle
J^{(1)}_{\mu_{i_1}}(x_{i_1})
J^{(1)}_{\mu_{i_2}}(x_{i_2}) .....
J^{(1)}_{\mu_{i_{\mid p \mid}}}(x_{i_{\mid p \mid}})
J^{(I_{j_1})}_{\nu_{j_1}}(y_{j_1})
J^{(I_{j_2})}_{\nu_{j_2}}(y_{j_2}) .....
J^{(I_{j_k})}_{\nu_{j_{\mid p \mid}}}(y_{j_{\mid p \mid}})
\Big\rangle_{0 \; c}^\theta \]
\[
= \; \left[ \sum_{\pi \in {\cal P}_n} \prod_{p \in \pi}
\Big\langle
J^{(1)}_{\mu_{i_1}}(x_{i_1})
J^{(1)}_{\mu_{i_2}}(x_{i_2}) .....
J^{(1)}_{\mu_{i_{\mid p \mid}}}(x_{i_{\mid p \mid}})
\Big\rangle_{0 \; c}^\theta \right]\]
\[
\times
\left[ \sum_{\pi^\prime \in {\cal P}_{k}} \prod_{p^\prime \in \pi^\prime}
\Big\langle
J^{(I_{j_1})}_{\nu_{j_1}}(y_{j_1})
J^{(I_{j_2})}_{\nu_{j_2}}(y_{j_2}) .....
J^{(I_{j_k})}_{\nu_{j_{\mid p\prime \mid}}}
(y_{j_{\mid p^\prime \mid}}) \Big\rangle_{0 \; c}^\theta \right] \]
\begin{equation}
= \; \Big\langle J^{(1)}_{\mu_1}(x_1) J^{(1)}_{\mu_2}(x_2) .....
J^{(1)}_{\mu_n}(x_n) \Big\rangle \;\;
\Big\langle J^{(I_1)}_{\nu_1}(y_1)
J^{(I_2)}_{\nu_2}(y_2) ..... J^{(I_k)}_{\nu_k}(y_k) \Big\rangle_0^\theta \; .
\end{equation}
{}From this the tensor product structure of the Hilbert space follows
easily. $\Box$ \vskip3mm
\noindent
It was pointed out in Equations (6.64), (6.65)
that the dependence on the gauge field of the fully connected
correlations cancels entirely.
Hence the vector currents in the 'massless' sector of the Hilbert space
obey the same algebra as in a system of uncoupled fermions; this is the
well-known level 1 representation of the
$\mbox{SU(N)}_L\times \mbox{SU(N)}_R$ current
(Kac-Moody) algebra (see for instance \cite{goddard}).

%
%

\chapter{The generalized Sine Gordon model}
The Cartan currents can be bosonized
in the massive model as well. This gives rise to a generalized
Sine Gordon model that will be identified in the first section of this
chapter. Furthermore I will show that the expansion in
terms of the fermion masses
converges if a space-time cutoff is introduced. It will
be argued in Section 7.3 that the known methods to remove
the cutoff fail. The spectrum of the model will be
discussed semiclassically in Section 7.4, and a Witten-Veneziano
formula will be shown to hold in 7.5 in this approximation.
%
%
\section{Definition of the model}
In Chapter 6 it was shown that it is possible
to find a common
bosonization of the vector currents together with
the chiral densities which show up in
the mass perturbation series.
The task of this section is to identify
the bosonic model in which the Cartan currents of the massive model
are bosonized by the prescription (6.37). For the one flavor case ($N=1$)
this is provided by the Sine Gordon model \cite{coleman1}.
Thus for $N>1$ one expects
a generalization.

As has been outlined in Chapter 3.4 the strategy for the construction of the
massive model is to sum
up the expansion of the mass term of the action. If one is
interested in generating functionals for Cartan currents, the formula
corresponding to (3.37) reads
\[
F(A^{(I)}_\mu) := \frac{1}{Z}  \Bigg\langle
\exp \left( - S_M [ \overline{\psi}, \psi ] \right) \;
\exp \left( i e \sum_{I=1}^N \Big( A^{(I)}_\mu, J^{(I)}_\mu \Big) \right)
\Bigg\rangle_0^\theta
\]
\[
= \; \frac{1}{Z} \sum_{n=0}^\infty
\frac{(-1)^n}{n!} \Bigg\langle  \Big( S_M [ \overline{\psi}, \psi ] \Big)^n \;
\exp \left( i e \sum_{I=1}^N \Big( A^{(I)}_\mu, J^{(I)}_\mu \Big) \right)
\Bigg\rangle_0^\theta
\]
\[
\; = \frac{1}{Z} \sum_{n=0}^\infty
\frac{1}{n!} \Bigg\langle
\left( \sum_{b=1}^N m^{(b)} \int_\Lambda d^2 x \; t(x) \Big[
\overline{\psi}^{(b)}(x) P_+ \psi^{(b)}(x) +
\overline{\psi}^{(b)}(x) P_- \psi^{(b)}(x) \Big] \right)^n \]
\begin{equation}
\exp \left( i e \sum_{I=1}^N \Big( A^{(I)}_\mu, J^{(I)}_\mu \Big) \right)
\Bigg\rangle_0^\theta  \; .
\end{equation}
Of course also $Z$ has to be expanded in that way.
Obviously all the expansion coefficients are of the form of the
generalized generating functional $E(n_b,m_b;A^{(b)})$ (c.f. (6.1))
which was entirely bosonized in Section 6.2.
Thus one can insert the bosonization prescriptions (6.37), (6.38) and obtain
\[
F(A^{(I)}_\mu) =
\frac{1}{Z} \sum_{n=0}^\infty
\frac{1}{n!} \Bigg\langle \;
\Bigg( \frac{1}{2\pi} \sum_{b=1}^N m^{(b)} c^{(b)} \int_\Lambda
d^2 x \; t(x) \;
\]
\[
\Bigg[
: e^{- i 2 \sqrt{\pi} \sqrt{\frac{\pi}{\pi\!+\!gN}} U_{1 b} \Phi^{(1)}(x)}
:_{M^{(1)}} \; \prod_{I=2}^N
: e^{- i 2 \sqrt{\pi} U_{I b} \Phi^{(I)}(x)}
:_{M^{(I)}} e^{+ i \frac{\theta}{N}}
\]
\[
+ \; : e^{+ i 2 \sqrt{\pi} \sqrt{\frac{\pi}{\pi +gN}} U_{1 b} \Phi^{(1)}(x)}
:_{M^{(1)}} \; \prod_{I=2}^N
: e^{+ i 2 \sqrt{\pi} U_{I b} \Phi^{(I)}(x)}
:_{M^{(I)}} e^{- i \frac{\theta}{N}} \; \Bigg] \Bigg)^n
\]
\[
\exp \left( - i e \Bigg[ \frac{1}{\sqrt{\pi\!+\!gN}}
\Big( A^{(1)}_\mu, \varepsilon_{\nu \mu} \partial_\nu \Phi^{(1)} \Big)  +
\sum_{I=2}^N \frac{1}{\sqrt{\pi}}
\Big( A^{(I)}_\mu, \varepsilon_{\nu \mu} \partial_\nu \Phi^{(I)} \Big) \Bigg]
\right) \Bigg\rangle_{\{K^{(I)}\}} \; =
\]
\begin{equation}
\Bigg\langle \!\exp \left( - S_{int} [ \Phi^{(I)} ]
- i e \Bigg[ \frac{1}{\sqrt{\pi\!+\!gN}}
\Big( A^{(1)}_\mu, \varepsilon_{\nu \mu} \partial_\nu \Phi^{(1)} \Big)  +
\sum_{I=2}^N \frac{1}{\sqrt{\pi}}
\Big( A^{(I)}_\mu, \varepsilon_{\nu \mu} \partial_\nu \Phi^{(I)} \Big) \Bigg]
\right)\!\Bigg\rangle_{\!\{K^{(I)}\}} .
\end{equation}
In the last step I summed up the formal expansion in the quark
masses in the bosonized model. The convergence of the series will be
proven in the next section justifying the expansion. The part
of the action $S_{int} [\Phi^{(I)}]$ that describes the
interaction of the fermions
can be read off as
\[
S_{int} [ \Phi^{(I)} ] \; := \;
- \frac{1}{\pi} \sum_{b=1}^N m^{(b)} c^{(b)} \int_\Lambda d^2 x \; t(x)
\]
\begin{equation}
: \cos \left(
2 \sqrt{\pi} \sqrt{\frac{\pi}{\pi\!+\!gN}} U_{1 b} \Phi^{(1)}(x)
+ 2\sqrt{\pi} \sum_{I=2}^N U_{I b} \Phi^{(I)}(x) - \frac{\theta}{N} \right) :
\; .
\end{equation}
The Wick ordering of the cosine is understood in the way it is defined
in the
perturbation expansion (7.2).

I conclude this section with displaying the classical Lagrangian
${\cal L}_{GSG}$
which corresponds to the newly defined model. It can be read off
from the $K^{(I)}$ (see (6.35) and (6.36)) and $S_{int}$
\[
{\cal L}_{GSG} =
+ \frac{1}{2} \sum_{I=1}^N \partial_\mu \Phi^{(I)} \partial_\mu \Phi^{(I)}
+ \frac{1}{2} \Big( \Phi^{(1)} \Big)^2 \; \frac{e^2 N}{\pi+gN}
\]
\begin{equation}
- \; \frac{1}{\pi} \sum_{b=1}^N m^{(b)} c^{(b)}
\cos \left(
2 \sqrt{\pi} \sqrt{\frac{\pi}{\pi\!+\!gN}} U_{1 b} \Phi^{(1)}
+ 2\sqrt{\pi} \sum_{I=2}^N U_{I b} \Phi^{(I)} - \frac{\theta}{N} \right)
\; .
\end{equation}
The model which is described by this Lagrangian will be refered to as the
{\it Generalized Sine Gordon model} (GSG).

At this point I draw another lesson that recovers a property
of the $\theta$-vacuum in QCD.
\vskip3mm
\noindent
{\bf Lesson 3 :}
\vskip3mm
\noindent
{\it Physics does not depend on $\theta$ if at least one of the
fermion masses vanishes.}
\vskip3mm
\noindent
This property of the QCD $\theta$-vacuum was discussed in Section 2.2.
In $\mbox{QED}_2$ it can be seen by the following arguments.

Without loss of generality $m^{(2)}$ can be set to zero. Using
\begin{equation}
U_{N 1} \; = \; \frac{1}{\sqrt{2}} \; \; \; , \; \; \;
U_{N 2} \; = \; \frac{-1}{\sqrt{2}} \; \; \; , \; \; \;
U_{N b} \; = \; 0 \; \; \; \; \mbox{for} \; \; 3 \leq b \leq N \; \; ,
\end{equation}
for $n \geq 2$ (compare (B.17)), one obtains for the interaction term (7.3)
\[
\sum_{b=1}^N m^{(b)} c^{(b)}
: \cos \left(
2 \sqrt{\pi} \sqrt{\frac{\pi}{\pi\!+\!gN}} U_{1 b} \Phi^{(1)}(x)
+ 2\sqrt{\pi} \sum_{I=2}^N U_{I b} \Phi^{(I)}(x) - \frac{\theta}{N} \right) :
\]
\[
= \; m^{(1)} c^{(1)} : \cos \Bigg(
2 \sqrt{\pi} \sqrt{\frac{\pi}{\pi\!+\!gN}} U_{1 b} \Phi^{(1)}(x)
+ 2\sqrt{\pi} \sum_{I=2}^{N-1} U_{I b} \Phi^{(I)}(x)
\]
\[
+ \; 2\sqrt{\pi} \frac{1}{\sqrt{2}}\Phi^{(N)}(x)
- \frac{\theta}{N} \Bigg) :
\]
\begin{equation}
+ \;
\sum_{b=3}^N m^{(b)} c^{(b)}
: \cos \left(
2 \sqrt{\pi} \sqrt{\frac{\pi}{\pi\!+\!gN}} U_{1 b} \Phi^{(1)}(x)
+ 2\sqrt{\pi} \sum_{I=2}^{N-1} U_{I b} \Phi^{(I)}(x) -
\frac{\theta}{N} \right) :
\; .
\end{equation}
Since $\Phi^{(N)}$ is a massless field and shows up only in
the first term on the right hand side of (7.6) it can be shifted by a constant
in order to change $\theta$. If none of the masses vanishes, $\Phi^{(N)}$
enters the interaction term twice but with different sign, as can
be seen from (7.5). The value of $\theta$ cannot be changed then, and
physics depends on it.

%
%
\section{Convergence of the mass perturbation series}
In this section it is proven that the mass perturbation series converges
if a space-time cutoff $\Lambda$ is imposed. The proof follows
mainly the strategy developed by Fr\"ohlich
for the one flavor case \cite{frohlich}, \cite{frohlich2}.
At several points some extra work has to be done, and it turns out
that the equations for more than one flavor become rather monstrous.
In other words, this section is of a more technical nature.

\noindent
The interaction term under consideration can be rewritten as (compare (7.3))
\begin{equation}
U_\Lambda \; := \; - 2 \sum_{b=1}^N \beta^{(b)} \int_{\Lambda} d^2x \;
t(x)\; : \cos \left( 2 \sqrt{\pi} \sum_{I=1}^N U_{I b} \Phi^{(I)}(x) \; - \;
\frac{\theta}{N} \right) :_{K^W} \; .
\end{equation}
$\Lambda$ denotes a finite rectangle in $\mbox{I\hspace{-0.62mm}R}^2$
and $t$ is some test function with
$\sup_{x \in \Lambda} \; \leq \; 1$.
$\beta^{(b)}$ are real positive
coefficients (compare (7.3), (6.51)).
Wick ordering is with respect to
\begin{equation}
K^W \; := \; \mbox{diag} \left(
\frac{\pi}{\pi + gN} \frac{1}{-\triangle+\frac{e^2N}{\pi + gN}} \; , \;
\frac{1}{-\triangle +1}\; , \; . \; . \; . \; . \; , \;
\frac{1}{-\triangle +1} \right) \; .
\end{equation}
The covariance for the expectation values reads
\begin{equation}
K^\mu \; := \; \mbox{diag} \left(
\frac{\pi}{\pi + gN} \frac{1}{-\triangle+\frac{e^2N}{\pi + gN}} \; , \;
\frac{1}{-\triangle +\mu}\; , \; . \; . \; . \; . \; , \;
\frac{1}{-\triangle +\mu} \right) \; ,
\end{equation}
and the limit $\mu \; \longrightarrow \; 0$ is taken in the end. Note the
factor $\pi/(\pi+gN)$ in front of the first entry of $K^\mu$ and $K^W$.
It was included in the covariance in order to remove $\sqrt{\pi/(\pi+gN)}$
from the first argument of the cosine (compare (7.3) and (7.7)).
It will be proven
\[
e^{-U_\Lambda} \; \; \; \mbox{is integrable with respect to} \; \;
d\mu_{K^0}[\Phi] \; \; ,
\]
\begin{equation}
\mbox{ and the mass perturbation series converges.}
\end{equation}
where the covariance $K^0$ formally denotes the limit $\mu \; \longrightarrow
\; 0$ taken in the end, and the exponential function is understood in the
sense of its expansion. In particular one has to prove that the series
converges. Since the proof is rather lengthy, I decided to divide
it into several steps.
\vskip3mm
\noindent
{\bf Step 1 :} Convenient notation
\vskip3mm
\noindent
The diagonal covariance (7.9) can easily be interpreted from a quantum field
theoretical point of view, but is less suitable for the proof below.
Define new fields $\varphi^{(b)}\; , \; \; b \; = \; 1,2, \; ... \; , N \;$ by
\begin{equation}
\varphi^{(b)} \; := \; \sum_{I=1}^N U_{I b} \; \Phi^{(I)} \; .
\end{equation}
Using the orthogonality of the matrix $U$ (compare Appendix B.3) one
immediately finds that (7.10) is equivalent to showing that
\begin{equation}
{\cal E} \; := \;
\exp \left( \;  2 \sum_{b=1}^N \beta^{(b)} \int_{\Lambda} d^2x \;
t(x)\; : \cos \left( 2 \sqrt{\pi} \varphi^{(b)}(x) \; - \;
\frac{\theta}{N} \right) :_{Q^W} \; \right) \; ,
\end{equation}
is integrable with respect to $d\mu_{Q^0}[\varphi]$ where
\begin{equation}
Q^\mu_{a b} \; := \; \sum_{I=1}^N U_{Ia} U_{Ib} K^\mu_{II} \; \; ,
\end{equation}
and
\begin{equation}
Q^W_{a b} \; := \; \sum_{I=1}^N U_{Ia} U_{Ib} K^W_{II} \; \; .
\end{equation}
\vskip3mm
\noindent
{\bf Step 2 :} Dependence on $\theta$
\vskip3mm
\noindent
Define
\begin{equation}
\chi^{(b)}_\pm [ C ] \; := \; \int_{\Lambda} d^2x \; t(x) \;
: e^{\pm i2 \sqrt{\pi} \; \varphi^{(b)}(x)} :_{C} \; \; ,
\end{equation}
and
\begin{equation}
\sigma_\pm [ C ] \; := \; \sum_{a=1}^N \; \beta^{(a)} \;
\chi^{(a)}_\pm [ C ] \; ,
\end{equation}
where $C$ is some covariance. Thus
\[
{\cal E} = \; \sum_{n=0}^\infty \; \frac{1}{n!}
\left\langle \left[ e^{-i\frac{\theta}{N}} \sigma_+[Q^W] \; +
\;e^{+i\frac{\theta}{N}} \sigma_-[Q^W] \right]^n \right\rangle_{Q^\mu}
\]
\[
= \; \sum_{n=0}^\infty \; \frac{1}{n!}
\sum_{q=0}^n \; { n \choose q} e^{-i\frac{\theta}{N} q}
e^{+i\frac{\theta}{N}(n-q)}
\left\langle \left[ \sigma_+[Q^W] \right]^q
\left[ \sigma_-[Q^W] \right]^{n-q} \right\rangle_{Q^\mu}
\]
\[ = \; \sum_{n=0}^\infty \; \frac{1}{n!} \frac{1}{2} \Bigg\{
\sum_{q=0}^n \; { n \choose q} e^{-i\frac{\theta}{N}2 q}
e^{+i\frac{\theta}{N}(n-q)}
\left\langle \left[ \sigma_+[Q^W] \right]^q
\left[ \sigma_-[Q^W] \right]^{n-q} \right\rangle_{Q^\mu}
\]
\[
+  \;\sum_{r=0}^n \; { n \choose n-r} e^{-i\frac{\theta}{N}r}
e^{+i\frac{\theta}{N}(n-r)}
\left\langle \left[ \sigma_+[Q^W] \right]^r
\left[ \sigma_-[Q^W] \right]^{n-r} \right\rangle_{Q^\mu} \Bigg\}
\]
\begin{equation}
= \; \sum_{n=0}^\infty \; \frac{1}{n!}
\sum_{q=0}^n \; { n \choose q} \cos\Big(\frac{\theta}{N}(n-2q)\Big)
\left\langle \left[ \sigma_+[Q^W] \right]^q
\left[ \sigma_-[Q^W] \right]^{n-q} \right\rangle_{Q^\mu} \; .
\end{equation}
In the last step I made use of
${ n \choose n-r} = { n \choose r}$,
performed a transformation of the summation index $q := n - r$
in the second sum, and applied
\begin{equation}
\left\langle \left[ \sigma_+[Q^W] \right]^q
\left[ \sigma_-[Q^W] \right]^{n-q} \right\rangle_{Q^\mu}
\; = \; \left\langle \left[ \sigma_+[Q^W] \right]^{n-q}
\left[ \sigma_-[Q^W] \right]^q \right\rangle_{Q^\mu} \; .
\end{equation}
The latter can be seen to hold from a transformation
$\varphi \; \longrightarrow \; - \varphi$ which transforms
$\sigma_\pm[Q^W] \; \longrightarrow \; \sigma_\mp[Q^W]$ but
leaves the expectation value invariant. Thus ${\cal E}$
can always be bounded by the $\theta = 0$ expression
\begin{equation}
\left\langle \exp \left( \;  2 \sum_{b=1}^N \beta^{(b)} \int_{\Lambda} d^2x \;
t(x)\; : \cos \left( 2 \sqrt{\pi} \varphi^{(b)}(x) \;
\right) :_{Q^W} \; \right) \right\rangle \; .
\end{equation}
All the expectation values involved are real, since
the coefficients $\beta^{(b)}$ and also the Gaussian
integrals (see (A.21)) are real.
\vskip3mm
\noindent
{\bf Step 3 :} cosh-bound
\vskip3mm
\noindent
The following estimate can be seen to hold if
the cosh is expressed in terms of exponentials
\[
\left\langle
\exp \left( \; 2 \sum_{b=1}^N \beta^{(b)} \int_{\Lambda} d^2x \;
t(x)\; : \cos \left( 2 \sqrt{\pi} \varphi^{(b)}(x) \right) :_{Q^W} \; \right)
\right\rangle_{Q^\mu} \]
\[ \leq \;
\left\langle \prod_{b=1}^N
2 \cosh\left( \; 2 \beta^{(b)} \int_{\Lambda} d^2x \;
t(x)\; : \cos \left( 2 \sqrt{\pi} \varphi^{(b)}(x)
\right) :_{Q^W} \; \right) \;
\right\rangle_{Q^\mu}
\]
\[
:= \; \left\langle \prod_{b=1}^N
2 \left( \sum_{n_b=0}^\infty \frac{1}{(2 n_b)!}
\Big(\beta^{(b)}\Big)^{2n_b} \;
\left[\chi^{(b)}_+[Q^W] + \chi^{(b)}_-[Q^W]\right]^{2n_b} \right)
\right\rangle_{Q^\mu}
\]
\begin{equation}
= \; 2^N \prod_{a=1}^N \sum_{n_a=0}^\infty
\frac{(\beta^{(a)})^{2 n_a}}{(2 n_a)!}
\prod_{b=1}^N
\sum_{q_b=0}^{2n_b} {2 n_b \choose q_b}
E^\mu(\{q,n\}) \; ,
\end{equation}
where
\begin{equation}
E^\mu(\{q,n\}) \; := \;
\left\langle \prod_{b=1}^N
\left[\chi^{(b)}_+ [ Q^W ] \right]^{q_b}
\left[\chi^{(b)}_- [ Q^W ] \right]^{2n_b-q_b}
\right\rangle_{Q^\mu} \; .
\end{equation}
\vskip3mm
\noindent
{\bf Step 4 :} Change of Wick ordering
\vskip3mm
\noindent
Change of Wick ordering means (compare Appendix A.4)
\begin{equation}
: e^{i2\sqrt{\pi} \; \varphi^{(b)}(x)} :_{Q^{old}} \; \; = \; \;
: e^{i2\sqrt{\pi} \; \varphi^{(b)}(x)} :_{Q^{new}} \;
\lim_{z\rightarrow 0}
e^{\frac{1}{2} 4 \pi [ Q^{old}_{bb}(z) - Q^{new}_{bb}(z) ]} \; .
\end{equation}
Now I set
\begin{equation}
Q^{old} \; := \; Q^W \; \; \; \; \; \; , \; \; \; \; \; \;
Q^{new} \; := \; Q^\mu \; .
\end{equation}
One has to evaluate
\[
Q^{old}_{bb}(z) - Q^{new}_{bb}(z) \; = \;
\sum_{I=1}^N U_{Ib} U_{Ib} \Big[ K^W_{II}(z) - K^\mu_{II}(z) \Big]
\]
\begin{equation}
= \; \sum_{I=1}^N U_{Ib} U_{Ib} (1-\delta_{I1})
\Big[ \frac{1}{4\pi}\ln(\mu^2) + O(z^2) \Big] \; ,
\end{equation}
where I used (7.8), (7.9) and (A.43) in the last step. Thus one obtains
\begin{equation}
: e^{i2\sqrt{\pi} \; \varphi^{(b)}(x)} :_{Q^W} \; = \;
: e^{i2\sqrt{\pi} \; \varphi^{(b)}(x)} :_{Q^\mu} \; \;
\Big( \mu \Big)^{\; \sum_{I=2}^N (U_{Ib})^2} \; .
\end{equation}
Inserting this in (7.21) one ends up with
\begin{equation}
E^\mu(\{q,n\}) \; = \;
\prod_{a=1}^N
\Big( \mu \Big)^{\; 2n_{a} \sum_{I=2}^N (U_{Ia})^2}
\left\langle \prod_{b=1}^N
\left[\chi^{(b)}_+ [ Q^\mu ] \right]^{q_b}
\left[\chi^{(b)}_- [ Q^\mu ] \right]^{2n_b-q_b}
\right\rangle_{Q^\mu} \; .
\end{equation}
\vskip3mm
\noindent
{\bf Step 5 :} Inverse conditioning
\vskip3mm
\noindent
In this step the inverse conditioning formula (B.22) from Appendix B.4
is applied for
\begin{equation}
C^1 \; := \; Q^\mu \; \; \; \; \; \; , \; \; \; \; \; \;
C^2 \; := \; Q^W \; .
\end{equation}
Thus one has to check $C^1 \geq C^2$. Obviously
\[
\Big( \; f \; , \; \Big[ C^1 - C^2 \Big] \; f \; \Big) \; = \;
\Big( \; f \; , \; U^T \; \Big[ K^\mu - K^W \Big] \; U \; f \; \Big)
\]
\begin{equation}
= \; \Big( \; g \; , \; \mbox{diag} (0, 1, \; ... \; 1)
\Big[\frac{1}{-\triangle + \mu^2} - \frac{1}{-\triangle + 1}\Big]
\; g \; \Big) \; \geq 0 \; ,
\end{equation}
for $\mu^2 \leq 1$ and inverse conditioning can be applied.
$g$ was defined as $g = U f$ and $f$ denotes an arbitrary vector of
test functions. Formula (B.22) gives
\[
\left\langle \prod_{b=1}^N
\left[\chi^{(b)}_+ [ Q^\mu ] \right]^{q_b}
\left[\chi^{(b)}_- [ Q^\mu ] \right]^{2n_b-q_b}
\right\rangle_{Q^\mu} \; \]
\begin{equation}
\leq \; \left\langle \prod_{b=1}^N
\left[\chi^{(b)}_+ [ Q^W ] \right]^{q_b}
\left[\chi^{(b)}_- [ Q^W ] \right]^{2n_b-q_b}
\right\rangle_{Q^W} \;
e^{4 \pi \sum_{b=1}^N 2n_b \lambda^{(b)}} \; \; .
\end{equation}
for $\mu^2 \leq 1$. One has to compute
\[
\lambda^{(b)} \; = \; \lim_{z\rightarrow 0} \frac{1}{2}
\Big[ C^1_{bb}(z) - C^2_{bb}(z) \Big]  \; = \;
\lim_{z\rightarrow 0} \frac{1}{2} \sum_{I=1}^N \Big( U_{Ib} \Big)^2 \;
\Big[ K^\mu_{II}(z) - K^W_{II}(z) \Big]
\]
\begin{equation}
= \; -\frac{1}{2} \; \sum_{I=2}^N \Big( U_{Ib} \Big)^2 \frac{1}{4\pi}
\ln \Big( \mu^2 \Big) \; ,
\end{equation}
where (7.27), (7.13), (7.14) and (A.43) from the
appendix were used in the last step.
Thus the power of $\mu$ that emerges from the change of Wick ordering
in (7.26) is eaten up by the power that comes from inverse conditioning, as
can be seen from (7.29). Thus the combination
of changing the Wick ordering and inverse conditioning leads to an
upper bound for ${\cal E}$ which does not depend on $\mu$ any longer
\[
\left\langle \prod_{b=1}^N
2 \cosh\left( \; 2 \beta^{(b)} \int_{\Lambda} d^2x \;
t(x)\; : \cos \left( 2 \sqrt{\pi} \varphi^{(b)}(x)
\right) :_{Q^W} \; \right) \;
\right\rangle_{Q^W}
\]
\begin{equation}
= \; 2^N \prod_{a=1}^N \sum_{n_a=0}^\infty
\frac{(\beta^{(a)})^{2 n_a}}{(2 n_a)!}
\prod_{b=1}^N
\sum_{q_b=0}^{2n_b} {2 n_b \choose q_b}
\left\langle \prod_{c=1}^N
\left[\chi^{(c)}_+ [ Q^W ] \right]^{q_c}
\left[\chi^{(c)}_- [ Q^W ] \right]^{2n_c-q_c}
\right\rangle_{Q^W} .
\end{equation}
\vskip3mm
\noindent
{\bf Step 6 :} Conditioning
\vskip3mm
\noindent
Define
$M^2 \; := \; \mbox{min}\big( 1 , e^2 N/(\pi + gN)\big)$
which implies
\begin{equation}
\frac{1}{-\triangle + M^2} \; \geq \;
\frac{1}{-\triangle + \frac{e^2}{\pi+gN}} \; \; \; \; \; \; , \; \; \; \; \;
\frac{1}{-\triangle + M^2} \; \geq \; \frac{1}{-\triangle + 1} \; .
\end{equation}
Define
\begin{equation}
K^M \; := \; \frac{1}{-\triangle + M^2} \; \mbox{diag} \left(
\frac{\pi}{\pi + gN} \; , \;
1 \; , \; . \; . \; . \; . \; , \;
1 \right) \; ,
\end{equation}
and
\begin{equation}
Q^M \; := \; U^T K^M U \; = \;
\frac{1}{-\triangle + M^2} \; \alpha \; .
\end{equation}
The matrix $\alpha$ is defined as (use (B.20) in the second step)
\begin{equation}
\alpha_{ab} \; := \;
\frac{\pi}{\pi\!+\!gN} U_{1a} U_{1b}
+  \sum_{I=2}^N U_{Ia} U_{Ib}  \; = \;
\frac{\pi}{\pi\!+\!gN} \frac{1}{N} \; + \;
\delta_{ab} - \frac{1}{N} \; = \;
\delta_{ab} - \frac{g}{\pi\!+\!gN} \; .
\end{equation}
Inspecting (7.32) and (7.34) immediately shows
(the argument is similar to (7.28))
that
$Q^M \; \geq \; Q^W$. Thus Corollary B.1
(Equation (B.30) in the conditioning appendix) can be applied for
$C^1 = Q^M$ and $C^2 = Q^W$, and the upper bound (7.31) is replaced by
\[
\left\langle \prod_{b=1}^N
2 \cosh\left( \; 2 \beta^{(b)} \int_{\Lambda} d^2x \;
t(x)\; : \cos \left( 2 \sqrt{\pi} \varphi^{(b)}(x)
\right) :_{Q^M} \; \right) \;
\right\rangle_{Q^M}
\]
\begin{equation}
= \; 2^N \prod_{a=1}^N  \sum_{n_a=0}^\infty
\frac{(\beta^{(a)})^{2 n_a}}{(2 n_a)!}
\prod_{b=1}^N
\sum_{q_b=0}^{2n_b} {2 n_b \choose q_b}
\left\langle \prod_{c=1}^N
\left[\chi^{(c)}_+ [ Q^M ] \right]^{q_c}
\left[\chi^{(c)}_- [ Q^M ] \right]^{2n_c-q_c}
\right\rangle_{Q^M} .
\end{equation}
\vskip3mm
\noindent
{\bf Step 7 :} Dirichlet boundary conditions
\vskip3mm
\noindent
Define a new covariance
\begin{equation}
Q^{M,S}_{ab} \; := \; \frac{1}{-\triangle_S + M^2} \;
\alpha_{ab} \; ,
\end{equation}
where $\triangle_S$ is the Laplace operator with zero Dirichlet data
on the circle $\partial S$, as is discussed in Appendix B.6.
Using (B.34) and Formula (7.37) one immediately infers
\begin{equation}
Q^M \; \geq \; Q^{M,S} \; ,
\end{equation}
and the inverse conditioning formula (B.22) can be applied.
\[
\lambda^{(b)} \; = \; \lim_{x\rightarrow y} \frac{1}{2}
\left[ Q_{bb}^M(x,y) - Q_{bb}^{M,S}(x,y) \right] \; =
\]
\begin{equation}
\Big( 1\!-\!\frac{g}{\pi\!+\!gN} \Big) \lim_{x\rightarrow y} \frac{1}{2}
\left[ \frac{1}{-\triangle\!+\!M^2}(x,y) -
\frac{1}{-\triangle_S\!+\!M^2}(x,y) \right] \; \leq \;
\Big( 1\!-\!\frac{g}{\pi\!+\!gN} \Big)\tilde{\omega} \; .
\end{equation}
In the first step the explicit form (7.35) of
$\alpha_{ab}$ was used. In the second step the bound (B.35)
was included. Defining now
\begin{equation}
\omega \; := \; \exp \left( 4 \pi \Big( 1 - \frac{g}{\pi + gN} \Big)
\tilde{\omega}
\right)
\end{equation}
one concludes from the inverse conditioning formula
\[
\left\langle \prod_{b=1}^N
\left[\chi^{(b)}_+ [ Q^M ] \right]^{q_b}
\left[\chi^{(b)}_- [ Q^M ] \right]^{2n_b-q_b}
\right\rangle_{Q^M} \;
\]
\begin{equation} \leq \;
\omega^{\sum_{a=1}^N 2 n_a} \; \left\langle \prod_{b=1}^N
\left[\chi^{(b)}_+ [ Q^{M,S} ] \right]^{q_b}
\left[\chi^{(b)}_- [ Q^{M,S} ] \right]^{2n_b-q_b}
\right\rangle_{Q^{M,S}} \; .
\end{equation}
Inserting this into (7.36) the upper bound for ${\cal E}$ now reads
\begin{equation}
\left\langle \prod_{b=1}^N
2 \cosh\left( \; 2 \omega\beta^{(b)} \int_{\Lambda} d^2x \;
t(x)\; : \cos \left( 2 \sqrt{\pi} \varphi^{(b)}(x)
\right) :_{Q^{M,S}} \; \right) \;
\right\rangle_{Q^{M,S}} \; .
\end{equation}
Using (B.33) one concludes (similar to (7.28))
\begin{equation}
Q^{0,S} \; \geq \; Q^{M,S} \; .
\end{equation}
Conditioning then gives the new upper bound
\[
\left\langle \prod_{b=1}^N
2 \cosh\left( \; 2 \omega \beta^{(b)} \int_{\Lambda} d^2x \;
t(x)\; : \cos \left( 2 \sqrt{\pi} \varphi^{(b)}(x)
\right) :_{Q^{0,S}} \; \right) \;
\right\rangle_{Q^{0,S}} \; =
\]
\begin{equation}
2^N \prod_{a=1}^N \sum_{n_a=0}^\infty
\frac{(\omega \beta^{(a)})^{2 n_a}}{(2 n_a)!}
\prod_{b=1}^N
\sum_{q_b=0}^{2n_b} {2 n_b \choose q_b}
\left\langle \prod_{c=1}^N
\left[\chi^{(c)}_+ [ Q^{0,S} ] \right]^{q_c}
\left[\chi^{(c)}_- [ Q^{0,S} ] \right]^{2n_c-q_c}
\right\rangle_{Q^{0,S}} \; .
\end{equation}
\vskip3mm
\noindent
{\bf Step 8 :} Reduction to neutral contributions
\vskip3mm
\noindent
Since $\chi^{(b)}_+[C] = \overline{\chi^{(b)}_-[C]}$ one can bound the
expectation values in (7.44) by their neutral contributions :
\[
\left\langle \prod_{b=1}^N
\left[\chi^{(b)}_+ [ Q^{0,S} ] \right]^{q_b}
\left[\chi^{(b)}_- [ Q^{0,S} ] \right]^{2n_b-q_b}
\right\rangle_{Q^{0,M}}  \]
\begin{equation}
\leq \; \left\langle \prod_{b=1}^N
\Bigg| \chi^{(b)}_+ [ Q^{0,S} ] \Bigg|^{q_b}
\Bigg| \chi^{(b)}_- [ Q^{0,S} ] \Bigg|^{2n_b-q_b}
\right\rangle_{Q^{0,S}}
=
\left\langle \prod_{b=1}^N
\bigg[\chi^{(b)}_+ [ Q^{0,S} ] \; \chi^{(b)}_- [ Q^{0,S} ] \bigg]^{n_b}
\right\rangle_{Q^{0,S}} \; .
\end{equation}
Using
\begin{equation}
\sum_{q_b = 1}^{2n_b} { 2n_b \choose q_b} \; = \; 2^{2 n_b} \; ,
\end{equation}
one ends up with (insert (7.45) in (7.44))
\begin{equation}
{\cal E} \; \leq \;
2^N \prod_{a=1}^N \sum_{n_a=0}^\infty
\frac{( 2 \omega \beta^{(a)})^{2 n_a}}{(2 n_a)!}
\left\langle \prod_{b=1}^N
\bigg[\chi^{(b)}_+ [ Q^{0,S} ] \chi^{(b)}_- [ Q^{0,S} ] \bigg]^{n_b}
\right\rangle_{Q^{0,S}} \; .
\end{equation}
\vskip3mm
\noindent
{\bf Step 9 :} Explicit evaluation
\vskip3mm
\noindent
\[
\left\langle \prod_{b=1}^N
\bigg[\chi^{(b)}_+ [ Q^{0,S} ] \chi^{(b)}_- [ Q^{0,S} ] \bigg]^{n_b}
\right\rangle_{Q^{0,S}} \;
\]
\[ = \;
\int_{\Lambda} \prod_{a=1}^N \left[ \prod_{i=1}^{n_a}
d^2x_{i}^{(a)} d^2y_{i}^{(a)} t(x_{i}^{(a)}) t(y_{i}^{(a)}) \right]
\]
\[ \times
\left\langle \prod_{b=1}^N \prod_{j=1}^{n_b}
\left[ : e^{+i 2\sqrt{\pi}\varphi^{(b)}(x_{j}^{(b)})}:_{Q^{0,S}}:
: e^{-i 2\sqrt{\pi}\varphi^{(b)}(y_{j}^{(b)})}:_{Q^{0,S}}
\right] \right\rangle_{Q^{0,S}} \; \leq
\]
\[
\int_{\Lambda} \prod_{a=1}^N \left[ \prod_{i=1}^{n_a}
d^2x_{i}^{(a)} d^2y_{i}^{(a)} \right]
\left\langle \prod_{b=1}^N \prod_{j=1}^{n_b}
\left[ : e^{+i 2\sqrt{\pi}\varphi^{(b)}(x_{j}^{(b)})}:_{Q^{0,S}}:
: e^{-i 2\sqrt{\pi}\varphi^{(b)}(y_{b}^{(b)})}:_{Q^{0,S}}
\right] \right\rangle_{Q^{0,S}}
\]
\[
= \; \int_{\Lambda} \prod_{a=1}^N \left[ \prod_{i=1}^{n_a}
d^2x_{i}^{(a)} d^2y_{i}^{(a)} \right] \exp \Bigg\{
-4 \pi \frac{1}{2} \sum_{b,c=1}^N \sum_{j=1}^{n_b} \sum_{l=1}^{n_c}
\Bigg[ \Big(1\!-\!\delta_{bc}\delta_{jl}\Big)
\]
\[
\times\bigg(
Q_{bc}^{0,S}(x_j^{(b)},x_l^{(c)})+Q_{bc}^{0,S}(y_j^{(b)},y_l^{(c)})
\bigg)
-Q_{bc}^{0,S}(x_j^{(b)},y_l^{(c)})-Q_{bc}^{0,S}(y_j^{(b)},x_l^{(c)}) \Bigg]
\Bigg\}
\]
\[
= \; \int_{\Lambda} \prod_{a=1}^N \left[ \prod_{i}^{n_a}
d^2x_{i}^{(a)} d^2y_{i}^{(a)} \right]
\]
\[
\times \prod_{b=1}^N e^{ -2\pi\big(1-\frac{g}{\pi+gN}\big)
\sum_{j=1}^{n_b}\sum_{l=1}^{n_b}\Big[\big(1-\delta_{jl}\big)
\big(C^{0,S}(x_j^{(b)},x_l^{(b)})+C^{0,S}(y_j^{(b)},y_l^{(b)})\big)
-2C^{0,S}(x_j^{(b)},y_l^{(b)}) \Big] }
\]
\begin{equation}
\times\prod_{c\neq d}^N e^{ 2\pi\frac{g}{\pi+gN}
\sum_{k=1}^{n_c} \sum_{h=1}^{n_d} \Big[
C^{0,S}(x_k^{(c)},x_h^{(d)})+
C^{0,S}(y_k^{(c)},y_h^{(d)})-
C^{0,S}(x_k^{(c)},y_h^{(d)})-
C^{0,S}(y_k^{(c)},x_h^{(d)})\Big] } \; .
\end{equation}
In the first step $\sup_{x \in \Lambda} t(x) \; \leq \; 1$ was used to
remove the test function $t$. Then the Gaussian integral was solved, and
finally $Q^{0,S}_{ab}(x,y) \; = \; C^{0,S}(x,y) \alpha_{ab}$ (compare
(7.37) and (7.35)) was inserted.
Using the explicit form (B.36) for the propagator $C^{0,S}$ one obtains
\[
e^{ -2\pi\big(1-\frac{g}{\pi+gN}\big)
\sum_{i=1}^{n_a}\sum_{j=1}^{n_a}\Big[\big(1\!-\!\delta_{ij}\big)
\big(C^{0,S}(x_i^{(a)},x_j^{(a)})+C^{0,S}(y_i^{(a)},y_j^{(a)})\big)
-2C^{0,S}(x_i^{(a)},y_j^{(a)}) \Big] }
\]
\[
= \; \Bigg| \frac{ \prod_{i<j}^{n_a}
\big( \tilde{x}^{(a)}_i - \tilde{x}^{(a)}_j \big)
\big( \tilde{y}^{(a)}_i - \tilde{y}^{(a)}_j \big)
\big( \hat{x}^{(a)}_i - \hat{x}^{(a)}_j \big)
\big( \hat{y}^{(a)}_i - \hat{y}^{(a)}_j \big)}{ \prod_{i,j=1}^{n_a}
\big( \tilde{x}^{(a)}_i - \tilde{y}^{(a)}_j \big)
\big( \hat{x}^{(a)}_i - \hat{y}^{(a)}_j \big)} \Bigg|^{[1-\frac{g}{\pi+gN}]}
\]
\begin{equation}
\times
\Bigg| \frac{\prod_{i,j=1}^{n_a}
\big( \hat{x}^{(a)}_i - \tilde{y}^{(a)}_j \big)
\big( \tilde{x}^{(a)}_i - \hat{y}^{(a)}_j \big)}{\prod_{i,j=1}^{n_a}
\big( \hat{x}^{(a)}_i - \tilde{x}^{(a)}_j \big)
\big( \tilde{y}^{(a)}_i - \hat{y}^{(a)}_j \big)} \Bigg|^{[1-\frac{g}{\pi+gN}]}
\times
\Bigg| \prod_{i=1}^{n_a}
\big( \hat{x}^{(a)}_i - \tilde{x}^{(a)}_i \big)
\big( \tilde{y}^{(a)}_i - \hat{y}^{(a)}_i \big) \Bigg|^{[1-\frac{g}{\pi+gN}]}
\; .
\end{equation}
For the definition of $\tilde{x}$ and $\hat{x}$ see (B.37) and (B.38).
The third factor on the right hand side can be bounded due to the
geometrical setting (see Figure B.1)
\begin{equation}
\Bigg| \prod_{i=1}^{n_a}
\big( \hat{x}^{(a)}_i - \tilde{x}^{(a)}_i \big)
\big( \tilde{y}^{(a)}_i - \hat{y}^{(a)}_i \big) \Bigg|^{[1-\frac{g}{\pi+gN}]}
\; \leq \; \left( 8^{[1-\frac{g}{\pi+gN}]} \right)^{2n_a} \; ,
\end{equation}
where I made use of (B.39).
Evaluating explicitely the last term in (7.48) gives $(a \neq b)$
\[
e^{ 2\pi\frac{g}{\pi+gN}
\sum_{i=1}^{n_a} \sum_{j=1}^{n_b} \Big[
C^{0,S}(x_i^{(a)},x_j^{(b)})+
C^{0,S}(y_i^{(a)},y_j^{(b)})-
C^{0,S}(x_i^{(a)},y_j^{(b)})-
C^{0,S}(y_i^{(a)},x_j^{(b)})\Big] }
\]
\[
= \;
\Bigg| \frac{\prod_{i=1}^{n_a}\prod_{j=1}^{n_b}
\big( \tilde{x}^{(a)}_i - \tilde{y}^{(b)}_j \big)
\big( \tilde{y}^{(a)}_i - \tilde{x}^{(b)}_j \big)
\big( \hat{x}^{(a)}_i - \hat{y}^{(b)}_j \big)
\big( \hat{y}^{(a)}_i - \hat{x}^{(b)}_j \big)}
{\prod_{i=1}^{n_a}\prod_{j=1}^{n_b}
\big( \tilde{x}^{(a)}_i - \tilde{x}^{(b)}_j \big)
\big( \tilde{y}^{(a)}_i - \tilde{y}^{(b)}_j \big)
\big( \hat{x}^{(a)}_i - \hat{x}^{(b)}_j \big)
\big( \hat{y}^{(a)}_i - \hat{y}^{(b)}_j \big)}
\Bigg|^{\frac{1}{2} \frac{g}{\pi+gN}} \]
\begin{equation}
\times
\Bigg|
\frac{\prod_{i=1}^{n_a}\prod_{j=1}^{n_b}
\big( \hat{x}^{(a)}_i - \tilde{x}^{(b)}_j \big)
\big( \hat{y}^{(a)}_i - \tilde{y}^{(b)}_j \big)
\big( \tilde{x}^{(a)}_i - \hat{x}^{(b)}_j \big)
\big( \tilde{y}^{(a)}_i - \hat{y}^{(b)}_j \big)}
{\prod_{i=1}^{n_a}\prod_{j=1}^{n_b}
\big( \hat{x}^{(a)}_i - \tilde{y}^{(b)}_j \big)
\big( \hat{y}^{(a)}_i - \tilde{x}^{(b)}_j \big)
\big( \tilde{x}^{(a)}_i - \hat{y}^{(b)}_j \big)
\big( \tilde{y}^{(a)}_i - \hat{x}^{(b)}_j \big)}
\Bigg|^{\frac{1}{2} \frac{g}{\pi+gN}} \; .
\end{equation}
Putting things together (use (7.48)-(7.51))
\[
\left\langle \prod_{b=1}^N
\left[\chi^{(b)}_+ [ Q^{0,S} ] \chi^{(b)}_- [ Q^{0,S} ] \right]^{n_b}
\right\rangle_{Q^{0,S}} \;
\]
\begin{equation}
\leq \;
\prod_{a=1}^N \left( 8^{[1-\frac{g}{\pi+gN}]} \right)^{2n_a} \;
\int_{\Lambda} \prod_{b=1}^N  \prod_{i=1}^{n_b}
d^2x_{i}^{(b)} d^2y_{i}^{(b)} \;
{\cal F}\Big( \{ x, y \} \Big) \; ,
\end{equation}
where I defined
\[
{\cal F}\Big( \{ x, y \} \Big) \; := \;
\prod_{a=1}^N
\Bigg| \frac{\prod_{i,j=1}^{n_a}
\big( \hat{x}^{(a)}_i - \tilde{y}^{(a)}_j \big)
\big( \tilde{x}^{(a)}_i - \hat{y}^{(a)}_j \big)}{\prod_{i,j=1}^{n_a}
\big( \hat{x}^{(a)}_i - \tilde{x}^{(a)}_j \big)
\big( \tilde{y}^{(a)}_i - \hat{y}^{(a)}_j \big)} \Bigg|^{[1-\frac{g}{\pi+gN}]}
\]
\[
\times \prod_{a=1}^N
\Bigg| \frac{ \prod_{i<j}^{n_a}
\big( \tilde{x}^{(a)}_i - \tilde{x}^{(a)}_j \big)
\big( \tilde{y}^{(a)}_i - \tilde{y}^{(a)}_j \big)
\big( \hat{x}^{(a)}_i - \hat{x}^{(a)}_j \big)
\big( \hat{y}^{(a)}_i - \hat{y}^{(a)}_j \big)}{ \prod_{i,j=1}^{n_a}
\big( \tilde{x}^{(a)}_i - \tilde{y}^{(a)}_j \big)
\big( \hat{x}^{(a)}_i - \hat{y}^{(a)}_j \big)} \Bigg|^{[1-\frac{g}{\pi+gN}]}
\]
\[
\times \prod_{a \neq b}^N
\Bigg| \frac{\prod_{i=1}^{n_a}\prod_{j=1}^{n_b}
\big( \tilde{x}^{(a)}_i - \tilde{y}^{(b)}_j \big)
\big( \tilde{y}^{(a)}_i - \tilde{x}^{(b)}_j \big)
\big( \hat{x}^{(a)}_i - \hat{y}^{(b)}_j \big)
\big( \hat{y}^{(a)}_i - \hat{x}^{(b)}_j \big)}
{\prod_{i=1}^{n_a}\prod_{j=1}^{n_b}
\big( \tilde{x}^{(a)}_i - \tilde{x}^{(b)}_j \big)
\big( \tilde{y}^{(a)}_i - \tilde{y}^{(b)}_j \big)
\big( \hat{x}^{(a)}_i - \hat{x}^{(b)}_j \big)
\big( \hat{y}^{(a)}_i - \hat{y}^{(b)}_j \big)}
\Bigg|^{\frac{1}{2} \frac{g}{\pi+gN}}
\]
\begin{equation}
\times \prod_{a \neq b}^N
\Bigg|
\frac{\prod_{i=1}^{n_a}\prod_{j=1}^{n_b}
\big( \hat{x}^{(a)}_i - \tilde{x}^{(b)}_j \big)
\big( \hat{y}^{(a)}_i - \tilde{y}^{(b)}_j \big)
\big( \tilde{x}^{(a)}_i - \hat{x}^{(b)}_j \big)
\big( \tilde{y}^{(a)}_i - \hat{y}^{(b)}_j \big)}
{\prod_{i=1}^{n_a}\prod_{j=1}^{n_b}
\big( \hat{x}^{(a)}_i - \tilde{y}^{(b)}_j \big)
\big( \hat{y}^{(a)}_i - \tilde{x}^{(b)}_j \big)
\big( \tilde{x}^{(a)}_i - \hat{y}^{(b)}_j \big)
\big( \tilde{y}^{(a)}_i - \hat{x}^{(b)}_j \big)}
\Bigg|^{\frac{1}{2} \frac{g}{\pi+gN}} \; .
\end{equation}
\vskip3mm
\noindent
{\bf Step 10 :} Cauchy's identity
\vskip3mm
\noindent
Define the following vectors
of complex space-time arguments
\begin{equation}
w^{(a)} := \left( \begin{array}{c}
\tilde{x}_1^{(a)} \\ . \\ . \\ . \\
\tilde{x}_{n^a}^{(a)} \\
\hat{y}_1^{(a)} \\ . \\ . \\ . \\
\hat{y}_{n^a}^{(a)}
\end{array} \right) \; , \;
z^{(a)} := \left( \begin{array}{c}
\tilde{y}_1^{(a)} \\ . \\ . \\ . \\
\tilde{y}_{n^a}^{(a)} \\
\hat{x}_1^{(a)} \\ . \\ . \\ . \\
\hat{x}_{n^a}^{(a)}
\end{array} \right) \; , \;
w^{(a,b)} := \left( \begin{array}{c}
\tilde{x}_1^{(a)} \\ . \\ . \\ . \\
\tilde{x}_{n^a}^{(a)} \\
\hat{y}_1^{(a)} \\ . \\ . \\ . \\
\hat{y}_{n^a}^{(a)} \\
\hat{x}_1^{(b)} \\ . \\ . \\ . \\
\hat{x}_{n^b}^{(b)} \\
\tilde{y}_1^{(b)} \\ . \\ . \\ . \\
\tilde{y}_{n^b}^{(b)} \end{array} \right) \; , \;
z^{(a,b)} := \left( \begin{array}{c}
\tilde{y}_1^{(a)} \\ . \\ . \\ . \\
\tilde{y}_{n^a}^{(a)} \\
\hat{x}_1^{(a)} \\ . \\ . \\ . \\
\hat{x}_{n^a}^{(a)} \\
\hat{y}_1^{(b)} \\ . \\ . \\ . \\
\hat{y}_{n^b}^{(b)} \\
\tilde{x}_1^{(b)} \\ . \\ . \\ . \\
\tilde{x}_{n^b}^{(b)} \end{array} \right) \; .
\end{equation}
Cauchy's identity (see \cite{deutsch}, compare (5.27))
\begin{equation}
\Bigg| { \; \atop { \mbox{det} \atop {\scriptstyle (i,j)} } }\Bigg(
\frac{1}{w_i - z_j} \Bigg) \Bigg| \; = \;
\Bigg|
\frac{\prod_{1 \leq i < j \leq n} (w_i - w_j )
( z_i - z_j ) }
{\prod_{i,j = 1}^n ( w_i - z_j ) } \Bigg| \; ,
\end{equation}
can be used to rewrite the integrand ${\cal F}$ in terms of
determinants. A straightforward computation gives
\[
\Bigg| { \; \atop { \mbox{det} \atop {\scriptstyle i,j=1,...,2n_a} } }\Bigg(
\frac{1}{w^{(a)}_i - z^{(a)}_j} \Bigg) \Bigg|
\]
\[
= \;
\Bigg| \frac{ \prod_{i<j}^{n_a}
\big( \tilde{x}^{(a)}_i - \tilde{x}^{(a)}_j \big)
\big( \tilde{y}^{(a)}_i - \tilde{y}^{(a)}_j \big)
\big( \hat{x}^{(a)}_i - \hat{x}^{(a)}_j \big)
\big( \hat{y}^{(a)}_i - \hat{y}^{(a)}_j \big)}{ \prod_{i,j=1}^{n_a}
\big( \tilde{x}^{(a)}_i - \tilde{y}^{(a)}_j \big)
\big( \hat{x}^{(a)}_i - \hat{y}^{(a)}_j \big)} \Bigg|
\]
\begin{equation}
\times
\Bigg| \frac{\prod_{i,j=1}^{n_a}
\big( \hat{x}^{(a)}_i - \tilde{y}^{(a)}_j \big)
\big( \tilde{x}^{(a)}_i - \hat{y}^{(a)}_j \big)}{\prod_{i,j=1}^{n_a}
\big( \hat{x}^{(a)}_i - \tilde{x}^{(a)}_j \big)
\big( \tilde{y}^{(a)}_i - \hat{y}^{(a)}_j \big)} \Bigg|
\; ,
\end{equation}
and $(a \neq b)$
\[
\Bigg|
{ \; \atop { \mbox{det} \atop {\scriptstyle i,j=1,...,2(n_a+n_b)} } }\Bigg(
\frac{1}{w^{(a,b)}_i - z^{(a,b)}_j} \Bigg) \Bigg|
\]
\[
= \;
\Bigg| { \; \atop { \mbox{det} \atop {\scriptstyle i,j=1,...,2n_a} } }\Bigg(
\frac{1}{w^{(a)}_i - z^{(a)}_j} \Bigg) \Bigg| \; \times \;
\Bigg| { \; \atop { \mbox{det} \atop {\scriptstyle i,j=1,...,2n_b} } }\Bigg(
\frac{1}{w^{(b)}_i - z^{(b)}_j} \Bigg) \Bigg|
\]
\[
\times
\Bigg| \frac{\prod_{i=1}^{n_a}\prod_{j=1}^{n_b}
\big( \tilde{x}^{(a)}_i - \tilde{y}^{(b)}_j \big)
\big( \tilde{y}^{(a)}_i - \tilde{x}^{(b)}_j \big)
\big( \hat{x}^{(a)}_i - \hat{y}^{(b)}_j \big)
\big( \hat{y}^{(a)}_i - \hat{x}^{(b)}_j \big)}
{\prod_{i=1}^{n_a}\prod_{j=1}^{n_b}
\big( \tilde{x}^{(a)}_i - \tilde{x}^{(b)}_j \big)
\big( \tilde{y}^{(a)}_i - \tilde{y}^{(b)}_j \big)
\big( \hat{x}^{(a)}_i - \hat{x}^{(b)}_j \big)
\big( \hat{y}^{(a)}_i - \hat{y}^{(b)}_j \big)}
\Bigg|
\]
\begin{equation}
\times
\Bigg|
\frac{\prod_{i=1}^{n_a}\prod_{j=1}^{n_b}
\big( \hat{x}^{(a)}_i - \tilde{x}^{(b)}_j \big)
\big( \hat{y}^{(a)}_i - \tilde{y}^{(b)}_j \big)
\big( \tilde{x}^{(a)}_i - \hat{x}^{(b)}_j \big)
\big( \tilde{y}^{(a)}_i - \hat{y}^{(b)}_j \big)}
{\prod_{i=1}^{n_a}\prod_{j=1}^{n_b}
\big( \hat{x}^{(a)}_i - \tilde{y}^{(b)}_j \big)
\big( \hat{y}^{(a)}_i - \tilde{x}^{(b)}_j \big)
\big( \tilde{x}^{(a)}_i - \hat{y}^{(b)}_j \big)
\big( \tilde{y}^{(a)}_i - \hat{x}^{(b)}_j \big)}
\Bigg| \; .
\end{equation}
Introducing the abbreviations
\begin{equation}
A(a,a) \; := \;
{ \; \atop { \mbox{det} \atop {\scriptstyle i,j=1,...,2n_a} } }\Bigg(
\frac{1}{w^{(a)}_i - z^{(a)}_j} \Bigg)  \; ,
\end{equation}
and
\begin{equation}
B(a,b) \; := \;
{ \; \atop { \mbox{det} \atop {\scriptstyle i,j=1,...,2(n_a+n_b)} } }\Bigg(
\frac{1}{w^{(a,b)}_i - z^{(a,b)}_j} \Bigg) \; ,
\end{equation}
one can write ${\cal F}$ in the handy form (compare (7.53))
\[
{\cal F} \; = \;
\left[\prod_{a=1}^N \; \Big| A(a,a) \Big|
\right]^{1 - \frac{g}{\pi+gN} - (N-1)\frac{g}{\pi+gN} }\;
\left[\prod_{a<b}^N \; \Big| B(a,b) \Big| \right]^{\frac{g}{\pi+gN}}
\]
\begin{equation}
= \;
\left[\prod_{a=1}^N \; \Big| A(a,a) \Big| \right]^{\frac{\pi}{\pi+gN}}\;
\left[\prod_{a<b}^N \; \Big| B(a,b) \Big| \right]^{\frac{g}{\pi+gN}} \; ,
\end{equation}
where I used
\begin{equation}
\Big| B(a,b) \Big| \; = \; \Big| B(b,a) \Big| \; ,
\end{equation}
which can be seen to hold from (7.54) and (7.59).
\vskip3mm
\noindent
{\bf Step 11 :} H\"older's inequality
\vskip3mm
\noindent
Let
\begin{equation}
\frac{1}{p} \; + \; \frac{1}{q} \; = \; 1 \; .
\end{equation}
Then the following chain of inequalities holds
\[
\int_{\Lambda} \prod_{b=1}^N \prod_{i=1}^{n_b}
d^2x_{i}^{(b)} d^2y_{i}^{(b)} \;
{\cal F}\Big( \{ x, y \} \Big)
\]
\[
\leq \;
\Bigg[ \int_{\Lambda} \prod_{a=1}^N \prod_{i=1}^{n_a}
d^2x_{i}^{(a)} d^2y_{i}^{(a)} \prod_{b=1}^N
\Big| A(b,b) \Big|^{\frac{\pi}{\pi+gN}p} \Bigg]^{\frac{1}{p}}
\]
\[
\times
\Bigg[ \int_{\Lambda} \prod_{c=1}^N \left[ \prod_{j=1}^{n_c}
d^2x_{j}^{(c)} d^2y_{j}^{(c)} \right] \prod_{d<e}^N
\Big| B(d,e) \Big|^{\frac{g}{\pi+gN}q} \Bigg]^{\frac{1}{q}}
\]
\[
\leq \;
\Bigg[ \int_{\Lambda} \prod_{a=1}^N \left[ \prod_{i=1}^{n_a}
d^2x_{i}^{(a)} d^2y_{i}^{(a)} \right] \prod_{a=1}^N
\Big| A(a,a) \Big|^{\frac{\pi}{\pi+gN}p} \Bigg]^{\frac{1}{p}}
\]
\begin{equation}
\times
\prod_{a<b}^N
\left[ \int_{\Lambda} \left[
\prod_{i=1}^{n_a} d^2x_{i}^{(a)} d^2y_{i}^{(a)}
\prod_{j=1}^{n_b} d^2x_{j}^{(b)} d^2y_{j}^{(b)} \right]
\Big| B(a,b) \Big|^{\frac{g}{\pi+gN}q(N-1)} \right]^{\frac{1}{q(N-1)}} \; .
\end{equation}
In the first step the usual H\"older inequality was used,
and in the second step I applied Corollary B.2
(Equations (B.48) - (B.51)) proven in Appendix B.7.

To apply the bounds on the integrals over the determinants
obtained in Appendix B.8, the exponents in (7.63) have to
obey the following inequalities
\begin{equation}
\frac{\pi}{\pi+gN} \; p \; \; <  \; \; 1 \; ,
\end{equation}
and
\begin{equation}
\frac{g}{\pi+gN}\; q \; (N-1) \; \; < \; \; 1 \; .
\end{equation}
Using (7.62) both restrictions can be rewritten in terms of $q$
leading to
\begin{equation}
\frac{1}{q} \; \in \;
\frac{g}{\pi + gN} \; \Big( (N-1) \; , \; N \Big) \; .
\end{equation}
Since the interval on the right hand side of (7.66) is not empty,
such a $q$ can be found and fixes $p$ via (7.62).
(B.57) then implies
\begin{equation}
\int_{\Lambda} \left[ \prod_{i=1}^{n_a}
d^2x_{i}^{(a)} d^2y_{i}^{(a)} \right]
\Big| A(a,a) \Big|^{\frac{\pi}{\pi+gN}p} \; \leq \;
(2n_a)! \bigg[ \Xi\Big(\frac{\pi}{\pi+gN}p\Big)\bigg]^{2n_a} \; ,
\end{equation}
and
\[
\int_{\Lambda} \left[
\prod_{i=1}^{n_a} d^2x_{i}^{(a)} d^2y_{i}^{(a)}
\prod_{j=1}^{n_b} d^2x_{j}^{(b)} d^2y_{j}^{(b)} \right]
\Big| B(a,b) \Big|^{\frac{g}{\pi+gN}q(N-1)} \;
\]
\begin{equation}
\leq \;
(2n_a + 2n_b)! \bigg[ \Xi \Big(\frac{g}{\pi+gN}q(N-1)\Big)
\bigg]^{2(n_a+n_b)} \; .
\end{equation}
The constants $\Xi(..)$ do not depend on the numbers of arguments
$2n_a\; , \; \; 2(n_a + n_b)$.
The bounds (7.67) and (7.68) establish
\begin{equation}
\Big| A(a,a) \Big|^{\frac{\pi}{\pi + gN}}  \;  \in \;
L^{p}(\mbox{I\hspace{-0.62mm}R}^{4n_a})
\; \; \; \; \;  \mbox{and} \; \; \; \; \;
\Big| B(a,b) \Big|^{\frac{g}{\pi + gN}q}\;  \in \;
L^{N-1}(\mbox{I\hspace{-0.62mm}R}^{4(n_a+n_b)}) \; ,
\end{equation}
which are necessary conditions for the application of H\"older's inequality
and the Corollary B.2 in (7.63).
Define
\begin{equation}
\xi \; := \; \max \bigg\{ \Xi \Big(\frac{\pi}{\pi+gN}p\Big) \; , \;
\Xi \Big(\frac{g}{\pi+gN}q(N-1)\Big) \bigg\} \; .
\end{equation}
Thus
\[
\int_{\Lambda} \prod_{b=1}^N \prod_{i=1}^{n_b}
d^2x_{i}^{(b)} d^2y_{i}^{(b)}  \;
{\cal F}\Big( \{ x, y \} \Big)
\]
\[ \leq \; \bigg[ \prod_{a=1}^N (2n_a)! \bigg]^{\frac{1}{p}} \;
\bigg[ \prod_{b<c}^N(2n_b+2n_c)! \bigg]^{\frac{1}{q(N-1)}} \;
\prod_{d=1}^N \Big( \xi \Big)^{\frac{1}{p} 2n_d } \;
\prod_{e<f}^N \Big( \xi \Big)^{\frac{1}{q(N-1)}(2n_e + 2n_f)}
\]
\begin{equation}
= \; \bigg[ \prod_{a=1}^N (2n_a)! \bigg]^{\frac{1}{p}} \;
\bigg[ \prod_{b<c}^N(2n_b+2n_c)! \bigg]^{\frac{1}{q(N-1)}} \;
\prod_{d=1}^N \Big( \xi \Big)^{2n_d} \; ,
\end{equation}
where I used
\begin{equation}
\sum_{a<b}^N \Big( n_a + n_b \Big) \; =  \; (N-1) \sum_{a=1}^N n_a \; ,
\end{equation}
and (7.62) in the last step. Finally (see (7.52))
\[
\prod_{a=1}^N \frac{1}{(2n_a)!} \;
\left\langle \prod_{b=1}^N
\left[\chi^{(b)}_+ [ Q^{0,S} ] \chi^{(b)}_- [ Q^{0,S} ] \right]^{n_b}
\right\rangle_{Q^{0,S}} \;
\]
\begin{equation}
\leq \;
\prod_{a=1}^N
\left(\xi \;  8^{[1-\frac{g}{\pi+gN}]}
\right)^{2n_a} \;
\prod_{b=1}^N \Big[(2n_b)!\Big]^{\frac{1}{p}-1} \;
\prod_{c<d}^N \Big[(2n_c+2n_d)!\Big]^{\frac{1}{q(N-1)}} \; .
\end{equation}
Using
\[
\prod_{a=1}^N \Big[(2n_a)!\Big]^{\frac{1}{p}-1} \;
\prod_{b<c}^N \Big[(2n_b+2n_c)!\Big]^{\frac{1}{q(N-1)}}
\; = \;
\prod_{a=1}^N \Big[(2n_a)!\Big]^{\frac{-1}{q}} \;
\prod_{b<c}^N \Big[(2n_b+2n_c)!\Big]^{\frac{1}{q(N-1)}}
\]
\[
 = \;
\left[\prod_{a<b}^N \frac{(2n_a+2n_b)!}{(2n_a)!(2n_b)!}
\right]^{\frac{1}{q(N-1)}}
\; = \; \left[\prod_{a<b}^N { 2n_a+2n_b \choose  2n_a }
\right]^{\frac{1}{q(N-1)}}
\]
\begin{equation}
\leq \;
\left[\prod_{a<b}^N 2^{ (2n_a+2n_b)}
\right]^{\frac{1}{q(N-1)}} \; = \;
\prod_{a=1}^N 2^{ 2n_a \frac{1}{q}}
\end{equation}
one concludes
\begin{equation}
\prod_{a=1}^N \frac{1}{(2n_a)!} \;
\left\langle \prod_{b=1}^N
\left[\chi^{(b)}_+ [ Q^{0,S} ] \chi^{(b)}_- [ Q^{0,S} ] \right]^{n_b}
\right\rangle_{Q^{0,S}} \; \leq \;
\prod_{a=1}^N \left(\xi \;  8^{[1-\frac{g}{\pi+gN}]} \; 2^{\frac{1}{q}}
\right)^{2n_a} \; .
\end{equation}
Inserting this into the series (7.47) the final bound is obtained as
\begin{equation}
{\cal E} \; \leq \;
2^N \prod_{a=1}^N \; \sum_{n_a=0}^\infty \left(
\frac{\beta^{(a)}}{r} \right)^{2 n_a}
\; ,
\end{equation}
with
\begin{equation}
r \; := \;
\bigg[ 2 \omega \xi \;  8^{[1-\frac{g}{\pi+gN}]} \; 2^{\frac{1}{q}}
\bigg]^{-1} \; .
\end{equation}
Clearly the series converges if
\begin{equation}
\beta^{(a)} \; < \; r\; \; , \; \; \; \; \; \; \forall \; \;
a \; = \; 1\; , \; ... \; , \; N \; \; \; ,
\end{equation}
and (7.10) then holds. $ \; \; \; \Box$

%
%
\section{Remarks on the mass perturbation}
The convergence of the mass perturbation series in the
presence of the space-time cutoff $\Lambda$
is a nice result. In particular it establishes the existence
of the model for small quark masses and finite $\Lambda$. But
in order to extract
the physical spectrum one has to send $\Lambda$ to infinity
since it breaks translation invariance. For the
$N=1$ flavor case Fr\"ohlich and Seiler \cite{seilerfroh} using the
Cluster Expansion, were able to remove $\Lambda$ nonperturbatively.
Below it will be shown for the first few terms of the expansion,
that for $N=1$ it is even possible to send
$\Lambda \rightarrow \infty$ termwise.
For $N>1$ it turns out that the known methods to remove $\Lambda$
do not work. Of course it should be possible to remove $\Lambda$
nonperturbatively. This would either require an adaption of the
cluster expansion, or even some new techniques as will be argued in the
end of this section.

Below I will discuss the problem with taking the termwise limit,
as it shows up when one tries to compute the masses of the
particles that correspond to the Cartan currents.
In order to extract the self energies one has to compute the
fully connected two point functions of the  Cartan currents. The
generating functional $W[\alpha]$ for connected correlation functions
is given by
\begin{equation}
W [\alpha] \; := \; \ln \Big( Z[\alpha] \Big) \; ,
\end{equation}
where
\begin{equation}
Z[\alpha] \; := \; \lim_{\mu \rightarrow 0} \;
\left\langle e^{
i \sum_{I=1}^N
\big( \Phi^{(I)}, \alpha^{(I)} \big) }
\;
e^{-S_{int} } \right\rangle_{K^\mu} \; .
\end{equation}
The interaction term $S_{int}$ is given by (see (7.3))
\[
S_{int} [ \Phi^{(I)} ] \; := \;
- \frac{1}{2\pi} \sum_{a=1}^N m^{(a)} c^{(a)} \int_{\Lambda} d^2 x \;
\]
\begin{equation}
\bigg[ \prod_{I=1}^N
: e^{- i 2 \sqrt{\pi} \omega{(I)} U_{I a} \Phi^{(I)}(x)}
:_{M^{(I)}} \; e^{+ i \frac{\theta}{N}} \; + \;
 \prod_{I=1}^N
: e^{+ i 2 \sqrt{\pi} \omega{(I)} U_{I a} \Phi^{(I)}(x)}
:_{M^{(I)}} \; e^{- i \frac{\theta}{N}} \bigg] \; .
\end{equation}
I introduced
\begin{equation}
\omega^{(1)} \; := \; \sqrt{\frac{\pi}{\pi+gN}}
\; \; \; \; , \; \; \; \;
\omega^{(I)} \; := \; 1 \; \; \; \mbox{for} \; \; \; 2 \leq I \leq N \; ,
\end{equation}
to account for the factors in the Wick ordered exponential
properly (compare (7.2)). The test function $t(x)$ that shows up in (7.2)
was replaced by the characteristic function of the finite rectangle $\Lambda$
in space time. The
covariance $K^\mu$ reads (compare (6.35) and (6.36))
\begin{equation}
K^\mu \; := \; \mbox{diag} \left(
\frac{1}{-\triangle+m_d^2} \; , \;
\frac{1}{-\triangle +\mu}\; , \; . \; . \; . \; . \; , \;
\frac{1}{-\triangle +\mu} \right) \; ,
\end{equation}
where I introduced
\begin{equation}
m_d \; := \; e \sqrt{\frac{N}{\pi + gN}} \; ,
\end{equation}
for the dynamically generated mass.
The Wick ordering masses $M^{(I)}$ that enter $S_{int}$ are fixed to
\begin{equation}
M^{(1)} \; := \; m_d
\; \; \; \; , \; \; \; \;
M^{(I)} \; := \; 1 \; \; \; \mbox{for} \; \; \; 2 \leq I \leq N \; .
\end{equation}
Finally I remark that the sources $\alpha^{(I)}$
in (7.80) have to be chosen neutral
\begin{equation}
\int \; d^2x \; \alpha^{(I)}(x) \; = \; 0
\; \; \; \; \; \mbox{for} \; \; \; 1 \leq I \leq N
\; .
\end{equation}
Expansion of $Z[\alpha]$ up to third order in the fermion masses gives
\begin{equation}
Z[\alpha] \; = \; Z^{(0)}[\alpha] \; + \; Z^{(1)}[\alpha] \; + \;
\frac{1}{2} Z^{(2)}[\alpha] \; + \; O(m^3) \; ,
\end{equation}
where
\begin{equation}
Z^{(0)}[\alpha] \; := \; \lim_{\mu \rightarrow 0} \;
\left\langle
e^{+i \sum_{I=1}^N
\big(\Phi^{(I)}, \alpha^{(I)} \big) } \right\rangle_{K^\mu}
\; ,
\end{equation}
and
\[
Z^{(1)}[\alpha] \; := \;
\frac{1}{2\pi} \sum_{a=1}^N m^{(a)} c^{(a)} \; \int_{\Lambda} d^2x \;
\lim_{\mu \rightarrow 0} \; \Bigg\langle
e^{+i \sum_{I=1}^N \big(\Phi^{(I)}, \alpha^{(I)} \big) }\;
\]
\begin{equation}
\bigg[ \prod_{I=1}^N
: e^{- i 2 \sqrt{\pi} \omega{(I)} U_{I a} \Phi^{(I)}(x)}
:_{M^{(I)}} \; e^{+ i \frac{\theta}{N}} \; + \;
 \prod_{I=1}^N
: e^{+ i 2 \sqrt{\pi} \omega{(I)} U_{I a} \Phi^{(I)}(x)}
:_{M^{(I)}} \; e^{- i \frac{\theta}{N}} \bigg] \Bigg\rangle_{K^\mu} \; ,
\end{equation}
and finally
\[
Z^{(2)}[\alpha] \; := \;
\frac{1}{(2\pi)^2}
\sum_{a_1,a_2 = 1}^N \bigg[
\prod_{i=1}^2 m^{(a_i)} c^{(a_i)} \int_{\Lambda} d^2x_i \bigg] \;
\lim_{\mu \rightarrow 0} \; \Bigg\langle
e^{ + i \sum_{I=1}^N \big(\Phi^{(I)}, \alpha^{(I)} \big) } \; \prod_{j=1}^2
\]
\begin{equation}
\bigg[ \prod_{I=1}^N
: e^{- i 2 \sqrt{\pi} \omega{(I)} U_{I a_j} \Phi^{(I)}(x_j)}
:_{M^{(I)}} e^{+ i \frac{\theta}{N}} \; +
 \prod_{I=1}^N
: e^{+ i 2 \sqrt{\pi} \omega{(I)} U_{I a_j} \Phi^{(I)}(x_j)}
:_{M^{(I)}} e^{- i\frac{\theta}{N}} \bigg] \; \Bigg\rangle_{K^\mu} \; .
\end{equation}
$Z^{(0)}$ can be evaluated easily (see (A.21) and (A.38)) and gives
\begin{equation}
Z^{(0)} \; = \; e^{ -\frac{1}{2}
\big( \alpha^{(1)} ,
C_{m_d} \alpha^{(1)} \big) } \;
\prod_{I=2}^N e^{ -\frac{1}{2}
\big( \alpha^{(I)} ,
C_0 \alpha^{(I)} \big) } \; ,
\end{equation}
where the massive covariance $C_{m_d}$ is
given by (A.33) and the massless covariance $C_0$ by (A.40).

The expectation values that show up in $Z^{(1)}$ factorize with
respect to the flavors, and give rise to
\[
\left\langle
e^{ + i \big(\Phi^{(1)}, \alpha^{(1)} \big) }
: e^{\pm i 2 \sqrt{\pi} \sqrt{\frac{\pi}{\pi\!+\!gN}} U_{1 a} \Phi^{(1)}(x)}
:_{m_d} \right\rangle_{C_{m_d}}
\]
\begin{equation}
\times \; \prod_{I=2}^N \; \lim_{\mu \rightarrow 0} \;
\left\langle
e^{ + i \big(\Phi^{(I)}, \alpha^{(I)} \big) }
: e^{\pm i 2 \sqrt{\pi} U_{I a} \Phi^{(I)}(x)}
:_1 \right\rangle_{C_\mu} \; .
\end{equation}
For $N \geq 2$ these terms vanish. The reason for this is that
for $I=2$ the neutrality condition is never fulfilled. In
particular, since $\alpha$ was chosen neutral (see (7.86)),
$U_{2 a}$ would have to vanish for neutrality. But $U_{2 a}$
can only assume the values $1/\sqrt{(N-1 + (N-1)^2)}$ and
$-(N-1)/\sqrt{(N-1 + (N-1)^2)}$ as can be seen from (B.17).
Hence neutrality is violated and the expectation values (7.92)
are all equal to zero and thus $Z^{(1)}$ vanishes. The situation is
different for the $N=1$ case, since no massless particles are involved
then. The result for $Z_1^{(1)}$
(to distinguish the $N=1$ flavor result from the general $Z^{(1)}$
an extra subscript 1 was added)
is given by
\[
Z_1^{(1)} [\alpha] \; = \; Z_1^{(0)} [\alpha] \; \frac{1}{2\pi} m c
\int_{\Lambda} d^2 x
\]
\begin{equation}
\left[ e^{ + 2 \sqrt{\pi} \sqrt{\frac{\pi}{\pi+g}}
\big( \delta(x), C_{m_d} \alpha \big) + i\theta} \; + \;
e^{ - 2 \sqrt{\pi} \sqrt{\frac{\pi}{\pi+g}}
\big( \delta(x), C_{m_d} \alpha  \big) - i\theta} \right] \; .
\end{equation}
All flavor indices are suppressed.
The convolution with the $\delta$-functional is understood
as $\Big(\delta(x),t\Big) = t(x)$.

\noindent
The expectation values that enter $Z^{(2)}$ also factorize and
are given by
\[
e^{+i2\frac{\theta}{N}} \; \lim_{\mu \rightarrow 0}
\prod_{I=1}^N\left\langle e^{i \big( \Phi^{(I)} , \alpha^{(I)} \big)}
:e^{-i2 \sqrt{\pi} \omega^{(I)} U_{Ia} \Phi^{(I)}(x)}:_{M^{(I)}}
:e^{-i2 \sqrt{\pi} \omega^{(I)} U_{Ib} \Phi^{(I)}(y)}:_{M^{(I)}}
\right\rangle_{K^\mu_{II}}
\]
\[
+ \; \lim_{\mu \rightarrow 0}
\prod_{I=1}^N\left\langle e^{i \big( \Phi^{(I)} , \alpha^{(I)} \big)}
:e^{-i2 \sqrt{\pi} \omega^{(I)} U_{Ia} \Phi^{(I)}(x)}:_{M^{(I)}}
:e^{+i2 \sqrt{\pi} \omega^{(I)} U_{Ib} \Phi^{(I)}(y)}:_{M^{(I)}}
\right\rangle_{K^\mu_{II}}
\]
\[
+ \; \lim_{\mu \rightarrow 0}
\prod_{I=1}^N\left\langle e^{i \big( \Phi^{(I)} , \alpha^{(I)} \big)}
:e^{+i2 \sqrt{\pi} \omega^{(I)} U_{Ia} \Phi^{(I)}(x)}:_{M^{(I)}}
:e^{-i2 \sqrt{\pi} \omega^{(I)} U_{Ib} \Phi^{(I)}(y)}:_{M^{(I)}}
\right\rangle_{K^\mu_{II}}
\]
\begin{equation}
+e^{-i2\frac{\theta}{N}} \lim_{\mu \rightarrow 0}
\prod_{I=1}^N\left\langle e^{i \big( \Phi^{(I)} , \alpha^{(I)} \big)}
:e^{+i2 \sqrt{\pi} \omega^{(I)} U_{Ia} \Phi^{(I)}(x)}:_{M^{(I)}}
:e^{+i2 \sqrt{\pi} \omega^{(I)} U_{Ib} \Phi^{(I)}(y)}:_{M^{(I)}}
\right\rangle_{K^\mu_{II}} \; .
\end{equation}
Since the $\alpha^{(I)}$ are neutral, the neutrality is again determined
by the $U_{Ia}\; , \; \; 2 \leq I \leq N$. For the first and the
last term in (7.94) the condition for nonvanishing contributions reads
\begin{equation}
U_{Ia} \; + \; U_{Ib} \; \stackrel{!}{=} \; 0 \; \; \; \; \;
\forall \; \; I \; = \; 2, \; . . . \; N \; .
\end{equation}
Define $N$ vectors $\vec{h}^{(a)}\; \; \; a = 1,2, ... N$ each
with $N-1$ entries
\begin{equation}
\vec{h}^{(a)}_I \; := \; U_{Ia} \; \; , \; \;
\; \; I \; = \; 2,3\, \; ... \; n \; \; .
\end{equation}
Thus the neutrality condition reads
\begin{equation}
\vec{h}^{(a)} \; + \; \vec{h}^{(b)} \; \stackrel{!}{=} \; 0 \; .
\end{equation}
Inspecting (B.17) shows that for $N > 2$ none of the
vectors $\vec{h}^{(a)}$ is the negative of another one, and
(7.97) can never be fulfilled then. For the following I restrict myself to the
case $N \geq 3$ and hence the first and the last terms in (7.94) vanish.

The neutrality condition for the second and the third term in (7.94)
reads
\begin{equation}
\vec{h}^{(a)} \; - \; \vec{h}^{(b)} \; \stackrel{!}{=} \; 0 \; .
\end{equation}
Inspecting (B.17) again shows that for none of the
vectors $\vec{h}^{(a)}$ is equal to any other, and
(7.98) only has the trivial solution $a = b$.
Thus for $N \geq 3$ one obtains
\[
Z^{(2)}[\alpha] \; = \; Z^{(0)}[\alpha] \; \frac{2}{(2\pi)^2}
\sum_{a=1}^N \Big( m^{(a)} c^{(a)} \Big)^2 \int_{\Lambda} d^2x \; d^2y
\]
\begin{equation}
e^{ + 2 \sqrt{\pi} \sqrt{\frac{\pi}{\pi+gN}} U_{1a}
\big( \delta(x)-\delta(y), C_{m_d} \alpha^{(1)} \big)}
\prod_{I=2}^N
e^{ + 2 \sqrt{\pi} U_{I a}
\big( \delta(x)-\delta(y), C_0 \alpha^{(I)} \big)}
\; \rho(x-y) \; ,
\end{equation}
where
\[
\rho(x-y) \; := \;
\lim_{\mu \rightarrow 0}
\prod_{I=1}^N\left\langle
:e^{-i2 \sqrt{\pi} \omega^{(I)} U_{Ia} \Phi^{(I)}(x)}:_{M^{(I)}}
:e^{+i2 \sqrt{\pi} \omega^{(I)} U_{Ia} \Phi^{(I)}(y)}:_{M^{(I)}}
\right\rangle_{K^\mu_{II}}
\]
\[
= \; e^{4\pi \frac{\pi}{\pi+gN} \frac{1}{N} C_{m_d} (x-y) } \;
\prod_{I=2}^N e^{4\pi \big( U_{Ia} \big)^2 C_0 (x-y)}
\]
\begin{equation}
= \;
e^{ \frac{2}{N} \frac{\pi}{\pi+gN} \mbox{K}_0(m_d|x-y|)}
\left( \frac{1}{(x-y)^2} \right)^{\frac{N-1}{N}}
\left( \frac{e^{2\gamma}}{4} \right)^{-\frac{N-1}{N}} \; .
\end{equation}
In the last step I inserted (A.33) and (A.40) for the covariances
and used (B.20) to remove the $U_{Ia}$, which shows that $\rho$ does not
depend on $a$.

Again the result for $N=1$ flavor is different, and I quote it for later
reference
\[
Z^{(2)}_1[\alpha] \; = \; Z_1^{(0)}[\alpha] \; \frac{1}{(2\pi)^2}
\Big( m c \Big)^2 \int_{\Lambda} d^2x \; d^2y
\]
\[
\bigg[
2e^{ + 2 \sqrt{\pi} \sqrt{\frac{\pi}{\pi+g}}
\big( \delta(x)-\delta(y), C_{m_d} \alpha \big)}
+ e^{ + 2 \sqrt{\pi} \sqrt{\frac{\pi}{\pi+g}}
\big( \delta(x)+\delta(y), C_{m_d} \alpha \big)}e^{i2\theta}
\]
\begin{equation}
+ \; e^{ - 2 \sqrt{\pi} \sqrt{\frac{\pi}{\pi+g}}
\big( \delta(x)+\delta(y), C_{m_d} \alpha \big)}e^{-i2\theta}\bigg]
\; \rho_1(x-y) \; ,
\end{equation}
and
\begin{equation}
\rho_1(x-y) \; := \;  e^{ 2 \frac{\pi}{\pi+g} \mbox{K}_0(m_d|x-y|)} \; .
\end{equation}
Putting things together one obtains the general structure
\begin{equation}
Z[\alpha] \; = \; Z^{(0)}[\alpha] \bigg[ 1 + \tilde{Z}^{(1)}[\alpha]
+ \frac{1}{2} \tilde{Z}^{(2)}[\alpha] + O(m^3) \bigg]\; ,
\end{equation}
which is correct for all $N$, since $Z^{(0)}$ always factorizes. The
$\tilde{Z}^{(I)}[\alpha]$ can be read off from (7.99) ((7.93), (7.101)
for $N=1$).
Thus the expansion of
$W[\alpha]$ up to third order in the fermion masses reads
\begin{equation}
W[\alpha] \; = \; \ln\Big(Z^{(0)}[\alpha]\Big)
+ \tilde{Z}^{(1)}[\alpha]
+ \frac{1}{2} \tilde{Z}^{(2)}[\alpha]
- \frac{1}{2} \Big( \tilde{Z}^{(1)}[\alpha]\Big)^2
+ O(m^3) \; .
\end{equation}
Before the more involved case of more than one flavors will be
attacked, I discuss $N=1$. The generating functional then reads
\[
W[\alpha] \; = \; -\frac{1}{2} \Big( \alpha, C_{m_d} \alpha \Big)
\]
\[
\; + \; \frac{1}{2\pi} m c
\int_{\Lambda} d^2 x \;
\left[ e^{ + 2 \sqrt{\pi} \sqrt{\frac{\pi}{\pi+g}}
\big( \delta(x), C_{m_d} \alpha \big) + i\theta} \; + \;
e^{ - 2 \sqrt{\pi} \sqrt{\frac{\pi}{\pi+g}}
\big( \delta(x), C_{m_d} \alpha  \big) - i\theta} \right]
\]
\[
+ \; \frac{1}{2}\frac{1}{(2\pi)^2}
\Big( m c \Big)^2 \int_{\Lambda} d^2x \; d^2y
\bigg[
2e^{ + 2 \sqrt{\pi} \sqrt{\frac{\pi}{\pi+g}}
\big( \delta(x)-\delta(y), C_{m_d} \alpha \big)}
\]
\[
+ \; e^{ + 2 \sqrt{\pi} \sqrt{\frac{\pi}{\pi+g}}
\big( \delta(x)+\delta(y), C_{m_d} \alpha \big)}e^{i2\theta}
+ e^{ - 2 \sqrt{\pi} \sqrt{\frac{\pi}{\pi+g}}
\big( \delta(x)+\delta(y), C_{m_d} \alpha \big)}e^{-i2\theta}\bigg]
\]
\begin{equation}
\times
\Big[ \rho_1(x-y)-1 \Big] \; \; \;
+ \; \; \; O(m^3) \; .
\end{equation}
The connected two point function is given by
\[
G(w,z) \; := \; \left\langle \Phi(w) \Phi(z) \right\rangle_c
\; = \; - \frac{\delta^2}{\delta \alpha(w) \; \delta \alpha(z)}
W[\alpha] \; \Bigg|_{\alpha = 0}
\]
\[
= \; C_{m_d} (w\!-\!z)  \; - \; \frac{mc}{2\pi} \frac{4\pi^2}{\pi\!+\!g} \;
\int_\Lambda d^2 x \;
C_{m_d}(x\!-\!w)C_{m_d}(x\!-\!z) \; 2 \cos(\theta)
\]
\[
- \; \frac{1}{2} \left(\frac{mc}{2\pi}\right)^2 \frac{4\pi^2}{\pi+g} \;
\int_\Lambda d^2 x d^2y \; \Bigg\{
2\Big[C_{m_d}(x\!-\!w)-C_{m_d}(y\!-\!w)\Big]
\Big[C_{m_d}(x\!-\!z)-C_{m_d}(y\!-\!z)\Big]
\]
\[
+ 2 \cos(2\theta) \Big[C_{m_d}(x\!-\!w)+C_{m_d}(y\!-\!w)\Big]
\Big[C_{m_d}(x\!-\!z)+C_{m_d}(y\!-\!z)\Big] \Bigg\} \]
\begin{equation}
\times \;
\Big[ \rho_1(x-y)-1 \Big] \; \; + \; \; O(m^3) \; .
\end{equation}
To obtain the Fourier transform of the propagator I insert
\begin{equation}
C_{m_d}(\xi) \; = \; \frac{1}{2\pi} \int d^2p \; \hat{C}_{m_d}(p) \;
e^{ip\xi} \; = \;
\frac{1}{(2\pi)^2} \int d^2p \; \frac{1}{p^2+m_d^2} \; e^{ip\xi} \; ,
\end{equation}
and end up with
\[
G(w,z) \; = \; \frac{1}{2\pi} \int d^2p \; \hat{C}_{m_d}(p) \; e^{ip(w-z)}
\]
\[
- \; \frac{mc}{2\pi} \frac{4 \pi^2}{\pi\!+\!g} \; \frac{1}{(2\pi)^2}
\int d^2p d^2q \; \hat{C}_{m_d}(p) \hat{C}_{m_d}(q)
\int_\Lambda d^2 x \; e^{ip(x-w)}e^{iq(x-z)}
\; 2 \cos(\theta)
\]
\[
- \; \frac{1}{2} \left(\frac{mc}{2\pi}\right)^2 \frac{4\pi^2}{\pi\!+\!g} \;
\frac{1}{(2\pi)^2} \int d^2p d^2q \; \hat{C}_{m_d}(p) \hat{C}_{m_d}(q)
\int_{\Lambda^\prime} d^2 x d^2\xi \; e^{ix(p+q)}
\]
\[
\times \bigg\{
2 [1\!- e^{-ip\xi}- e^{-iq\xi}\!+\!e^{i\xi(p+q)}]
+ \; 2 \cos(2\theta) [1\!+\!e^{-ip\xi}+\!e^{-iq\xi}\!+\!e^{i\xi(p+q)}] \bigg\}
\]
\begin{equation}
\times\Big[ \rho_1(\xi)-1 \Big] \; \; + \; \;  O(m^3) \; .
\end{equation}
The transformation $x-y := \xi$ of the arguments was performed in the last
step, which changes the rectangle $\Lambda$ into some other finite
area $\Lambda^\prime$.
Inserting the expansion (A.6) in (7.102), one
finds
\begin{equation}
\rho_1(\xi) \; \sim \; \left( \frac{1}{(\xi)^2} \right)^{\frac{\pi}{\pi+g}}
\; \; \; \; \; \; \mbox{for} \; \; \; \xi \; \rightarrow \; 0 \; \; ,
\end{equation}
which shows that the short distance singularity of $\rho_1$ is
integrable for $g > 0$. Furthermore since $\mbox{K}_0$ approaches zero
exponentially, the Fourier transform of
\begin{equation}
\rho_1(\xi) \; -\; 1\; ,
\end{equation}
exists as can be seen from (7.102).
This allows to send the cutoff to infinity.
Using (B.4) one obtains for the propagator in momentum space
\[
\hat{G}(p) \; = \; \hat{C}_{m_d}(p)
\; - \;
mc \frac{4\pi^2}{\pi\!+\!g} \;
\Big( \hat{C}_{m_d}(p) \Big)^2 \; 2 \cos(\theta)
\; - \; \left(mc \right)^2 \frac{4\pi^2}{\pi\!+\!g} \;
\Big(\hat{C}_{m_d}(p) \Big)^2
\]
\begin{equation}
\times \bigg\{
[ 2 + 2 \cos(2\theta)] \widehat{(\rho_1\!-\!1)}(0) \; -
[ 2 - 2 \cos(2\theta)] \widehat{(\rho_1\!-\!1)}(p) \bigg\} \; +  O(m^3) \; .
\end{equation}
$\hat{G}(p)$ can now be inverted easily and the self energy can be
computed.

The situation for $N>1$ is different, since the term $Z^{(1)}$ linear in
the fermion masses vanishes due to the neutrality condition (see (7.92)).
Inserting (7.91) and (7.99) into (7.104) gives
\[
W[\alpha] \; = \; -\frac{1}{2} \Big( \alpha^{(1)}, C_{m_d} \alpha^{(1)} \Big)
-\frac{1}{2} \sum_{I=2}^N \Big( \alpha^{(I)}, C_0 \alpha^{(I)} \Big)
\]
\[
+\; \frac{1}{(2\pi)^2}
\sum_{a=1}^N \Big( m^{(a)} c^{(a)} \Big)^2 \int_{\Lambda} d^2x \; d^2y
\]
\begin{equation}
e^{ + 2 \sqrt{\pi} \sqrt{\frac{\pi}{\pi+gN}} U_{1a}
\big( \delta(x)-\delta(y), C_{m_d} \alpha^{(1)} \big)}
\prod_{I=2}^N
e^{ + 2 \sqrt{\pi} U_{I a}
\big( \delta(x)-\delta(y), C_0 \alpha^{(I)} \big)}
\; \rho(x-y) + \; O(m^3) \; .
\end{equation}
The connected two point function for $\Phi^{(1)}$ is given by
\[
G^{(1)}(w-z) \; := \;
\left\langle \Phi^{(1)}(w) \Phi^{(1)}(z) \right\rangle_c \; = \;
C_{m_d}(w-z) \;
\]
\[
- \; \frac{1}{\pi+gN} \frac{1}{N}
\sum_{a=1}^N \Big( m^{(a)} c^{(a)} \Big)^2 \int_{\Lambda} d^2x \; d^2y
\]
\begin{equation}
\times \bigg[ C_{m_d}(x\!-\!z) - C_{m_d}(y\!-\!z) \bigg]
\bigg[ C_{m_d}(x\!-\!w) - C_{m_d}(y\!-\!w) \bigg] \;
\rho(x-y) \; + \; O(m^3) \; .
\end{equation}
Inserting (7.107) one obtains
\[
G^{(1)}(w-z)
\; = \; \frac{1}{2\pi} \int d^2p \; \hat{C}_{m_d}(p) \; e^{ip(w-z)}
\]
\[
- \; \frac{1}{\pi+gN} \frac{1}{N}
\sum_{a=1}^N \Big( m^{(a)} c^{(a)} \Big)^2 \; \frac{1}{(2\pi)^2}
\int d^2p d^2q \; \hat{C}_{m_d}(p) \hat{C}_{m_d}(q)
\int_{\Lambda^\prime} d^2 x d^2\xi \; e^{ix(p+q)}
\]
\begin{equation}
2 [1 - e^{-ip\xi}- e^{-iq\xi} + e^{i\xi(p+q)}] \rho(\xi) \; +  O(m^3) \; .
\end{equation}
The properties of $\rho$ have to be discussed.
Inserting the expansion (A.6) one
obtains from (7.100)
\begin{equation}
\rho(\xi) \; \sim \; \left( \frac{1}{\xi^2} \right)^{
1 - \frac{1}{N}\big(1 - \frac{\pi}{\pi+gN} \big)}
\; \; \; \; \; \; \mbox{for} \; \; \; \xi \; \rightarrow \; 0 \; \; .
\end{equation}
This shows that the short distance singularity is
integrable also for $N > 1$. Unfortunately there remains an infrared problem.
Since $\mbox{K}_0$ approaches 0 for large argument, $\rho$ behaves as
\begin{equation}
\rho(\xi) \; \sim \; \left( \frac{1}{\xi^2} \right)^{1 - \frac{1}{N}}
\; \; \; \; \; \; \mbox{for} \; \; \; \xi \; \rightarrow \; \infty \; \; ,
\end{equation}
as can be seen from (7.100).
This implies that the cutoff cannot be removed properly
in the expression for $G^{(1)}$.
The analysis of the propagators for $\Phi^{(I)} \; , \; \; I > 2$
can be taken over by replacing $C_{m_d} \; \rightarrow \; C_0$.
The infrared problem remains, and one
has to conclude that the propagators in momentum space, and
thus the self energies cannot be computed termwise.

The reason why the cutoff $\Lambda$ cannot be removed termwise
is related to the
fact that the fermion determinant in infinite volume behaves as
$m^2 \ln(m)$ for small mass $m$ (compare Sections 4.1, 4.2).
Thus the power series expansion
of the determinant is only correct for finite cutoff $\Lambda$.
If one could
somehow reorder or sum up the expansion to extract the $m^2 \ln(m)$ behaviour
it might be possible to send $\Lambda$ to infinity also for $N>1$.
%
%
\section{Semiclassical approximation}
Since the extraction of physical results from the mass perturbation
with the $\Lambda-$cutoff present is rather
problematic, one can try to learn something from a semiclassical
approximation of the Lagrangian
\[
{\cal L}_{GSG} =
\frac{1}{2} \sum_{I=1}^N \partial_\mu \Phi^{(I)} \partial_\mu \Phi^{(I)}
+ \frac{1}{2} \Big( \Phi^{(1)} \Big)^2 \; \frac{e^2 N}{\pi+gN}
\]
\begin{equation}
- \; \frac{1}{\pi} \sum_{b=1}^N m^{(b)} c^{(b)}
\cos \left(
2 \sqrt{\pi} \sqrt{\frac{\pi}{\pi\!+\!gN}} U_{1 b} \Phi^{(1)}
+ 2\sqrt{\pi} \sum_{I=2}^N U_{I b} \Phi^{(I)} - \frac{\theta}{N} \right)
\; .
\end{equation}
The coefficients $c^{(b)}$ are given by (c.f. (6.51))
\begin{equation}
c^{(b)} = \Big(\frac{M^{(1)} e^\gamma}{2}
\Big)^{\frac{\pi}{\pi+gN} \frac{1}{N}} \;
\prod_{I=2}^N \Big(\frac{M^{(I)} e^\gamma}{2} \Big)^{(U_{Ib})^2} \; .
\end{equation}
The $M^{(I)}$ are the masses that are used for Wick ordering. They are
still free parameters. A natural choice is to Wick order
the fields with respect to their own mass.
Such a restriction can be used to fix the $M^{(I)}$ in the end.
To simplify the rather involved structure I consider a special case defined by
\vskip3mm
\noindent
{\bf 1: }
\begin{equation}
m^{(b)} \; = \; m \; \; \; \; \; \; \; \forall \; b = 1,2 \; ... \; N \; \; .
\end{equation}
{\bf 2:}
\begin{equation}
M^{(I)} \; = \; M \; \; \; \; \; \; \; \forall \; I = 2,3 \; ... \; N \; \; .
\end{equation}
The first condition assumes equal masses for all fermion fields.
The fields
$\Phi^{(I)}\; , \; I>1$ are treated symmetrically by the Lagrangian then.
The second restriction thus makes use of the fact that only $\Phi^{(1)}$
plays an extra role in ${\cal L}_{GSG}$.
Hence it makes sense to Wick order those fields with respect to the same
mass which is expressed by (7.120). This implies for the
constants $c^{(b)}$
\begin{equation}
c^{(b)} \; = \; \Big(\frac{M^{(1)} e^\gamma}{2}
\Big)^{\frac{\pi}{\pi+gN} \frac{1}{N}} \;
\Big(\frac{M e^\gamma}{2} \Big)^{1-\frac{1}{N}} \; =: \; c
\; \; \; \; \; \; \forall \; b = 1,2 \; ... \; N \; \; ,
\end{equation}
where I made use of (B.20) to remove the $U_{Ib}$ and thus the dependence
on $b$.

Thus the simplified Lagrangian reads
\begin{equation}
{\cal L}_{GSG} =
\frac{1}{2} \sum_{I=1}^N \partial_\mu \Phi^{(I)} \partial_\mu \Phi^{(I)}
+ V( \Phi^{(I)} ) \; ,
\end{equation}
where the potential $V( \Phi^{(I)} )$ is given by
\begin{equation}
\frac{1}{2} \Big( \Phi^{(1)} \Big)^2 \; \frac{e^2 N}{\pi+gN} \; - \;
\frac{1}{\pi} m c \sum_{b=1}^N
\cos \left(
2 \sqrt{\frac{\pi}{N}} \sqrt{\frac{\pi}{\pi\!+\!gN}} \Phi^{(1)}
+ 2\sqrt{\pi} \sum_{I=2}^N U_{I b} \Phi^{(I)} - \frac{\theta}{N} \right)
\; .
\end{equation}
It is rather instructive to plot the potential in the $N=2$ case.
In order to point out the interesting features of the potential,
I have chosen the numerical values
$g = 0, \; e^22/(\pi+g2) = 1, \; mc/\pi = 0.8, \; \theta/2 = 0$
for Figure 7.1 (next page).

Obviously there are infinitely many degenerate minima.
For the semiclassical approximation one has to find those
minima $\Phi^{(I)}_0$ of $V( \Phi^{(I)} )$, i.e. one has to solve the
equations
\[
0 \; \stackrel{!}{=} \; \frac{\partial}{\partial \Phi^{(1)}}
V( \Phi^{(I)} ) \Bigg|_{\Phi^{(I)} = \Phi^{(I)}_0} \; =
\]
\begin{equation}
\frac{e^2 N}{\pi\!+\!gN} \Phi^{(1)}_0 + \frac{2mc}{\sqrt{\pi\!+\!gN}}
\sum_{b=1}^N \frac{1}{\sqrt{N}}
\sin \left( 2 \sqrt{\frac{\pi}{N}}
\sqrt{\frac{\pi}{\pi\!+\!gN}} \Phi^{(1)}_0
+ 2\sqrt{\pi} \sum_{I=2}^N U_{I b} \Phi^{(I)}_0 - \frac{\theta}{N} \right)
\; ,
\end{equation}
and for $J = 2,3, \; ... \;N$
\[
0 \; \stackrel{!}{=} \; \frac{\partial}{\partial \Phi^{(J)}}
V( \Phi^{(I)} ) \Bigg|_{\Phi^{(I)} = \Phi^{(I)}_0} \; =
\]
\begin{equation}
\frac{2mc}{\sqrt{\pi}} \sum_{b=1}^N U_{J b}
\sin \left( 2\sqrt{\frac{\pi}{N}} \sqrt{\frac{\pi}{\pi\!+\!gN}} \Phi^{(1)}_0
+ 2\sqrt{\pi} \sum_{I=2}^N U_{I b} \Phi^{(I)}_0 - \frac{\theta}{N} \right)
\; .
\end{equation}
\newpage
{}.
\vskip120mm
\noindent
{\bf Figure 7.1 :}
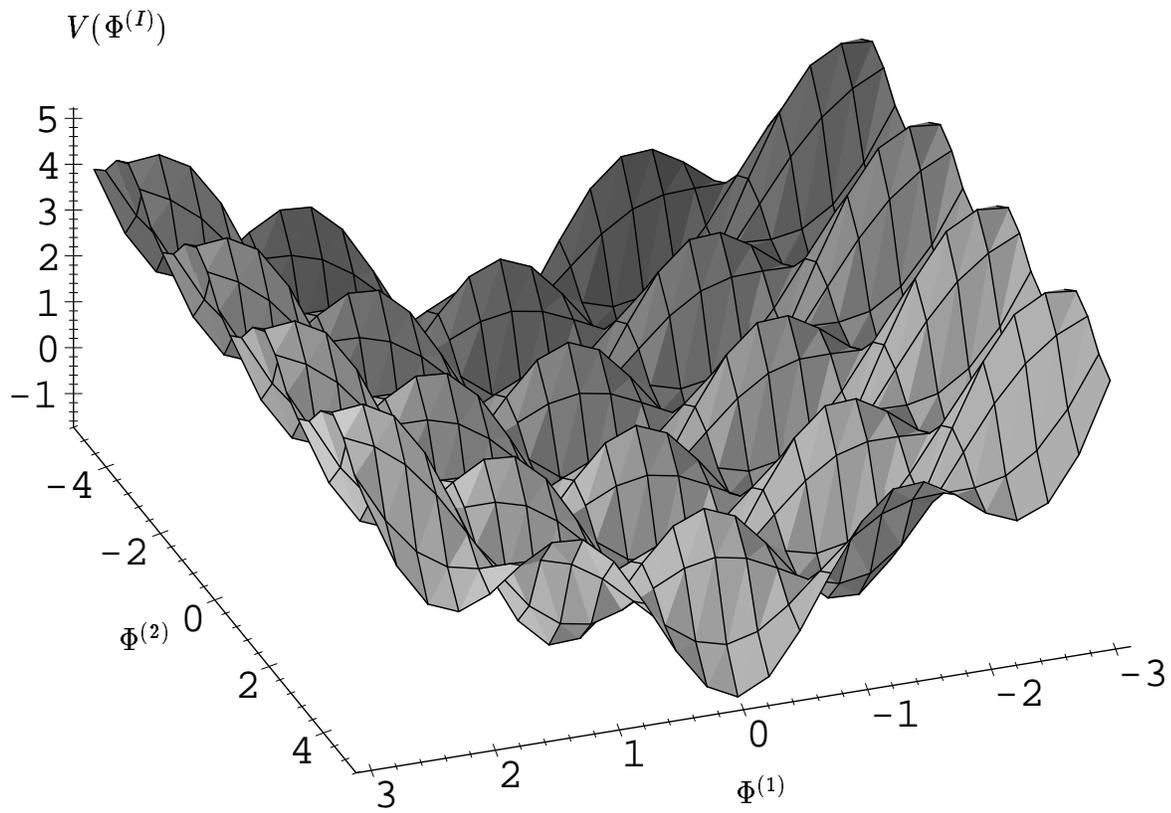
Plot of the potential $V(\Phi^{(I)})$ defined in Equation (7.123) for
$N=2$ flavors. The values of the parameters are given in the text.
\vskip3mm
\noindent
Again one can interpret the lines of $U$ (fixed $J$ in $U_{Jb}$ ) as
vectors $\vec{r}^{\;(J)}$ (see (B.17)) and denote
Equations (7.125) as products of two vectors
\begin{equation}
\vec{r}^{\;(J)} \cdot \vec{s} \; \stackrel{!}{=} \; 0 \; \; \; \; \; \; \; \;
\forall \; \; J \; = \; 2,3\; ... \; N \; ,
\end{equation}
where the entries of the vector $\vec{s}$ are given by
\begin{equation}
s_b :=
\frac{2mc}{\sqrt{\pi}} \sin \left( 2 \sqrt{\pi} \sqrt{\frac{\pi}{\pi\!+\!gN}}
\frac{1}{\sqrt{N}} \Phi^{(1)}_0
+ 2\sqrt{\pi} \sum_{I=2}^N U_{I b} \Phi^{(I)}_0 - \frac{\theta}{N} \right)
\; .
\end{equation}
Equation (7.126) has the only solution (see Appendix B.3)
\begin{equation}
\vec{s} \; = \; \lambda \; ( 1,1, \; ... \; 1 )
\; \; \; \; \; \; \; \lambda \; \in \; I\!\!R \; .
\end{equation}
Thus the set of equations (7.125) is equivalent to
\begin{equation}
2 \sqrt{\pi} \sqrt{\frac{\pi}{\pi\!+\!gN}}
\frac{1}{\sqrt{N}} \Phi^{(1)}_0
+ 2\sqrt{\pi} \sum_{I=2}^N U_{I b} \Phi^{(I)}_0 - \frac{\theta}{N}  \; = \;
\arcsin \left( \frac{ \lambda \sqrt{\pi}}{2m c} \right) \; ,
\end{equation}
for all $b = 1,2, ... N$.
Equation (7.124) can be used to express $\lambda$ in terms of $\Phi^{(1)}_0$,
\begin{equation}
\lambda  = - \sqrt{\frac{N}{\pi}} \frac{e^2}{\sqrt{\pi + gN}}
\Phi^{(1)}_0 \; .
\end{equation}
Inserting this one obtains for the set of equations (7.129)
\begin{equation}
\sum_{I=2}^N U_{Ib} \Phi^{(I)}_0 = \frac{1}{2\sqrt{\pi}}
\left[ \frac{\theta}{N} - \arcsin\left( \frac{e^2}{2mc}
\sqrt{\frac{N}{\pi\!+\!gN}}
\Phi^{(1)}_0 \right) \right] -
\frac{1}{\sqrt{N}}\sqrt{\frac{\pi}{\pi\!+\!gN}} \Phi^{(1)}_0  \; ,
\end{equation}
for all $b=1,2, ... N$.
Multiplication with $U_{Jb}$ and summing over $b$ gives
\begin{equation}
\sum_{I=2}^N \delta_{JI} \Phi^{(I)}_0 =
\delta_{J1} \left\{\frac{\sqrt{N}}{2\sqrt{\pi}}
\left[\frac{\theta}{N} - \arcsin\left(
\frac{e^2}{2mc} \sqrt{\frac{N}{\pi\!+\!gN}}
\Phi^{(1)}_0 \right) \right] - \sqrt{\frac{\pi}{\pi\!+\!gN}} \Phi^{(1)}_0
\right\} \; ,
\end{equation}
where I used the orthogonality of $U$ and $\sum_b U_{Jb} = \delta_{J1}
\sqrt{N}$. In the last expression the equations for the determination
of the minima are decoupled and can be solved easily.
For the case $2 \leq J \leq N$ one obtains
the naive solution
\begin{equation}
\Phi_0^{(J)} \; = \; 0 \; \; \; \; \; \; \; \; \forall \; \; J \; =
\; 2,3, \; ... \; N \; .
\end{equation}
Of course there exists an infinite countable set of solutions
since one can always shift the argument of the cosine in (7.123)
by integer multiples
of $2\pi$
giving rise to
\begin{equation}
2 \sqrt{\pi} \sum_{I=2}^N U_{I b} \Phi^{(I)}_0 = n_b 2 \pi \; ,\; \; \; \;
n_b \; \in \; \mbox{Z\hspace{-1.35mm}Z}
\; \; ,\; \; \; \; \forall \; \; b \; =
\; 1,2, \; ... \; N \; .
\end{equation}
Using the orthogonality of $U$,
the last expression can be inverted and one ends up with
\begin{equation}
\Phi^{(I)}_0 \; = \; \sqrt{\pi} \sum_{b=1}^N U_{Ib} \; n_b \; \; \; \; \;
\; I \; = \; 2,3,\; ... \; N \; .
\end{equation}
The $\Phi^{(1)}_0$ coordinate of the minimum has to fulfill the
equation that emerges from (7.124) when inserting (7.134)
\begin{equation}
\frac{1}{\sqrt{N}}\sqrt{\frac{\pi}{\pi\!+\!gN}} \Phi^{(1)}_0
\; = \;
\frac{1}{2\sqrt{\pi}}
\left[ \frac{\theta}{N} -
\arcsin\left( \frac{e^2}{2mc} \sqrt{\frac{N}{\pi\!+\!gN}}
\Phi^{(1)}_0 \right) \right] \; .
\end{equation}
Obviously this is a trivial modification of the transcendental equation
that determines the minimum in the one flavor case. The corresponding
potential is
\begin{equation}
V^{(1)} ( \Phi^{(1)} ) :=
\frac{1}{2} \Big( \Phi^{(1)} \Big)^2 \; \frac{e^2 N}{\pi+gN} \; - \;
\frac{1}{\pi} m c N
\cos \left(
2 \sqrt{\frac{\pi}{N}} \sqrt{\frac{\pi}{\pi\!+\!gN}} \Phi^{(1)}
- \frac{\theta}{N} \right)
\; .
\end{equation}
I plot the potential $V^{(1)} ( \Phi^{(1)} )$ for
$N=2, \; g=0, \; e^2 2/\pi = 1, \; mc/\pi = 0.8$ and
different values of
$\theta/2$.
\vspace*{\fill}
\newpage
\vspace*{18cm}
\noindent
{\bf Figure 7.2 :}
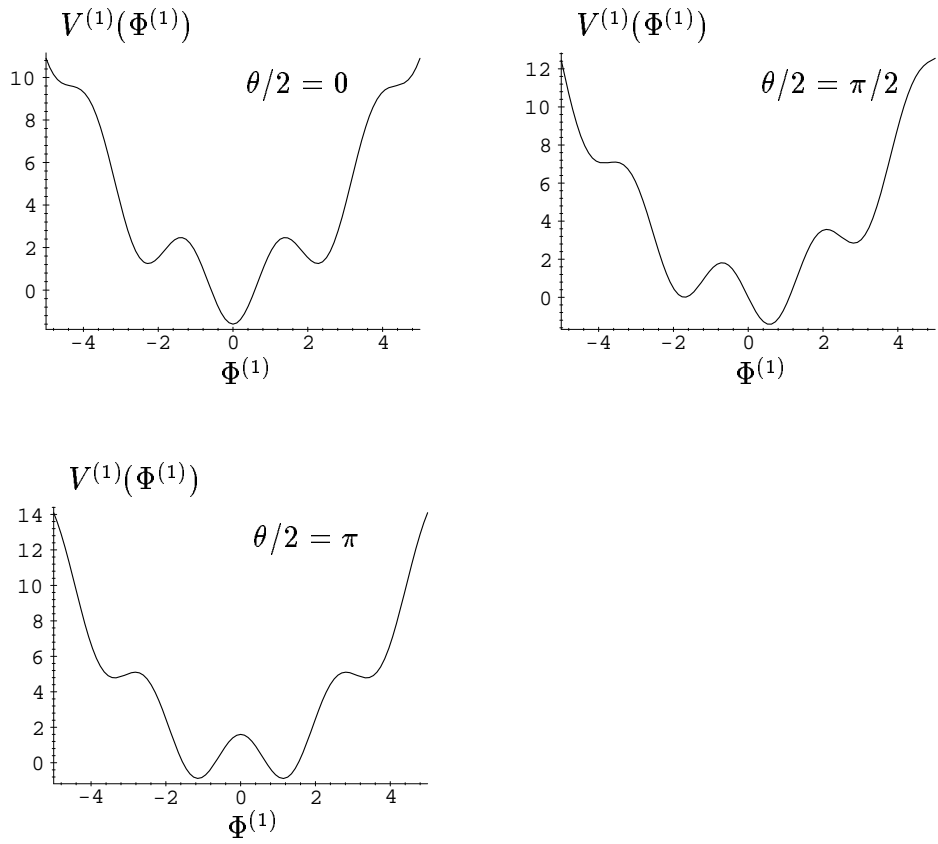
Plot of the potential $V^{(1)}(\Phi^{(1)})$ defined in (7.137) for
$N=2$ flavors and $\theta/2 = 0, \theta/2 = \pi/2$ and
$\theta/2 = \pi$. The values of the other paramters are given in the text.
\vskip3mm
\noindent
For $\theta/N \; = \; 0 \; \mbox{mod}(2\pi)$ there is one unique absolute
minimum and eventually (depending on the other parameters $e,mc,g$)
several local minima. Shifting now $\theta/N$ towards
$\pi \; \mbox{mod}(2\pi)$ the relative minima come down on one branch
of the potential leading to a degeneracy of the absolute minima
for $\theta/N \; = \; \pi \; \mbox{mod}(2\pi)$.

Thus the discussion of the position of the minima can be summed up as
follows.
For $2 \leq I \leq N$ the minima are given by (7.135), whereas
$\Phi^{(1)}_0$ has to be determined as solution of the
transcendental equation (7.136). There is an infinite set
of absolute minima due to (7.135), which gets doubled for $\theta = \pi/2$
as was shown in the discussion of $V^{(1)}(\Phi^{(1)})$.
Thus the vacuum structure of the Generalized Sine Gordon model is rather
different from the $N=1$ model, since the semiclassical vacuum is always
degenerate irrespective of the value of $\theta$. Of course in the quantized
theory this degeneracy vanishes, due to the Coleman theorem \cite{coleman2}
since it is related to a symmetry.

To evaluate the mass matrix of the effective theory around the minima,
the Hesse matrix
\begin{equation}
H_{I I^\prime} \; := \; \frac{ \partial^2 \; V (\Phi^{(J)})}
{\partial \Phi^{(I)} \partial \Phi^{(I^\prime)}} \; \;
\Bigg|_{\Phi^{(J)} = \Phi^{(J)}_0} \; ,
\end{equation}
has to be computed.
It can be evaluated easily
\[
H \; = \; \mbox{diag} \Big( \frac{e^2 N}{\pi\!+\!gN}, 0 , \; ... \; 0 \Big)
\; +
\]
\[
4 mc \; \tilde{\lambda} \; \sum_{b=1}^N \left[
\begin{array}{cccccc}
\frac{\pi}{\pi\!+\!gN} \frac{1}{N} &
\sqrt{\frac{\pi}{\pi\!+\!gN}} \sqrt{\frac{1}{N}} U_{2b} & . & . & . &
\sqrt{\frac{\pi}{\pi\!+\!gN}} \sqrt{\frac{1}{N}} U_{Nb} \\
\sqrt{\frac{\pi}{\pi\!+\!gN}} \sqrt{\frac{1}{N}} U_{2b} &
U_{2b} U_{2b} & . & . & . &
U_{2b} U_{Nb} \\
\sqrt{\frac{\pi}{\pi\!+\!gN}} \sqrt{\frac{1}{N}} U_{3b} &
U_{3b} U_{2b} & \; & \; & \; & \; \\
. & . & . & \; & \; & . \\
. & . & \; & . & \; & . \\
. & . & \; & \; & . & . \\
\sqrt{\frac{\pi}{\pi\!+\!gN}} \sqrt{\frac{1}{N}} U_{Nb} &
U_{Nb} U_{2b} & . & . & . & U_{Nb} U_{Nb}
\end{array} \right] \; =
\]
\\
\begin{equation}
\mbox{diag} \Big( \frac{e^2 N}{\pi\!+\!gN}
+ \frac{\pi}{\pi\!+\!gN} 4 mc \tilde{\lambda} \; , \;  4 mc \tilde{\lambda} \;
, \; . \; . \; . \; .\; . \; , \;
4 mc \tilde{\lambda} \Big) \; .
\end{equation}
Where the orthogonality of $U$ was used again.
$\tilde{\lambda}$ is defined as (compare (7.129))
\begin{equation}
\tilde{\lambda} \; := \; \cos \arcsin
\left( \frac{\lambda \sqrt{\pi}}{2mc} \right) \; = \;
\sqrt{1- \left( \frac{\lambda \sqrt{\pi}}{2mc} \right)^2} \; .
\end{equation}
The Hesse matrix comes out as a positive definite diagonal matrix. The
entries have to be interpreted as the squared masses of the fields $\Phi^{(I)}$
in an effective theory around the semiclassical vacua. The masses $m_I$
are given by
\begin{equation}
m_1 \; := \; \sqrt{  \; \left( \frac{e^2 N}{\pi} + 4 mc \right) } \;
\sqrt{ \frac{\pi}{\pi\!+\!gN}} \; \; \; \; \; \; \; \mbox{for} \; \; \;
\Phi^{(1)} \; ,
\end{equation}
and
\begin{equation}
m_I \; := \; \sqrt{ 4 mc } \;
\; \; \; \; \; \; \; \mbox{for} \; \; \;
\Phi^{(I)}\; , \; \; \; I \; = \; 2,3, \; ... \; N \; .
\end{equation}
It has to be pointed out that for vanishing fermion masses the
semiclassical approximation gives the correct result, and thus
is presumably rather good also for small masses.

%
%
\section{Witten-Veneziano formula}
The masses obtained in the semiclassical
approximation will now be used to test Witten-Veneziano formulas.
The Thirring coupling $g$ is set to zero in the following.
This is for two reasons. Firstly the Thirring term is not necessary
in the semiclassical approximation. Of course one could modify the
Witten-Veneziano formula to include a finite $g$. But the second
more physical reason shows why one should not do this.
Using the bosonization prescription (6.37) (set $g=0$)
the Thirring term (3.13) can be written as
\begin{equation}
\frac{1}{2} \frac{gN}{\pi}
\int d^2x \; \Big( \partial_\mu \Phi^{(1)}(x) \Big)
\Big( \partial_\mu \Phi^{(1)}(x) \Big) \; .
\end{equation}
This additional term causes an extra breaking of the symmetry of the scalar
fields $\Phi^{(I)}$, which has
to be distinguished from the breaking through the dynamically generated
mass $m_1 = e \sqrt{N/\pi}$. It even can be seen how (7.143) leads to the
extra factor attached to the $\Phi^{(1)}$ mass (compare (7.141)).
(7.143) is an extra contribution
to the kinetic term of $\Phi^{(1)}$. Normalizing the kinetic term canonically
gives exactly the factor $\sqrt{\pi/(\pi+gN)}$ in (7.141).

For $g=0$ the following Witten-Veneziano formula will be shown to hold:
\begin{equation}
m_1^2 \; - \;
\frac{1}{N-1} \sum_{I=2}^N \;m^2_I \; = \; \frac{4N}{(f^0_1)^2} \; P^0(0) \; .
\end{equation}
$f^0_1$ denotes the decay constant of the U(1)-pseudoscalar\footnote{
{}From my choice of the $\gamma$-algebra (see Appendix B.1) there follows
$J^{(I)}_{5 \mu} = -i \varepsilon_{\mu \nu} J^{(I)}_{\nu}$. Using (6.37)
one obtains the bosonization prescription $J^{(I)}_{5 \mu} \propto
\partial_\mu \Phi^{(I)}$ for the axial vector currents. This implies
that the properties (mass, decay constant) of the axial vector currents
and the vector currents are determined by the same field $\Phi^{(I)}$.
} particle in the
model with vanishing fermion masses.
Inserting the mass values (7.141) and (7.142) at $g=0$, one
finds that the left hand side of (7.144) reduces to
\begin{equation}
m_1^2 \; - \;
\frac{1}{N-1} \sum_{I=2}^N \;m^2_I \; \; = \; \; \frac{e^2 N}{\pi} \; \;
\stackrel{!}{=} \; \;
\frac{4N}{(f^0_1)^2} \; P^0(0) \; .
\end{equation}
In $\mbox{QED}_2$ the topological charge density $q(x)$ is given by
(see (3.21))
\begin{equation}
q(x) = \frac{e}{2\pi} F_{12}(x) \; .
\end{equation}
The topological susceptibility is defined to be (compare (2.46))
\begin{equation}
\chi_{top} = \int \langle q(x) q(0) \rangle d^2x = \frac{e^2}{(2\pi)^2}
\int \langle F_{12}(x) F_{12}(0) \rangle d^2x =
\frac{e^2}{2\pi} \hat{G}_{FF}(0) \; ,
\end{equation}
where $G_{FF}$ denotes the $F_{12}$ propagator, and $\hat{G}_{FF}(0)$
its Fourier transform at zero momentum. Since
\begin{equation}
F_{12}(x) = \varepsilon_{\mu \nu} \partial_\mu A_\nu(x) \; ,
\end{equation}
the propagator $G_{FF}$ thus is related to the gauge field propagator
$Q_{\mu \nu}$ (see (4.45)) and is given by
\begin{equation}
G_{FF} = - \varepsilon_{\mu \nu} \partial_\nu
Q_{\mu \rho} \varepsilon_{\rho \sigma} \partial_\sigma \; .
\end{equation}
Inserting $Q$ (4.45) and transforming to momentum space one obtains
\begin{equation}
\hat{G}_{FF}(p) = \frac{p^2}{p^2 + e^2\frac{N}{\pi}} =
1 - \frac{e^2\frac{N}{\pi}}{p^2 + e^2\frac{N}{\pi}} =
1 - \int_0^\infty \frac{d\rho(\mu^2)}{p^2 + \mu^2} \; .
\end{equation}
In the last step I made the spectral integral explicit. One nicely
sees that the spectral measure
\begin{equation}
d\rho(\mu^2) = \delta(\mu^2 - m_d^2) d\mu^2
\end{equation}
is `dominated' by the contribution of the U(1)-particle
(compare (2.53)).
{}From (7.147), (7.150) one immediately reads
off the contact term $P(0)$ in the spectral decomposition of
$\chi_{top}$
\begin{equation}
P^0(0) = \frac{e^2}{4\pi^2} \; .
\end{equation}
The last missing ingredient is the decay constant $f^0_1$ in
the massless model.
It is defined by
\begin{equation}
f^0_1 \; = \; m_d^{-2} \langle 0 \mid
\partial_\mu J^{(1)}_{5 \; \mu} \mid 1 \rangle \; ,
\end{equation}
where $\mid 1 \rangle$ is the state that corresponds to the U(1)-current.
In QCD this would be the state $\mid \eta^\prime \rangle$.
Using the definition of the U(1)-current (3.12) the anomaly
equation\footnote{Note that in my definition of the U(1)-current there is
an extra factor $1/\sqrt{N}$ which modifies (3.19) by this factor, compared
to the usual notation.} for $g=0$ takes the form (compare (3.19))
\begin{equation}
\partial_\mu J^{(1)}_{5 \; \mu} \; = \; 2 \sqrt{N} q \; ,
\end{equation}
which I insert in (7.153) to end up with
\begin{equation}
f^0_1=
m_d^{-2} \; 2 \sqrt{N} \langle 0 \mid q \mid 1 \rangle =
m_d^{-2} \;
2 \sqrt{N} \langle 0 \mid q \; \frac{i}{m} F_{12} \mid 0 \rangle \; .
\end{equation}
In the last step the state $| 1 \rangle$ was generated as
$Z^{\frac{1}{2}} F_{12} | 0 \rangle$ with the normalization condition
\begin{equation}
\hat{G}_{11} \; \stackrel{!}{=} \; \frac{1}{p^2 + m_d^2}
\; \; + \; \; \mbox{contact term} \; ,
\end{equation}
giving rise to $Z = -\frac{1}{m_d^2}$.
One ends up with
\begin{equation}
f_1^0 \; = \; i \frac{1}{\sqrt{\pi}} \; .
\end{equation}
Insertion of (7.152) and (7.157) in (7.145) gives an identity.
This explicit computation shows that Equation (7.144) holds in the
semiclassical approximation of
$\mbox{QED}_2$. In particular one finds that the
contributions of the mass perturbation cancel on the
left hand side of (7.144). Thus one can draw Lesson 4.
\vskip3mm
\noindent
{\bf Lesson 4 :}
\vskip3mm
\noindent
{\it The masses of the particles that correspond to the Cartan currents
obey the Witten-Veneziano formula} (7.144).
\vskip3mm
\noindent
It has to be remarked that (7.144) is also
a verification of the original form of the Witten-Veneziano formula,
because the topological susceptibility of the quenched theory reduces
to the contact term. It is not true, however, that the topological
susceptibility appearing in the formula expresses a property of the long
distance fluctuations of the topological density.

%
%

\chapter{Confinement in the massless model}
In this chapter the problem of confinement in the massless model
with $g=0$ will
be adressed.
I consider a generalization of the
confinement criterion proposed by Fredenhagen and Marcu \cite{fredenhagen}.
The original proposal (for lattice-QCD) is to study a sequence of dipole
states
\begin{equation}
| \Phi_{\vec{x},\vec{y}} \rangle \; = \;
\sum_{i,\alpha,\beta} \psi_{i , \alpha}(\vec{x}) \; U_{\alpha \beta}
\Big( {\cal C}_{\vec{x},\vec{y}} \Big) \; \overline{\psi}_{i,\beta}(\vec{y})
\; | 0 \rangle \; .
\end{equation}
$i$ is a flavor index, and $\alpha$ and $\beta$ are Dirac indices.
$U_{\alpha \beta} \Big( {\cal C}_{\vec{x},\vec{y}} \Big)$ is the path ordered
integral of the gauge field along the path ${\cal C}_{\vec{x},\vec{y}}$
which connects the points in space $\vec{x}$ and $\vec{y}$. It is chosen as a
rectangle in the upper Euclidean time half plane
(compare Figure 8.1 for the 2d case).

If quark fragmentation occurs for $|\vec{x}-\vec{y}| \rightarrow \infty$,
the transition probability of $| \Phi_{\vec{x},\vec{y}} \rangle$ into
hadronic states (including the vacuum) should go to 1. In particular
since all hadronic states are local excitations of the vacuum one expects
\begin{equation}
\lim_{|\vec{x}-\vec{y}| \rightarrow \infty}
\frac{| \langle 0 | \Phi_{\vec{x},\vec{y}} \rangle |^2}{
|| \Phi_{\vec{x},\vec{y}} ||^2} \; = \; \; \mbox{const} \; \neq \; 0 \; .
\end{equation}
If the limit (8.2) is zero, this is an indication that the sequence of
dipole states becomes orthogonal to all hadron states and therefore
approximates an isolated quark. In \cite{fredenhagen} it is discussed
that the denominator in (8.2) should be replaced by a Wilson loop if
the order parameter is computed in the continuum
in order to avoid singularities from quark fields at coinciding points.
It is known from experiment that a quark-antiquark system is confined.
Thus one has a guideline how to construct a proper
confinement criterium for QCD which is our best theory
for strong interactions. In $\mbox{QED}_2$ there is no such hint. Thus I
consider the following, more general order parameter. Define
for $0 \leq l \leq \mbox{N}$
\begin{equation}
\rho^{(l)} \; := \;
\frac{\Big| \left\langle N^{(l)}(L) \right\rangle^\theta_0 \Big|^2}{
\left\langle W^{(l)}(L) \right\rangle^\theta_0} \; ,
\end{equation}
where
\begin{equation}
N^{(l)}(L) \; := \; \sum_{\{\alpha_a,\beta_a\}}\prod_{b=1}^l \left[
\psi_{\alpha_b}^{(b)}(-L,0) \; U\Big({\cal C}(L)\Big)\;
\overline{\psi}_{\beta_b}^{(b)}(+L,0) \right] \; ,
\end{equation}
with the gauge field transporter $U\Big({\cal C}(L)\Big)$ defined as
\begin{equation}
U\Big({\cal C}(L)\Big) \; := \; \exp
\left( i \int_{{\cal C}(L)} \; e A_\mu(x) \; dx_\mu \right) \; .
\end{equation}
Finally $W^{(m)}(L)$ denotes the Wilson loop
\begin{equation}
W^{(l)}(L) \; := \;  \; \exp
\left( i l \int_{{\cal C}^W(L)} \; e A_\mu(x) \; dx_\mu \right) \; .
\end{equation}
The contours ${\cal C}(L)$ for the gauge field transporter
and ${\cal C}^W(L)$ for the Wilson loop are given in Figure 8.1.
\begin{center}
\unitlength0.6cm
\begin{picture}(19.5,11.5)
\put(1.5,8){\vector(0,1){2}}
\put(1.5,8){\vector(0,-1){1}}
\put(12.5,8){\vector(0,1){2}}
\put(12.5,8){\vector(0,-1){4}}
\put(0.95,8.3){\mbox{L}}
\put(11.6,6.8){\mbox{2L}}
\thicklines
\put(2,7){\line(0,1){3}}
\put(2,10){\line(1,0){6}}
\put(8,7){\line(0,1){3}}
\put(5.2,10.0){\line(-1,1){0.6}}
\put(5.2,10.0){\line(-1,-1){0.6}}
\put(2.3,6.8){\mbox{(-L,0)}}
\put(6.25,6.8){\mbox{(L,0)}}
\put(2,7){\circle*{0.2}}
\put(8,7){\circle*{0.2}}
\put(13,4){\line(0,1){6}}
\put(13,10){\line(1,0){6}}
\put(13,4){\line(1,0){6}}
\put(19,4){\line(0,1){6}}
\put(16.2,10.0){\line(-1,1){0.6}}
\put(16.2,10.0){\line(-1,-1){0.6}}
\put(15.8,4.0){\line(1,1){0.6}}
\put(15.8,4.0){\line(1,-1){0.6}}
\put(13.3,6.8){\mbox{(-L,0)}}
\put(17.25,6.8){\mbox{(L,0)}}
\put(13,7){\circle*{0.2}}
\put(19,7){\circle*{0.2}}
\end{picture}
\end{center}
\vskip-10mm
\noindent
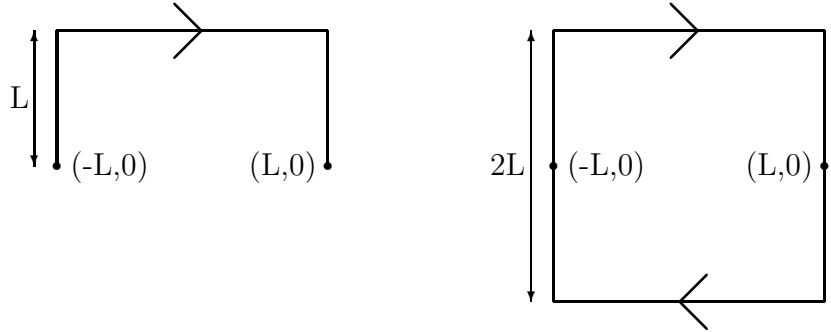
{\bf Figure 8.1 :} The contours for the gauge field transporter
and the Wilson loop.
\vskip3mm
\noindent
The contour integrals showing up in (8.5) and (8.6) can be
rewritten in terms of scalar products (insert ${\cal C}(L)$
and ${\cal C}^W(L)$ for ${\cal K}\;$)
\begin{equation}
\int_{{\cal K}(L)} \; A_\mu(x) \; dx_\mu \; = \;
\Big(A_\mu, j_{{\cal K} \mu}^{(L)} \Big) \; ,
\end{equation}
where the current $j_{\cal K}$ has its support on the contour ${\cal K}$.
In particular the current for the
gauge field transporter is given by
\begin{equation}
j^{(L)}(x) = \left( \begin{array}{l}
\theta(x_1 + L) \theta(L-x_1) \;
\delta(x_2 - L) \\ \; \\
\theta(x_2) \theta(L-x_2) \; [
\delta(x_1 + L) - \delta(x_1 - L) ]
\end{array} \right) \; .
\end{equation}
The current for the Wilson loop reads
\begin{equation}
j_W^{(L)}(x) = \left( \begin{array}{l}
\theta(x_1 + L) \theta(L-x_1) \; [
\delta(x_2 - L) - \delta(x_2 + L) ] \\ \; \\
\theta(x_2 + L) \theta(L-x_2) \; [
\delta(x_1 + L) - \delta(x_1 - L) ]
\end{array} \right) \; .
\end{equation}
Later I will need the Fourier transform of the currents given by
(compare (B.3))
\begin{equation}
\hat{j}^{(L)}(p) \; = \;
\frac{1}{\pi} \sin(p_1 L) \left( \begin{array}{c}
\frac{e^{-ip_2 L}}{p_1} \\ \; \\
\frac{1 - e^{-ip_1 L}}{p_2}
\end{array} \right) \; ,
\end{equation}
and
\begin{equation}
\hat{j}_W^{(L)}(p) \; = \;
\frac{2i}{\pi} \sin(p_1 L) \sin(p_2 L) \left( \begin{array}{c}
- \frac{1}{p_1} \\ \; \\
+ \frac{1}{p_2}
\end{array} \right) \; .
\end{equation}
I start with the evaluation of the Wilson loop.
\begin{equation}
\left\langle W^{(l)}(L) \right\rangle^\theta_0 \; = \;
\int d\mu_Q[A] \; \exp
\left( i l e \Big( A , j^{(L)}_{W} \Big) \right) \; = \;
\exp
\left( -\frac{1}{2} l^2 e^2 \Big( j^{(L)}_{W} , Q \;
j^{(L)}_{W} \Big) \right)  \; .
\end{equation}
The covariance $Q$ for the gauge field is given by (4.45).
Fourier transforming
it and inserting (8.11) gives for the Wilson loop
\begin{equation}
\left\langle W^{(l)}(L) \right\rangle^\theta_0 \; = \;
\exp
\left( -\frac{1}{2} l^2 e^2 \frac{4}{\pi^2} \;
\int d^2p \; \frac{1}{p^2 + m_d^2} \; \frac{p^2}{p_1^2 p_2^2} \;
\sin^2 (p_1 L) \sin^2 (p_2 L)
\right) \; ,
\end{equation}
where the dynamically generated mass $m_d$ reads $(g=0)$
\begin{equation}
m_d \; := \; e \sqrt{\frac{N}{\pi}} \; .
\end{equation}
Before one starts to evaluate the numerator of (8.3) one first has to
check which case of the
$\theta$-prescription has to be applied (compare (5.53)).
According to the $\mbox{U(1)}_A$ transformation defined in (5.50)
one finds that
\begin{equation}
\prod_{a=1}^l \left[
\psi_1^{(a)}(-L,0) \; U\Big({\cal C}(L)\Big)\;
\overline{\psi}_2^{(a)}(L,0) \right] \; ,
\end{equation}
remains invariant,
and the $\theta$-prescription reduces to the naive
expectation value as can be seen from (5.53). Of course this is only one of
the terms showing up in the sum for $N^{(l)}(L)$. It will turn out that it
is the crucial contribution that determines the confinement behaviour.
Thus I start my analysis with this expression,
and discuss the other terms later.
\[
\left\langle \prod_{a=1}^l \left[
\psi_1^{(a)}(-L,0) \; U\Big({\cal C}(L)\Big)\;
\overline{\psi}_2^{(a)}(L,0) \right] \right\rangle^\theta_0
\]
\begin{equation}
= \;
\int d\mu_Q[A] \; \exp
\left( i l e \Big( A , j^{(L)} \Big) \right) \; \prod_{a=1}^l G_{12} \Big(
(-L,0),(+L,0); A \Big) \; .
\end{equation}
Using (4.53) to rewrite $A_\mu$ in terms of the scalar field $\varphi$
one obtains (insert (4.58) for the propagator)
\[
\left\langle \prod_{a=1}^l \left[
\psi_1^{(a)}(-L,0) \; U\Big({\cal C}(L)\Big)\;
\overline{\psi}_2^{(a)}(L,0) \right] \right\rangle^\theta_0
\]
\[
= \;
\frac{1}{(2\pi)^l} \left( \frac{-1}{2L} \right)^l
\int d\mu_{\tilde{Q}}[\varphi]
\exp \bigg( - l e \Big( \varphi, \delta(-L,0) - \delta(L,0) -
i \varepsilon_{\mu \nu} \partial_\nu j^{(L)}_\mu \Big) \bigg)
\]
\[
= \; \frac{1}{(2\pi)^l} \left( \frac{-1}{2L} \right)^l
\exp \left( \frac{1}{2} l^2 e^2 \Big( [ \delta(-L,0) - \delta(L,0) ],
\tilde{Q} [ \delta(-L,0) - \delta(L,0) ] \Big) \right)
\]
\[
\times
\exp \left( - \frac{1}{2} l^2 e^2
\Big( \varepsilon_{\mu \nu} \partial_\nu j^{(L)}_\mu \; , \; \tilde{Q} \;
 \varepsilon_{\rho \sigma} \partial_\sigma j^{(L)}_\rho \Big) \; \right)
\]
\begin{equation}
\times
\exp \bigg( \; -\frac{i}{2} l^2 e^2 \; 2 \mbox{Re}
\Big( \varepsilon_{\mu \nu} \partial_\nu j^{(L)}_\mu \; , \; \tilde{Q} \;
[ \delta(-L,0) - \delta(L,0) ] \; \Big) \; \bigg) \; .
\end{equation}
The convolution with the $\delta$-functional $\delta(x)$ with support
at the space-time point $x$ is understood as $\big(t,\delta(x)\big) =
t(x)$.
Again the scalar products appearing in (8.17)
will be rewritten as momentum space integrals.
The Fourier transform of the $\delta$-functional is given by (compare (B.3))
\begin{equation}
[ \; \hat{\delta}(-L,0) - \hat{\delta}(L,0) \; ] (p) \; = \;
 \frac{i}{\pi} \sin(p_1 L) \; .
\end{equation}
Using this and (8.8) and Equation (4.54) for the covariance
$\tilde{Q}$, the term that mixes the current $j^{(L)}$
and the $\delta$-functionals can be
written as
\[
\Big( \varepsilon_{\mu \nu} \partial_\nu j^{(L)}_\mu \; , \; \tilde{Q} \;
[ \delta(-L,0) - \delta(L,0) ] \; \Big)
\]
\begin{equation}
= \;
 \frac{1}{\pi^2} \int d^2p \frac{1}{p^2 + m_d^2} \; \frac{1}{p^2} \;
\sin^2(p_1L) \;
\bigg[ \frac{p_2}{p_1}e^{ip_2 L} -
\frac{p_1}{p_2}\Big( 1 - e^{ip_2 L}\Big) \bigg] \; = \; 0 \; .
\end{equation}
In the last step I used that the integrand is odd in $p_1$
to show that this term vanishes. Rewriting the other two
exponents in (8.17) in terms of momentum space integrals, one ends up with
\[
\left\langle \prod_{a=1}^l \left[
\psi_1^{(a)}(-L,0) \; U\Big({\cal C}(L)\Big)\;
\overline{\psi}_2^{(a)}(L,0) \right] \right\rangle^\theta_0
\]
\[
= \;
\frac{1}{(2\pi)^l} \left( \frac{-1}{2L} \right)^l
\exp \left( \frac{1}{2} l^2 e^2 \frac{1}{\pi^2} \int d^2 p \;
\frac{1}{p^2 + m_d^2} \; \frac{1}{p^2} \; \sin^2(p_1 L) \right)
\]
\begin{equation}
\times
\exp \left( - \frac{1}{2}  l^2 e^2 \frac{1}{\pi^2} \int d^2 p \;
\frac{1}{p^2 + m_d^2} \; \frac{1}{p^2} \; \sin^2(p_1 L)
\left[ \frac{p_2^2}{p_1^2} + 2\frac{p^2}{p_2^2}\Big(1-\cos(p_2L)\Big)
\right]\right) \; .
\end{equation}
Considering the  contribution
coming from (8.16) alone, one obtains
\[
\rho_{12}^{(l)}(L)
\; := \;
\frac{\Big|
\left\langle \prod_{a=1}^l \left[
\psi_1^{(a)}(-L,0) \; U\Big({\cal C}(L)\Big)\;
\overline{\psi}_2^{(a)}(L,0) \right] \right\rangle^\theta_0
\Big|^2}{
\left\langle W^{(l)}(L) \right\rangle^\theta_0}
\]
\begin{equation}
= \;
\frac{1}{(2\pi)^{2l}} \left( \frac{1}{2L} \right)^{2l}
\exp \left( l^2 I_1  \; + \; l^2 I_2 \; + \; l^2 I_3 \right) \; ,
\end{equation}
with $I_1$ from the Wilson loop result (8.13)
\begin{equation}
I_1 \; := \; e^2 \frac{2}{\pi^2} \;
\int d^2p \;
\frac{1}{p^2 + m_d^2} \; \frac{p^2}{p_1^2 p_2^2} \;
\sin^2 (p_1 L) \sin^2 (p_2 L) \;
\end{equation}
and $I_2$ from (8.20)
\begin{equation}
I_2 \; := \; e^2 \frac{1}{\pi^2} \int d^2 p \;
\frac{1}{p^2 + m_d^2} \; \frac{1}{p^2} \; \sin^2(p_1 L) \; ,
\end{equation}
and finally $I_3$ which stems from the second exponent in (8.20)
\begin{equation}
I_3 \; := \;
- e^2 \frac{1}{\pi^2} \int d^2 p \;
\frac{1}{p^2 + m_d^2} \; \frac{1}{p^2} \; \sin^2(p_1 L)
\left[ \frac{p_2^2}{p_1^2} + 2\frac{p^2}{p_2^2}\Big(1-\cos(p_2L)\Big)
\right] \; .
\end{equation}
Using trigonometric identities and the symmetry of some terms in the
integrands under the interchange $p_1 \leftrightarrow p_2$ one can show
\begin{equation}
I_1 \; + \; I_3 \; = \; I_2 \; + \; I_R \; ,
\end{equation}
where $I_R$ is given by
\[
I_R \; = \;
\frac{e^2}{\pi^2} \int d^2p \; \frac{1}{p^2 + m_d^2}
\; \frac{1}{p_2^2} \; \Big( \sin^2(p_2 L) - 2 \sin^2(p_2 L/2) \Big)
\]
\begin{equation}
- \; 2 \frac{e^2}{\pi^2} \int d^2p \; \frac{1}{p^2 + m_d^2}
\; \frac{1}{p_2^2} \; \cos(p_1 2L) \;
\Big( \sin^2(p_2 L) -  \sin^2(p_2 L/2) \Big) \; .
\end{equation}
Using Formula (B.9) from Appendix B.1, one can solve
$I_2$ (insert $m_d$)
\begin{equation}
I_2 \; := \; \frac{1}{N} \left[
\ln (2 L) \; + \; \ln \left( \frac{e}{2} \sqrt{\frac{N}{\pi}} \right)
\; + \; \mbox{K}_0 \left( 2L e  \sqrt{\frac{N}{\pi}} \right) \; + \;
\gamma \right] \; .
\end{equation}
$I_R$ was not solved explicitely, but in the Appendix B.1
the behaviour
\begin{equation}
I_R \; = \; \frac{1}{N} \; + \; O(e^{-L}) \; \; \; \; \; \;
\mbox{for} \; \; L\; \rightarrow \; \infty \; ,
\end{equation}
is established (see (B.12)).
Thus when putting things together, one concludes
\[
\rho_{12}^{(l)}(L)\; = \;
\frac{1}{(2\pi)^{2l}} \; \left( \frac{1}{4 L^2} \right)^l \;
\left( 4 L^2 \right)^{\frac{l^2}{N}} \;
\left( \frac{e^2}{4} \frac{N}{\pi} e^{2\gamma+1} \right)^{\frac{l^2}{N}} \;
\exp \left( O(e^{-L}) \right)
\]
\begin{equation}
= \; \left( \frac{1}{4 L^2} \right)^{l\big(1 - \frac{l}{N} \big)} \;
\frac{1}{(2\pi)^{2l}} \;
\left( \frac{e^2}{4} \frac{N}{\pi} e^{2\gamma+1} \right)^{\frac{l^2}{N}} \;
\exp \left( O(e^{-L}) \right) \; .
\end{equation}
Inspecting (8.17)-(8.20) one finds that
\begin{equation}
\prod_{a=1}^l \left[
\psi_2^{(a)}(-L,0) \; U\Big({\cal C}(L)\Big)\;
\overline{\psi}_1^{(a)}(L,0) \right] \; ,
\end{equation}
gives the same result as was obtained for (8.17). Thus taking
into account all possible combinations of the terms (8.17) and (8.30)
in $\rho^{(l)}(L)$ one obtains an extra factor 4 compared to (8.29).
Multiplying $\rho_{12}^{(l)}(L)$ with this extra factor
4, gives already the final result, since all other possible contributions
vanish with an extra power of $1/L$
for $L \rightarrow \infty$. This can be seen as follows.
Consider e.g. ($l\geq 2$)
\[
\left\langle \psi_2^{(1)}(-L,0) \; U\Big({\cal C}(L)\Big)\;
\overline{\psi}_1^{(1)}(L,0) \prod_{a=2}^l \left[
\psi_1^{(a)}(-L,0) \; U\Big({\cal C}(L)\Big)\;
\overline{\psi}_2^{(a)}(L,0) \right] \right\rangle^\theta_0
\]
\[
= \;
\int d\mu_Q[\varphi] \; \exp
\left( i l e \Big( \varphi , \varepsilon_{\mu\nu} \partial_\nu j^{(L)}_\mu
\Big) \right)
\]
\[
\times
G_{21} \Big(
(-L,0),(+L,0); \varphi \Big) \prod_{a=2}^l
G_{12} \Big( (-L,0),(+L,0); \varphi \Big)
\]
\[
= \;
\frac{1}{(2\pi)^{2l}} \left(\frac{-1}{2L}\right)^{l} \;
d\mu_Q[\varphi] \; \exp
\bigg( i l e \Big( \varphi , \varepsilon_{\mu\nu} \partial_\nu j^{(L)}_\mu
\Big) \bigg)
\]
\begin{equation}
\times
\exp \bigg(- [l-2] e \Big( \varphi , \delta(-L,0) - \delta(L,0) \Big)
\bigg) \; ,
\end{equation}
where I used the exponential dependence (4.58) on the
external field $\varphi$ of the propagator in the last step.
Comparing (8.17) and (8.31) one finds that the latter expression has
only the factor $l-2$ in front of the exponent containing the
$\delta$-functionals. After evaluation of the functional
integrals this amounts to an extra factor ($l \geq 2$)
\begin{equation}
\exp \left( [ -l^2 + (l-2)^2 ] \; \frac{I_2}{2} \right) \; \sim \;
\left( \frac{1}{2L} \right)^{\frac{2}{N}(l-1)} \; .
\end{equation}
Thus all terms of the type (8.31)
where the spinor indices do not all assume the same value
are suppressed by the extra factor (8.32).
Finally there are some more possible contributions that were not discussed
so far. They show up only for $l=\mbox{N}$ when clustering can be violated in
principle and a different case of the $\theta$-prescription (5.53) has to be
used. A typical contribution of that type is
\[
\left\langle \prod_{a=1}^N \left[
\psi_1^{(a)}(-L,0) \; U\Big({\cal C}(L)\Big)\;
\overline{\psi}_1^{(a)}(L,0) \right] \right\rangle^\theta_0
\]
\begin{equation}
= \; e^{i\theta} \; \lim_{\tau \rightarrow \infty}
\left\langle {\cal O}^a_-(y + \hat{\tau}) \prod_{a=N}^l \left[
\psi_1^{(a)}(-L,0) \; U\Big({\cal C}(L)\Big)\;
\overline{\psi}_1^{(a)}(L,0) \right] \right\rangle_0  \; ,
\end{equation}
where I already inserted the $\theta$-prescription (5.52).
Due to the vanishing diagonal entries of the propagator (see (4.58))
the $\psi^{(a)}_1(-L,0)$ and $\overline{\psi}^{(a)}_1(L,0)$ can
only contract to ${\cal O}^a_-$, and the terms
\begin{equation}
\exp \bigg(\pm e \Big( \varphi , \delta(-L,0) - \delta(L,0) \Big)
\bigg) \; ,
\end{equation}
cannot emerge. As for the term (8.31) one concludes that (8.33)
acquires an extra factor
\begin{equation}
\left( \frac{1}{2L} \right)^{\frac{l^2}{2}} \; .
\end{equation}
The final result thus reads
\[
\rho^{(l)}(L) \;
\]
\begin{equation}
= \;
\left( \frac{1}{4 L^2} \right)^{l\big(1 - \frac{l}{N} \big)}
\frac{4}{(2\pi)^{2l}}
\left( \frac{e^2}{4} \frac{N}{\pi} e^{2\gamma+1} \right)^{\frac{l^2}{N}}
\exp \left( O(e^{-L}) \right) \; \left[
1 + O\left( \left( \frac{1}{2L} \right)^{\frac{2}{N}(l-1)} \right) \right] .
\end{equation}
Obviously the damping factor can only be switched off by setting
$l \; = \; \mbox{N}$, and one ends up
with\footnote{Up to an overall factor the result (8.37)
for $\mbox{N}=1$ flavor can be found in \cite{alonso}.}
\begin{equation}
\rho^{(l)} = \lim_{L \rightarrow \infty} \rho^{(l)}(L) =
\left\{ \begin{array}{cl}
\frac{4}{(2\pi)^{2N}} \Big(\frac{e^2 N}{4 \pi}\Big)^{N}
e^{N(2\gamma +1)} & \mbox{for}\;\; l = N \\
0 & \mbox{for}\;\; l < N  \; \; .
\end{array} \right.
\end{equation}
Thus in the N-flavor model an arrangement of
N `quarks' is bound by a confining force to an arrangement of N
`antiquarks'.

A generalization of the order parameter (8.3)
shows that an operator of $\mbox{N}+l$ with
$0 < l < \mbox{N}$ behaves like the product of only $l$ quarks\footnote{
In general products of the type (8.39) with $n$ factors
always behaves as the $l =  \mbox{N mod(N)}$ expression (8.3).}. Define
\begin{equation}
\rho^{(N+l)} \; := \;
\frac{\Big| \left\langle N^{(N+l)}(L) \right\rangle^\theta_0 \Big|^2}{
\left\langle W^{(N+l)}(L) \right\rangle^\theta_0} \; ,
\end{equation}
where
\[
N^{(N+l)}(L) \; := \;
\sum_{\{\alpha_a,\beta_a,\alpha^\prime_a,\beta^\prime_a\}}
\Bigg\{ \prod_{b=1}^N \left[
\psi_{\alpha_b}^{(b)}(-L,0) \; U\Big({\cal C}(L)\Big)\;
\overline{\psi}_{\beta_b}^{(b)}(+L,0) \right] \]
\begin{equation}
\times \; \prod_{c=1}^l \left[
\psi_{\alpha^\prime_c}^{(c)}(-L,0) \; U\Big({\cal C}(L)\Big)\;
\overline{\psi}_{\beta^\prime_c}^{(c)}(+L,0) \right] \Bigg\} \; ,
\end{equation}
$W^{(N+l)}(L)$ is obtained by replacing $l \rightarrow \mbox{N}+l$ in (8.6).
Again I start with considering a special arrangement of the
spinor indices
\[
\left\langle \prod_{b=1}^N \left[
\psi_{1}^{(b)}(-L,0) \; U\Big({\cal C}(L)\Big)\;
\overline{\psi}_{2}^{(b)}(+L,0) \right]
\prod_{c=1}^l \left[
\psi_{1}^{(c)}(-L,0) \; U\Big({\cal C}(L)\Big)\;
\overline{\psi}_{2}^{(c)}(+L,0) \right]
\right\rangle_0^\theta
\]
\[
= \;
\int d\mu_{\tilde{Q}}[\varphi]
\exp \bigg( i [N+l] e \Big( \varphi,
\varepsilon_{\mu \nu} \partial_\nu j^{(L)}_\mu \Big) \bigg)
\]
\[
\times \prod_{b=1}^l \left[ G_{12}\Big((-L,0),(L,0);\varphi\Big)-
G_{12}\Big((-L,0),(L,0);\varphi\Big) \right] \]
\begin{equation}
\times \;
\prod_{c=l+1}^N G_{12}\Big((-L,0),(L,0);\varphi\Big) \; = \; 0 \; ,
\end{equation}
where I used the factorization with respect to the flavors, and rewrote
the expectation values
$\langle \psi^{(a)}_1(-L,0) \overline{\psi}_2^{(a)}(L,0)
\psi^{(a)}_1(-L,0) \overline{\psi}_2^{(a)}(L,0) \rangle_0 $ for fixed
$a$ in terms
of propagators. Analyzing other possible arrangements of the spinor indices
one finds that the crucial terms take the form
\[
\left\langle \prod_{b=1}^N \left[
\psi_{1}^{(b)}(-L,0) \; U\Big({\cal C}(L)\Big)\;
\overline{\psi}_{1}^{(b)}(+L,0) \right]
\prod_{c=1}^l \left[
\psi_{2}^{(c)}(-L,0) \; U\Big({\cal C}(L)\Big)\;
\overline{\psi}_{2}^{(c)}(+L,0) \right]
\right\rangle_0^\theta
\]
\[
= \;
\int d\mu_{\tilde{Q}}[\varphi]
\exp \bigg( i [N+l] e \Big( \varphi,
\varepsilon_{\mu \nu} \partial_\nu j^{(L)}_\mu \Big) \bigg)
\]
\[
\prod_{b=1}^l \left[ - G_{12}\Big((-L,0),(L,0);\varphi\Big)
G_{21}\Big((-L,0),(L,0);\varphi\Big) \right]
\prod_{c=l+1}^N G_{12}\Big((-L,0),(L,0);\varphi\Big)
\]
\[
= \; (-1)^N \frac{1}{(2\pi)^{N+l}} \left( \frac{1}{2L} \right)^{N+l}
\int d\mu_{\tilde{Q}}[\varphi]
\exp \bigg( i [N+l] e \Big( \varphi,
\varepsilon_{\mu \nu} \partial_\nu j^{(L)}_\mu \Big) \bigg)
\]
\[
\times \exp \bigg( - [N-l] e \Big( \varphi, \delta(-L,0) - \delta(L,0) \bigg)
\]
\begin{equation}
= \; (-1)^N \frac{1}{(2\pi)^{N+l}} \left( \frac{1}{2L} \right)^{N+l}
\exp \left( \frac{1}{2} [N-l]^2 I_2  \; + \; \frac{1}{2} [N+l]^2 I_3
\right) \; .
\end{equation}
Considering again the contributions that stem from terms (8.41)
alone, one obtains the behaviour
\[
\sim \; \left( \frac{1}{2L} \right)^{2[N+l]}
\exp \left( [N-l]^2 I_2  \; + \; [N+l]^2 \Big( I_1 + I_3 \Big)
\right)
\]
\[
\sim \; \left( \frac{1}{2L} \right)^{2[N+l]}
\exp \left( \Big( [N-l]^2 + [N+l]^2 \Big) I_2 \right)
\]
\begin{equation}
\sim \; \left( \frac{1}{2L}
\right)^{2[N+l] - \frac{1}{N}\big([N-l]^2 + [N+l]^2\big) } \; = \;
\left( \frac{1}{4L^2}
\right)^{l\big(1-\frac{l}{N}\big)} \; .
\end{equation}
Arguments similar to the discussion for the simpler case
$\rho^{(l)}$ show that terms that are not of the type (8.41)
acquire extra powers of $1/L$, and thus do not influence the general
behaviour.
One ends up with
\begin{equation}
\rho^{(N+l)} = \lim_{L \rightarrow \infty} \rho^{(N+l)}(L) =
\left\{ \begin{array}{cl}
\mbox{const} \; \; \neq 0 & \; \; \; \mbox{for}\;\; l = 0,N \\
0 & \; \; \; \mbox{for}\;\; 0 < l < N  \; \; \; .
\end{array} \right.
\end{equation}
The physical interpretation suggested by this behaviour of the
Fredenhagen-Marcu order parameters (8.37) and (8.43) is the following:
The model has N
distinct superselection sectors labeled by a charge $Q$ that is defined
only modulo N. To obtain a state in the sector of charge $Q=n$,
$n<\mbox{N}$, one applies an operator consisting of $n$ `quarks' and $n$
antiquarks, separated by distance $2L$ and takes the limit $L\to\infty$.

A rather curious result is obtained, when one computes $\rho^{(l)}(L)$
defined in (8.3) in the model with finite $g$, i.e. with the
Thirring term coupled. One finds
\begin{equation}
\lim_{L \rightarrow \infty} \rho^{(l)}(L) =
\left\{ \begin{array}{cl}
\infty & \; \; \; \mbox{for}\;\; l = N \\
0 & \; \; \; \mbox{for}\;\; 0 < l < N  \; \; \; .
\end{array} \right.
\end{equation}
This result might be related to problems with OS-positivity.
It turns out that one recovers a finite constant
(and thus the problems with OS-positivity vanish),
when the auxiliary field is transported along the contours as well.
In particular $U\Big({\cal C}(L)\Big)$ defined in (8.5) is replaced by
\begin{equation}
\exp
\left( i \int_{{\cal C}(L)} \; \Big[ e A_\mu(x) + \sqrt{g} h_\mu(x) \Big]
\; dx_\mu \right) \; ,
\end{equation}
and similar for the Wilson loop (8.6).

%
%

\chapter*{Summary}
\addcontentsline{toc}{chapter}{Summary}
Using the method of Euclidean path integrals, $\mbox{QED}_2$ with
mass and N flavors of fermions has been
investigated. In order to use the explicitely known determinant for
massless fermions, the expectation functional was expanded with
respect to the fermion masses. It has been argued that it does not make
sense to expand the determinant directly since all involved terms
behave $\propto m^2 \ln m$ in infinite volume.
A Thirring term has been included in order
to make the short distance singularities which show up in the mass
perturbation integrable. It can be produced by
an auxiliary field which couples in the same way as the gauge field does.
Using the Gaussian behaviour of fermion
determinant and action, the formally defined path integral was given a
mathematically precise meaning in terms of Gaussian functional integrals.

Evaluation of a general ansatz allows the identification of operators
that violate clustering in the massless model.
It turned out that the cluster decomposition
property is violated by operators that are singlets under
$\mbox{U}(1)_V \times \mbox{SU(N)}_L \times \mbox{SU(N)}_R$
but transform nontrivially under $\mbox{U}(1)_A$. The nontrivial
transformation properties under $\mbox{U}(1)_A$ were used to decompose
the expectation functional into clustering $\theta$-vacua. The
original vacuum state was shown to be a mixture of the $\theta$-vacua.

A generalized generating functional was used to bosonize the currents
corresponding to a Cartan subalgebra of U(N) together with the chiral
densities. It was shown that for vanishing fermion masses the
Cartan currents can be bosonized in terms of one massive and N-1
massless scalar fields. It was demonstrated that no bosonization in terms
of local scalar fields exists for the whole set of $\mbox{N}^2$
currents corresponding to all generators of U(N). Nevertheless it was
possible to show that the Hilbert space for all $\mbox{N}^2$
classically conserved currents is a tensor product of the Hilbert space
for the U(1)-current (which is the Fock space
of a massive free field) with the Hilbert space of $\mbox{N}^2-1$ currents
constructed out of free massless fermions.

Summing up the mass perturbation series,
the Cartan currents were boson\-ized also for nonvanishing fermion
masses. The corresponding scalar theory turned out to be a
generalization of the Sine-Gordon model. The mass perturbation
series was shown to converge when imposing a space-time cutoff. By
evaluating explicitely the first few terms of the series it was
demonstrated that removing the cutoff termwise is only possible
for the one flavor model. It was argued that the
correct treatment (sum up the series and remove the cutoff nonperturbatively)
requires some new mathematical methods. Since the space time cutoff spoils
translation invariance which is necessary for the computation of
self energies, one is reduced to a semiclassical approximation in
order to compute the mass spectrum of the bosonized currents.
Nevertheless for
vanishing fermion masses the semiclassical approximation is exact,
and thus is expected to give good results for small masses.

A generalization of the Fredenhagen Marcu order parameter was
evaluated in order to investigate the confinement properties of the massless
model. It turned out that in the N flavor model an arrangement of N
quarks is bound by a confining force to an arrangement of N antiquarks.

So far for the construction of $\mbox{QED}_2$. The whole investigation
was motivated by a critical survey of three topics from QCD which are
closely related to each other. Namely the construction of
$\theta$-vacua from topologically nontrivial sectors, the U(1)-problem
and Witten-Veneziano formulas. Those three subjects can be modelled
rather well in $\mbox{QED}_2$. The idea is to circumvent poorly
defined concepts like the superposition of topologically nontrivial sectors
to a $\theta$-vacuum, and to learn from the construction summarized above.
This enterprise lead to the following four lessons for $\mbox{QED}_2$.
\vskip5mm
\noindent
{\bf Lesson 1 :} (page 51)
\vskip2mm
\noindent
{\it The structure
of the vacuum functional that has been suggested
within the instanton picture is recovered.}
\vskip2mm
\noindent
In particular
only operators with chirality
$2 \mbox{N} \nu \; , \; \nu \in
\mbox{Z\hspace{-1.35mm}Z}$ have nonvanishing vacuum expectation values,
as has been claimed by 't Hooft for QCD.
\vskip5mm
\noindent
{\bf Lesson 2 :} (page 66)
\vskip2mm
\noindent
{\it The axial U(1)-symmetry is not a symmetry on the physical
Hilbert space, and there is no  U(1)-problem for $QED_2$.}
\vskip2mm
\noindent
The same should be true for QCD since it is doubtful if the generator for
the U(1)-axial symmetry really exists on the physical Hilbert space.
\vskip5mm
\noindent
{\bf Lesson 3 :} (page 74)
\vskip2mm
\noindent
{\it Physics does not depend on $\theta$ if at least one of the
fermion masses vanishes.}
\vskip2mm
\noindent
This property is commonly believed to hold for QCD as well.
\vskip5mm
\noindent
{\bf Lesson 4 :} (page 102)
\vskip2mm
\noindent
{\it The masses of the particles that correspond to the Cartan currents
obey a Witten-Veneziano type formula.}
\vskip2mm
\noindent
Witten-Veneziano formulas were also derived for QCD (see the discussion in
Section 2.3 for their status).
\vskip5mm
\noindent
To sum up, several interesting insights into the three `mysteries'
as they show up in $\mbox{QED}_2$ have been obtained. It is hoped that
this helps to come to a better understanding of their QCD counterparts.

%
%
\chapter*{Acknowledgements}
\addcontentsline{toc}{chapter}{Acknowledgements}
\vskip5mm
\noindent
I start my acknowledgements with thanking Dr. Erhard Seiler, my advisor at
the Max Planck Institute in Munich. He always had an open ear for discussions
and taught me to ask the right questions. Consulting him was always a great
help and it was a pleasure to learn from him.

I also thank Professor Chistian Lang from the Institute for
Theoretical Physics of the University of Graz, my second advisor
who supported me in many ways.
I had several valuable
discussions with him that often forced me to think more carefully about
physical problems
that seemed already solved.

Finally I would like to thank my colleagues in Munich and Graz.
Namely Dr. Gerhard Buchalla, Dipl. Phys. Wolf M\"osle and
Dipl. Phys. Doreen Wackeroth for
helping me with more phenomenologically oriented questions,
thus keeping my feet
on the ground of `real physics'.
I thank Dr. Max Niedermaier for trying to show me the heaven of
mathematical rigor, and Prof. Peter Breitenlohner,
Dr. Bernd Br\"ugman, Dr. Holger Ewen and
Doz. Helmut Gausterer for discussions and
helping me with problems of more general nature.

%
%
\begin{appendix}
\chapter{The field theory appendix}
This appendix contains the ingredients from (two dimensional)
Euclidean quantum
field theory I am going to use. All the material is well known,
but distributed over various textbooks. The appendix summarizes the
formulas and fixes the notation.

%
%
\section{Propagators in two dimensions}
In this section the expressions for various two dimensional
Euclidean propagators that will be used in the main part are sumarized.\\
\vskip3mm
\noindent
{\bf Free massless bosons: }\\
The defining equation for the Green's functions reads
\begin{equation}
-\triangle C_0(x-y) = \delta(x-y) \; ,
\end{equation}
and is solved by
\begin{equation}
C_0(x) = -\frac{1}{4\pi} \ln(\mu^2 x^2) \; .
\end{equation}
$\mu^2$ is an arbitrary constant (compare the appendix on Wick ordering).
The above solution is understood in the sense of distributions.
After smearing with a test function $t$
\begin{equation}
(-\triangle C_0,t) \; \equiv \; (C_0, -\triangle t) \; = \;
t(0) \; = \; (\delta,t) \; .
\end{equation}
It can be found in e.g. \cite{gelf}. It has to be remarked that
a massless scalar field $\varphi$ in two dimensions does not define
a proper Wightman field theory \cite{coleman2},
but $\partial_\mu \varphi$ does.
Only the latter will be used here, and one has to take derivatives
of the formal propagator (A.2) which remove the dependence of the
results on $\mu$.
\vskip3mm
\noindent
{\bf Free massive bosons: } \\
The defining equation reads
\begin{equation}
\Big(-\triangle + m^2 \Big) C_m(x-y) \; = \; \delta(x-y) \; .
\end{equation}
It can be solved by Fourier transformation
\begin{equation}
C_m(x) \; = \; \frac{1}{(2\pi)^2}\int d^2k \frac{e^{ikx}}{k^2 + m^2} \; = \;
\frac{1}{2\pi}\mbox{K}_0(m|x|) \; .
\end{equation}
The momentum space integral is to be interpreted in distributional
sense again (see \cite{gelf}) and can be found in Appendix B.2.
It has the following short and long distance behaviour.
\begin{equation}
C_m(x) = \left\{ \begin{array}{cl}
- \frac{1}{2\pi}
\left( \ln \Big( \frac{m|x|}{2} \Big) + \gamma + O(x^2) \right)
& \mbox{for} \; x \rightarrow 0 \\ \; & \; \\
\frac{1}{2\pi} \left( \frac{\pi}{2m|x|} \right)^\frac{1}{2} e^{-m|x|}
\left( 1 + O \Big(\frac{1}{x} \Big) \right)
& \mbox{for} \; x \rightarrow \infty
\end{array} \right.
\end{equation}
\vskip3mm
\noindent
{\bf Free massless fermions:} \\
The fermion propagator can be constructed from the boson propagator.
It has to obey
\begin{equation}
\gamma_\mu \partial_\mu G^o(x-y) \; = \; \delta(x-y) \; .
\end{equation}
Using $\not{\!\partial}\not{\!\partial} = \triangle$ and the boson
propagator one finds the solution
\begin{equation}
G^o(x) \; = \; - \; \gamma_\mu \partial_\mu \; C_0(x) \; = \; \frac{1}{2\pi}
\frac{\gamma_\mu x_\mu}{x^2} \; .
\end{equation}
\vskip3mm
\noindent
{\bf Massless fermions in an external field:} \\
The Green's function equation reads
\begin{equation}
\gamma_\mu \Big( \partial_\mu -iB_\mu (x) \Big) G(x,y;B) \; = \;
\delta(x-y) \; .
\end{equation}
The solution is related to the free propagator $G^o(x)$
\begin{equation}
G(x,y;B) \; = \;  G^o(x-y) e^{i[\Phi(x) - \Phi(y)]}  \; ,
\end{equation}
where
\begin{equation}
\Phi(x) \; = \; -\int d^2z D(x-z) \Big( \partial_\mu B_\mu(z) + i \gamma_5
\varepsilon_{\mu \nu} \partial_\mu B_\nu(z) \Big) \; .
\end{equation}
By direct evaluation it is easy to show that
$\partial_\mu \Phi(x) = B_\mu(x)$.
It has to be remarked that the inversion of the Laplace operator
requires some mild regularity and falloff properties of the external field
$B_\mu$.
In particular its Fourier transform at zero momentum has to vanish.
This corresponds to zero winding (compare Section 3.3). \\
In the main part I will only work with transverse fields obeying
$\partial_\mu B_\mu(x) =0$. In this case $\Phi$ reduces to
\begin{equation}
\Phi(x) = - i \gamma_5 \int d^2z D(x-z) \Big( + i \gamma_5
\varepsilon_{\mu \nu} \partial_\mu B_\nu(z) \Big) \; =
 i \gamma_5 \frac{\varepsilon_{\mu \nu} \partial_\mu}{\triangle} B_\nu(x) \; .
\end{equation}
$e^{i[\Phi(x) - \Phi(y) ]}$ can be evaluated easily and gives the diagonal
matrix in spin space
\begin{equation}
e^{i[\Phi(x) - \Phi(y) ]} \; = \; \mbox{diag} \Big(
e^{ - \big[ \chi(x) - \chi(y) \big]} \; , \;
e^{ + \big[ \chi(x) - \chi(y) \big]}
\Big) \; ,
\end{equation}
where I defined
\begin{equation}
\chi(x) :=
\frac{\varepsilon_{\mu \nu} \partial_\mu}{\triangle} B_\nu \; \; ,
\; \; \tilde{x} := x_1 + ix_2 \; .
\end{equation}
Performing the matrix multiplication one ends up with
\begin{equation}
G(x,y;B) =
\frac{1}{2\pi} \frac{1}{(x-y)^2} \left( \begin{array}{cc}
0 & e^{-[\chi(x)-\chi(y)]} \; \overline{(\tilde{x}-\tilde{y})} \\
e^{+[\chi(x)-\chi(y)]} \; (\tilde{x}-\tilde{y}) & 0
\end{array} \right) \; .
\end{equation}

%
%
\section{Gaussian measures}
The aim of this appendix is to give a taste of the mathematics of Gaussian
functional integrals, and to introduce
the notations used in the main part.
A nice introduction to the topic can be found in \cite{roepstorff},
the mathematical details are discussed in \cite{glimm}.

Gaussian measures are measures on the space of tempered
distributions
${\cal S}^\prime(\mbox{I\hspace{-0.62mm}R}^d)$, the dual of the
Schwartz space ${\cal S}(\mbox{I\hspace{-0.62mm}R}^d)$.
Measureable sets can be constructed by considering {\it
cylinder sets} $Z$ defined in the following way.
Let $t_1,  ... , t_n$
be a fixed set of test functions in
${\cal S}(\mbox{I\hspace{-0.62mm}R}^d)$
and $B$ a Borel set in $\mbox{I\hspace{-0.62mm}R}^n$. The set
\begin{equation}
Z := \left\{ \; \varphi \in {\cal S}^\prime(\mbox{I\hspace{-0.62mm}R}^d)
\; \Big| \; (\varphi(t_1), ... , \varphi(t_n)) \in B \; \right\} \; ,
\end{equation}
is called a cylinder set generated by $t_1,  ... , t_n$, with
basis $B$.
Equation (A.16) already shows that measures on the cylinder sets
can be defined by making use of the measurability of $B$.

The second ingredient for the construction of Gaussian measures
are {\it covariance operators}. A continuous, positive, linear map $C$ from
${\cal S}(\mbox{I\hspace{-0.62mm}R}^d)$ to
${\cal S}(\mbox{I\hspace{-0.62mm}R}^d)$ which is nondegenerate
\begin{equation}
\Big( t, C t \Big)_{L^2} \; = \; 0 \; \; \; \; \; \;
\mbox{only for} \; \; t \; = \; 0 \; ,
\end{equation}
is called a covariance operator.

Now one can define the {\it Gaussian measure with covariance $C$ }
of a cylinder set $Z$ by
\begin{equation}
\mu_C[Z] \; := \; \int_{\varphi \in B} \; d\mu_C[\varphi] \; ,
\end{equation}
\begin{equation}
d\mu_C[\varphi] \; := \; \frac{1}{\sqrt{\mbox{det}(2\pi \tilde{C})}} \;
\exp\left( \;  -\frac{1}{2} \sum_{i,j =1}^n \;
\alpha_i \Big( \tilde{C}^{-1} \Big)_{i,j}\alpha_j \; \right) \;
\prod_{l=1}^n d\alpha_l
\; ,
\end{equation}
where
\begin{equation}
\tilde{C}_{i,j} \; := \; \Big( t_i , C t_j \Big)_{L^2} \; ,
\end{equation}
and $\prod_{l=1}^n d\alpha_l$ denotes Lebesgue measure which has to be
integrated over $B$ the basis of $Z$.

So far for the constructive aspects of Gaussian measures. To show
that (A.18) defines a proper measure on all of
${\cal S}^\prime(\mbox{I\hspace{-0.62mm}R}^d)$, some
more work has to be done. It has to be established
that the cylinder sets can be extended to a Boolean $\sigma$-algebra.
Furthermore it has to be shown that $\mu_C[Z]$ does not depend on the
choice of the generating elements $t_1,  ... , t_n$ or the basis $B$.
Finally it has to be checked that $\mu_C[Z]$ obeys the properties
( $\sigma$-additivity, regularity, ..... ), that allow
to extend it to a measure on all of
${\cal S}^\prime(\mbox{I\hspace{-0.62mm}R}^d)$. Most of this material
can be found in the very explicit books of Gelfand and Shilow (Vilenkin)
\cite{gelf}.

I finish this section by quoting the two `holy formulas' of
Gaussian integration
\begin{equation}
\int \; d\mu_C[\varphi] \; e^{ \pm i \varphi(t)} \; = \;
e^{ - \frac{1}{2}(t,C t) }  \; ,
\end{equation}
\begin{equation}
\int \; d\mu_C[\varphi] \; e^{ \pm \varphi(t)} \; = \;
e^{ + \frac{1}{2}(t,C t) }  \; .
\end{equation}
Those two results can be obtained easily from (A.18)-(A.20)
since it is sufficient
to consider the cylinder set with generating element $t$ and basis
$B \; = \; \mbox{I\hspace{-0.62mm}R}$. The Equations (A.21), (A.22) reduce to
ordinary Gaussian integrals then.

%
%

\section{Finite action is zero measure}
In this appendix I discuss a toy example which illustrates that
field configurations with finite action have measure zero. In fact this
is a well known feature which e.g. follows
from the properties of Gaussian measures discussed in
\cite{colella}.
The formulation here I borrow from \cite{seilertoy}.

The system describes infinitely many uncoupled harmonic oscillators.
The action is given by
\begin{equation}
S [\{b_j\}] \; := \;
\frac{1}{2} \; \sum_{n=1}^\infty \; b_n^2 \; \; \; \; , \; \;
b_n \; \in \; \mbox{I\hspace{-0.62mm}R} \; .
\end{equation}
The measure on the space of series $\{b_j\}_{j=1}^\infty$
is defined as the product of normalized Gaussian measures, symbolically
\begin{equation}
d \mu[\{b_j\}] \; := \; \; ` \; \exp \Big( \; -S[\{b_j\}]\; \Big) \;
\prod_{n=1}^\infty
\frac{db_n}{\sqrt{2\pi}} \; \mbox{'} \; \; \; \;  ,
\end{equation}
and expectation values of operators $O$ acting on
$\{b_j\}_{j=1}^\infty$ are defined as
\begin{equation}
\langle \; O \; \rangle \; := \; \int \; d \mu[\{b_j\}] \; O[\{b_j\}]\; \; .
\end{equation}
{}From the normalization in (A.24) it follows that
\begin{equation}
\langle \; 1 \; \rangle \; = \; 1 \; ,
\end{equation}
and hence the measure is a proper probability measure.
Now one uses a special observable given by
\begin{equation}
O_N[\{b_j\}] \; := \; \exp \left( \; - \frac{\lambda}{2} \;
\sum_{n=1}^N \; b_n^2 \; \right)\; \; \; \; , \; \; \lambda > 0 \; ,
\end{equation}
and considers the limit of $N \in \mbox{I\hspace{-0.58mm}N}$ going to infinity.
The evaluation of the expectation value only makes use of Gaussian integrals
\[
\lim_{N\rightarrow \infty} \; \langle \; O_N \; \rangle \; = \;
\lim_{N\rightarrow \infty} \;
\int \; \prod_{n=1}^\infty \frac{db_n}{\sqrt{2\pi}}
\exp \left( \; - \; \frac{1+ \lambda}{2} \;
\sum_{n=1}^N \; b_n^2  \; - \;
\frac{1}{2} \;
\sum_{n=N+1}^\infty \; b_n^2 \right) \]
\begin{equation}
= \; \lim_{N \rightarrow \infty} \;
\prod_{n=1}^N \frac{1}{\sqrt{1+\lambda}} \; = \; 0 \; .
\end{equation}
Using Fatou's lemma (see Vol.1 of \cite{reed})
one obtains
\begin{equation}
\langle \; O_\infty \; \rangle \; := \;
\langle \lim_{N \rightarrow \infty} \; O_N \; \rangle \; \leq
\;\lim_{N \rightarrow \infty} \; \langle \; O_N \; \rangle \; = \; 0\; ,
\end{equation}
one has to conclude from
(A.25) that
$O_\infty [\{b_j\}] \; = \; 0$ almost everywhere,
and from the special choice (A.27) for the observable then follows
\begin{equation}
S[\{b_j\}] \; = \; \infty \; \; \; \; \; \; \mbox{almost everywhere} \; \; .
\end{equation}
This concludes the toy model discussion. When investing a little bit of
time, the same can be shown for the Gaussian integrals of the last section
as well. One could define the observable $O_N$ to be
\begin{equation}
O_N [ \varphi ] \; := \; \exp
\left( - \frac{\lambda}{2} \Big( \varphi, P_N C^{-1} P_N \; \varphi \Big)
\right) \; ,
\end{equation}
where $P_N$ is the projector on the first $N$ eigenvectors of the
covariance operator $C$.
Then one can essentially repeat the arguments given above.

%
%
\section{Wick ordering and massless particles}
In this appendix Wick ordering of massive as well as massless
bosons is discussed, and the neutrality condition for the massless case
is proven.
An introduction to the topic can be found in \cite{glimm}.

First I consider two dimensional, Euclidean, massive fields with covariance
\begin{equation}
C_m := \left( -\triangle + m^2 \right)^{-1} \; .
\end{equation}
The corresponding Green's function which I denote by the same symbol but
an extra space-time argument is given by (compare the propagator appendix)
\begin{equation}
C_m(x-y) \; = \; \frac{1}{2\pi} \mbox{K}_0(m|x-y|) \; .
\end{equation}
The Wick ordering of an exponential with respect to mass $M$ is defined
as
\begin{equation}
: e^{i\varphi(f)} :_M \; \; \equiv \; \frac{e^{i\varphi(f)}}
{\langle e^{i\varphi(f)} \rangle_{C_M}} \; = \;
e^{i\varphi(f) + \frac{1}{2}(f,C_M f)} \; .
\end{equation}
Here $\varphi$ is a real scalar field with covariance $C_m$ and $f$
denotes a test function in Schwartz space
${\cal S} (\mbox{I\hspace{-0.62mm}R} )$.
With this definition one has for example
\[
\Big\langle : e^{i\varphi(f)} :_M e^{i\varphi(g)} :_M \Big\rangle_{C_m} \; = \;
\exp \left( -\frac{1}{2}(f,C_m f) - (f,C_m g) - \frac{1}{2}(g,C_m g) \; +
\right. \]
\begin{equation}
\left. \frac{1}{2}(f,C_M f) + \frac{1}{2}(g,C_M g) \right) \; .
\end{equation}
Usually massive fields are normal ordered with respect to their own
mass which simplyfies the last equation to
\begin{equation}
\Big\langle : e^{i\varphi(f)} :_m e^{i\varphi(g)} :_m \Big\rangle_{C_m} \; = \;
\exp \Big( - (f,C_m g) \Big) \; .
\end{equation}
The test functions $f$ and $g$ can be replaced by $\delta$-sequences
leading to e.g.
\begin{equation}
\langle : e^{i\varphi(x)} :_m e^{i\varphi(y)} :_m \rangle_{C_m} \; = \;
\exp \left( - C_m(x-y) \right) \; .
\end{equation}
For massless particles the strategy is to Wick order with respect to a
given fixed mass $M$, to take the expectation value with respect to
$C_m$ and to perform the limit $m \rightarrow 0$ in the end
\begin{equation}
\Big\langle : e^{i\varphi(f)} :_M e^{i\varphi(g)} :_M \Big\rangle_{C_{m=0}}
\; := \;
\lim_{m \rightarrow 0}
\Big\langle : e^{i\varphi(f)} :_M e^{i\varphi(g)} :_M \Big\rangle_{C_{m}} \; .
\end{equation}
In the massless limit the
{\it neutrality condition} \cite{coleman1}
has to be obeyed in order to obtain
nonvanishing expectation values.
\vskip3mm
\noindent
{\bf Lemma A.1 :} Neutrality condition
\vskip3mm
\noindent
For test functions $t_j \; , \; j = 1, \; .. \; , n$
\[
\lim_{m \rightarrow 0} \;
\langle \prod_{j=1}^{n} : e^{i \varphi(t_j)} :_M  \rangle_{C_{m}}
\]
\begin{equation}
= \; \left\{ \begin{array}{cc}
e^{+\frac{1}{2} \sum_{i=1}^n \big(t_i,C_M t_i \big)}
e^{-\frac{1}{2} \sum_{i \neq j} \big(t_i,C_0 t_j \big)}
& \mbox{for} \; \sum_{j=1}^n q_j \; = \; 0 \\ \; & \; \\
0 & \mbox{for} \; \sum_{j=1}^n q_j \; \neq \; 0 \\
\end{array} \right. \; ,
\end{equation}
where
\begin{equation}
C_0(x) \; = \; -\frac{1}{4\pi}\left( \ln(x^2) + 2\gamma - \ln4 \right) \; ,
\end{equation}
and
\begin{equation}
q_j \; := \; \int \; d^2x \; t_j(x) \; \; .
\end{equation}
\vskip3mm
\noindent
{\bf Proof :}
\vskip3mm
\noindent
\begin{equation}
\Big\langle \prod_{j=1}^{n} : e^{i \varphi(t_j)} :_M
\Big\rangle_{C_{m}}
\; = \; e^{ \frac{1}{2} \sum_{i=1}^n \big( t_i,C_M t_i\big)
-\frac{1}{2} \sum_{i,j} \big(t_i,C_m t_j\big)} \; .
\end{equation}
{}From the propagator appendix I use the short distance
($\Leftrightarrow$ small mass)
behaviour (A.6) of the massive boson propagator
\begin{equation}
C_m(z) = -\frac{1}{4\pi} \ln(z^2)
-\frac{1}{4\pi} \left( 2\gamma + \ln\Big(\frac{m^2}{4}\Big) +
O\Big(m^2 z^2\Big) \right) \; .
\end{equation}
This implies
\[
\Big(t_i,C_m t_j \Big) = \int \; d^2x\; d^2y \;  C_m(x-y) \; t_i(x) \; t_j(y)
\]
\begin{equation}
= \;\Big(t_i,C_0 t_j \Big) \; - \;
\frac{1}{4\pi} \ln\Big(m^2 \Big) \; q_i \; q_j \; + \; O(m^2)
\; .
\end{equation}
Inserting this into (A.42) gives
\[
\Big\langle \prod_{j=1}^{n} : e^{i\varepsilon_j \varphi(x_j)} :_M \;
\Big\rangle_{C_{m}}
= \; e^{ \frac{1}{2} \sum_{i=1}^n \big( t_i,C_M t_i\big)}
e^{-\frac{1}{2} \sum_{i,j} \big(t_i,C_0 t_j\big)}
\]
\begin{equation}
\times m^{\frac{1}{4\pi}\left( \sum_{i=1}^n q_i \right)^2} e^{O(m^2)} \; ,
\end{equation}
which immediately leads to the desired result.$\; \Box$
\vskip3mm
\noindent
Inserting $\delta$-sequences one can find another formulation
\cite{frohlich}.
\vskip3mm
\noindent
{\bf Lemma A.2 :}
\vskip3mm
\noindent
For pairwise disjoint space time arguments $x_j \; , \; j = 1, \; .. \; , n$
and real constants $\varepsilon_j \; , \; j = 1, \; .. \; , n$
\[
\lim_{m \rightarrow 0} \;
\langle \prod_{j=1}^{n} : e^{i\varepsilon_j \varphi(x_j)} :_M  \rangle_{C_{m}}
\]
\begin{equation}
= \; \left\{ \begin{array}{cc}
\left(\frac{1}{M}\right)^{\frac{1}{4\pi}\sum_{j=1}^n \varepsilon_j^2}
e^{-\frac{1}{2} \sum_{i \neq j} \varepsilon_i \varepsilon_j
C_0(x_i-x_j)}
& \mbox{for} \; \sum_{j=1}^n \varepsilon_j \; = \; 0 \\ \; & \; \\
0 & \mbox{for} \; \sum_{j=1}^n \varepsilon_j \; \neq \; 0 \; \; .\\
\end{array} \right.
\end{equation}
\vskip3mm
\noindent
{\bf Proof :}
\vskip3mm
\noindent
For the proof one only has to insert some $\delta$-sequence and to use
\begin{equation}
\lim_{z\rightarrow 0} \left( C_M(z) \; - \; C_0(z) \right) \; = \;
- \frac{1}{4\pi} \; \ln \Big( M^2 \Big) \; ,
\end{equation}
which directly follows from
the short distance behaviour (A.43) of massive propagators. $\;  \Box$
\vskip3mm
\noindent
Usually massless bosons are Wick ordered with respect to mass $M = 1$,
which makes the extra power of $1/M$ in (A.46) equal to one.
The propagator $C_0(x)$ can be rewritten as
$-1/4\pi \ln(\mu^2 x^2)$ which coincides with the expression given
in the propagator appendix.

The neutrality condition which was presented as an algebraic identitiy
is a nice consistency check of the formalism for massless
scalar fields in two dimensions.
The Lagrangian is invariant under
\begin{equation}
\varphi(x) \; \longrightarrow \; \varphi(x) + c \; ,
\end{equation}
where $c$ is some constant.
The expectation values considered in (A.39) formally transform like
\begin{equation}
\Big\langle \prod_{j=1}^{n} : e^{i \varphi(t_j)} :_M
\Big\rangle_{C_{m=0}} \; \longrightarrow \;
\Big\langle \prod_{j=1}^{n} : e^{i \varphi(t_j)} :_M
\Big\rangle_{C_{m=0}} \;
\exp \left( i c \sum_{j=1}^n q_j \right) \; .
\end{equation}
If now the neutrality condition were not there, the symmetry would be
broken, which is not possible since continuous symmetries
cannot be broken in two dimensions \cite{coleman2}.
Thus one can consider the neutrality condition as a direct consequence of
Coleman's theorem.

%
%
\chapter{The technical appendix}
The appendix B is a summary of notational conventions
and of formulas that can not be found
the literature. Since they are of technical nature I did not include
them into the main part.

%
%
\section{Notational conventions}
{\bf $\gamma$-algebra :} \\
\vskip3mm \noindent
It is convenient to  use the following representation of the
2d Euclidean \\ $\gamma$-matrices which makes the fermion
propagator antidiagonal
\begin{equation}
\gamma_1 =  \left( \begin{array}{cc}
0 & 1 \\ 1 & 0 \end{array} \right) \; \; , \; \;
\gamma_2 =  \left( \begin{array}{cc}
0 & -i \\ i & 0 \end{array} \right) \; \; , \; \;
\gamma_5 = i \gamma_2 \gamma_1 =  \left( \begin{array}{cc}
1 & 0 \\ 0 & -1 \end{array} \right) \; .
\end{equation}
They obey the commutation relations $\{\gamma_\mu , \gamma_\nu \} =
2 \delta_{\mu \nu} \: ; \: \mu,\nu = 1,2,5 $.
Chiral projectors $P_\pm$ are defined as
\begin{equation}
P_\pm := \frac{1}{2} \Big( 1 \pm \gamma_5 \Big) \; .
\end{equation}
{\bf Fourier transform:} \\
\vskip3mm \noindent
I use a symmetric normalization of the $1/2\pi$ factors
which gives for the Fourier transform in two dimensions
\begin{equation}
\hat{f} (p) := \frac{1}{2\pi} \int d^2x \; f(x) e^{- ipx} \; \; \; \; ,
\; \; \; \; f(x) = \frac{1}{2\pi} \int d^2p \; \hat{f}(p) e^{ipx} \; ,
\end{equation}
and formally (correct when smeared with test functions)
\begin{equation}
\delta(x) = \frac{1}{(2\pi)^2} \int d^2p \; e^{ipx} \; .
\end{equation}

%
%

\section{Some integrals}
The following integrals are used in the main part. They all can be evaluated
after a transformation to polar coordinates $r,\varphi$. It turned out
that integrating over $\varphi$ first is simpler. For some of the integrals
partial integration in $r$ is necessary to bring them into a form such
that they can be found in the integral tables
\cite{abra}, \cite{bronstein}, \cite{rys} and \cite{prud}. In all cases
it was possible to cross-check the formulas. For properties of special
functions I use \cite{magnus}.
\vskip3mm
\noindent
${\bf I_1 : }$
\begin{equation}
I_1 \; := \; \int d^2p \; \frac{1}{p^2 + \lambda^2} e^{ipx}\; =
2\pi \mbox{K}_0(\lambda |x|) \; ,
\end{equation}
where $\mbox{K}_0$ is the modified Bessel function
(see \cite{magnus} p. 66). Here one should remark, that $I_1$ is not
absolutely convergent, but converges conditionally for $x \neq 0$.
Thus some regularization procedure has to be applied. In particular
$(p^2-\lambda^2)^{-1}$ can be replaced by $(p^2-\lambda^2)^{-\alpha}$
which gives an absolutely convergent integral for
$\alpha > 1$ which can be solved explicitely.
The result (B.5) is then obtained by analytic continuation to $\alpha = 1$
(see Vol. I of \cite{gelf} for details).
%
%
\vskip3mm
\noindent
${\bf I_2 : }$
\[
I_2 \; := \; \int d^2p \; \frac{1}{(p^2 + \lambda^2) p^2}
\Big(1 - \cos(px)\Big) \]
\begin{equation}
= \;
\frac{2\pi}{\lambda^2}\Bigg( \ln|x| + \mbox{K}_0\Big(\lambda |x|\Big)
+ \ln\Big(\frac{\lambda}{2}\Big) + \gamma \Bigg) \; ,
\end{equation}
where $\gamma = 0.577216..$ denotes Euler's constant. \\
%
%
\vskip3mm
\noindent
${\bf I_3 : }$
\[ I_3 \; := \; \int d^2p e^{-2\frac{|p|}{n}} \frac{1}{p^2}
\Big(1 - \cos(px)\Big) \]
\[
= \;
2\pi \ln \Bigg( \frac{n}{4} \Bigg[ \frac{2}{n}
+ \sqrt{ \Big(\frac{2}{n}\Big)^2 + x^2} \Bigg] \Bigg) \]
\begin{equation}
= \;
2\pi \Bigg( \ln|x| + \ln\Big(\frac{n}{4}\Big) + O\Big(\frac{1}{n}\Big)
\; \Bigg) \; \;
\mbox{for} \; n \rightarrow \infty \; .
\end{equation}
%
%
\vskip3mm
\noindent
${\bf I_4 : }$
\[ I_4 \; := \;
\int d^2p \; e^{-2\frac{|p|}{n}} \frac{1}{p^2 + \lambda^2} \]
\[
= \;- 2\pi \Bigg( \cos\Big(\frac{2}{n}\lambda\Big)
\mbox{Ci}\Big(\frac{2}{n}\lambda\Big) \;
+ \; \sin\Big(\frac{2}{n}\lambda\Big)
\mbox{si}\Big(\frac{2}{n}\lambda\Big) \Bigg)  \]
\begin{equation}
= \;
2\pi \Bigg( -\gamma - \ln\Big(\frac{2}{n}\lambda\Big) + O\Big(\frac{1}{n}\Big)
\; \Bigg) \; \; \; \;
\mbox{for} \; n \rightarrow \infty \; .
\end{equation}
Here $\mbox{si}(x)$ and $\mbox{Ci}(x)$ denote the sine and the
cosine integral (see \cite{magnus} p. 347).
%
%
\vskip3mm
\noindent
${\bf I_5 : }$
\[
I_5 \; := \; \int d^2p \; \frac{1}{(p^2 + \lambda^2) p^2}
\sin^2(px)
\]
\begin{equation}
= \;
\frac{\pi}{\lambda^2}
\Bigg( \ln\Big(2|x|) + \mbox{K}_0\Big(\lambda 2|x|\Big)
+ \ln\Big(\frac{\lambda}{2}\Big) + \gamma \Bigg) \; .
\end{equation}
%
%
\vskip3mm
\noindent
${\bf I_6 : }$
\[
I_6 \; = \;
\int d^2p \; \frac{1}{p^2 + \lambda^2}
\; \frac{1}{p_2^2} \; \Big( \sin^2(p_2 L) - 2 \sin^2(p_2 L/2) \Big)
\]
\begin{equation}
- \; 2 \int d^2p \; \frac{1}{p^2 + \lambda^2}
\; \frac{1}{p_2^2} \; \cos(p_1 2L) \;
\Big( \sin^2(p_2 L) -  \sin^2(p_2 L/2) \Big) \; .
\end{equation}
For this integral the large $L$ behaviour is of interest.
In both terms the
$p_2$ integral can be solved using the formula
\begin{equation}
\int_0^\infty \; dx \; \frac{\sin^2(bx)}{x^2(x^2+z^2)} \; = \;
\frac{\pi}{4z^3} \left( e^{-bz} \; + 2bz -1 \right) \;
\end{equation}
which can be found in \cite{prud}.
Some terms then cancel each other and the remaining integrals either
vanish exponentially with $L$, or give a constant.
One ends up with
\begin{equation}
I_6 \; = \; \frac{\pi}{\lambda^2} \; + \; O\Big(e^{-L}) \; \; \; \;
\; \mbox{for} \; \; L \; \rightarrow \; \infty \; \; .
\end{equation}

%
%

\section{Some matrices}
In this appendix I discuss the properties of some matrices acting
in flavor space.
\begin{equation}
R := \left( \begin{array}{ccccc}
N-1 & -1 & . & . & -1 \\
-1 & N-1 & -1 &   & .  \\
 . & -1  & . & . & .  \\
 . &     & . & . & -1 \\
-1 & .   & . & -1 & N-1
\end{array} \right) \; .
\end{equation}
$R$ has the following set of orthonormal eigenvectors
\[
\vec{r}^{\;(1)} = c^{(1)}\left( \begin{array}{c}
1 \\ 1 \\ . \\ .\\  . \\ . \\ 1 \end{array}  \right) \; , \;
\vec{r}^{\;(2)} = c^{(2)}\left( \begin{array}{c}
1 \\ . \\ . \\ .\\  . \\ 1 \\ -(N-1) \end{array}   \right) \; ,
\]
\begin{equation}
\vec{r}^{\;(3)} = c^{(3)}\left( \begin{array}{c}
1 \\ . \\ . \\ .\\  1 \\ -(N-2) \\ 0 \end{array}   \right)
\; , ........ \; , \;
\vec{r}^{\;(N)} = c^{(N)}\left( \begin{array}{c}
\; 1 \\ -1 \\ 0 \\ .\\  . \\ 0 \\ 0 \end{array}   \right) \; ,
\end{equation}
where the normalization constants are given by
\[
c^{(1)} = \frac{1}{\sqrt{N}} \; , \;
c^{(2)} = \frac{1}{\sqrt{N-1+(N-1)^2}} \; , \]
\begin{equation}
c^{(3)} = \frac{1}{\sqrt{N-2+(N-2)^2}} \; , ...... \; , \;
c^{(N)} = \frac{1}{\sqrt{2}} \; .
\end{equation}
Obviously the eigenvalues $e^{(I)} \; , \; I = 1 ... N$ are given by
\begin{equation}
e^{(1)} = 0 \; , \; e^{(2)} = N \; , \;
e^{(3)} = N \; , ........ \;  , \; e^{(N)} = N \; .
\end{equation}
This implies that $R$ can be diagonalized by the matrix $U$
constructed out of the vectors $\vec{r}^{\;(I)} \; , \; I = 1 ... N$
\begin{equation}
U := \Big( \vec{r}^{\;(1)},\vec{r}^{\;(2)}, .......
,\vec{r}^{\;(N)} \Big)^T \; .
\end{equation}
\begin{equation}
U R U^T = \mbox{diag}\Big( 0 , N , ....... N \Big) \; .
\end{equation}
$U$ is an orthogonal matrix
\begin{equation}
U^T = U^{-1} \; .
\end{equation}
Finally I denote the useful identity
\begin{equation}
\sum_{I=2}^{N} U_{Ia} U_{Ib} = \delta_{ab} - \frac{1}{N} \; .
\end{equation}
Using the fact that $U_{1a} = 1/\sqrt{N}$ for $ a = 1,2, .., N$ and the
orthogonality of $U$ one obtains
\begin{equation}
\delta_{ab} = \sum_{I=1}^{N} U_{Ia} U_{Ib} =
\sum_{I=2}^{N} U_{Ia} U_{Ib} + \frac{1}{N} \; ,
\end{equation}
which implies the quoted identity.

%
%

\section{Inverse conditioning}
In this appendix the inverse conditioning formula \cite{frohlich2}
adapted to the case of several flavors is proven.
\vskip3mm
\noindent
{\bf Lemma B.1 :} Inverse conditioning
\vskip3mm
\noindent
Let $C^1, C^2$ be covariances that obey $C^1 \geq C^2 \geq 0$ as quadratic
forms. Then the following inequality holds
\[
\left\langle \prod_{b=1}^N \left[
\prod_{i_b}^{q_b} : e^{i2\sqrt{\pi} \varphi^{(b)}(x^{(b)}_{i_b})}:_{C^1}
\prod_{j_b}^{2n_b-q_b} : e^{-i2\sqrt{\pi} \varphi^{(b)}(y^{(b)}_{j_b})}:_{C^1}
\right] \right\rangle_{C^1}
\]
\begin{equation}
\leq \; e^{4\pi\sum_{b=1}^N 2n_b \lambda^{(b)}}
\left\langle \prod_{b=1}^N \left[
\prod_{i_b}^{q_b} : e^{i2\sqrt{\pi} \varphi^{(b)}(x^{(b)}_{i_b})}:_{C^2}
\prod_{j_b}^{2n_b-q_b} : e^{-i2\sqrt{\pi} \varphi^{(b)}(y^{(b)}_{j_b})}:_{C^2}
\right] \right\rangle_{C^2} \; ,
\end{equation}
where
\begin{equation}
\lambda^{(b)} \; := \; \lim_{z \rightarrow 0} \;
\frac{1}{2} \Big[ C^1_{bb}(z) \; - \; C^2_{bb}(z) \Big] \; .
\end{equation}
\vskip3mm
\noindent
{\bf Proof : }
\vskip3mm
\noindent
Denote by $\varphi(\delta_n(x))$ the convolution of the field
$\varphi$ with a $\delta$-sequence $\delta_n$ peaked at $x$. This allows
to write the left hand side of (B.22) as
\[
\lim_{n\rightarrow \infty}
\left\langle \prod_{b=1}^N \left[
\prod_{i_b}^{q_b} : e^{i2\sqrt{\pi}
\varphi^{(b)}(\delta_n(x^{(b)}_{i_b}))}:_{C^1}
\prod_{j_b}^{2n_b-q_b} : e^{-i2\sqrt{\pi}
\varphi^{(b)}(\delta_n(y^{(b)}_{j_b}))}:_{C^1}
\right] \right\rangle_{C^1}
\]
\[
= \; \lim_{n\rightarrow \infty}
\left\langle e^{i2\sqrt{\pi} \varphi(f_n)} \right\rangle_{C^1} \;
e^{\frac{1}{2} 4\pi \sum_{b=1}^N 2n_b
\big(\delta_n(\xi),C^1_{bb} \delta_n(\xi)\big)}
\]
\[
= \; \lim_{n\rightarrow \infty} e^{-\frac{1}{2} 4\pi \big( f_n, C^1 f_n \big)}
e^{\frac{1}{2} 4\pi \sum_{b=1}^N 2n_b
\big(\delta_n(\xi),C^2_{bb} \delta_n(\xi)\big)}
e^{\frac{1}{2} 4\pi \sum_{b=1}^N 2n_b
\big(\delta_n(\xi), [C^1_{bb}-C^2_{bb}] \delta_n(\xi)\big)}
\]
\[
\leq \;
\lim_{n\rightarrow \infty} e^{-\frac{1}{2} 4\pi \big( f_n, C^2 f_n \big)}
e^{\frac{1}{2} 4\pi \sum_{b=1}^N 2n_b
\big(\delta_n(\xi),C^2_{bb} \delta_n(\xi)\big)}
e^{\frac{1}{2} 4\pi \sum_{b=1}^N 2n_b
\big(\delta_n(\xi), [ C^1_{bb}-C^2_{bb} ] \delta_n(\xi)\big)}
\]
\begin{equation}
= \;
e^{4\pi\sum_{b=1}^N 2n_b \lambda^{(b)}}
\left\langle \prod_{b=1}^N \left[
\prod_{i_b}^{q_b} : e^{i2\sqrt{\pi} \varphi^{(b)}(x^{(b)}_{i_b})}:_{C^2}
\prod_{j_b}^{2n_b-q_b} : e^{-i2\sqrt{\pi} \varphi^{(b)}(y^{(b)}_{j_b})}:_{C^2}
\right] \right\rangle_{C^2} \; .
\end{equation}
$f_n$ denotes the vector composed from the sum over all
$\delta$-sequences $\delta_n$ peaked at the various space-time arguments.
In the last step I used
\begin{equation}
\lambda^{(b)} \; = \; \lim_{n \rightarrow \infty} \; \frac{1}{2}
\Big(\delta_n(\xi), \Big[ C^1_{bb} - C^2_{bb} \Big] \delta_n(\xi) \Big) \; ,
\end{equation}
which coincides with (B.23). $\xi$ denotes some dummy space-time argument. \\
$\Box$

%
%

\section{Conditioning}
In this section the conditioning formula
\cite{frohlich2}, \cite{guerra} will be proven.
Again the result is quoted in the form which is suitable for
the N-flavor case.
\vskip3mm
\noindent
{\bf Lemma B.2 :} Conditioning
\vskip3mm
\noindent
Let $C^1, C^2$ be two covariances that obey $ C^1 \geq C^2 \geq 0$
as quadratic forms. Then the conditioning formula holds
\[
\left\langle \prod_{b=1}^N \; \exp \left( -2 \sum_{b=1}^N \beta^{(b)}
\epsilon^{(b)}
\int_{\Lambda} d^2 x \; t(x)
: \cos\Big[ 2 \sqrt{\pi} \varphi^{(b)}(x) \Big] :_{C^2} \right)
\right\rangle_{C^2}
\]
\begin{equation}
\leq \;
\left\langle \prod_{b=1}^N \; \exp \left( -2 \sum_{b=1}^N \beta^{(b)}
\epsilon^{(b)}
\int_{\Lambda} d^2 x \; t(x)
: \cos\Big[ 2 \sqrt{\pi} \varphi^{(b)}(x) \Big] :_{C^1} \right)
\right\rangle_{C^1}
\; ,
\end{equation}
where $\epsilon^{(b)} \in \{-1,+1\}$ arbitrary but fixed.
\vskip3mm
\noindent
{\bf Proof :}
\vskip3mm
\noindent
The proof makes use of Jensen's inequality (see e.g. \cite{simonphi})
\begin{equation}
\int \; d\mu[\varphi] \; \exp \Big( F (\varphi) \Big) \; \geq \;
\exp \left( \; \int \; d\mu[\varphi] \; F (\varphi) \; \right) \; \; ,
\end{equation}
whenever $\int \; d\mu[\varphi] \; \exp \big( F (\varphi) \big) \; < \infty$.
Introducing new fields $\theta_1$ with covariance $C^1 - C^2$ and
$\theta_2$ with covariance $C^2$, the right hand side of (B.26) can
be written as
\[
\left\langle \prod_{b=1}^N \; e^{ -2 \sum_{b=1}^N \beta^{(b)} \epsilon^{(b)}
\int_{\Lambda} d^2 x \; t(x) \;
: \cos\big[ 2 \sqrt{\pi} \big( \theta_1^{(b)}(x) +
\theta_2^{(b)}(x)\big)\big] :_{C^1-C^2,C^2} } \right\rangle_{C^1-C^2,C^2}
\]
\[
\geq \; \left\langle \prod_{b=1}^N \;
e^{ -2 \int d\mu_{C^1-C^2}[\theta_1]
\sum_{b=1}^N \beta^{(b)} \epsilon^{(b)}
\int_{\Lambda} d^2 x \; t(x)
: \cos\big[ 2 \sqrt{\pi} \big( \theta_1^{(b)}(x) +
\theta_2^{(b)}(x) \big) \big] :_{C^1-C^2,C^2} } \right\rangle_{C^1-C^2,C^2}
\]
\begin{equation}
= \;
\left\langle \prod_{b=1}^N \; e^{ -2 \sum_{b=1}^N \beta^{(b)} \epsilon^{(b)}
\int_{\Lambda} d^2 x \; t(x) \;
: \cos\big[ 2 \sqrt{\pi} \theta_2^{(b)}(x)  \big] :_{C^2} }
\right\rangle_{C^2} \; ,
\end{equation}
where Jensen's inequality for $\theta_1$ was used in the first step.
The second step made use of
\[
\int d\mu_{C^1-C^2}[\theta_1] \;
\int_{\Lambda} d^2 x \; t(x) \;
: \cos\Big[ 2 \sqrt{\pi} \Big( \theta_1^{(b)}(x) +
\theta_2^{(b)}(x) \Big) \Big] :_{C^1-C^2,C^2}
\]
\begin{equation}
= \;
\int_{\Lambda} d^2 x \; t(x) \;
: \cos\Big[ 2 \sqrt{\pi} \theta_2^{(b)}(x) \Big] :_{C^2} \; ,
\end{equation}
which can be seen to hold from the definition
of Wick ordering (A.34). \\
$\Box$
\vskip3mm
\noindent
{}From Lemma B.2 one easily reads off the following corollary by
expressing the cosh in terms of exponentials and inserting the
corresponding values for $\epsilon^{(b)} \in \{-1,+1\}$.
\vskip3mm
\noindent
{\bf Corollary B.1 :}
\vskip3mm
\noindent
For $C_1 \geq C_2$ the following inequality holds
\[
\left\langle \prod_{b=1}^N \; 2 \cosh \left( -2 \sum_{b=1}^N \beta^{(b)}
\int_{\Lambda} d^2 x \; t(x)
: \cos\Big[ 2 \sqrt{\pi} \varphi^{(b)}(x) \Big] :_{C^2} \right)
\right\rangle_{C^2}
\]
\begin{equation}
\leq \;
\left\langle \prod_{b=1}^N \; 2 \cosh \left( -2 \sum_{b=1}^N \beta^{(b)}
\int_{\Lambda} d^2 x  \; t(x)
: \cos\Big[ 2 \sqrt{\pi} \varphi^{(b)}(x) \Big] :_{C^1} \right)
\right\rangle_{C^1}
\; .
\end{equation}

%
%

\section{Dirichlet boundary conditions}
In this appendix formulas for covariances with
Dirichlet boundary
conditions are collected.
They all are discussed in the proof for one flavor
by Fr\"ohlich \cite{frohlich}, \cite{frohlich2}.

By scaling one may choose for the space-time cutoff $\Lambda$ a unit square,
and by Euclidean invariance of the measure one may suppose that
$\Lambda$ is centered at (1,0) with sides parallel to the
coordinate axes. Let $S$ be the disc of radius 2 centered at
(0,0). The geometry is illustrated in Figure B.1.
\vskip95mm
\noindent
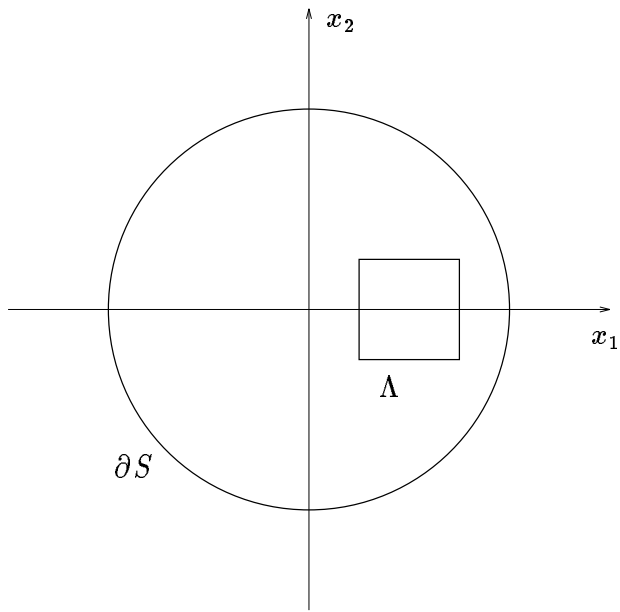
{\bf Figure B.1 :} The boundary $\partial S$ and the space-time
region $\Lambda$.
\vskip3mm
\noindent
As can be evaluated easily
\begin{equation}
\mbox{dist} \; ( \Lambda, \partial S ) \; = \; 2 - \frac{\sqrt{10}}{2}
\; > \; 0 \; .
\end{equation}
Let $\triangle_S$ be the Laplacian on $L^2(S,d^2x)$ with
zero Dirichlet data on $\partial S$. Because of zero Dirichlet data
on the boundary, $-\triangle_S$ is strictly positive, and hence
gives rise to a proper covariance $\Big( -\triangle_S )^{-1}$.
Furthermore
\begin{equation}
\frac{1}{ -\triangle_S + M^2} \; ,
\end{equation}
is a proper covariance operator as well.
Obviously as an operator on $L^2(S,d^2x)$
\begin{equation}
\frac{1}{ -\triangle_S } \; \geq \; \frac{1}{ -\triangle_S + M^2} \; ,
\end{equation}
for some real mass $M$, and finally
\begin{equation}
\frac{1}{ -\triangle + M^2} \; \geq \; \frac{1}{ -\triangle_S + M^2} \; ,
\end{equation}
which follows from the fact that $\triangle$ is not strictly positive.
Furthermore since
$\mbox{dist} \; ( \Lambda, \partial S ) \; > \; 0 \;$
\begin{equation}
\sup_{x,y \in \Lambda} \left[ \frac{1}{-\triangle + M^2}(x,y) -
\frac{1}{-\triangle_S + M^2}(x,y) \right] \; \leq \; \tilde{\omega} \; < \;
\infty \; .
\end{equation}
Using the method of image charges
one can construct an explicit representation of the
Green's function $( -\triangle_S )^{-1}(x,y) \; =: \; C^{0,S}(x,y)$
\begin{equation}
\Big( -\triangle_S )^{-1}(x,y) \; = \;
-\frac{1}{4\pi} \left( \;
\ln\Big| \tilde{x} - \tilde{y}\Big| \; + \;
\ln\Big| \hat{x} - \hat{y}\Big| \; - \;
\ln\Big| \hat{x} - \tilde{y}\Big| \; - \;
\ln\Big| \tilde{x} - \hat{y}\Big| \; \right)\; ,
\end{equation}
where
\begin{equation}
\tilde{x} \; := \; x_1 \; + \; ix_2 \; ,
\end{equation}
denotes the complex coordinate already encountered in (A.14), and
\begin{equation}
\hat{x} \; \; := \; \; 4\; / \; \overline{\tilde{x}} \; ,
\end{equation}
is the reflection of $\tilde{x}$ at the circle $\partial S$.

Finally
\begin{equation}
\Big| \tilde{x} \; - \; \hat{x} \Big| \; < \; 8 \; \; ,
\end{equation}
for $x \; \in \; \Lambda$. This can easily be seen to hold from the
definition of $\tilde{x}, \; \hat{x}$ and the relative position
of $\Lambda $ and $\partial S$ (compare Figure B.1).

%
%

\section{A generalized H\"older inequality}
In this section a generalization of H\"older's inequality is proven
(This generalization is an exercise in \cite{dunford}.).
Furthermore I infer a corollary
that is needed in the main text.
\vskip3mm
\noindent
{\bf Lemma B.3 : } (generalized H\"older inequality)
\vskip3mm
\noindent
Let for positive numbers $0 < q_1,q_2, ... q_n < 1$
\begin{equation}
\frac{1}{q_1} + \frac{1}{q_2} + ....... + \frac{1}{q_n} \; = \; 1 \; ,
\end{equation}
and
\begin{equation}
f_i(x) \in L^{q_i}(\mbox{I\hspace{-0.62mm}R}^D)
\; \; , \; \; i = 1,2,...,n \; \; ,
\end{equation}
where the number of dimensions $D$ is a positive but arbitrary integer.
Then
\begin{equation}
\prod_{i=1}^{n} f_i(x) \; \in \; L^1(\mbox{I\hspace{-0.62mm}R}^D) \; ,
\end{equation}
and
\begin{equation}
\int d^Dx \prod_{i=1}^{n} |f_i(x)|  \; \leq \;
\prod_{i=1}^{n} \|f_i\|_{q_i} \; ,
\end{equation}
where
\begin{equation}
\|f\|_{q} \; := \; \left( \int d^Dx |f_i(x)|^q \right)^\frac{1}{q} \; .
\end{equation}
\vskip3mm
\noindent
{\bf Proof:} (by induction)
\vskip3mm
\noindent
{\bf i:}
For $n=2$ the claim reduces to the usual H\"older inequality. \\
{\bf ii:}
Let the lemma be true for $n-1$. \\
{\bf iii:} $n:$ \\
$q_i \in \mbox{I\hspace{-0.62mm}R} \; , \; i = 1,2,...,n$ obeying (B.40) and
$f_i \in  L^{q_i}(\mbox{I\hspace{-0.62mm}R}^D) \; , \;
i = 1,2,...,n$ are given.
Define
\begin{equation}
\frac{1}{p} \; := \; \frac{1}{q_n} \; \; \; \; \;   , \; \; \; \; \;
\frac{1}{q} \; := \;
\frac{1}{q_1} + \frac{1}{q_2} + ....... + \frac{1}{q_{n-1}} \; .
\end{equation}
Obviously
\begin{equation}
\frac{1}{p} + \frac{1}{q} = 1 \; \; \; \; \;  , \; \; \; \; \;
\frac{1}{q_1/q} + ....... + \frac{1}{q_{n-1}/q} = 1 \; .
\end{equation}
{}From $f_i(x) \in L^{q_i}(\mbox{I\hspace{-0.62mm}R}^D) \; , \;
i = 1,...,n-1 \; $ there follows
$| f_i(x) |^q \in L^{q_i/q}(\mbox{I\hspace{-0.62mm}R}^D) \; , \;
i = 1,...,n-1\; .$ Thus the
assumption for $n-1$ and the usual H\"older inequality can be applied to
finish the proof,
\[
\int d^Dx \prod_{i=1}^{n} |f_i(x)| \;
\leq \;
\left[\int d^Dx \prod_{i=1}^{n-1} |f_i(x)|^q \right]^{\frac{1}{q}} \;
\left[\int d^Dx |f_n(x)|^p \right]^{\frac{1}{p}} \;
\leq \]
\begin{equation}
\prod_{i=1}^{n-1} \left[ \int d^Dx |f_i(x)|^{q q_1/q}
\right]^{\frac{1}{q q_1/q}} \;
\left[\int d^Dx |f_n(x)|^p \right]^{\frac{1}{p}} \; = \;
\prod_{i=1}^{n} \|f_i\|_{q_i} \; .
\end{equation}
$\Box$
\vskip3mm
\noindent
Lemma B.3 allows to prove a corollary that will be used in the
main part.
\vskip3mm
\noindent
{\bf Corollary B.2 :}
\vskip3mm
\noindent
Denote by $f^{(a,b)}\; , \; \; 1 \leq a < b \leq N$
functions depending on the coordinates
\begin{equation}
z_1^{(a)} \; , \; . \; . \; . \; , \;  z_{n_a}^{(a)}
\; \; \; , \; \; \;
z_1^{(b)} \; , \; . \; . \; . \; , \;  z_{n_b}^{(b)} \; \; ,
\end{equation}
where each $z_j^{(c)}$ is itself $D$-dimensional. Assume
\begin{equation}
f^{(a,b)} \; \in \; L^{N-1}
\Big( \mbox{I\hspace{-0.62mm}R}^{D ( n_a + n_b)} \; , \;
\prod_{i=1}^{n_a}d^Dz_i^{(a)} \prod_{j=1}^{n_b} d^Dz_j^{(b)} \Big) \; \; ,
\end{equation}
for
$ 1 \leq a < b \leq N$. Then
\begin{equation}
\prod_{a<b}^N \; f^{(a,b)} \; \in \; L^1
\Big( \mbox{I\hspace{-0.62mm}R}^{D \sum_{a=1}^N  n_a } \; , \;
\prod_{b=1}^N \prod_{j=1}^{n_b} d^Dz_j^{(b)} \Big) \; \; ,
\end{equation}
and
\begin{equation}
\int \prod_{a=1}^N \prod_{j=1}^{n_a} d^Dz_j^{(a)}
\bigg| \prod_{b<c}^N  \; f^{(b,c)} \bigg| \; \leq \;
\prod_{a<b}^N \Bigg[ \int \;
\prod_{i=1}^{n_a} d^Dz_i^{(a)} \prod_{j=1}^{n_b} d^Dz_j^{(b)} \;
\bigg| f^{(a,b)} \bigg|^{N-1} \Bigg]^{\frac{1}{N-1}} \; .
\end{equation}
\vskip3mm
\noindent
{\bf Proof :}
\vskip3mm
\noindent
The statement will be proven by direct construction. For
convenience I introduce the notations
\begin{equation}
\Big|\!\Big|\!\Big| f^{(a,b)} \Big|\!\Big|\!\Big|_b \; := \;
\Bigg[ \int \;
\prod_{j=1}^{n_b} d^Dz_j^{(b)} \;
\bigg| f^{(a,b)} \bigg|^{N-1} \Bigg]^{\frac{1}{N-1}} \; ,
\end{equation}
and
\begin{equation}
\Big|\!\Big|\!\Big| f^{(a,b)} \Big|\!\Big|\!\Big| \; := \;
\Bigg[ \int \;
\prod_{i=1}^{n_a} d^Dz_i^{(a)} \; \prod_{j=1}^{n_b} d^Dz_j^{(b)} \;
\bigg| f^{(a,b)} \bigg|^{N-1} \Bigg]^{\frac{1}{N-1}} \; .
\end{equation}
The latter is of course the usual H\"older norm $|\!| f |\!|_{N-1}$.
{}From (B.49) and Definition (B.52) there follows immediately
that
\begin{equation}
\Big|\!\Big|\!\Big| f^{(a,b)} \Big|\!\Big|\!\Big|_b
\; \in \; L^{N-1}
\Big( \mbox{I\hspace{-0.62mm}R}^{D n_a } \; , \;
\prod_{i=1}^{n_a}d^Dz_i^{(a)} \Big) \; \; .
\end{equation}
Then
\[
\int \prod_{a=1}^N \prod_{i=1}^{n_a} d^Dz_i^{(a)}
\bigg| \prod_{b<c}^N  \; f^{(b,c)} \bigg|
\]
\[
= \;
\int \prod_{a=1}^{N-1} \prod_{i=1}^{n_a} d^Dz_i^{(a)}
\prod_{b<c}^{N-1} \bigg| \; f^{(b,c)} \bigg| \;
\int \prod_{j=1}^{n_N} d^Dz_j^{(N)}
\prod_{d=1}^{N-1} \bigg| \; f^{(d,N)} \bigg|
\]
\[
\leq \;
\int \prod_{a=1}^{N-1} \prod_{i=1}^{n_a} d^Dz_i^{(a)} \;
\prod_{b<c}^{N-1} \bigg| \; f^{(b,c)} \bigg| \;
\prod_{d=1}^{N-1}
\Big|\!\Big|\!\Big| f^{(d,N)} \Big|\!\Big|\!\Big|_N
\]
\[
= \;
\int \prod_{a=1}^{N-2} \prod_{i=1}^{n_a} d^Dz_i^{(a)}
\prod_{b<c}^{N-2} \bigg| \; f^{(b,c)} \bigg| \;
\int \prod_{j=1}^{n_{N-1}} d^Dz_j^{(N-1)}
\prod_{d=1}^{N-2} \bigg| \; f^{(d,N-1)} \bigg|
\prod_{e=1}^{N-1}
\Big|\!\Big|\!\Big| f^{(e,N)} \Big|\!\Big|\!\Big|_N
\]
\[
\leq \;
\int \prod_{a=1}^{N-2} \prod_{i=1}^{n_a} d^Dz_i^{(a)}
\prod_{b<c}^{N-2} \bigg| \; f^{(b,c)} \bigg| \;
\prod_{d=1}^{N-2}
\Big|\!\Big|\!\Big| f^{(d,N-1)} \Big|\!\Big|\!\Big|_{N-1}
\Big|\!\Big|\!\Big| f^{(d,N)} \Big|\!\Big|\!\Big|_N
\times
\Big|\!\Big|\!\Big| f^{(N-1,N)} \Big|\!\Big|\!\Big|
\]
\[
\leq \;
\int \prod_{a=1}^{N-3} \prod_{i=1}^{n_a} d^Dz_i^{(a)}
\prod_{b<c}^{N-3} \bigg| \; f^{(b,c)} \bigg| \;
\prod_{d=1}^{N-3}
\Big|\!\Big|\!\Big| f^{(d,N-2)} \Big|\!\Big|\!\Big|_{N-2}
\Big|\!\Big|\!\Big| f^{(d,N-1)} \Big|\!\Big|\!\Big|_{N-1}
\Big|\!\Big|\!\Big| f^{(d,N)} \Big|\!\Big|\!\Big|_N
\]
\[
\times
\Big|\!\Big|\!\Big| f^{(N-2,N-1)} \Big|\!\Big|\!\Big|
\Big|\!\Big|\!\Big| f^{(N-2,N)} \Big|\!\Big|\!\Big|
\Big|\!\Big|\!\Big| f^{(N-1,N)} \Big|\!\Big|\!\Big|
\]
\[
\begin{array}{c}
. \\
. \\
. \\
. \\
{}.
\end{array}
\]
\[
\leq \;
\int \prod_{a=1}^{2} \prod_{i=1}^{n_a} d^Dz_i^{(a)}
\bigg| \; f^{(1,2)} \bigg| \;
\prod_{b=1}^2
\prod_{c=3}^{N}
\Big|\!\Big|\!\Big| f^{(b,c)} \Big|\!\Big|\!\Big|_c
\times
\prod_{d=3}^{N}
\prod_{e=d+1}^{N}
\Big|\!\Big|\!\Big| f^{(d,e)} \Big|\!\Big|\!\Big|
\]
\[
\leq \;
\int \prod_{i=1}^{n_1} d^Dz_i^{(1)}
\prod_{a=2}^{N}
\Big|\!\Big|\!\Big| f^{(1,a)} \Big|\!\Big|\!\Big|_a
\times
\prod_{b=2}^{N}
\prod_{c=b+1}^{N}
\Big|\!\Big|\!\Big| f^{(b,c)} \Big|\!\Big|\!\Big|
\]
\begin{equation}
= \; \prod_{a<b}^{N}
\Big|\!\Big|\!\Big| f^{(a,b)} \Big|\!\Big|\!\Big| \; .
\end{equation}
In the chain of inequalities above, I successively applied
the generalized H\"older inequality (B.43) with
\begin{equation}
\frac{1}{q_i} \; = \; \frac{1}{N-1} \; \; , \; \; \; \;  i \; = \; 1 \; , \;
. \; . \; . \; N-1 \; \; \; \; .
\end{equation}
$\Box$

%
%
\section{Bound on integrals over Cauchy determinants}
The following bound on integrals over Cauchy determinants can be found
in \cite{frohlich2}. I quote it for the convenience of the reader.
\vskip3mm
\noindent
{\bf Lemma B.4 :}
\vskip3mm
\noindent
For $\alpha \; < \; 1$ the following bound holds
\begin{equation}
\int_{\Lambda} \left[ \prod_{i=1}^{n}
d^2x_{i} d^2y_{i}  \right]
\Bigg|
{ \; \atop { \mbox{det} \atop {\scriptstyle i,j=1,...2n} } }\Bigg(
\frac{1}{w_i - z_j} \Bigg)
\Bigg|^\alpha \; \leq \;
(2n)! \bigg[ \Xi\Big(\alpha \Big)\bigg]^{2n} \; ,
\end{equation}
where
\begin{equation}
w := \left( \begin{array}{c}
\tilde{x}_1 \\ . \\ . \\ . \\
\tilde{x}_{n} \\
\hat{y}_1 \\ . \\ . \\ . \\
\hat{y}_{n}
\end{array} \right) \; , \;
z := \left( \begin{array}{c}
\tilde{y}_1 \\ . \\ . \\ . \\
\tilde{y}_{n} \\
\hat{x}_1 \\ . \\ . \\ . \\
\hat{x}_{n}
\end{array} \right) \; \; .
\end{equation}
For the definitions of $\tilde{x}$ and $\hat{x}$ see (B.37) and (B.38).
\vskip3mm
\noindent
{\bf Proof :}
\vskip3mm
\noindent
\[
\int_{\Lambda} \left[ \prod_{i=1}^{n}
d^2x_{i} d^2y_{i}  \right]
\Bigg|
{ \; \atop { \mbox{det} \atop {\scriptstyle i,j=1,...2n} } }\Bigg(
\frac{1}{w_i - z_j} \Bigg)
\Bigg|^\alpha \;
\]
\[
= \;
\int_{\Lambda} \left[ \prod_{i=1}^{n}
d^2x_{i} d^2y_{i}  \right]
\Bigg| \sum_{\pi(2n)} \mbox{sign}(\pi) \; \prod_{j=1}^{2n}
\frac{1}{w_j - z_{\pi(j)}} \Bigg|^\alpha
\]
\[
\leq \;
\sum_{\pi(2n)}
\int_{\Lambda} \left[ \prod_{i=1}^{n}
d^2x_{i} d^2y_{i}  \right]
\prod_{j=1}^{2n} \Bigg| \frac{1}{w_j - z_{\pi(j)}} \Bigg|^\alpha
\]
\begin{equation}
\leq \;
\sum_{\pi(2n)} \bigg[ \Xi\Big(\alpha \Big)\bigg]^{2n} \; = \;
(2n)! \bigg[ \Xi\Big(\alpha \Big)\bigg]^{2n} \; .
\end{equation}
In the last step I used that
$\prod_{j=1}^{2n} \Bigg| \frac{1}{w_j - z_{\pi(j)}} \Bigg|^\alpha$
is $\prod_{i=1}^{n} d^2x_{i} d^2y_{i}$ integrable for $\alpha < 1$
and
\begin{equation}
\int_{\Lambda} \left[ \prod_{i=1}^{n}
d^2x_{i} d^2y_{i}  \right]
\prod_{j=1}^{2n} \Bigg| \frac{1}{w_j - z_{\pi(j)}} \Bigg|^\alpha
\; \leq \; \bigg[ \Xi\Big(\alpha \Big)\bigg]^{2n} \; ,
\end{equation}
for some constant $\Xi\Big(\alpha \Big)$, independent of the
permutation $\pi$ (see \cite{frohlich2} for details).

\end{appendix}

%
%

\end{document}